\documentclass[10pt]{amsart}
\usepackage{amsmath, amsthm, amssymb, bm}
\usepackage{amsfonts}
\usepackage{mathrsfs}
\usepackage{euscript}
\usepackage{graphicx,color}
\usepackage{amscd}
\usepackage{geometry}
\usepackage[dvipsnames]{xcolor}
\usepackage[colorlinks=true,linkcolor=red,citecolor=blue]{hyperref}
\usepackage{color}
\usepackage{epsfig}
\usepackage{mathtools}
\usepackage{verbatim}
\usepackage{slashed}
\usepackage{tikz}
\usepackage{cleveref}
\usepackage{physics}
\usepackage{bbm}
\usepackage{graphics}
\usepackage{tikz-cd}
\usepackage[pagewise]{lineno}
\usepackage{graphics}

\usepackage{sansmath}
\usetikzlibrary{shadings,intersections} 

\usepackage{enumitem}

\usepackage{slashed}
\title[On Kerr black hole formation and a New Approach toward Penrose Inequality]{On Kerr black hole formation \\ with Complete Apparent Horizon \\ and a New Approach toward Penrose Inequality}
\date{\today}

\author{Xinliang An} 
\address{Department of Mathematics, National University of Singapore, Singapore 119076}
\email{matax@nus.edu.sg}

\author{Taoran He}
\address{Department of Mathematics, National University of Singapore, Singapore 119076}
\email{taoran\textunderscore he@u.nus.edu}

\theoremstyle{plain}
\newtheorem{lemma}{Lemma}[section]
\newtheorem{definition}[lemma]{Definition}
\newtheorem{proposition}[lemma]{Proposition}
\newtheorem{theorem}[lemma]{Theorem}
\newtheorem{corollary}[lemma]{Corollary}

\newtheorem{remark}{Remark}

\numberwithin{equation}{section}
\DeclareMathOperator{\err}{Err}
\DeclareMathOperator{\lot}{l.o.t.}
\DeclareMathOperator{\spn}{span}
\DeclareMathOperator{\sgn}{sgn}

\DeclareFontFamily{U}{mathx}{}
\DeclareFontShape{U}{mathx}{m}{n}{<-> mathx10}{}
\DeclareSymbolFont{mathx}{U}{mathx}{m}{n}
\DeclareMathAccent{\widehat}{0}{mathx}{"70}
\DeclareMathAccent{\widecheck}{0}{mathx}{"71}

\begin{document}

\newcommand{\ub}{\underline{u}}
\newcommand{\Ub}{\underline{U}}
\newcommand{\Cb}{\underline{C}}
\newcommand{\Lb}{\underline{L}}
\newcommand{\Lh}{\hat{L}}
\newcommand{\Lbh}{\hat{\Lb}}
\newcommand{\phib}{\underline{\phi}}
\newcommand{\Phib}{\underline{\Phi}}
\newcommand{\Db}{\underline{D}}
\newcommand{\Dh}{\hat{D}}
\newcommand{\Dbh}{\hat{\Db}}
\newcommand{\omb}{\underline{\omega}}
\newcommand{\omh}{\hat{\omega}}
\newcommand{\ombh}{\hat{\omb}}
\newcommand{\Pb}{\underline{P}}
\newcommand{\chib}{\underline{\chi}}
\newcommand{\chih}{\hat{\chi}}
\newcommand{\chibh}{\hat{\chib}}
\def\hch{\hat{\chi}}
\def\hchb{\hat{\chib}}
\newcommand{\kb}{\overline{\kappa}}
\newcommand{\kbb}{\overline{\underline{\kappa}}}
\newcommand{\vsgmb}{\underline{\varsigma}}
\newcommand{\ud}{\underline}
\newcommand{\alb}{\underline{\alpha}}
\newcommand{\zeb}{\underline{\zeta}}
\newcommand{\beb}{\underline{\beta}}
\newcommand{\etb}{\underline{\eta}}
\newcommand{\Mb}{\underline{M}}
\newcommand{\oth}{\hat{\otimes}}
\newcommand{\tR}{\tilde{R}}
\newcommand{\fb}{\underline{f}}
\newcommand{\tf}{\widetilde{f}}
\newcommand{\tfb}{\widetilde{\fb}}
\newcommand{\mD}{\mathcal{D}}
\newcommand{\Ga}{\Gamma}
\newcommand{\Gac}{\check{\Ga}}

\def\a {\alpha}
\def\b {\beta}
\def\ab {\alphab}
\def\bb {\betab}
\def\nab {\nabla}
\def\ssnab {\slashed{\nabla}}

\def\ub {\underline{u}}
\def\th {\theta}
\def\Lb {\underline{L}}
\def\Hb {\underline{H}}
\def\chib {\underline{\chi}}
\def\chih {\hat{\chi}}
\def\chibh {\hat{\underline{\chi}}}
\def\omegab {\underline{\omega}}
\def\etab {\underline{\eta}}
\def\betab {\underline{\beta}}
\def\alphab {\underline{\alpha}}
\def\Psib {\underline{\Psi}}
\def\hot{\widehat{\otimes}}
\def\Phib {\underline{\Phi}}
\def\thb {\underline{\theta}}
\def\vth {\vartheta}
\def\t {\tilde}
\def\st {\tilde{s}}
\def\tH {\widetilde{H}}
\def\tHb {\widetilde{\Hb}}
\def\tS {\widetilde{S}}

%%%%%%%%
%%%%%%%%
\def\d {\delta}
\def\f {\frac}
\def\i {\infty}
\def\l {\bigg(}
\def\r {\bigg)}
\def\S {S_{u,\underline{u}}}
\def\o{\omega}
\def\O{\Omega}

\def\od{\omega^{\dagger}}
\def\ombd{\underline{\omega}^{\dagger}}
\def\K{K-\frac{1}{|u|^2}}
\def\ut{\frac{1}{|u|^2}}
\def\Kb{K-\frac{1}{(u+\underline{u})^2}}
\def\M{\mathcal}
\def\p{\psi}

\def\D{\Delta}
\def\T{\Theta}
\def\s{S_{u',\underline{u}'}}
\def\Hu{H_u^{(0,\underline{u})}}
\def\Hbu{\underline{H}_{\underline{u}}^{(u_{\infty},u)}}
\def\ee{(\eta,\underline{\eta})}

\def\at{a^{\f12}}
\def\sigmac{\check{\sigma}}
\def\p{\psi}
\def\q{\underline{\psi}}
\def\ls{\leq}
\def\ls{\lesssim}
\def\oo{\Omega\mbox{tr}\chib-\frac{2}{u}}
\def\om{\omega}
\def\Om{\Omega}
%%%%%%%%
%%%%%%%%
\renewcommand{\div}{\mbox{div}\,}
\renewcommand{\curl}{\mbox{curl}\,}
\newcommand{\trchb}{\mbox{tr} \chib}
\def\trch{\mbox{tr}\chi}

\newcommand{\Ls}{{\mathcal L} \mkern-10mu /\,}
\newcommand{\eps}{{\epsilon} \mkern-8mu /\,}
%%%%%%%%%
%%%%%%%%%

\renewcommand{\tr}{\mbox{tr}}

\newcommand{\xib}{\underline{\xi}}
\newcommand{\psib}{\underline{\psi}}
\newcommand{\rhob}{\underline{\rho}}
\newcommand{\thetab}{\underline{\theta}}
\newcommand{\gammab}{\underline{\gamma}}
\newcommand{\nub}{\underline{\nu}}
\newcommand{\lb}{\underline{l}}
\newcommand{\mub}{\underline{\mu}}
\newcommand{\Xib}{\underline{\Xi}}
\newcommand{\Thetab}{\underline{\Theta}}
\newcommand{\Lambdab}{\underline{\Lambda}}
\newcommand{\vphb}{\underline{\varphi}}

\newcommand{\ih}{\hat{i}}
\newcommand{\ui}{u_{\infty}}
\newcommand{\shb}{L^2_{sc}(\underline{H}_{\ub}^{(u_{\infty},u)})}
\newcommand{\sh}{L^2_{sc}(H_{u}^{(0,\ub)})}
\newcommand{\Rb}{\underline{\mathcal{R}}}
\newcommand{\tc}{\widetilde{\tr\chib}}

\newcommand{\tcL}{\widetilde{\mathscr{L}}}

\newcommand{\sRic}{Ric\mkern-19mu /\,\,\,\,}
\newcommand{\sL}{{\cal L}\mkern-10mu /}
\newcommand{\sLh}{\hat{\sL}}
\newcommand{\sg}{g\mkern-9mu /}
\newcommand{\seps}{\epsilon\mkern-8mu /}
\newcommand{\sd}{d\mkern-10mu /}
\newcommand{\sR}{R\mkern-10mu /}
\newcommand{\snab}{\nabla\mkern-13mu /}
\newcommand{\sdiv}{\mbox{div}\mkern-19mu /\,\,\,\,}
\newcommand{\scurl}{\mbox{curl}\mkern-19mu /\,\,\,\,}
\newcommand{\slap}{\mbox{$\triangle  \mkern-13mu / \,$}}
\newcommand{\sGamma}{\Gamma\mkern-10mu /}
\newcommand{\somega}{\omega\mkern-10mu /}
\newcommand{\somb}{\omb\mkern-10mu /}
\newcommand{\spi}{\pi\mkern-10mu /}
\newcommand{\sJ}{J\mkern-10mu /}
\renewcommand{\sp}{p\mkern-9mu /}
\newcommand{\su}{u\mkern-8mu /}

\newcommand{\red}{\textcolor{red}}
\newcommand{\blue}{\textcolor{blue}}
\newcommand{\pur}{\textcolor{purple}}
\newcommand{\gre}{\textcolor{OliveGreen}}

\newcommand{\wc}{\widecheck}
\newcommand{\parallelsum}{\mathbin{\!/\mkern-5mu/\!}}
\newcommand{\prtheta}{\f{\partial}{\partial \vartheta}}
\newcommand{\prv}{\f{\partial}{\partial v}}
\newcommand{\pr}{\partial}
\newcommand{\intM}{{}^{(int)}\mathcal{M}}
\newcommand{\intL}{{}^{(int)}\mathcal{L}}
\newcommand{\pulM}{{}^{(P)}\mathcal{M}}
\newcommand{\Int}{{}^{(int)}}
\newcommand{\Ext}{{}^{(ext)}}
\newcommand{\pul}{{}^{(pul)}}
\newcommand{\glo}{{}^{(glo)}}
\newcommand{\tran}{{}^{(tran)}}
\newcommand{\MM}{\mathcal{M}}
\newcommand{\TT}{\mathcal{T}}
\newcommand{\LL}{\mathcal{L}}
\newcommand{\HH}{\mathcal{H}}
\newcommand{\JJ}{\mathfrak{J}}
\newcommand{\PP}{\mathcal{P}}
\newcommand{\DD}{\mathcal{D}}
\newcommand{\fd}{\mathfrak{d}}
\renewcommand{\SS}{\mathcal{S}}
\newcommand{\bfD}{\mathbf{D}}
\newcommand{\bfg}{\mathbf{g}}
\newcommand{\bfR}{\mathbf{R}}
\newcommand{\bfK}{\mathbf{K}}
\newcommand{\KK}{\mathcal{K}}
\newcommand{\tu}{\widetilde{u}}
\newcommand{\tub}{\widetilde{\ub}}
\newcommand{\Tub}{\widetilde{\underline{U}}}
\newcommand{\Tu}{\widetilde{U}}
\newcommand{\tX}{\widetilde{X}}
\newcommand{\ubrtp}{\ub, r, \theta, \varphi}
\newcommand{\ubtp}{\ub, \theta, \varphi}
\newcommand{\tp}{\theta, \varphi}
\newcommand{\rthe}{r, \theta}
\newcommand{\ainft}{a_{\infty}}
\newcommand{\minft}{m_{\infty}}
\newcommand{\te}{\tilde{e}}
\newcommand{\fepub}{\f{\epsilon_0}{\ub^{1+\delta_{dec}}}}
\newcommand{\tfepub}{\f{\epsilon_0}{\tub^{1+\delta_{dec}}}}
\newcommand{\ths}{\theta_{*}}
\newcommand{\rs}{r_{*}}
\newcommand{\fs}{f_{*}}
\newcommand{\bs}{b_{*}}
\newcommand{\La}{\Lambda}
\newcommand{\la}{\lambda}
\newcommand{\tla}{\widetilde{\lambda}}
\newcommand{\De}{\Delta}
\newcommand{\de}{\delta}
\newcommand{\tDe}{\widetilde{\Delta}}
\newcommand{\hb}{\underline{h}}
\newcommand{\thth}{\th^1, \th^2}
\newcommand{\tth}{\tilde{\theta}}
\newcommand{\tnab}{\widetilde{\nab}}
\newcommand{\Sis}{\Sigma^*}
\newcommand{\AH}{\mathcal{AH}}
\newcommand{\tujp}{\langle \tu \rangle}
\newcommand{\les}{\lesssim}
\newcommand{\Rs}{R_{\ast}}
\newcommand{\ov}{\overline}
\newcommand{\ms}{\mathbb{S}^2}
\newcommand{\nabr}{\mathring{\nab}}
\newcommand{\cf}{c_{\infty}}
\newcommand{\Ric}{\textbf{Ric}}
\def\si{\sigma}
\def\Si{\Sigma}
\def\II{\mathscr{I}}
\def\atrch{{}^{(a)}\trch}
\def\atrchb{{}^{(a)}\trchb}
\def\dual{{}^*}
\renewcommand{\c}{\cdot}
%\bfseries%% font

\begin{abstract}

Arising from admissible extended scale-critical short-pulse initial data, we show that $3+1$ dimensional Einstein vacuum equations admit dynamical Kerr black hole formation solutions. Our hyperbolic arguments combine the scale-critical gravitational-collapse result by An--Luk with the recent breakthrough by Klainerman--Szeftel on proving nonlinear Kerr stability with small angular momentum, which requires us to perform various specific coordinate changes and frame transformations. Furthermore, allowing large spacetime angular momentum, with new elliptic arguments and precise leading order calculations, we also solve the apparent horizon in Kerr black hole formation spacetimes (including Klainerman--Szeftel's Kerr stability spacetimes) and conduct an exploration, detailing the emergence, evolution, asymptotics and final state of the apparent horizon. Building on our analysis, without time symmetric assumption, we then put forward a new mathematical framework and prove both the dynamical Penrose inequality and the spacetime Penrose inequality in our black-hole formation spacetimes and in the perturbative regime of subextremal Kerr black holes. Collectively, without assuming any symmetry, we extend Christodoulou's celebrated trapped surface formation theorem to a black hole formation result.

\end{abstract}

\maketitle

\tableofcontents
%\newpage

\section{Introduction}
The mathematical theory of black hole formation is a compelling subject, showing a profound intersection of nonlinear wave equations, geometric analysis, and physical perspectives from general relativity and astrophysics. The mathematical description of the whole formation process demands detailed mathematical studies of various dynamical stages, i.e., its emergence, evolution, asymptotics and final state. Thought experiments on this motivated Penrose \cite{Penrosenaked} to raise his perspective of the Penrose inequality in 1973. Meanwhile, rigorous mathematical proof of the black hole formation remained to be a very challenging problem, that Wheeler suggested to Christodoulou in late 1960s. For spherically symmetric Einstein-scalar field system, Christodoulou had a series of celebrated results on this topic in 1990s. In 2008, Christodoulou published a breakthrough work \cite{Chr:book}. With no symmetry assumption, he showed that, \textit{a trapped surface (lying inside the black hole region) can form dynamically in the evolution of Einstein vacuum spacetimes.}  

The work \cite{A-L} by An--Luk and the work \cite{An:scale} by An later extend \cite{Chr:book} to the scale critical regime and the emergence of the trapped region and its boundary (apparent horizon) is studied. Our results here further extend \cite{Chr:book,A-L} and address the subsequent evolution, asymptotic behavior and final state of the corresponding gravitational collapse. In particular, we connect our arguments to nonlinear Kerr stability with small angular momentum, the main conclusion of which was recently proved in a breakthrough work \cite{KS:main} by Klainerman--Szeftel. Based on all above, we further solve the entire apparent horizon with detailed information and in perturbative Kerr regime we prove the Penrose inequality. 
\vspace{2mm}

In this paper we study the Einstein vacuum equations (EVEs)
\begin{equation}\label{Intro:EVE eqn}
    \Ric(\bfg)=0
\end{equation}
for a $3+1$ dimensional Lorentzian manifold $(\MM, \bfg)$.
Bridging hyperbolic and elliptic approaches, we show that: \textit{Arising from admissible characteristic initial data, the Einstein vacuum equations admit  black-hole formation solutions: In the process of gravitational collapse, a whole black hole region with its boundary (apparent horizon) emerges from a spacetime point; Subsequently, it grows in evolution, and settles down to a slowly-rotating Kerr black hole. Moreover, the apparent horizon is spacelike in its initial stage, achronal during its evolution, asymptotically null and converging to the event horizon in its final stage. Furthermore, in this process the Penrose inequality is also proved to be true.}
\vspace{2mm}

As mentioned above, Christodoulou invented the short-pulse method in \cite{Chr:book}. Later, the first author and Luk \cite{A-L} extended Christodoulou's result \cite{Chr:book} by relaxing the size of required initial data and established the existence of solutions for Einstein vacuum equations up to the ``center'' of gravitational collapse. This scale-critical existence result further allows the first author \cite{An:scale} to solve the boundary of the trapped region, that is the so-called apparent horizon and it emerges from a spacetime point.

An important feature in the proofs of \cite{Chr:book} and \cite{A-L} is that the spacetime region of study has a short characteristic length in the outgoing direction. Via \cite{A-L} by An--Luk and \cite{An:AH,An:scale} by the first author, the initial stage of black hole formation is depicted in details, however, its later evolution is still unrevealing, which requires a global in-time-result.

In \cite{An-He} we made one step further and a new concept called the \textit{null comparison principle} was introduced by us. After conducting proof based on it and verifying this null comparison principle, in this paper we show that our solved apparent horizon is achronal.

After this period, the spacetime metric enters a regime, which is close to a slowly-rotating Kerr metric. The mechanism of Kerr stability then comes into play. The breakthrough of nonlinear Kerr stability with small angular momentum is achieved in a series of works \cite{KS:Kerr1,KS:Kerr2,KS:main} by Klainerman--Szeftel, and companion papers in \cite{GKS} by Giorgi--Klainerman--Szeftel and \cite{Shen} by Shen. We also hope to mention two striking proofs of nonlinear Schwarzschild stability with the absence of angular momentum, i.e., \cite{K-S} by Klainerman--Szeftel, and \cite{D-H-R-T} by Dafermos--Holzegel--Rodnianski--Taylor, and an earlier important achievement \cite{H-V} by Hintz--Vasy for proving the nonlinear stability of Kerr-de Sitter spacetimes. For the contributions of many previous works, we refer to the references in these aforementioned papers. 

\begin{figure}[htp]
    \centering

\tikzset{every picture/.style={line width=0.75pt}} %set default line width to 0.75pt        

\tikzset{every picture/.style={line width=0.75pt}} %set default line width to 0.75pt        

\begin{tikzpicture}[x=0.75pt,y=0.75pt,yscale=-1,xscale=1]
%uncomment if require: \path (0,415); %set diagram left start at 0, and has height of 415

%Straight Lines [id:da04148539083938896] 
\draw [color={rgb, 255:red, 0; green, 0; blue, 0 }  ,draw opacity=1 ] [dash pattern={on 0.84pt off 2.51pt}]  (372.93,14.81) -- (447.93,88.21) ;
%Straight Lines [id:da6200668403455578] 
\draw [color={rgb, 255:red, 0; green, 0; blue, 0 }  ,draw opacity=1 ]   (261.93,54.01) -- (375.53,167.21) ;
%Curve Lines [id:da3789953131295283] 
\draw [color={rgb, 255:red, 0; green, 0; blue, 0 }  ,draw opacity=1 ]   (261.93,54.01) .. controls (303.87,45.43) and (324.53,37.77) .. (372.93,14.81) ;
%Straight Lines [id:da03823923770601123] 
\draw    (228.17,329.63) -- (447.37,88.73) ;
\draw [shift={(447.6,88.48)}, rotate = 312.3] [color={rgb, 255:red, 0; green, 0; blue, 0 }  ][line width=0.75]      (0, 0) circle [x radius= 1.34, y radius= 1.34]   ;
%Straight Lines [id:da8594658011416159] 
\draw    (372.6,15.08) -- (375.2,167.48) ;
%Straight Lines [id:da8073915065973601] 
\draw    (300,91.48) -- (372.6,15.08) ;
%Straight Lines [id:da011586745020532696] 
\draw    (226,61.48) -- (352,193.08) ;
%Curve Lines [id:da814717229444376] 
\draw    (226,61.48) .. controls (243.68,67.91) and (260.37,69.33) .. (276.25,67.18) .. controls (311.56,62.38) and (342.86,39.9) .. (372.16,15.45) ;
\draw [shift={(372.6,15.08)}, rotate = 320.12] [color={rgb, 255:red, 0; green, 0; blue, 0 }  ][line width=0.75]      (0, 0) circle [x radius= 2.01, y radius= 2.01]   ;
%Straight Lines [id:da6142804922901302] 
\draw    (226,61.48) -- (261.93,54.01) ;
%Straight Lines [id:da7952380756222553] 
\draw    (226.01,62.49) -- (228.17,329.63) ;
\draw [shift={(226,61.48)}, rotate = 89.54] [color={rgb, 255:red, 0; green, 0; blue, 0 }  ][line width=0.75]      (0, 0) circle [x radius= 2.01, y radius= 2.01]   ;
%Straight Lines [id:da26620688080587096] 
\draw  [dash pattern={on 0.84pt off 2.51pt}]  (297.87,45.43) -- (309.2,57.84) -- (309.2,57.84) ;
%Straight Lines [id:da3325060155806061] 
\draw  [dash pattern={on 0.84pt off 2.51pt}]  (309.2,57.84) -- (374.16,122.8) ;
\draw [shift={(309.2,57.84)}, rotate = 45] [color={rgb, 255:red, 0; green, 0; blue, 0 }  ][fill={rgb, 255:red, 0; green, 0; blue, 0 }  ][line width=0.75]      (0, 0) circle [x radius= 1.34, y radius= 1.34]   ;
%Straight Lines [id:da8708184444504157] 
\draw    (374.87,203.77) -- (362.48,190.15) ;
\draw [shift={(361.13,188.67)}, rotate = 47.71] [color={rgb, 255:red, 0; green, 0; blue, 0 }  ][line width=0.75]    (6.56,-1.97) .. controls (4.17,-0.84) and (1.99,-0.18) .. (0,0) .. controls (1.99,0.18) and (4.17,0.84) .. (6.56,1.97)   ;
%Straight Lines [id:da9818525816414958] 
\draw    (381.87,197.43) -- (369.48,183.81) ;
\draw [shift={(368.13,182.33)}, rotate = 47.71] [color={rgb, 255:red, 0; green, 0; blue, 0 }  ][line width=0.75]    (6.56,-1.97) .. controls (4.17,-0.84) and (1.99,-0.18) .. (0,0) .. controls (1.99,0.18) and (4.17,0.84) .. (6.56,1.97)   ;
%Straight Lines [id:da8685904482451177] 
\draw    (387.53,190.77) -- (375.15,177.15) ;
\draw [shift={(373.8,175.67)}, rotate = 47.71] [color={rgb, 255:red, 0; green, 0; blue, 0 }  ][line width=0.75]    (6.56,-1.97) .. controls (4.17,-0.84) and (1.99,-0.18) .. (0,0) .. controls (1.99,0.18) and (4.17,0.84) .. (6.56,1.97)   ;

% Text Node
\draw (371.16,-2.64) node [anchor=north west][inner sep=0.75pt]  [font=\scriptsize]  {$i^{+}$};
% Text Node
\draw (418.76,43.76) node [anchor=north west][inner sep=0.75pt]  [font=\scriptsize]  {$\mathscr{I}^{+}$};
% Text Node
\draw (316.76,23.96) node [anchor=north west][inner sep=0.75pt]  [font=\scriptsize]  {$\mathcal{A}$};
% Text Node
\draw (344.36,45.56) node [anchor=north west][inner sep=0.75pt]  [font=\scriptsize]  {$\mathcal{H}^{+}$};
% Text Node
\draw (283.76,65.83) node [anchor=north west][inner sep=0.75pt]  [font=\scriptsize]  {$\mathcal{AH}$};
% Text Node
\draw (285.96,49.69) node [anchor=north west][inner sep=0.75pt]  [font=\tiny]  {$M_{\widetilde{\underline{u}}}$};
% Text Node
\draw (376.76,78.36) node [anchor=north west][inner sep=0.75pt]  [font=\scriptsize]  {$\mathcal{T}$};
% Text Node
\draw (206.76,56.76) node [anchor=north west][inner sep=0.75pt]  [font=\scriptsize]  {$O$};
% Text Node
\draw (331.16,92.36) node [anchor=north west][inner sep=0.75pt]  [font=\scriptsize]  {$\widetilde{\underline{H}}_{\widetilde{\underline{u}}}$};
% Text Node
\draw (249.76,173.16) node [anchor=north west][inner sep=0.75pt]  [font=\scriptsize]  {Minkowski};
% Text Node
\draw (276.76,107.83) node [anchor=north west][inner sep=0.75pt]  [font=\scriptsize]  {$ \begin{array}{l}
\text{short}\ \\
\ \ \ \ \ \ \text{pulse}
\end{array}$};
% Text Node
\draw (411.76,138.49) node [anchor=north west][inner sep=0.75pt]  [font=\scriptsize]  {Kerr\ stability};
% Text Node
\draw (381.43,203.16) node [anchor=north west][inner sep=0.75pt]  [font=\scriptsize]  {incoming\ radiation};

\end{tikzpicture}

    \caption{Process of Kerr Black Hole Formation}
    \label{fig:KerrBHformation1}
\end{figure}
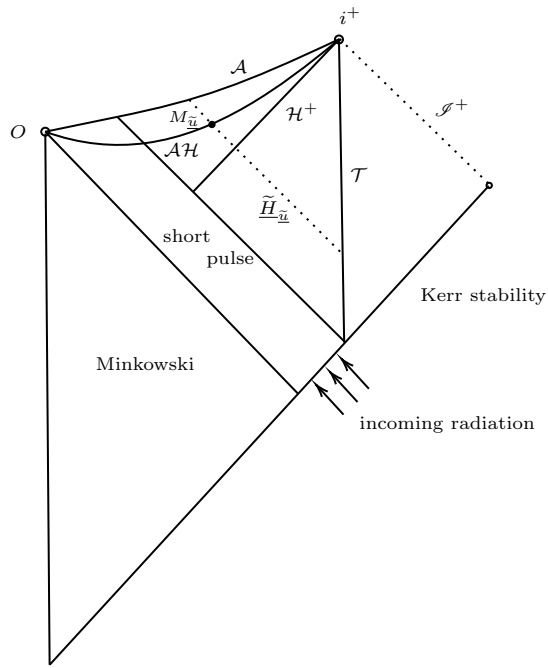

Back to this paper, the aim here is to establish the global picture as portrayed in \Cref{fig:KerrBHformation1} and to prove Kerr black hole formation, via connecting the scale-critical black hole emergence results in \cite{A-L,An:scale} to the main theorem of nonlinear Kerr stability in Klainerman--Szeftel \cite{KS:main}.

We achieve our goal by the following steps:
\vspace{1mm}

We first extend \cite{A-L,An:scale} and introduce the admissible Kerr black hole formation initial data along an outgoing null hypersurface, such that, after the short-pulse region, the solved spacetime metric near the initial outgoing null hypersurface and along a later incoming null hypersurface enters the perturbative Kerr regime, and we further verify the required initial layer conditions in \cite{KS:main}. We can then apply the main result in \cite{KS:main} and obtain the global-in-time solutions for EVEs \eqref{Intro:EVE eqn}.

 Next we proceed to solve the boundary of the trapped hole region, i.e., the apparent horizon. In the short-pulse regime, in \cite{An:AH} by the first author and in \cite{An-Han} by An--Han, the corresponding apparent horizon is identified by solving the marginally outer trapped surface (MOTS) along each incoming null hypersurface. For solving the apparent horizon in Kerr black hole-formation spacetimes, two obstructions arise:
 \begin{enumerate}
     \item The spacetime solved in \cite{KS:main} is not foliated by incoming null cones. It is sliced by  slightly timelike hypersurfaces.
     \item In perturbative Kerr regime, new elliptic arguments are required to extract the precise information of Kerr geometry.
 \end{enumerate}
To overcome these obstructions, we adopt the Pretorius--Israel type coordinates and construct incoming null cones with precise leading behaviors.

Utilizing new elliptic techniques and precise geometric information of our constructed incoming null cones with Pretorius--Israel type coordinates, we are able to establish the existence and uniqueness of MOTS along each incoming null cone. These new elliptic arguments rely on a direct derivation of the $C^{1,\a}$ a priori estimate via employing \textit{the quasi-conformal mapping method} and \textit{the Leray--Schauder fixed point theorem}.

We further provide a comprehensive analysis of the apparent horizon, and prove that: \textit{the formed apparent horizon emerges from a spacetime point, is smooth (except at that point), is asymptotically null and eventually approaches the event horizon.} By validating our null comparison principle formulated in \cite{An-He}, we also show that our solved apparent horizon is \textit{(locally) achronal.}\footnote{This means that the apparent horizon locates in $\MM\backslash J^-(\II^+)$, i.e., the complement of causal past of the future null infinity.}

Building on our aforementioned analysis, in this paper we proceed to put forward a new hyperbolic approach to study the Penrose inequality. In particular, we prove the desired Penrose inequality in our black hole formation spacetimes and in perturbative Kerr regime. Our approach aligns with Penrose's original perspective in \cite{Penrosenaked} by considering the hyperbolic evolution of Einstein's equations and we also provide a rigorous new mathematical framework: \textit{For the spacetime Penrose inequality, our approach is as starting with a spacelike initial data set, that contains a MOTS within the perturbed Kerr regime, we first generate the corresponding spacetime, which solves the $3+1$ Einstein vacuum equations. Adapting to the initial MOTS, within the solved spacetime we then construct suitable incoming and outgoing null cones as foliations. Thereafter, we solve the MOTS along each incoming null cone. By assembling all these MOTS, we identify the apparent horizon, and ultimately demonstrate its smoothness and achronality to prove the Penrose inequality.}

\subsection{Main Results} 
 With no symmetry assumption, this paper investigates the entire process of black hole formation for $3+1$ Einstein vacuum equations. As portrayed in \Cref{fig:KerrBHformation1}, we demonstrate the following physical picture: \textit{Initially, a central singularity and a tiny trapped region emerge in the course of gravitational collapse. The boundary of the trapped region (apparent horizon) also forms and expands along the spacetime evolution. Going through a transition period, the spacetime geometry evolves into a regime, which is close to a slowly-rotating Kerr metric. The mechanism of nonlinear Kerr stability then kicks in. Eventually, the spacetime metric settles down to the Kerr value with small angular momentum.}

In more formal words, the first main result of this paper can be summarized as
\begin{theorem}\label{Main Thm1}
    With admissible characteristic initial data, the Einstein vacuum equations \eqref{Intro:EVE eqn} admit Kerr black hole formation solutions. Each solution processes a complete apparent horizon, originating from an emerging spacetime center point, being spacelike in the short-pulse region, being asymptotically null and converging to the event horizon as the advanced time tends to the timelike infinity.
\end{theorem}

In later sections, we call our admissible characteristic initial data as \textit{admissible Kerr black hole formation initial data}. Its precise definition is stated in \Cref{Subsec:admiKerr}. There we impose open set conditions. Smooth examples of these initial data are given in \Cref{Apx:example} and they allow general dynamical asymptotics, which are used in \cite{KS:main} by Klainerman--Szeftel.

Our proof of \Cref{Main Thm1} consists of two main components: the hyperbolic part and the elliptic part. In the hyperbolic part, we combine and extend the scale-critical short-pulse result \cite{A-L} by the first author and Luk with the main results of the nonlinear Kerr stability for small angular momentum \cite{KS:main} by Klainerman and Szeftel. Notably, the general covariant modulated (GCM) admissible spacetimes constructed in \cite{KS:main} are not foliated by null cones, but instead by slightly timelike hypersurfaces, which is a feature necessitated by the spin of Kerr solutions. In our elliptic part, our marginally outer trapped surface (MOTS) will be solved along each incoming null hypersurface. We hence construct the precise (incoming) null hypersurfaces in these GCM admissible spacetimes. We adopt a Pretorius--Israel type coordinate system in these perturbed Kerr spacetimes. Pretorius and Israel \cite{P-I} introduced these coordinates in exact Kerr spacetimes. Here we generalize their coordinates into our dynamical spacetimes. Leveraging the characteristic method based on specific Pretorius--Israel coordinates, 
we solve the eikonal equation for the optical function $\tub$ and construct the incoming null cones $\tHb_{\tub}$ explicitly in perturbative Kerr spacetimes. This approach enables us to capture the precise leading terms of the corresponding outgoing null expansion, which is a critical component requested in our elliptic argument.

We then solve the equation of MOTS along each newly constructed incoming null hypersurface. The equation we are dealing with is a quasilinear elliptic equation with quadratic growth of the gradient. Contrast to the approach in \cite{An:AH,An-He}, where a modified Bochner formula is designed to capture the Schwarzschild geometry, in this paper we adopt new elliptic arguments via using the quasi-conformal mapping, which equip us with the flexibility to deal with Kerr geometry. 
To prove the existence of MOTS $M_{\tub}$ along each incoming null cone $\tHb_{\tub}$, instead of relying on the continuity method adopted in \cite{An:AH,An-He}, we apply the Leray--Schauder fixed point theorem here with our established a priori estimates and with a crucial utilization of the Miranda--Talenti type inequality on $\ms$. Meanwhile, the uniqueness of MOTS along each incoming null cone $\tHb_{\tub}$, and the smoothness, the asymptotics of the apparent horizon $\AH=\cup_{\tub} M_{\tub}$ are obtained via analyzing the linearized equation. Notably, despite that in our hyperbolic part we apply the main results of \cite{KS:main} by Klainerman--Szeftel, where they proved the nonlinear Kerr stability with small angular momentum, our elliptic arguments described above remain valid for the perturbative spacetimes of subextremal Kerr with full range, as long as the desired Kerr stability holds.
\vspace{3mm}

In \cite{An-He} we introduced a new concept called the \textit{null comparison principle} and we proved that \textit{if the null comparison principle holds true with respect to the MOTS along the incoming null cone, then the corresponding apparent horizon must be piecewise spacelike or piecewise null.} Based on the precise calculation for the leading terms of the outgoing null expansion and relying on analytic properties of the linearized operator for the MOTS equation, in this paper we verify the null comparison principle for every MOTS solved as mentioned above. Hence, our apparent horizon is \textit{(locally) achronal for all time}. This further implies our area increasing law of the MOTS. We summarize these properties as
\begin{theorem}\label{Main Thm2}
    The apparent horizon $\AH=\cup_{\tub} M_{\tub}$ constructed in this paper is either piecewise spacelike or piecewise (outgoing) null. Moreover, the area of MOTS $M_{\tub}$ is non-decreasing as $\tub$ grows.
\end{theorem}
\begin{remark}
    For the apparent horizon constructed in \Cref{Main Thm1}, verifying the associated null comparison principle demands the detailed information from controlling the derivatives of the location of $\AH$, i.e., requiring the smallness of its gradient and Hessian. This information has been derived and fully utilized in this paper via our method of solving MOTSs along incoming null hypersurfaces. Compared to finding the MOTS along a spacelike hypersurface, due to the blow-up nature of the corresponding Jang's equation, it is usually hard to show the smallness of the corresponding second fundamental form and curvatures of MOTS.
\end{remark}
Equipped with \Cref{Main Thm1} and \Cref{Main Thm2}, we march to fulfill a perspective of Penrose, i.e., the \emph{Penrose inequality} in Kerr black hole formation spacetimes. In 1973, Penrose \cite{Penrosenaked} proposed this inequality to explore the relation between the surface area of the black hole and the total mass of the spacetime. 

\begin{figure}[htp]
    \centering

\tikzset{every picture/.style={line width=0.75pt}} %set default line width to 0.75pt        

\begin{tikzpicture}[x=0.75pt,y=0.75pt,yscale=-0.6,xscale=0.8]%uncomment if require: \path (0,300); %set diagram left start at 0, and has height of 300

%Straight Lines [id:da4156818570001959] 
\draw    (226.78,219.69) -- (226.78,-5.8) ;
%Curve Lines [id:da8682928903352223] 
\draw    (264,214.03) .. controls (230.54,157.91) and (226.49,124.45) .. (226.78,-5.8) ;
%Shape: Ellipse [id:dp034564106250814186] 
\draw   (264.74,209.16) .. controls (269.42,199.36) and (293.93,193.58) .. (319.48,196.26) .. controls (345.02,198.93) and (361.93,209.03) .. (357.25,218.83) .. controls (352.57,228.63) and (328.06,234.4) .. (302.52,231.73) .. controls (276.97,229.06) and (260.06,218.95) .. (264.74,209.16) -- cycle ;
%Curve Lines [id:da4164359253093157] 
\draw    (393.23,-6.83) .. controls (393.23,107.46) and (380.06,182.17) .. (357.25,218.83) ;
%Straight Lines [id:da14245119979959275] 
\draw    (393.23,-6.83) -- (393.23,219.17) ;
%Shape: Ellipse [id:dp37038899224321153] 
\draw   (227.92,-0.38) .. controls (221.05,-21.88) and (252.12,-41.88) .. (297.32,-45.05) .. controls (342.51,-48.22) and (384.73,-33.36) .. (391.6,-11.86) .. controls (398.47,9.64) and (367.39,29.64) .. (322.19,32.81) .. controls (277,35.98) and (234.79,21.12) .. (227.92,-0.38) -- cycle ;
%Shape: Ellipse [id:dp0735962877251326] 
\draw   (231.28,231.92) .. controls (216.14,212.37) and (239,190.63) .. (282.35,183.36) .. controls (325.69,176.09) and (373.12,186.05) .. (388.26,205.59) .. controls (403.41,225.14) and (380.55,246.89) .. (337.2,254.16) .. controls (293.85,261.43) and (246.43,251.47) .. (231.28,231.92) -- cycle ;
%Straight Lines [id:da9249598285412924] 
\draw  [dash pattern={on 0.84pt off 2.51pt}]  (424.5,-11.34) -- (512,220.25) ;
%Straight Lines [id:da1168388200898598] 
\draw  [dash pattern={on 0.84pt off 2.51pt}]  (189.33,-12.33) -- (118.5,216.25) ;
%Shape: Ellipse [id:dp6138619577221648] 
\draw  [dash pattern={on 0.84pt off 2.51pt}] (160.28,182.43) .. controls (227.43,157.29) and (350.88,153.33) .. (436,173.59) .. controls (521.13,193.86) and (535.7,230.67) .. (468.55,255.82) .. controls (401.4,280.96) and (277.95,284.92) .. (192.83,264.66) .. controls (107.7,244.39) and (93.13,207.58) .. (160.28,182.43) -- cycle ;
%Shape: Ellipse [id:dp539996290781753] 
\draw  [dash pattern={on 0.84pt off 2.51pt}] (190.27,2.64) .. controls (180.12,-25.02) and (224.12,-50.64) .. (288.55,-54.57) .. controls (352.99,-58.5) and (413.45,-39.25) .. (423.61,-11.59) .. controls (433.77,16.07) and (389.77,41.69) .. (325.33,45.62) .. controls (260.9,49.55) and (200.43,30.3) .. (190.27,2.64) -- cycle ;
%Shape: Ellipse [id:dp12142620520817193] 
\draw  [dash pattern={on 0.84pt off 2.51pt}] (199.1,17.9) .. controls (244.05,0.67) and (329.27,-2.08) .. (389.44,11.76) .. controls (449.61,25.59) and (461.95,50.77) .. (417,68) .. controls (372.05,85.23) and (286.83,87.98) .. (226.66,74.14) .. controls (166.49,60.31) and (154.15,35.13) .. (199.1,17.9) -- cycle ;
%Straight Lines [id:da32058807011400314] 
\draw    (240.33,121) -- (214.8,202.45) ;
%Straight Lines [id:da32784758227134814] 
\draw    (376,125) -- (405.3,202.95) ;
%Curve Lines [id:da28865001437710025] 
\draw  [dash pattern={on 0.84pt off 2.51pt}]  (234.8,138.5) .. controls (253.8,104.45) and (366.8,111.95) .. (381.8,140.45) ;
%Curve Lines [id:da02675561391186787] 
\draw    (234.8,138.5) .. controls (237.3,176.95) and (381.3,175.95) .. (381.8,140.45) ;

% Text Node
\draw (475.76,96.59) node [anchor=north west][inner sep=0.75pt]  [font=\scriptsize]  {$\mathscr{I}^{+}$};
% Text Node
\draw (313.46,281.9) node [anchor=north west][inner sep=0.75pt]  [font=\small]  {$\Sigma $};
% Text Node
\draw (334.17,-77.11) node [anchor=north west][inner sep=0.75pt]  [font=\scriptsize]  {$Kerr( a_{\infty } ,m_{\infty })$};
% Text Node
\draw (469.58,228.31) node [anchor=north west][inner sep=0.75pt]  [font=\scriptsize]  {$m$};
% Text Node
\draw (398.91,-17.19) node [anchor=north west][inner sep=0.75pt]  [font=\scriptsize]  {$m_{\infty }$};
% Text Node
\draw (306.05,209.62) node [anchor=north west][inner sep=0.75pt]  [font=\scriptsize]  {$A$};
% Text Node
\draw (398.02,104.13) node [anchor=north west][inner sep=0.75pt]  [font=\scriptsize]  {$\mathcal{H}^{+}$};
% Text Node
\draw (358.68,95.67) node [anchor=north west][inner sep=0.75pt]  [font=\scriptsize]  {$\mathcal{AH}$};
% Text Node
\draw (300.41,-24.56) node [anchor=north west][inner sep=0.75pt]  [font=\scriptsize]  {$A_{\infty }$};
% Text Node
\draw (448.91,29.4) node [anchor=north west][inner sep=0.75pt]  [font=\scriptsize]  {$M_{B}\left(\tilde{u}\right)$};
% Text Node
\draw (289.91,93.9) node [anchor=north west][inner sep=0.75pt]  [font=\scriptsize]  {$A_{M}\left(\widetilde{\underline{u}}\right)$};
% Text Node
\draw (200.91,141.4) node [anchor=north west][inner sep=0.75pt]  [font=\scriptsize]  {$\underline{H}_{\widetilde{\underline{u}}}$};
\end{tikzpicture}

    \caption{A New Approach Toward Penrose Inequality}
    \label{fig:penroseargument}
\end{figure}
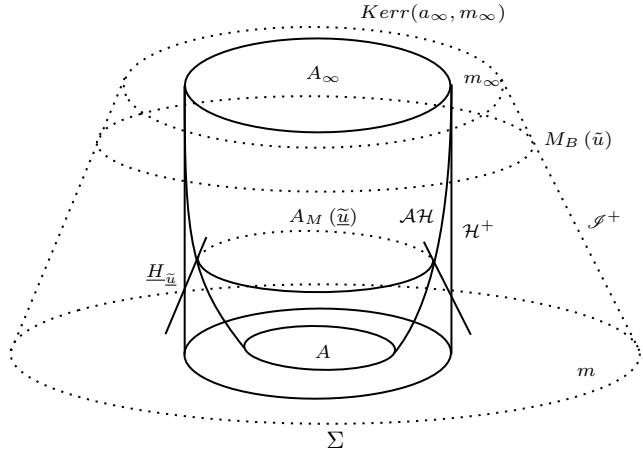

Building upon our analysis, as shown in \Cref{fig:penroseargument}, in this paper we put forward a new mathematical framework and prove both the dynamical Penrose inequality and the spacetime Penrose inequality in our black hole formation spacetimes and in the perturbative regime of Kerr black holes. We note that our arguments do not use the condition that the spacetime angular momentum $a$ is small and hence are applicable for large $a$ scenario.

We first state our result for the dynamical Penrose inequality.
\begin{theorem}[Dynamical Penrose Inequality]\label{Main Thm3:Penrose ineq 1}
    The dynamical Penrose inequality holds true in our Kerr black hole formation spacetimes and in Klainerman--Szeftel's Kerr stability spacetimes. Specifically, letting $A_M(\tub)$ be the area of MOTS $M_{\tub}$ and defining $M_B(\tu)$ (with $\tu$ being an outgoing optical function) as the Bondi mass along the future null infinity, we have the following inequality
    \begin{equation}\label{Intro:penrose ineq1}
        M_B(\tu)\ge m_{\infty}\ge \sqrt{\f{m_{\infty }( m_{\infty } +\sqrt{m_{\infty }^{2} -a_{\infty }^{2}})}{2}} \ge \sqrt{\f{A_M(\tub)}{16\pi}}.
    \end{equation}
    Here $m_{\infty}, a_{\infty}$ are the final mass, final angular momentum of the dynamical spacetime with $|a_{\infty}/m_{\infty}|\ll 1$.

     If assuming nonlinear Kerr stability in full subextremal regime, we have that \eqref{Intro:penrose ineq1} still holds with $|a_{\infty}/m_{\infty}|<1$.
\end{theorem}

The logic of our approach for proving the above Penrose inequality is as follows:  As portrayed in \Cref{fig:penroseargument}, we start from  considering a spacelike initial slice $\Si$, which has the ADM mass $m$. Assume a MOTS with area $A$ is embedded in $\Si$. The main body of current paper is to construct the complete apparent horizon $\AH$ and to establish the desired black hole area law along it. In this article we first prove that for the MOTS $M_{\tub}$, which is a spacelike section of the apparent horizon, its area is non-decreasing for all $\tub$. Given the desired nonlinear Kerr stability conclusions for the full subextremal range, we further verify that the induced metric along $\AH$  eventually stabilizes to that of the horizon for a stationary Kerr black hole with the final mass $m_{\infty}$ and the final angular momentum $a_{\infty}$ with $|a_{\infty}/m_{\infty}|<1$. For the exact Kerr metric, the surface area of the MOTS lying along the horizon can be explicitly calculated and it is
\begin{equation*}
    A_{\infty } =8\pi m_{\infty }\left( m_{\infty } +\sqrt{m_{\infty }^{2} -a_{\infty }^{2}}\right).
\end{equation*}
Since part of the initial mass would be radiated away along with gravitational waves during the gravitational collapse,\footnote{It is usually referred as the Bondi mass loss.} along the future null infinity $\II^+$ it holds that $m_{\infty}\le M_B(\tu)\le m$. 
Therefore, collecting all these, we hence obtain a string of inequalities, i.e.,
\begin{equation}\label{Intro:ineq Penrose ineq}
    A\le A_M(\tub) \le A_{\infty}\le 16\pi m_{\infty}^2\le 16\pi [M_B(\tu)]^2\le 16\pi m^2.
\end{equation}
This gives rise to \eqref{Intro:penrose ineq1}.
\vspace{2mm}

\begin{figure}[ht]
    \centering
  \tikzset{every picture/.style={line width=0.75pt}} %set default line width to 0.75pt        

\begin{tikzpicture}[x=0.75pt,y=0.75pt,yscale=-0.8,xscale=1]

%Shape: Parallelogram [id:dp2362012717073685] 
\draw   (222.23,130.88) -- (501.84,130.88) -- (382.01,203.52) -- (102.4,203.52) -- cycle ;
%Curve Lines [id:da22345420634118107] 
\draw    (263.16,165.51) .. controls (251.44,147.68) and (243.44,130.48) .. (237.3,94.44) ;
%Shape: Ellipse [id:dp17129628744944514] 
\draw   (263.81,160.93) .. controls (267.96,151.71) and (289.65,146.27) .. (312.26,148.78) .. controls (334.88,151.3) and (349.85,160.81) .. (345.7,170.03) .. controls (341.56,179.25) and (319.87,184.68) .. (297.25,182.17) .. controls (274.64,179.66) and (259.66,170.15) .. (263.81,160.93) -- cycle ;
%Curve Lines [id:da43002599336704006] 
\draw    (367.44,96.27) .. controls (364.24,136.48) and (356.24,150.48) .. (345.7,170.03) ;
%Straight Lines [id:da813248823391504] 
\draw    (242.2,77.97) -- (214.64,162.88) ;
%Straight Lines [id:da6694153647289117] 
\draw    (362.3,81.73) -- (390.24,166.08) ;
%Curve Lines [id:da7583160071210381] 
\draw  [dash pattern={on 0.84pt off 2.51pt}]  (237.3,94.44) .. controls (254.12,62.4) and (354.16,69.45) .. (367.44,96.27) ;
%Curve Lines [id:da06897526645907093] 
\draw    (237.3,94.44) .. controls (239.52,130.62) and (366.99,129.68) .. (367.44,96.27) ;
%Straight Lines [id:da6493769222632892] 
\draw    (284.9,89.85) -- (263.16,165.51) ;
%Straight Lines [id:da38789705642209127] 
\draw    (317.4,91.35) -- (346.74,167.08) ;
%Curve Lines [id:da12050348467573113] 
\draw  [dash pattern={on 0.84pt off 2.51pt}]  (215.8,162.42) .. controls (238.24,126.93) and (371.69,134.75) .. (389.4,164.46) ;
%Curve Lines [id:da5089876251320691] 
\draw    (215.8,162.42) .. controls (218.76,202.51) and (388.81,201.46) .. (389.4,164.46) ;

% Text Node
\draw (265.36,210.56) node [anchor=north west][inner sep=0.75pt]  [font=\small]  {$( \Sigma ,g,k)$};
% Text Node
\draw (348.72,164.12) node [anchor=north west][inner sep=0.75pt]  [font=\small]  {$M_{0}$};
% Text Node
\draw (333.72,89.12) node [anchor=north west][inner sep=0.75pt]  [font=\small]  {$M_{\widetilde{\underline{u}}}$};

\end{tikzpicture}

    \caption{Spacetime Penrose Inequality}
    \label{fig:Riem penrose ineq}
\end{figure}
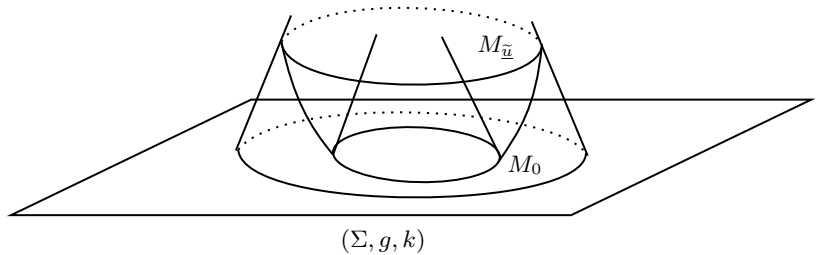

In the Riemannian setting, when restricting \eqref{Intro:ineq Penrose ineq} to the initial data on $\Si$, we would expect the spacetime Penrose inequality
\begin{equation*}
    A\le 16 \pi m^2.
\end{equation*}
Here as in \Cref{fig:Riem penrose ineq}, $A$ represents the area of the MOTS $M_0$ along $\Si$ and  $m$ denotes the ADM mass of the initial data set $(\Si, g, k)$. Note that in our constructed spacetimes, we do not impose the time symmetric condition. With the aid of time symmetry (near the MOTS), namely, $k\equiv 0$, the general Riemannian Penrose $A\le 16\pi m^2$ inequality has been proved by Huisken--Ilmanen \cite{H-I} via the inverse mean curvature flow and by Bray \cite{Bray} via the conformal flow.

In this current paper, as portrayed in \Cref{fig:Riem penrose ineq}, with initial data set $(\Si, g, k)$ in perturbative Kerr regime, we evolve Einstein vacuum equations and obtain the same spacetime picture as in \Cref{fig:penroseargument}. Employing the aforementioned logic, we thus prove the following spacetime Penrose inequality with no time symmetric assumption.

\begin{theorem}[Spacetime Penrose Inequality]\label{Main Thm3:Penrose ineq 2}
    Within the perturbative regime of slowly rotating Kerr black holes, assume that we have a spacelike initial data set $(\Si, g, k)$ with a perturbed MOTS $M_0$, that is close to the MOTS in the exact Kerr. Let $m$ be the ADM mass of $(\Si, g)$ and denote $A$ to be the area of the MOTS $M_0$. Then we have the following Penrose inequality
    \begin{equation}\label{Intro:penrose ineq}
        m\ge \sqrt{\f{A}{16\pi}},
    \end{equation}
   with equality if and only if the Gauss curvature of the MOTS $M_0$ is pointwise equal to $\f{1}{4m^2}$.

  Moreover, assuming the validity of nonlinear Kerr stability for the full subextremal range, we still have \eqref{Intro:penrose ineq} with the aforementioned rigidity in the perturbative regime of all subextremal Kerr black holes.
\end{theorem}

\subsection{Key Points and New Ingredients of the Paper}
\subsubsection{Admissible Kerr Black Hole Formation Initial Data}\label{Introsec:Extended ID}
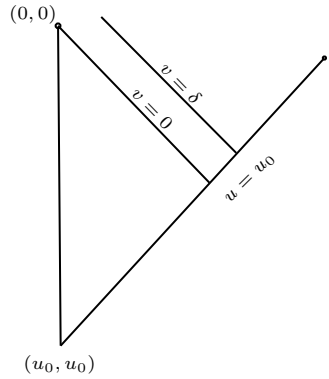
\begin{figure}[htp]
    \centering
   \tikzset{every picture/.style={line width=0.75pt}} %set default line width to 0.75pt        

\begin{tikzpicture}[x=0.75pt,y=0.75pt,yscale=-0.6,xscale=0.6]

%Straight Lines [id:da3713363955240051] 
\draw [color={rgb, 255:red, 0; green, 0; blue, 0 }  ,draw opacity=1 ]   (281.93,14.38) -- (394.3,128.45) ;
%Straight Lines [id:da41801556136918605] 
\draw    (248.17,290) -- (467.37,49.1) ;
\draw [shift={(467.6,48.85)}, rotate = 312.3] [color={rgb, 255:red, 0; green, 0; blue, 0 }  ][line width=0.75]      (0, 0) circle [x radius= 1.34, y radius= 1.34]   ;
%Straight Lines [id:da5964874586864704] 
\draw    (246,21.85) -- (372,153.45) ;
%Straight Lines [id:da9422549115992721] 
\draw    (246.01,22.86) -- (248.17,290) ;
\draw [shift={(246,21.85)}, rotate = 89.54] [color={rgb, 255:red, 0; green, 0; blue, 0 }  ][line width=0.75]      (0, 0) circle [x radius= 2.01, y radius= 2.01]   ;

% Text Node
\draw (202.76,1.63) node [anchor=north west][inner sep=0.75pt]  [font=\scriptsize]  {$( 0,0)$};
% Text Node
\draw (214.76,295.4) node [anchor=north west][inner sep=0.75pt]  [font=\scriptsize]  {$( u_{0} ,u_{0})$};
% Text Node
\draw (380.78,165.77) node [anchor=north west][inner sep=0.75pt]  [font=\scriptsize,rotate=-315]  {$u=u_{0}$};
% Text Node
\draw (313.52,68.86) node [anchor=north west][inner sep=0.75pt]  [font=\scriptsize,rotate=-46.12]  {$v=0$};
% Text Node
\draw (338.02,47.36) node [anchor=north west][inner sep=0.75pt]  [font=\scriptsize,rotate=-46.12]  {$v=\delta $};

\end{tikzpicture}

    \caption{Admissible Characteristic Initial Data}
    \label{fig:admisID}
\end{figure}

To connect the scale-critical short-pulse result \cite{A-L} to the stability of Kerr \cite{KS:main}, we introduce and prescribe admissible characteristic initial data along outgoing null hypersurface $H_{u_0}=\qty{u=u_0}$ up to the center as shown in \Cref{fig:admisID}. Later we call these initial data as the \textit{admissible Kerr black hole formation initial data}. 

Specifically, define $S_{u, v}$ to be the intersected 2-sphere for level sets of $u$ and $v$ in double null coordinates.
\begin{itemize}
    \item Along $H_{u_0}\cap\qty{u_0 \le v \le \de}$ with $0<\de\ll 1$, we prescribe the scale-critical short-pulse data that are consistent with \cite{A-L}. Note that the shear tensor $\chih$ is large along $H_{u_0}\cap\qty{0 \le v \le \de}$ and it is of size $\f{A^{\f12}}{|u_0|}$ with $A=\de^{-1}\gg 1$.

    \item To guarantee that the initial metric on $S_{u_0, \de}$ (at the end of the short pulse) is sufficiently close to the Schwarzschild metric with mass $m_0=1/4$, we further impose 
    \begin{equation}\label{intro:integral bound}
     \int_{0}^{\de} |u_0|^2|\chih|^2(u_0, v, \th^1, \th^2) dv=4m_0+\lot
 \end{equation}
and
\begin{equation*}
    |\chih|+\de|\a|\lesssim \f{\de^2 A^{\f12}}{|u_0|} \quad \text{on} \quad S_{u_0, \de}.
\end{equation*}
Here $\a\approx \pr_v \chih$ is the most singular curvature component and its precise definition is provided in \eqref{def curvatures} in \Cref{Subsec:framdecomp}.
    
As a consequence, we can then propagate the enhanced bounds for $\chih$ and $\a$ along the incoming null direction and get
\begin{equation*}
    |\chih|+\de|\a|\lesssim \f{\de^2 A^{\f12}}{|u|} \qquad \text{along} \quad \Hb_{\de}^{[u_0, -\de A]}\coloneqq\qty{u_0\le u\le -\de A, \ v=\de}.
\end{equation*}
The estimates of other anomalous geometric quantities can also be improved accordingly. These imply that the characteristic data along the incoming null hypersurface $\Hb_{\de}^{[u_0, -\de A]}$ are sufficiently close to the Schwarzschild value with mass parameter $m_0$.

\item  After the short pulse, along $H_{u_0}\cap \qty{v\ge \de}$ we prescribe asymptotic-to-Kerr data, which satisfy the decay assumptions for initial data as required in \cite{KS:main}. Typically, we let
\begin{equation*}
    |\chih|+r|\a|\lesssim \f{\epsilon_1}{r^{\f52+\de_B}}.
\end{equation*}
Here $r$ is the area radius for $S_{u, v}$. Parameters $\de_B, \epsilon_1$ obey $0<\de_B\ll 1$ and $\de A^{\f12}\ll \epsilon_1\ll 1$.
\end{itemize}
We note that smooth examples of these admissible Kerr black hole formation initial data are provided in \Cref{Apx:example}. In particular, the initial data at later stage can be freely prescribed in this construction.

\vspace{2mm}

 We now take our null hypersurfaces  $\{ u_0\le u\le -\de A, \ v=\de\}$ and $\{u=u_0,\  \de\le v<\infty\}$ as initial hypersurfaces. To verify the initial conditions in \cite{KS:main} by Klainerman--Szeftel, we here construct the so called initial data layers in the strip-like neighborhood of both $\{ u_0\le u\le -\de A, \ v=\de\}$ and $\{u=u_0, v_0\le v<\infty\}$. This process is carried out as follows:
 \begin{itemize}
     \item By a standard local argument (such as in \cite{Luk}), we can solve for the region $\{ u_0 \le u\le -\de A, \ \de \le v\le \de+d\}$.
     \item To obtain the strip-type neighborhood of the outgoing initial null hypersurface $\{u=u_0, v_0\le v<\infty\}$, we prove a semi-global existence result in this unbounded region. For this step, we employ a similar approach as in \cite{L-Z,Shen:Minkext}, utilizing the $r^p$-weighted estimates and a last slice argument. 
 \end{itemize} 
 We point out that the double null foliation is employed in \cite{A-L} for the short-pulse region. In \cite{KS:main} for studying the Kerr stability, totally different foliations and frames are used, which are referred to as the principal geodesic (PG) structures. To combine both results, we need to construct the associated PG structures in the solved initial data layers and carry out \textit{nine} detailed changes of coordinates and null frames throughout the entire paper.
\vspace{2mm}

Therefore, applying the main theorem in \cite{KS:main}, we derive a global spacetime region in the future of our admissible characteristic initial data, and the spacetime are $\epsilon_0$-close to\footnote{In our proof, we choose $\epsilon_0=\epsilon_1^{\f12}$ with $\epsilon_1$ measuring the size of perturbations in the initial data layer.} and eventually converges to a nearby slowly rotating Kerr black hole $Kerr(a_{\infty}, m_{\infty})$ with final mass $m_{\infty}$ and final angular momentum $a_{\infty}$ satisfying $|a_{\infty}/m_{\infty}|\ll 1$. We also want to mention that the spacetime and the apparent horizon constructed in this paper are both smooth. In \cite{K-U:gluing}, a gluing argument is used and the spacetime obtained exhibits only $C^2$ regularity.

\subsubsection{Pretorius--Israel Type Coordinate System in Perturbed Kerr spacetimes}\label{Introsubsubsec:PI coordinate}
In \cite{KS:main} Klainerman--Szeftel constructed the interior region $\intM$ with incoming principal geodesic (PG) structures. They use the coordinate system $(\ub, r, \th, \varphi)$, which approximates the Eddington--Finkelstein (EF) coordinates in Kerr spacetime. A notable feature of their PG structures is that $\ub$ is not necessarily an optical function and the level set of $\ub$ could be a bit timelike. In our elliptic arguments, we solve the MOTS along each incoming null hypersurface. In order to obtain these incoming null hypersurfaces and the corresponding MOTSs in $\intM$ with precise leading order behaviors, inspired by \cite{P-I} and \cite{D-L:Kerrint}, we construct Pretorius--Israel type coordinates in the following steps. Although $|a_{\infty}|\lesssim \epsilon_0$ in our Kerr black hole formation spacetime, we point out that our construction here is also valid for large $a_{\infty}$ case.
\begin{itemize}
    \item We first note that the original Pretorius--Israel coordinates are derived in the exact Kerr spacetime and strongly rely on the specific structures of Kerr metric, which fail to be directly applicable to our obtained perturbed Kerr spacetimes. In this current article, we instead start from solving the eikonal equation 
\begin{equation}\label{Intro:incoming eikonal eqn}
    \bfg^{\mu \nu}\pr_{\mu} \tub \pr_{\nu} \tub=0
\end{equation}
by the method of characteristics. To initialize $\tub$ appropriately, we appeal to the incoming optical function $\tub_0=\tub_0(\ub,r,\th)=\ub+\fs(r,\th)$ for the exact Kerr solution $Kerr(a_{\infty}, m_{\infty})$.\footnote{The precise definitions and the construction of $\tub_0$ and $\fs$ are given in \Cref{Appendix:coordinate in Mint}.} Then we set $\tub=\tub_0$ along $r=r_0$ with $r_0\approx |u_0|$ being a large positive constant.

\item For $x=(\ub, r, \th, \varphi)\in \intM$ and the momentum vector $p=(p_{\ub}, p_{r}, p_{\th}, p_{\varphi})$, we proceed to define 
   \begin{equation*}
       F(x, p)\coloneqq\bfg^{\a \b} (\pr_{\a} \tub_0+p_{\a}) (\pr_{\b}\tub_0+p_{\b}).
   \end{equation*}
   Then for $\hb=\tub-\tub_0$, we can transfer \eqref{Intro:incoming eikonal eqn} into the form 
   \begin{equation*}
       F(x, D\hb)=0 \qquad \text{with} \quad D\hb\coloneqq(\pr_{\ub} \hb, \pr_{r} \hb, \pr_{\th} \hb, \pr_{\varphi} \hb).
   \end{equation*}
  From the classical theory of first-order nonlinear PDEs, we further convert $F(x, D\hb)=0$ to the below Hamiltonian system\footnote{Interested readers are referred to Chapter 3 of \cite{Evans} for more details.}
   \begin{equation}\label{Intro:Eqn: Character for hb}
      \left\{ \begin{aligned}
           &\f{d}{ds} p_{\a}=D_{x^{\a}}F(x, p), \\
           &\f{d}{ds} x^{\a}=-D_{p_{\a}} F(x, p)=-2g^{\a \b}(\pr_{\b} \tub_0+p_{\b}), \\
       \end{aligned} \right.
   \end{equation}
with initial data $x(0)=(\ub, r_0, \th, \varphi), \ p(0)=\mu_+ Dr(\ub, r_0, \th, \varphi)$. Here $\mu_+$ is the larger root of the quadratic equation\footnote{Here $\bfD$ is the spacetime gradient operator and $\c$ stands for the spacetime inner product.}
\begin{equation*}
    \l (\bfD r\cdot \bfD r)\mu^2+ 2(\bfD \tub_0 \cdot \bfD r)\mu+\bfD \tub_0\cdot \bfD \tub_0 \r\Big|_{r=r_0}=0.
\end{equation*}
Based on the above system, we have that the solution to \eqref{Intro:incoming eikonal eqn} along the characteristic curve $x=x(t)$ satisfies
\begin{align*}
    \hb(x(s))=\hb(x(0))-\int_{0}^{s} D_{p_{\a}} F(x(s'), p(s')) p_{\a} (s') ds', \quad D\hb(x(s))=p(x(s)).
\end{align*}

\item Owing to the smallness of $\mu_+$ and $p(0)$, we solve \eqref{Intro:Eqn: Character for hb} in $\intM=\qty{r_{\HH}\le r\le r_0}$. Together with hyperbolic estimates derived in \cite{KS:main}, we obtain that
\begin{equation*}
    \tub=\tub_0+O(\fepub),
\end{equation*}
 where $\epsilon_0$ represents the size of perturbation and $0<\de_{dec}\ll 1$. 
\end{itemize}
\vspace{3mm}

Next we move to construct an incoming geodesic foliation in $\intM$, that is adapted to our incoming optical function $\tub$. Introduce $\t{r}(r,\th)\coloneqq e^{\kappa \rs(r,\th)}+r_{+,\infty}$ with $r_{\pm,\infty}\coloneqq m_{\infty}\pm\sqrt{m_{\infty}^2-a^2_{\infty}}$ and $\kappa\coloneqq\f{r_{+,\infty}-r_{-,\infty}}{2m_{\infty}r_{+,\infty}}$. Define $\tHb_{\tub}$ to be the constant $\tub$ hypersurface and set $\tS_{\tub, \t{r}}$ to be the level set of $\t{r}$ along  $\tHb_{\tub}$.
\begin{itemize}
    \item Setting $\te_3=-\bfD \tub$,  we introduce a new coordinate system $(\tub, \t{r}, \tth^1, \tth^2)$ with the adapted null frame $\{\te_1, \te_2, \te_3, \te_4 \}$, such that
\begin{equation*}
    \te_1, \te_2\in T\tS_{\tub, \t{r}} \qquad \text{and} \qquad \te_3(\tth^a)=0 \quad \text{with} \quad a=1, 2.
\end{equation*}
This allows us to obtain the boundedness of geometric quantities with respect to the new null frame $\{\te_1, \te_2, \te_3, \te_4 \}$ from hyperbolic estimates demonstrated in \cite{KS:main}, with the aid of transformation formulas in \cite{KS:Kerr1}.
\item For the outgoing null expansion $\tr \t{\chi}=\bfg(\bfD_{\te^a} \te_4, \te_a)$, a more precise asymptotic behavior is established in this paper. First we have
\begin{equation}\label{Intro:est for tr tch1}
    \tr{\t{\chi}}=
    \tr{\t{\chi}}_0 +O(\tfepub).
\end{equation}
Here $\tr{\t{\chi}}_0$ represent the corresponding outgoing null expansions in the exact Kerr spacetime $Kerr(\ainft, \minft)$.

\item In \Cref{Appendix:Null expansion}, we further analyze $\tr{\t{\chi}}_0$ and establish that\footnote{We use the notation $A\sim B$ if there exists some constant $C\ge 1$ such that $C^{-1}\le A/B\le C$.}
\begin{equation}\label{Intro:est for tr tch0}
    \tr{\t{\chi}}_0= F\c(\t{r}-r_{+, \infty}) \quad \text{with} \quad F\sim1.
\end{equation}
Hence, this yields
\begin{equation}\label{Intro:est for tr tch}
    \tr{\t{\chi}}= 
    F\c(\t{r}-r_{+, \infty}) +O(\tfepub).
\end{equation}
 The leading order structure $F\c(\t{r}-r_{+, \infty})$ and $F$ being positive play vital roles in the proof for the existence of MOTS and the Penrose inequality, with more details explained in \Cref{Intro:sec:MOTS} and \Cref{Introsec:Penrose Inequality in Perturbative Kerr Regime}.
    \item The derivation of \eqref{Intro:est for tr tch0} is rather crucial and delicate. One may easily fail to get this desired form by just straightforward calculation, as the expression of $\tr{\t{\chi}}_0$ contains more than $10$ fractions that involve functions implicitly depending on $r, \th$ and  quadratic radical functions as well as elliptic integrals. To achieve \eqref{Intro:est for tr tch0}, we instead conduct a carefully designed indirect argument:
\begin{itemize}
   \item First, we analyze the asymptotic behavior of $\tr{\t{\chi}}_0$ in two different regimes $\t{r}\gg 1$ and $\t{r}$ near $(r_{+,\infty})^-$. This enables us to verify that $\tr{\t{\chi}}_0$ has the desired sign in both of these scenarios;
    \item Next, using the monotonicity of $\tr{\t{\chi}}_0$ along the outgoing null direction, which is from the Raychaudhuri equation, we show that $\tr{\t{\chi}}_0$ is positive for $\t{r}>r_{+,\infty}$ and is negative for $\minft\le \t{r}<r_{+,\infty}$;
    \item We then note that $\f{\tr{\t{\chi}}_0}{\t{r}-r_{+,\infty}}=F(\t{r},\th)$ only depends on $\t{r}, \th$ and it suffices for us to prove that
    \begin{equation*}
        F(r_{+,\infty},\th)=\lim\limits_{\t{r}\to r_{+,\infty}}\f{\tr{\t{\chi}}_0}{\t{r}-r_{+,\infty}}>0.
    \end{equation*}
    This significantly simplifies the computations, as most of terms in $F$ vanish in the limit as $\t{r}$ approaches $r_{+,\infty}$.
\end{itemize}
\end{itemize}
\vspace{2mm}

Additionally, to obtain a smooth apparent horizon originating from the short-pulse region and tending to the timelike infinity, we further construct a global optical function that smoothly transits from the level sets of $v$ in the short-pulse region to the level sets of $\tub$ in $\intM$. The associated incoming geodesic foliation is constructed in the transition region as well.
\begin{itemize}
   \item In the transition region, we have to relate the double-null coordinates $(u, v, \th^1, \th^2)$ used in the short pulse region with the later designed incoming geodesic coordinates $(\tub, \t{r}, \tth^1, \tth^2)$. This is achieved by evaluating the transition coefficients\footnote{The precise definition of transition coefficients are provided in \Cref{Lem:frame transform} in \Cref{Subsec:frametrans}, which allows us to transfer the geometric information in different coordinates and null frames.} $(f''',\fb''', \la''')$ between the associated previous short-pulse null frame $(\pul e_{\mu})$ and the later incoming geodesic null frame $(\te_{\mu})$ in the corresponding regions. More precisely, we have the following chains of relations 
    \begin{equation*}
        (\pul e_{\mu})\xrightarrow{( \Int f', \Int\fb', \Int\la')}\big((\Int e_0)_{\mu}\big)\xrightarrow{(f'', \fb'', \la'')}(\Int e_{\mu})\xrightarrow{(f,\fb,\la)} (\te_{\mu}).
    \end{equation*}
    Here $\big((\Int e_0)_{\mu}\big)$ and $(\Int e_{\mu})$ are adapted to the incoming PG coordinates  $(\ub_{\LL_0}, \Int r_{\LL_0}, \\ \Int\th_{\LL_0}, \Int\th_{\LL_0})$ in the incoming initial data layer and the incoming PG coordinates $(\ub, r, \th, \varphi)$ in $\intM$, respectively.
    \item Thanks to the smallness of the transition coefficients involved in the intermediate changes of frames, we prove that $(f''',\fb''', \la''')$ are small as well, which implies that $\pr_{\tub}v\sim 1, \pr_{\tth^a}v=O(\epsilon_0)$. This further enables us to use the implicit function theorem to express $\tub$ as a function of $v, \t{r}, \tth^1, \tth^2$ and to construct the desired transition null foliation.
\end{itemize}
\subsubsection{Existence and Uniqueness of MOTS Along Each Incoming Null Hypersurface}\label{Intro:sec:MOTS}
In this section, we aim to solve the unique MOTS along incoming null cones obtained from the preceding section. It is worth mentioning that even though $|a_{\infty}/m_{\infty}|\lesssim \epsilon_0$ in our setting, our elliptic arguments do not rely on the smallness of $a_{\infty}$ and can be applied to the full subextremal regime with $|a_{\infty}/m_{\infty}|<1$. We begin with stating the equation of MOTS and transferring it into a favorable form of quasilinear elliptic equation in local coordinates.
\begin{itemize}
    \item Following the deformation procedure as described in \cite{An-He}, along $\tHb_{\tub}$ setting the MOTS $M_{\tub}$ to be with coordinates $\{ \t{r}=R(\tth^1, \tth^2)\}\eqqcolon \tS_{\tub, R}$, the equation of MOTS can be obtained from $\trch'\big|_{M_{\tub}}\equiv 0$ as\footnote{For notational simplicity, we sometimes drop the dependence on $\tub$ if there is no danger of confusion.}
\begin{equation}\label{Intro:MOTS main equation}
   \begin{aligned}
    0=L(R, \tub)\coloneqq(\f{1}{2} f^{-1} \trch')|_{\tS_{\tub, R}}=
&\tDe_{\tS_{\tub, R}} R+\big(f^{-1}\tnab f+(\t{\eta}+\t{\zeta})\big)\cdot \tnab R-2f\t{\chibh}_{ac}\tnab^a R \tnab^c R\\&+(\te_3(f)-\f{1}{2}f\tr\t{\chib}-2\t{\omegab} f)|\tnab R|^2+\f{1}{2}f^{-1}\tr\t{\chi}.
\end{aligned}
\end{equation}
Here $f=[\te_3(\t{r})]^{-1}\sim -1$ and $\tDe$, $\tnab$ correspond to the Laplace--Beltrami operator and the induced covariant derivative on $\tS_{\ub, \t{r}}$, respectively. We also note the Ricci coefficients in \eqref{Intro:MOTS main equation} are defined with regard to the null frame $\{\te_1, \te_2, \te_3, \te_4 \}$ by
\begin{equation}\label{Intro:Riccidefine}
    \begin{aligned}
  \t{\chib}_{ab}=&\bfg(\bfD_{\te_a} \te_3, \te_b),  &\quad \tr\t{\chib}=&\de^{ab}\t{\chib}_{ab}, &\quad   \t{\chibh}_{ab}=&\t{\chib}_{ab}-\f12\de_{ab}\tr\t{\chib}, \\
  \t{\eta}_a=&-\frac 12 \bfg(\bfD_{\te_3} \te_a, \te_4), &\quad \t{\zeta}_a=&\frac 1 2 \bfg(\bfD_{\te_a} \te_4, \te_3), &\quad \t{\omegab}=&-\frac 14 \bfg(\bfD_{\te_3} \te_4, \te_3).
\end{aligned}
\end{equation}

We remark that in \cite{An:AH,An-He} we employ the modified Bochner formula to obtain a priori gradient estimates for equation similar to \eqref{Intro:MOTS main equation}. However, this approach is not applicable in this paper due to the complexity of Kerr geometry and the largeness of $a_{\infty}$.

 For instance, in this paper the geometric quantities in front of $\tnab R$ obey the upper bound
    \begin{equation}\label{Intro:Ricci bound}
        \Big|\Big(\tnab f, \t{\eta}, \t{\zeta}, \t{\chibh}, \te_3(f), \tr\t{\chib}+\f{2}{r} \Big) \Big|\lesssim \f{|a_{\infty}|}{r^2}+\tfepub,
    \end{equation}
and this bound lacks decay in $\tub$ due to the presence of the non-zero $a_{\infty}$; on the contrast, compared to the setting in \cite{An-He} there we have $a_{\infty}=0$. Meanwhile, in our main theorem, we allow $a_{\infty}$ to be not necessarily small and we need quantitative decay estimates for $\nab R$ to prove the desired Penrose inequality.

In this paper, to overcome these difficulties, we carry out a new method with quasi-conformal mappings and apply the Leray--Schauder fixed point theorem.

\item The estimates of geometric quantities adapted to $\{\te_1, \te_2, \te_3, \te_4 \}$ allow us to rewrite \eqref{Intro:MOTS main equation} in the following form
\begin{equation}\label{Intro:MOTS main equation new}
       L(R, \tub)=a^{ij}(\omega, R,  DR)D_{ij} R+b(\omega, R, DR)+c(\omega, R)(R-r_{+, \infty})=0,
    \end{equation}
    where $\omega$ is a local chart\footnote{As the Ricci coefficient $\omega=-\f14\bfg(\bfD_{4} e_3, e_4)$ does not appear in the equation of MOTS, we use $\omega$ to represent a local chart on $\mathbb{S}^2$ if there is no danger of confusion in this section.} on $\mathbb{S}^2$ and $a^{ij}(\omega, \t{r},  p), b(\omega, \t{r},  p), c(\omega, \t{r})$ are smooth functions satisfying
    \begin{enumerate}
        \item $a^{ij}(\omega, \t{r},  p)=a^{ji}(\omega, \t{r},  p)$;
        \item there exist constants $\mu\ge \nu>0$ and $C_0>c_0>0$ independent of $\epsilon_0, \tub, \t{r}, p$, such that
        \begin{align}
            \nu |\xi|^2 \le& a^{ij}(\omega, \t{r},  p) \xi_i \xi_j \le \mu |\xi|^2 \qquad \text{for any} \quad \xi\in \mathbb{R}^2, \label{Intro:property for a} \\ 
            |b(\omega, \t{r},  p)|\lesssim&  |p|^{2}+|p|+\tfepub, \qquad -C_0\le c(\omega, \t{r})\le -c_0. \label{Intro:property for b}
        \end{align}
    \end{enumerate}
Note that the estimate \eqref{Intro:property for b} is from our new observation about $\tr\t{\chi}$ in \eqref{Intro:est for tr tch}. There the principal part $F(\t{r}-r_{+, \infty})$ of \eqref{Intro:est for tr tch} is related to the zeroth-order term in \eqref{Intro:MOTS main equation new}, while the remainder part $O(\tfepub)$ of \eqref{Intro:est for tr tch} reflects in the corresponding term in $b(\omega, \t{r},  p)$. 

 We remark that the significance of \eqref{Intro:est for tr tch} and \eqref{Intro:property for b} can be viewed in below two perspectives: 
 \begin{itemize}
     \item On the one hand, the negativity of zeroth-order coefficient $c(\omega, \t{r})$ (which comes from the fact that $F\sim 1$ and $f\sim -1$) implies a favorable a priori $C^{0}$ estimate for the equation of MOTS and the desired invertibility of the linearized operator $\pr_R L$. 
     \item On the other hand, the remainder $O(\tfepub)$ manifests the asymptotic properties of MOTSs and the corresponding apparent horizon as $\tub\to \infty$. 
 \end{itemize}
\end{itemize}
\vspace{3mm}

In this article, we solve the equation of MOTS \eqref{Intro:MOTS main equation new} via a Leray--Schauder fixed point argument instead of the continuity method adopted in \cite{An:AH,An-Han,Annakedsingularity,An-He}. Our main ideas are as follows:
\begin{itemize}
    \item The Leray--Schauder fixed point theorem asserts that
    \begin{theorem}\label{intro:L-S fixed point thm}
    Let $M>0$ and $\mathcal{B}$ be a Banach space. Suppose $T$ is a compact mapping of a closed ball $\overline{B}_M\coloneqq\{x\in \mathcal{B}: \ \|x\|_{\mathcal{B}} \le M  \}$ into $\mathcal{B}$. If for all  $x\in \overline{B}_M$ and $\sigma\in [0, 1]$ satisfying $x=\sigma Tx$, it holds
    \begin{equation}\label{Intro:condition for L-S fixed point thm}
        \|x\|_{\mathcal{B}}< M.
    \end{equation}
  Then $T$ has a fixed point.
\end{theorem}
 To apply \Cref{intro:L-S fixed point thm}, for any given $S\in C^{1, \a}$ with $\a\in(0, 1)$, we consider the below linear elliptic equation
\begin{equation*}
         a^{ij}(\omega, S,  DS)D_{ij} R+b(\omega, S, DR)+c(\omega, S)(R-r_+)=0.
    \end{equation*}
In light of the classical theory of linear elliptic equations and the estimates in \eqref{Intro:property for a}, \eqref{Intro:property for b}, the equation above admits a unique solution $R\in C^{2, \a}$. This allows us to we define the compact map $T: C^{1, \a}\to C^{1,\a}$ by letting $R=T(S)$.

\item The condition \eqref{Intro:condition for L-S fixed point thm} is equivalent to the a priori $C^{1,\a}$ estimate for the equation
\begin{equation*}
         a^{ij}(\omega, R,  DR)D_{ij} R+\si b(\omega, R, DR)+c(\omega, R)(R-r_+)=0 \qquad \text{with} \quad \si\in[0, 1].
    \end{equation*}
Note that in reality, it is sufficient to derive the a priori $C^{1,\a}$ estimate for the case $\si=1$ of the above equation, which is exactly the equation of MOTS \eqref{Intro:MOTS main equation new}.
\end{itemize}

We then derive a priori estimates for the equation of MOTS following the scheme $C^{0}\to C^{1, \a}\to C^{2, \a}$. Note that the $C^0$ estimate is inferred directly from the negativity of $c(\omega, \t{r})$ in \eqref{Intro:property for b} and the maximum principle on $\mathbb{S}^2$. The crucial step is to derive the $C^{1, \alpha}$ estimate. In this article, we employ the below new elliptic arguments, invoking the quasi-conformal mapping in two dimensions.  
\begin{itemize}
\item Let
\begin{equation*}
    \a\coloneqq1+\f{\mu}{\nu}-\sqrt{(\f{\mu}{\nu})^2+\f{2\mu}{\nu}}\in (0, 1).
\end{equation*}
Given any $S\in C^{1,\a}$, we start from deriving a priori $C^{1, \alpha}$ estimates, which are independent of $S, DS$ and uniform in $\sigma\in[0, 1]$, for the equation\footnote{Finally we will choose $S=R$.}
\begin{equation*}
    \mathscr{L}(S)R+b(\omega, R, DR)-(1-\sigma)b(\omega, r_{+, \infty}, 0)=0.
\end{equation*}
Here
\begin{equation*}
     \mathscr{L}(S)R\coloneqq a^{ij}(\omega, S,  DS) D_{ij} R+c(\omega, S)(R-r_{+, \infty}).
\end{equation*}
From \eqref{Intro:property for b} we see that the zeroth order terms in $b(\o,\t{r},p)$ are small. It is worthwhile to mention that our elliptic arguments are quite robust and do not rely on the smallness of these terms. For future use, in the following, we will replace $\tfepub$ in \eqref{Intro:property for b} with $N$, where $N$ is allowed to be large in our consideration.

\item Assume first that $\pr_{\t{r}} b\le 0$ and denote $R_{\si}$ is the solution to
 \begin{equation}\label{Intro:eqn LSR}
     \mathscr{L}(S)R+b(\omega, R, DR)-(1-\sigma)b(\omega, r_{+, \infty}, 0)=0 \qquad \text{with} \quad \si\in[0, 1]. 
\end{equation}
Our strategy is to construct a finite sequence $0=\si_0<\si_1<\cdots<\si_n=1$, so that for each pair $(\si_i, \si_{i+1})$, we have the a priori bound for $\|R_{\si_i}-R_{\si_{i-1}}\|_{C^{1, \alpha}}$.
\begin{itemize}
    \item
  We first pick $\si_1, \si_2\in [0, 1]$ such that $|\si_1-\si_2|\le \de$ with $\de>0$ being determined later. Note that the assumption $\pr_{\t{r}} b\le 0$ validates the maximum principle for the equation of $R_{\si_1}-R_{\si_2}$, from which we infer that
\begin{equation*}
    |R_{\si_1}-R_{\si_2}|\lesssim |\si_1-\si_2|N. 
\end{equation*}
This together with \eqref{Intro:property for b} and \eqref{Intro:eqn LSR} gives
\begin{equation*}
     \begin{split}
          &\|\mathscr{L}(S)(R_{\sigma_1}-R_{\sigma_2})\|_{L^2}\\
          \lesssim& \|DR_{\sigma_1}\|_{L^4}^2+\|DR_{\sigma_1}\|_{L^2}+\|D(R_{\sigma_1}-R_{\sigma_2})\|_{L^4}^2+\|D(R_{\sigma_1}-R_{\sigma_2})\|_{L^2}+N  \\
\lesssim& F(\|DR_{\sigma_1}\|_{C^0})+\|R_{\sigma_1}-R_{\sigma_2}\|_{L^\infty} \|R_{\sigma_1}-R_{\sigma_2}\|_{H^2}+\|R_{\sigma_1}-R_{\sigma_2}\|_{L^\infty}^{\f12} \|R_{\sigma_1}-R_{\sigma_2}\|_{H^2}^{\f12}+N \\
\lesssim& F(\|DR_{\sigma_1}\|_{C^0})+|\sigma_1-\sigma_2|N \cdot\|R_{\sigma_1}-R_{\sigma_2}\|_{H^2}+\l |\sigma_1-\sigma_2|N \cdot\|R_{\sigma_1}-R_{\sigma_2}\|_{H^2}\r^{\f12}+N.
     \end{split}
 \end{equation*}
Here $F(x)=x^2+x$ and we use the interpolation inequality  $\|DR\|^2_{L^4}\lesssim \|R\|_{L^\infty} \|R\|_{H^2}$. Consequently, we deduce
\begin{equation*}
    \|\mathscr{L}(S)(R_{\sigma_1}-R_{\sigma_2})\|_{L^2}\lesssim F(\|DR_{\sigma_1}\|_{C^0})+N
\end{equation*}
provided that $|\si_1-\si_2|N\le \de N\ll 1 $ and we also obtain the below regularity estimate
\begin{equation}\label{Intro:regularity est}
        \|R_{\sigma_1}-R_{\sigma_2}\|_{H^2}\lesssim \|\mathscr{L}(S)(R_{\sigma_1}-R_{\sigma_2})\|_{L^2}+\|R_{\sigma_1}-R_{\sigma_2}\|_{L^2}.
\end{equation}
We emphasize that to get \eqref{Intro:regularity est} we cannot apply the classical regularity estimate for linear elliptic equations, as it depends on the modulus of continuity of $a^{ij}(\omega, S, DS)$ with respect to $\omega\in \mathbb{S}^2$, and thus involves $S$ and $DS$. To avoid this obstruction, we appeal to the Campanato condition and the sharp Miranda--Talenti inequality on $\mathbb{S}^2$, which only relies on the boundedness of the coefficients and the uniform ellipticity of the elliptic operator $\mathscr{L}(S)$. We also note that the Campanato condition is equivalent to the uniform ellipticity in two dimensions, but is more restrictive in higher dimensions.

In view of Sobolev embedding $L^{p}\hookrightarrow H^1$ for $1\le p<\infty$, we proceed and derive
\begin{equation}\label{intro:est LS R1-R2 Lp}
    \begin{aligned}
         &\|\mathscr{L}(S)(R_{\sigma_1}-R_{\sigma_2})\|_{L^p}\\\lesssim& \|DR_{\sigma_1}\|_{L^{2p}}^2+\|DR_{\sigma_1}\|_{L^{p}}+\|D(R_{\sigma_1}-R_{\sigma_2})\|_{L^{2p}}^2+\|D(R_{\sigma_1}-R_{\sigma_2})\|_{L^{p}}+N  \\
\lesssim&_p F(\|DR_{\sigma_1}\|_{C^0})+ \|R_{\sigma_1}-R_{\sigma_2}\|_{H^2}^2+\|R_{\sigma_1}-R_{\sigma_2}\|_{H^2}+N \\
\lesssim&  F(\|DR_{\sigma_1}\|_{C^0})+N.
    \end{aligned}
\end{equation}

  \item For a similar reason,  the classical $W^{2, p}$ estimate together with Sobolev embedding is inefficient to obtain the $C^{1, \alpha}$ estimate of $R_{\sigma_1}-R_{\sigma_2}$. Instead, we carry out the quasi-conformal mapping method and it allows us to derive the desired interior $C^{1, \alpha}$ estimates for the below linear elliptic equation in two variables 
\begin{equation}\label{Intro:linear eqn}
      au_{xx}+2bu_{xy}+cu_{yy}=f \qquad \text{in} \quad \Omega.
\end{equation}
These $C^{1, \alpha}$ estimates only depend on the boundedness and the uniform ellipticity of $a, b, c$. Specifically, for $f\in L^{2\b}(\O)$ with $\beta>\f{1}{1-\a}$, we derive
\begin{equation}\label{Intro:linear est}
         \|u\|_{C^{1, \a}(\O')}\le C(\gamma, \b, \Omega, \O') (\|u\|_{C^0(\O)}+\|f/\lambda\|_{L^{2\b}(\O)}),
    \end{equation}
where $\O'\subset \overline{\O'} \subset \O$, $\la, \La$ are the smaller and larger eigenvalues for the coefficient matrix of \eqref{Intro:linear eqn} and $\gamma=\sup\limits_{\O} (\La/\la)$. 

We remark that the classic quasi-conformal method presented in \cite{G-T} requires $\b=\infty$. In this current paper, for our later use we generalize this classic method and prove \eqref{Intro:linear est} while allowing $\beta$ to be a large but finite number.

\item Back to \eqref{intro:est LS R1-R2 Lp}, by virtue of \eqref{Intro:linear est}, we hence get
\begin{equation}\label{Intro:apriori}
    \|R_{\sigma_1}-R_{\sigma_2}\|_{C^{1, \alpha}}\lesssim \|DR_{\sigma_1}\|^2_{C^0}+\|DR_{\sigma_1}\|_{C^0}+N \qquad  \text{if} \quad  |\sigma_1-\sigma_2|\le \delta.
\end{equation}
Observe that $R_0\equiv r_{+, \infty}$ solves \eqref{Intro:eqn LSR} with $\si=0$. Employing \eqref{Intro:apriori} with the selections of $(\si_1, \si_2)=(n\de, (n-1)\de)$ for $n=1,2,\dots, \lfloor\f{1}{\delta}\rfloor$ and $(\si_1, \si_2)=(\lfloor\f{1}{\delta}\rfloor\de, 1)$, we arrive at the $C^{1, \alpha}$ for $R=R_1$:
\begin{equation*}
         \|R-r_{+, \infty}\|_{C^{1, \alpha}}\lesssim \max(N^{2^{CN}}, N). \\[2mm]
    \end{equation*}
\end{itemize}

    \item Next we turn to remove the assumption $\pr_{\t{r}} b\le 0$. Note that if $R$ is a $C^2$ solution to \eqref{Intro:MOTS main equation new}, then it can be verified easily that $\widetilde{R}=R$ also solves the below quasilinear elliptic equation
\begin{equation}\label{Intro:tR equation}
        \mathscr{L}(R)\widetilde{R}+\widetilde{b}(\omega, \widetilde{R}, D\widetilde{R})=0,
    \end{equation}
where
\begin{align*}
    \widetilde{b}(\omega, \t{r}, p)\coloneqq&\f{b(\omega, R, DR)}{|DR|^2+|DR|+N}\cdot (|p|^2+|p|+N).
\end{align*}
Since $\pr_{\t{r}} \widetilde{b}=0$, applying the previous result for the new elliptic equation \eqref{Intro:tR equation} and choosing $S=R$, our desired $C^{1, \alpha}$ estimate thus follows. By virtue of standard Schauder estimates with the selection of $N=\tfepub$, we further obtain
\begin{equation}\label{Intro:C2,a}
    \|R-r_{+, \infty}\|_{C^{2, \alpha}}\lesssim \tfepub.
\end{equation}

Incorporating with the Leray--Schauder fixed point theorem mentioned above, we obtain the existence of solutions to the equation of MOTS \eqref{Intro:MOTS main equation new}.
\end{itemize}
\vspace{2mm}

To prove the uniqueness of MOTS along $\tHb_{\tub}$, we examine the linearized operator of the equation $L(R, \tub)=0$ for the location of the MOTS. The linearized operator can be expressed as
\begin{equation*}
     \pr_R L(R, \tub)[W]=a^{ij} D_{ij}W+\l\pr_{p^i} a^{jk} D_{jk}R+\pr_{p^i} b\r \cdot D_i W+\l \pr_{\t{r}} a^{ij} D_{ij}R+\pr_r c(R-r_{+, \infty})+c  \r W.
\end{equation*}
\begin{itemize}
\item  In view of a priori estimates \eqref{Intro:C2,a} and \eqref{Intro:property for b}, for $R=R(\tub)$ solves $L(R, \tub)=0$, we have that the zeroth-order coefficient of $\pr_R L(R(\tub), \tub)$ obeys
\begin{equation*}
    \pr_{\t{r}} a^{ij} D_{ij}R+\pr_r c(R-r_{+, \infty})+c\le -c_0+O(\tfepub)\le -\f{c_0}{2}.
\end{equation*}
This indicates that $\pr_R L$ is invertible. Supposing that $R_1, R_2$ are two solutions to $L(R)=0$, then it yields
\begin{equation*}
    0=L(R_1)-L(R_2)=\int_{0}^1 \pr_R L(tR_1+(1-t)R_2) dt[R_1-R_2].
\end{equation*}
Note that the elliptic operator $\int_{0}^1 \pr_R L(tR_1+(1-t)R_2) dt$ shares similar properties with $\pr_R L(R, \tub)$. Hence, after applying the maximum principle on $\mathbb{S}^2$, we conclude that $R_1\equiv R_2$. This gives the uniqueness.
\begin{itemize}
    \item The invertibility of $\pr_R L$ at $R=R(\tub)$ comes from both the detailed structure of $\tr\t{\chi}_0$ established in \Cref{Appendix:Null expansion} and our desired $C^{1,\a}$ estimates via the quasi-conformal method.
\end{itemize}
\end{itemize}
\vspace{2mm}

\begin{remark}
    For the MOTS lying within the (incoming) null cones, the corresponding second order coefficient $a^{ij}(\omega, R, DR)$ as in \eqref{Intro:MOTS main equation new} for the equation of MOTS is actually independent of $DR$. Our new method allows the presence of $DR$. It is worthwhile to mention that in \cite{An-Han, Annakedsingularity} An-Han and An utilized that $a^{ij}=a^{ij}(\o, R)$ crucially. They went through the details of Moser's iteration and obtained the desired a priori estimates $C^0\to C^{0, \a}\to C^{1,\a}$. An important feature of our method in this paper is that our proof does not rely on this structure. Instead, we find that requiring the uniform ellipticity of $a^{ij}(\omega, R, DR)$ is sufficient to carry out our entire argument. Moreover, in perturbative Kerr regime we also derive more precise upper bounds for the a priori estimates.

    Our newly developed elliptic method in this paper  can be used to solve the MOTS along slightly timelike or spacelike hypersurfaces as well. For these cases, we need to note that the dependence of $a^{ij}(\omega, R, DR)$ on $DR$ may cause $a^{ij}(\omega, R, DR)$ to lose uniform ellipticity if $|DR|$ becomes too large. Hence, there we need to restrict our analysis to conditional solutions whose gradients satisfy appropriate bounds, which ensures that $a^{ij}(\omega, R, DR)$ remains uniformly elliptic.
\end{remark}
\begin{remark}
    In \cite{An:AH,An-He}, the modified Bochner formula was employed to deduce the precise a priori estimates for the equation of MOTS when the spacetime is very close and asymptotes to a Schwarzschild solution. However, that approach is not applicable in our current setting due to the complexity of Kerr geometry and the non-smallness of $a_{\infty}$. Specifically, the MOTS equation in our setting can be expressed as
 \begin{equation*}
     \De' R-\f{1}{R}|\nab' R|^2+O(a_{\infty})(|\nab' R|^2+|\nab' R|)+c(R-r_{+,\infty})+O(\tfepub)(|\nab' R|^2+|\nab' R|+1)=0.
 \end{equation*}
 Unlike in \cite{An:AH,An-He}, where $a_{\infty}$ is negligible, the coefficients in front of the term $|\nab' R|^2+ |\nab' R|$ here are of size $O(a_{\infty})$, not of size $O(\tfepub)$. And in this paper we allow $a_{\infty}$ not being small. This significantly complicates our arguments as compared to \cite{An:AH,An-He}.
\end{remark}

\subsubsection{Dynamics of Apparent Horizon}\label{Intro:sec:MOTSdynamic}
We proceed to discuss the properties of the apparent horizon $\AH\coloneqq\cup_{\tub>0} M_{\tub}$, which is composed of the MOTSs $M_{\tub}=\{\t{r}=R(\tub, \tth^1, \tth^2) \}\cap \tHb_{\tub}$. 
\begin{itemize}
    \item In the short-pulse region, by \cite{An:AH} we already have the smoothness of the apparent horizon $\AH\cap\qty{0<\tub<\tub_1}$ for the initial stage, with $\tub_1$ being the endpoint of short-pulse region. The smoothness of late stage of $\AH$ with $\tub\ge \tub_1$ follows by the same method as in \cite{An-He}, with the aid of the implicit function theorem and the invertibility of the linearized operator for the MOTS equation.
\item To investigate the asymptotics of the apparent horizon, we consider the tangential vector $X$ on $\AH=\{\t{r}=R(\tub, \tth^1, \tth^2) \}$, which is normal to the MOTSs $M_{\tub}$, as
\begin{equation*}
    X= e'_4-\f{e'_4(\t{r}-R)}{e'_3(\t{r}-R)}e'_3.
\end{equation*}
Here $\{e'_1, e'_2, e'_3, e'_4 \}$ is the null frame associated with the MOTS $M_{\tub}$. 

Applying the estimate for $R$ in \eqref{Intro:C2,a} and a similar estimate\footnote{This can be derived via differentiating $L(R(\tub), \tub)=0$ in $\tub$.} for $\pr_{\tub} R$, we then derive
\begin{align*}
    X(\tub)=2, \qquad e'_3(\t{r}-R)\sim -1, \qquad e'_4(\t{r}-R)O(\tfepub).
\end{align*}
This reveals that $\tub$ is an affine parameter of $X$ along $\AH$ and $X$ is an asymptotically null vector in the sense that
\begin{equation*}
    \bfg(X, X)=\f{4e'_4(\t{r}-R)}{e'_3(\t{r}-R)}=O(\tfepub)\to 0 \qquad \text{as} \quad \tub\to \infty.
\end{equation*}
Therefore, we conclude that the apparent horizon $\AH$ constructed in this paper is asymptotically null, i.e., the induced metric on $\AH$ is asymptotically degenerate.
\vspace{2mm}

Additionally, the derived estimate for $R$ in \eqref{Intro:C2,a} also allows us to deduce that the apparent horizon converges to the event horizon $\HH^+$ as $\tub\to\infty$. Denote $\Si_{\tub}$ as the intersection of the event horizon $\HH^+$ and the incoming null hypersurface $\tHb_{\tub}$. Set $\Si_{\tub}$ to be associated with the coordinates $\{\t{r}=R_{\HH^+}(\tub, \tth^1, \tth^2)  \}$ along $\tHb_{\tub}$. In view of the location of $\HH_+$ estimated in \cite{KS:main}, it follows that for all $\tub\ge\tub_1, \ (\tth^1, \tth^2)\in \ms$,
\begin{equation*}
|R(\tub, \tth^1, \tth^2)-R_{\HH^+}(\tub, \tth^1, \tth^2)|\lesssim
     \f{\sqrt{\epsilon_0}}{\tub^{1+\de_{dec}}}.
\end{equation*}
This reveals that the apparent horizon $\AH=\{\t{r}=R(\tub, \tth^1, \tth^2) \}$ approaches $\HH^+$ at the timelike infinity as $\tub\to \infty$.
\vspace{2mm}

\item Although we just showed that along $\AH$ the tangent vector $X$ normal to the MOTS $M_{\tub}$ is asymptotically null, it remains unclear whether $X$ (and thus $\AH$) could be timelike or achronal (spacelike or null). The achronality of the apparent horizon is always of significance as verifying it would immediately imply that the area of MOTS does not decrease along the apparent horizon toward the future. In \cite{An-He}, a new criterion named the null comparison principle was proposed by us to guarantee that the apparent horizon must be locally achronal (either piecewise spacelike or piecewise null). 

By analyzing the associated linearized operator for the equation of MOTS, together with obtained a priori estimates of $R$, we are able to validate the null comparison principle in the setting of this paper. It thus renders that our apparent horizon $\AH$ must be piecewise spacelike or piecewise null.
\vspace{2mm}

\item As a direct implication, we have that the area of MOTS $A_M(\tub)\coloneqq\text{Area}(M_{\tub})$ is non-decreasing along $\AH$ as $\tub$ grows, and $ \f{dA_M}{d\tub}=0$ along the null piece of $\AH$, while $\f{dA_M}{d\tub}>0$ along the spacelike piece of $\AH$. This monotonicity of $A_M(\tub)$ later plays a significant role in our proof for the Penrose inequality.
\end{itemize}
\subsubsection{Penrose Inequality in Kerr Black Hole Formation Spacetimes and in Perturbative Kerr Spacetimes}\label{Introsec:Penrose Inequality in Perturbative Kerr Regime}
An important part of this paper is to prove the Penrose inequality in perturbed Kerr spacetimes with no time symmetric conditions. Notably, this conclusion is extensively built on the main results in almost all sections of this paper. In the introduction, we discuss the hyperbolic part in \Cref{Introsec:Extended ID} and \Cref{Introsubsubsec:PI coordinate}, elliptic estimates in \Cref{Intro:sec:MOTS}, and dynamics of apparent horizon in \Cref{Intro:sec:MOTSdynamic}.

In our black hole formation spacetimes, we first establish the dynamical version of Penrose inequality, i.e.,
\begin{equation}\label{Intro:spacetimePenrose}
       M_B(\tu)\ge m_{\infty}\ge \f{\sqrt{r_+^2+a_{\infty}^2}}{2} \ge \sqrt{\f{A_M(\tub)}{16\pi}}.
\end{equation}
Here $\tu$ is the outgoing optical function and $M_B(\tu)$ represents the Bondi mass along the level set of $\tu$.
\begin{itemize}
    \item 
In light of the a priori estimate \eqref{Intro:C2,a}, the MOTS $M_{\tub}=\{\t{r}=R(\tub, \tth^1, \tth^2) \}\cap \tHb_{\tub}$ is sufficiently close to the 2-sphere $\tS_{\tub, r_{+, \infty}}$,\footnote{Recall that $r_{+, \infty}=m_{\infty}+\sqrt{m_{\infty}^2-a_{\infty}^2}$.} the area of which can be computed directly with hyperbolic estimates that are established in \Cref{Sec:Construct null cones} and \cite{KS:main}:
\begin{equation}\label{Intro:areaSur}
        \textrm{Area}\,(\tS_{\tub, r_{+,\infty}})=4\pi(r_{+, \infty}^2+a_{\infty}^2)+O(\sqrt{\fepub}).
    \end{equation}
Note that the induced metric on the MOTS $M_{\tub}$ satisfies
\begin{equation*}
        g'_{\tth^a \tth^b}=\bfg(\pr_{\tth^a}+\pr_{\tth^a} R \, \pr_{\t{r}}, \pr_{\tth^b}+\pr_{\tth^b} R\, \pr_{\t{r}})=\t{g}_{\tth^a \tth^b}
    \end{equation*}
with $\t{g}_{\tth^a \tth^b}$ being the induced metric on $\tS_{\tub, \t{r}}$. Hence, we can compare the areas of $M_{\tub}$ and $\tS_{\tub, r_{+, \infty}}$ as below
   \begin{equation*}
    \begin{aligned}
        \left| A_M(\tub)-\text{Area}\,(\tS_{\tub, r_{+,\infty}})\right| =&\left|\int_{\mathbb{S}^2} \l \sqrt{\det(\t{g})}\Big|_{\tS_{\tub, R}}-\sqrt{\det(\t{g})}\Big|_{\tS_{\tub, r_{+, \infty}}} \r d\tth^1 d\tth^2 \right|\\
        \lesssim& \sup\limits_{r, \tth^1, \tth^2} \left|\pr_{\t{r}} \l\sqrt{\det(\t{g})}\r\right|\sup\limits_{\tth^1, \tth^2} |R-r_{+,\infty}|\lesssim \tfepub.
    \end{aligned}
 \end{equation*}
 This together with \eqref{Intro:areaSur} gives
 \begin{equation}\label{Intro:limit AM}
     \lim\limits_{\tub\to \infty} A_M(\tub)=4\pi(r_{+, \infty}^2+a_{\infty}^2).
 \end{equation}
 A crucial conclusion in our \Cref{Subsec:area law} (see also discussions in \Cref{Intro:sec:MOTSdynamic}) is that $A_{M}(\tub)$ is non-decreasing in $\tub$ variable. Together with \eqref{Intro:limit AM}, we prove the right half inequality in \eqref{Intro:spacetimePenrose}
 \begin{equation}\label{Intro:rightPenrose}
     \f{\sqrt{r_{+,\infty}^2+a_{\infty}^2}}{2} \ge \sqrt{\f{A_M(\tub)}{16\pi}}. \\[2mm]
 \end{equation}
 
 \item To show the left half part of \eqref{Intro:spacetimePenrose}, we construct a double null foliation in the exterior region of our spacetime. The advantage of using the double null foliation is that it gives the Bondi mass loss formula, namely, $M_B(\tu)$ does not increase as $\tu$ grows. By the hyperbolic estimates provided in \cite{KS:main}, we also have that $M_B(\tu)\to m_{\infty}$ as $\tu\to \infty$. Along with \eqref{Intro:rightPenrose} and the below algebraic inequality 
 \begin{equation}\label{Basic ineq}
        r_{+,\infty}^2+a_{\infty}^2=(\minft+\sqrt{\minft^2-\ainft^2})^2+\ainft^2=2\minft^2+2\minft\sqrt{\minft^2-\ainft^2}\le 4\minft^2,
    \end{equation}
 we then prove \eqref{Intro:spacetimePenrose}.
 \end{itemize}
\vspace{2mm}

 We proceed to establish the spacetime Penrose inequality in perturbative Kerr regime:
\begin{equation}\label{Intro:riem penrose ineq1}
      m\ge \sqrt{\f{A}{16\pi}},
 \end{equation}
 where $m$ and $A$ represent the ADM mass and the area of the MOTS for the initial data set, respectively.
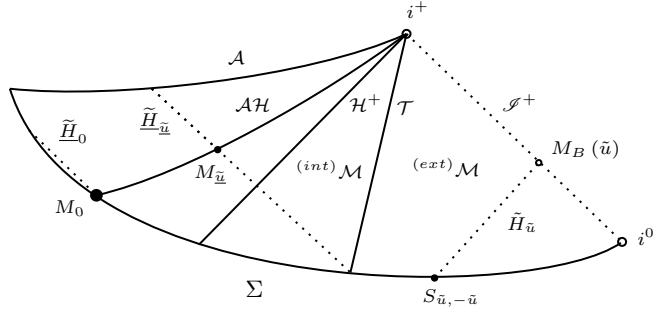
\begin{figure}[ht]
    \centering
\tikzset{every picture/.style={line width=0.75pt}} %set default line width to 0.75pt        

\begin{tikzpicture}[x=0.75pt,y=0.75pt,yscale=-1,xscale=1]

%Curve Lines [id:da599180705506545] 
\draw    (119.6,121.68) .. controls (153.43,229.94) and (382.89,228.73) .. (424.59,199.17) ;
\draw [shift={(425.2,198.72)}, rotate = 323.13] [color={rgb, 255:red, 0; green, 0; blue, 0 }  ][line width=0.75]      (0, 0) circle [x radius= 2.01, y radius= 2.01]   ;
%Straight Lines [id:da5569341305376619] 
\draw    (214,200.08) -- (317.6,94.08) ;
%Straight Lines [id:da9249598285412924] 
\draw  [dash pattern={on 0.84pt off 2.51pt}]  (317.6,94.08) -- (425.2,198.72) ;
%Curve Lines [id:da16323498668352543] 
\draw    (119.6,121.68) .. controls (172.53,127.25) and (269.03,116.19) .. (316.88,94.41) ;
\draw [shift={(317.6,94.08)}, rotate = 335.19] [color={rgb, 255:red, 0; green, 0; blue, 0 }  ][line width=0.75]      (0, 0) circle [x radius= 2.01, y radius= 2.01]   ;
%Curve Lines [id:da8665236965398172] 
\draw    (162.8,175.28) .. controls (208.4,164.88) and (277.6,124.08) .. (317.6,94.08) ;
\draw [shift={(162.8,175.28)}, rotate = 347.15] [color={rgb, 255:red, 0; green, 0; blue, 0 }  ][fill={rgb, 255:red, 0; green, 0; blue, 0 }  ][line width=0.75]      (0, 0) circle [x radius= 2.68, y radius= 2.68]   ;
%Straight Lines [id:da2514986180680092] 
\draw  [dash pattern={on 0.84pt off 2.51pt}]  (129.14,142.54) -- (162.8,175.28) ;
%Straight Lines [id:da2279706504090223] 
\draw  [dash pattern={on 0.84pt off 2.51pt}]  (190.42,121.66) -- (223.34,152.33) -- (289.6,214.08) ;
%Straight Lines [id:da03823923770601123] 
\draw  [dash pattern={on 0.84pt off 2.51pt}]  (331.6,216.88) -- (383.37,159.13) ;
\draw [shift={(383.6,158.88)}, rotate = 311.88] [color={rgb, 255:red, 0; green, 0; blue, 0 }  ][line width=0.75]      (0, 0) circle [x radius= 1.34, y radius= 1.34]   ;
\draw [shift={(331.6,216.88)}, rotate = 311.88] [color={rgb, 255:red, 0; green, 0; blue, 0 }  ][fill={rgb, 255:red, 0; green, 0; blue, 0 }  ][line width=0.75]      (0, 0) circle [x radius= 1.34, y radius= 1.34]   ;
%Straight Lines [id:da8594658011416159] 
\draw    (317.6,94.08) -- (289.6,214.08) ;
%Straight Lines [id:da27143287102420766] 
\draw  [dash pattern={on 0.84pt off 2.51pt}]  (190.42,121.66) -- (223.34,152.33) ;
\draw [shift={(223.34,152.33)}, rotate = 42.98] [color={rgb, 255:red, 0; green, 0; blue, 0 }  ][fill={rgb, 255:red, 0; green, 0; blue, 0 }  ][line width=0.75]      (0, 0) circle [x radius= 1.34, y radius= 1.34]   ;

% Text Node
\draw (316.16,76.36) node [anchor=north west][inner sep=0.75pt]  [font=\scriptsize]  {$i^{+}$};
% Text Node
\draw (431.36,190.36) node [anchor=north west][inner sep=0.75pt]  [font=\scriptsize]  {$i^{0}$};
% Text Node
\draw (363.76,122.76) node [anchor=north west][inner sep=0.75pt]  [font=\scriptsize]  {$\mathscr{I}^{+}$};
% Text Node
\draw (226.96,101.36) node [anchor=north west][inner sep=0.75pt]  [font=\scriptsize]  {$\mathcal{A}$};
% Text Node
\draw (287.53,123.02) node [anchor=north west][inner sep=0.75pt]  [font=\scriptsize]  {$\mathcal{H}^{+}$};
% Text Node
\draw (230.56,124.96) node [anchor=north west][inner sep=0.75pt]  [font=\scriptsize]  {$\mathcal{AH}$};
% Text Node
\draw (139.66,176.56) node [anchor=north west][inner sep=0.75pt]  [font=\scriptsize,xslant=-0.1]  {$M_{0}$};
% Text Node
\draw (210.16,158.56) node [anchor=north west][inner sep=0.75pt]  [font=\scriptsize]  {$M_{\widetilde{\underline{u}}}$};
% Text Node
\draw (388.16,146.16) node [anchor=north west][inner sep=0.75pt]  [font=\scriptsize]  {$M_{B}\left(\tilde{u}\right)$};
% Text Node
\draw (324.16,220.96) node [anchor=north west][inner sep=0.75pt]  [font=\scriptsize]  {$S_{\tilde{u} ,-\tilde{u}}$};
% Text Node
\draw (311.13,126.67) node [anchor=north west][inner sep=0.75pt]  [font=\scriptsize]  {$\mathcal{T}$};
% Text Node
\draw (235.76,216.56) node [anchor=north west][inner sep=0.75pt]  [font=\small]  {$\Sigma $};
% Text Node
\draw (260.85,156.1) node [anchor=north west][inner sep=0.75pt]  [font=\scriptsize]  {$^{( int)}\mathcal{M}$};
% Text Node
\draw (318.85,155.25) node [anchor=north west][inner sep=0.75pt]  [font=\scriptsize]  {$^{( ext)}\mathcal{M}$};
% Text Node
\draw (366.39,182.45) node [anchor=north west][inner sep=0.75pt]  [font=\scriptsize]  {$\tilde{H}_{\tilde{u}}$};
% Text Node
\draw (183.06,129.78) node [anchor=north west][inner sep=0.75pt]  [font=\scriptsize]  {$\widetilde{\underline{H}}_{\widetilde{\underline{u}}}$};
% Text Node
\draw (142.66,135.56) node [anchor=north west][inner sep=0.75pt]  [font=\scriptsize]  {$\widetilde{\underline{H}}_{0}$};

\end{tikzpicture}

    \caption{Proof of Spacetime Penrose Inequality}
    \label{Introfig:Penroseineq}
\end{figure}
\begin{itemize}
    \item 
Given the initial data triplet $(\Si, g, k)$, we first apply the result of Kerr stability \cite{KS:main},\footnote{The main conclusion of \cite{KS:main} shows the global existence of the spacetime in a sub-region. The existence of the full spacetime region is guaranteed by combining a standard interior local existence result and an exterior global existence result for the external region as in \cite{C-N,Shen:Kerr}.} and obtain the global existence of solutions to the EVEs \eqref{Intro:EVE eqn}. This solution $(\MM, \bfg)$ features a complete future null infinity $\II^+$ and contains a domain of outer communication, which converges to a nearby Kerr solution $Kerr(a_{\infty}, m_{\infty})$. The configuration of the spacetime $(\MM, \bfg)$ is illustrated in \Cref{Introfig:Penroseineq}. Here $\mathcal{A}$ serves as a spacelike inner boundary of $\MM$.

According to the barrier argument presented in \cite{A-Met,E}, in perturbative Kerr regime we infer that there exists a unique, smooth MOTS $M_0$ on $\Si$. Now we divide the spacetime $\MM$ into two parts, the interior region $\intM$ and the exterior region $\Ext\MM$, with the timelike boundary $\TT$ between these two parts chosen to be far away from $\mathcal{A}$, so that $\Int\Si=\Si\cap\intM$ contains $M_0$.

Within $\intM$, we then foliate the spacetime by incoming null hypersurfaces $\tHb_{\tub}$ ($\tub$ is the associated incoming optical function). The construction of $\tHb_{\tub}$ is similar to the procedure discussed in \Cref{Introsubsubsec:PI coordinate}. The only difference is that here we initialize $\tub$ along $\Int\Si\cup \TT$ so that $\tub=0$ on the MOTS $M_0$. Proceeding as discussed in \Cref{Intro:sec:MOTS}, \Cref{Intro:sec:MOTSdynamic} and the first part of this section, we solve the unique MOTS along each $\tHb_{\tub}$ and obtain a smooth apparent horizon $\AH$ consisting of MOTSs $M_{\tub}$ for all $\tub\ge0$. And we emphasize that the area of MOTS $M_{\tub}$ is non-decreasing in $\tub$ and converging to the value $4\pi(r_{+,\infty}^2+a_{\infty}^2)$ as $\tub\to\infty$. In particular, along $\Si$ the area $A$ of initial MOTS satisfies
\begin{equation}\label{Intro:penrose ineq right2}
    A\le 4\pi(r_{+,\infty}^2+a_{\infty}^2).
\end{equation}

\item In view of the algebraic calculation \eqref{Basic ineq}, it suffices to prove $m\ge m_{\infty}$. We achieve this by evaluating the limits of the Bondi mass at the spacelike infinity $i^0$ and at the timelike infinity $i^+$.
\begin{itemize}
    \item 
More precisely, we construct a double null foliation in $\Ext\MM$, with the outgoing optical $\tu$ initialized on $\II^+$ and the incoming optical function $\Ext\tub$ initialized on $\Ext\Si=\Si\cap\Ext\MM$ so that $\tH_{\tu}\cap \Si=\tS_{\tu, -\tu}$. Here $\tH_{\tu}$ is the level set of $\tu$ and $\tS_{\tu, \tub}$ denotes the intersection of $\tH_{\tu}$ and the constant $\Ext\tub$ hypersurface. Let $\{\te_1, \te_2, \te_3, \te_3 \}$ be the corresponding null frame. 

\item Define $m_H(\tu, \tub)$ to be the Hawking mass of the 2-sphere $\tS_{\tu, \tub}$. Using the hyperbolic estimates in \cite{KS:main}, we derive
\begin{align}
     \pr_{\tub} m_H=O(\t{r}^{-2}), \qquad m_H=m_{\infty}\l 1+O(\t{r}^{-1})+O(\tujp^{-\f12-\de_{dec}}) \r, \label{Intro:Eqn:mH}
\end{align}
where $\t{r}$ stands for the area radius of $\tS_{\tu, \tub}$, $\tujp\coloneqq\sqrt{|u|^2+1}$ and $0<\de_{dec}\ll 1$.

\item In the expansion of $m_H$ in \eqref{Intro:Eqn:mH}, letting $\t{r}\to \infty$ first and then setting $\tu\to \infty$, we hence deduce
\begin{equation*}
    M_B(\tu)=m_{\infty}\l 1+O(\tujp^{-\f12-\de_{dec}}) \r \qquad  \text{and} \qquad M_B(\infty)=m_{\infty}.
\end{equation*}
Here we use the definition of Bondi mass $M_B(\tu)\coloneqq\lim\limits_{\tub\to \infty} m_H(\tu, \tub)$.

\item On the other hand, integrating $\pr_{\tub} m_H$ backwards from $\infty$ to $\tub$, with \eqref{Intro:Eqn:mH} we obtain
\begin{align}\label{Intro:ineq Bondi}
      |m_H(\tu, \tub)-M_B(\tu)|\lesssim \f{1}{\t{r}}.
\end{align}
By virtue of \eqref{Intro:ineq Bondi} and the property $\lim\limits_{\tu\to-\infty} m_H(\tu, -\tu)=m$, we infer that $M_B(-\infty)=m$. Therefore, employing the Bondi mass loss formula, we arrive at 
\begin{equation*}
    m=M_B(-\infty)\ge M_B(\infty)=m_{\infty}.
\end{equation*}
Together with \eqref{Intro:penrose ineq right2} and \eqref{Basic ineq}, we hence conclude the desired spacetime Penrose inequality \eqref{Intro:riem penrose ineq1}.
\end{itemize}
\item  We further prove a rigidity result for our spacetime Penrose inequality when equality in \eqref{Intro:riem penrose ineq1} holds. As a direct consequence of $A=16\pi m^2$, we obtain a string of equalities 
\begin{equation*}
     A=A_M(\tub) =8\pi m_{\infty}(m_{\infty}+\sqrt{\minft^2-\ainft^2})=16\pi m_{\infty}^2=16\pi [M_B(\tu)]^2=16\pi m^2,
\end{equation*}
which implies $a_{\infty}=0, m_{\infty}=m$ and $A_M(\tub)\equiv A$. Based on the area increasing law of MOTS mentioned in \Cref{Introsec:Penrose Inequality in Perturbative Kerr Regime} via the null comparison principle, we must have that the apparent horizon $\AH$ is a non-expanding null horizon. Employing the associated transport equations along the outgoing null direction, we thus deduce that various geometric quantities along $\AH$ take their Schwarzschild values. In particular, the shear tensor satisfies $\chih'\equiv 0$ and the curvature component obeys $\rho'=-\f{1}{4m^2}$. Via incorporating with the Gauss equation, we further conclude that for the MOTS $M_0$ along initial spacelike hypersurface $\Si$, its the Gauss curvature is pointwise equal to $\f{1}{4m^2}$. Conversely, assuming that Gauss curvature of the MOTS $M_0$ is identically equal to $\f{1}{4m^2}$, utilizing the Gauss--Bonnet theorem, we easily get $A=16\pi m^2$.
\end{itemize}

\subsection{Other Related Works}
In our work, we employ the recent breakthrough of nonlinear Kerr stability. Its broader implications include the angular momentum memory effect along future null infinity \cite{An-He-Shen} by An--He--Shen, the detailed structures of null infinity and the resolution of supertranslation ambiguity \cite{KSW} by Klainerman--Shen--Wan, as well as the existence and smoothness of the event horizon \cite{Chen-K:EH} by Chen--Klainerman and \cite{Hintz} by Hintz.
We also remark earlier works to construct examples of black hole formation spacetimes with gluing to exact Kerr spacetimes, such as those by Li--Mei \cite{L-M}, Athanasiou--Lesourd \cite{At-Le} and Kehle--Unger \cite{K-U:gluing}. With the aid of the time-symmetric assumption (near the MOTS), Penrose inequality was proved by Huisken--Ilmanen in \cite{H-I} via the inverse mean curvature flow, and by Bray \cite{Bray} through the conformal flow. For recent advancements toward the Penrose inequality, we refer to the works and survey papers by Malec--Mars--Simon \cite{M-M-S}, Mars \cite{Mars}, Dain \cite{Dain}, Bray--Lee \cite{B-L}, Bray--Khuri \cite{B-K}, Khuri \cite{Khuri}, Brendle--Wang \cite{B-W}, Brendle--Hung--Wang \cite{B-H-W}, Han--Khuri--Weinstein--Xiong \cite{H-K-W-X}, Allen--Bryden--Kazaras--Khuri\cite{Allen:2025fhj} and references therein. 

\subsection{Structure of the Paper}
\begin{itemize}
    \item In \Cref{Sec:basic}, we state the basic geometric setup used in this work.
    \item In \Cref{Sec:Char Data}, we introduce the admissible Kerr black hole formation initial data and construct the associated initial layer. Then we conduct the frame transformation to the adapted principal geodesic (PG) foliation within the initial layer and verify the initial conditions required in \cite{KS:main}.
    \item In \Cref{Sec:Construct null cones}, we construct the incoming null hypersurfaces by solving the eikonal equation with precise Pretorius--Israel type coordinates. Then we demonstrate a smooth foliation transition between incoming null hypersurfaces in the short-pulse region and those in the region after the short pulse. We also derive detailed information for null expansions along these incoming null hypersurfaces.
    \item In \Cref{Sec:Existence}, we prove the existence and the uniqueness of MOTS along each incoming null hypersurface, with the aid of new a priori estimates for the equation of MOTS and the Leray--Schauder fixed point theorem.
    \item In \Cref{Sec:physical}, we establish global dynamics and properties of our constructed apparent horizon, by analyzing the linearized operator for the equation of MOTS and by verifying the null comparison principle. 
    \item In \Cref{Sec:penrose ineq}, without time symmetry condition we prove the Penrose inequality in Kerr black hole formation spacetime and perturbative Kerr regime.
\end{itemize}

\subsection{Notations and Conventions}
For being precise in our arguments, we list the following constants that will be used in this paper.

In the discussions of the initial data layer as in \Cref{Sec:Char Data},
\begin{itemize}
    \item The constants $m_0>0$ and $|a_0|< m_0$ are the mass and the angular momentum of the Kerr solution relative to which our initial perturbation is measured. In particular, we pick $m_0=1/4$ and $a_0=0$ in this article.
    \item The positive integer $N$ refers to the number of derivatives for our constructed perturbed initial data energy.
    \item The size of the perturbed initial data energy is measured by $\epsilon_1>0$.
    \item $\de>0$ corresponds to the characteristic length of the short-pulse region. 
    \item $A=\de^{-1}$ measures the size of $\chih$ prescribed in the short-pulse region.
    \item $u_0<0$ is tied to the initial outgoing hypersurface $u=u_0$.
    \item $r_0\approx |u_0|$ is related to the initialization of PG structures in the initial data layer. 
    \item $\de_B>0$ appears in the $r$-power for the initial decay estimates of $\a, \b$.
\end{itemize}
In the current paper, $-u_0,r_0>0$ are fixed, sufficiently large constants, $\de_{B}>0$ is a fixed, sufficiently small, universal constant. The smallness constants $\epsilon_1$ and $\de, A$ are chosen such that
\begin{equation*}
    0<\de A^{\f12}\ll \epsilon_1\ll 1. \\[2mm]
\end{equation*}

In the context of nonlinear Kerr stability in \Cref{Sec:Construct null cones},
\begin{itemize}
\item The constants $m_{\infty}>0$ and $|a_{\infty}|< m_0$ represent the final mass and the final angular momentum of the Kerr spacetime that our spacetime converges to.
    \item The integer $k_{small}>0$ refers to the number of derivatives that appears in the decay estimates for the solutions. 
    \item The size of the perturbation is measured by $\epsilon_0>0$.
    \item $\de_{dec}>0$ is tied to decay estimates in $u, \ub$ of Ricci coefficients and curvature components.
\end{itemize}
Here, we select $\de_{\HH}, \de_{dec}>0$ to be fixed, sufficiently small, universal constants. 

\vspace{2mm}
 Throughout this paper, given $A, B\ge0$, we denote $A\les B$ to stand for $A\leq CB$ with $C>0$ being a constant that is independent of constants $\epsilon_1, \epsilon_0$. 
 
 Also, for $A$ and $B$ with the same sign, we denote $A\sim B$ if there exists a constant $C\ge 1$ independent of constants $\epsilon_1, \epsilon_0$ such that $C^{-1}\le A/B \le C$.

\subsection{Acknowledgements}
The authors would like to thank Chao Li and Jingbo Wan for valuable conversations. XA is supported by MOE Tier 1 grants A-0004287-00-00, A-0008492-00-00 and MOE Tier 2 grant A-8000977-00-00.  TH acknowledges the support of NUS President Graduate Fellowship.

\section{Preliminaries}\label{Sec:basic}
\subsection{Null Pairs and Horizontal Structures}
Let $(\MM, \bfg)$ be a $3+1$ dimensional Lorentzian manifold. In this paper, we study the below Einstein vacuum equations (EVEs)
\begin{equation}\label{Eqn:EVE}
    \textbf{Ric}(\bfg)=0.
\end{equation}
 We consider a pair of normalized null vectors $(e_3, e_4)$ of $(\MM, \bfg)$ satisfying
\begin{equation*}
    \bfg(e_3, e_3)=\bfg(e_4, e_4)=0, \qquad \bfg(e_3, e_4)=-2.
\end{equation*}
In the tangent space, we further define the horizontal structure $\HH\coloneqq\qty{e_3, e_4}^{\perp}$ with respect to the null pair $(e_3, e_4)$. For each vector $X\in\HH$, we require $X$ to be perpendicular to both $e_3$ and $e_4$, i.e., 
\begin{equation*}
    \bfg(e_3, X)=\bfg(e_4, X)=0.
\end{equation*}
Due to the positive definiteness of $\bfg|_{\HH}$, we can find an orthonormal basis of $\HH$ and denote the basis as $\{ e_1, e_2\}$. For the rest of this paper, we refer to $(e_1, e_2, e_3, e_4)$ as the null frame at each point on $\MM$.

With $a,b,c$ denoting $1,2$, we proceed to define the left and right duals of horizontal 1-forms $\om$ and 2-covariant tensor-fields $U$ as
\begin{align*}
    \dual \om_a\coloneqq\in_{ab} \om^b, \qquad  \om\dual{}_a\coloneqq \om^b\in_{ba}, \qquad (\dual U)_{ab}\coloneqq\in_{ac} U^c{}_b, \qquad  ( U\dual)_{ab}\coloneqq U_{a}{}^c \in_{cb}.
\end{align*}
Here $\in_{ab}\coloneqq\f12\in$ with $\in$ being the volume form of $\MM$.\footnote{Note that we can choose $(e_1, e_2)$ such that $\in_{12}=\f12\in(e_1, e_2, e_3, e_4)=1$.}

Denote by $\SS_1$ the set of 1-forms on $\HH$, and by $\SS_2$ the set of symmetric traceless 2-tensors on $\HH$. Given $\xi, \eta\in \SS_1$, $U, V\in \SS_2$, we further define
\begin{align*}
    \xi\c\eta\coloneqq& \de^{ab}\xi_a\eta_b, &\quad \xi\wedge\eta\coloneqq& \in^{ab}\xi_a \eta_b=\xi\c\dual\eta, &\quad (\xi\hot\eta)_{ab}\coloneqq& \xi_a \eta_b+\xi_b\eta_a-\de_{ab} \xi\c\eta, \\
    (\xi\c U)_a\coloneqq& \de^{bc}\xi_b U_{ac}, &\quad U\wedge V\coloneqq& \in^{ab}U_{ac} V_{cb}, &\quad U\c V\coloneqq& U_{ab} V_{ab}.
\end{align*}

Next, we define the horizontal covariant derivative $\nab$ on $\HH$ by
\begin{equation*}
    \nab_X Y\coloneqq\bfD_X Y-\f12 \chib(X,Y)e_4-\f12\chi(X,Y)e_3 \qquad \text{for any} \quad X, Y\in \HH.
\end{equation*}
 Similarly, we define $\nab_3$ and $\nab_4$ to be the projections of spacetime covariant derivatives $\bfD_3$ and $\bfD_4$ onto $\HH$.\footnote{Here $\bfD_\mu\coloneqq\bfD_{e_{\mu}}$ with $\mu=1, 2, 3,4$.} More precisely, for all $X\in \HH$, there hold
 \begin{align*}
   \nab_3 X\coloneqq&\bfD_3 X-\f12 g(X, \bfD_3 e_3) e_4- \f12 g(X, \bfD_3 e_4 ) e_3,\\
 \nab_4 X\coloneqq&\bfD_4 X-\f12 g(X, \bfD_4 e_3) e_4-\f12 g(X, \bfD_4 e_4 ) e_3. 
 \end{align*}

For a given 1-form $\xi$ on $\HH$, we then define the following differential operators on $\HH$:
 \begin{equation*}
     \div \xi\coloneqq\de^{ab} \nab_a \xi_b, \qquad \curl \xi\coloneqq \de^{ab} \nab_a \xi_b, \qquad  (\nab\hot\xi)_{ab}\coloneqq \nab_a \xi_b+\nab_b\xi_a-\de_{ab}(\div \xi).
 \end{equation*}
 Given $U\in\SS_2$, we also denote $(\div U)_a\coloneqq \nab^{b}U_{ba}$.
 
\subsection{Frame Decomposition}\label{Subsec:framdecomp}
 With respect to an arbitrary null frame $(e_1, e_2, e_3, e_4)$ on $\MM$, we define the associated curvature components and Ricci coefficients. Let $\bfD$ and $\bfR$ be the spacetime covariant derivative and the Riemann curvature tensor. Denoting the indices $a, b$ to be $1,2$, we define null curvature components
 \begin{equation}\label{def curvatures}
\begin{aligned}
\a_{ab}&\coloneqq\mathbf{R}(e_a, e_4, e_b, e_4),\quad &   \ab_{ab}&\coloneqq\mathbf{R}(e_a, e_3, e_b, e_3),\\
\b_a&\coloneqq \frac 1 2 \mathbf{R}(e_a,  e_4, e_3, e_4) ,\quad& \bb_a &\coloneqq\frac 1 2 \mathbf{R}(e_a,  e_3,  e_3, e_4),\\
\rho&\coloneqq\frac 1 4 \mathbf{R}(e_4,e_3, e_4,  e_3),\quad& \sigma&\coloneqq\frac 1 4  \,^*\mathbf{R}(e_4,e_3, e_4,  e_3). 
\end{aligned}
\end{equation}
Here $\, ^*\bfR$ stands for the Hodge dual of $\bfR$. We also define Ricci coefficients as
\begin{equation}\label{def Ricci coefficients}
\begin{aligned}
\chi_{ab}\coloneqq&\mathbf{g}(\bfD_a e_4,e_b),&\quad \chib_{ab}\coloneqq&\mathbf{g}(\bfD_a e_3,e_b),\\
\xi_a\coloneqq&\frac 12 \mathbf{g}(\bfD_4 e_4,e_a), &\quad \xib_a\coloneqq&\frac 12 \mathbf{g}(\bfD_3 e_3,e_a),\\
\eta_a\coloneqq&-\frac 12 \mathbf{g}(\bfD_3 e_a,e_4),&\quad \etab_a\coloneqq&-\frac 12 \mathbf{g}(\bfD_4 e_a,e_3),\\
\omega\coloneqq&-\frac 14 \mathbf{g}(\bfD_4 e_3,e_4),&\quad \omegab\coloneqq&-\frac 14 \mathbf{g}(\bfD_3 e_4,e_3),\\
\zeta_a\coloneqq&\frac 1 2 \mathbf{g}(\bfD_a e_4,e_3).
\end{aligned}
\end{equation}

Owing to the potential lack of integrability of $\HH=\spn\{e_1, e_2 \}$, the null second fundamental forms $\chi$ and $\chib$ defined as above may not be symmetric.  Accordingly, we decompose $\chi, \chib$ into traceless, trace and anti-trace parts as
\begin{align*}
    \chi_{ab}=&\chih_{ab}+\f12\de_{ab}\tr\chi+\f12 \in_{ab}\atrch, \\
    \chib_{ab}=&\chibh_{ab}+\f12\de_{ab}\tr\chib+\f12 \in_{ab}\atrchb,
\end{align*}
where
\begin{equation*}
    \trch\coloneqq\de^{ab} \chi_{ab}, \qquad \trchb\coloneqq\de^{ab} \chib_{ab}, \qquad \atrch\coloneqq\in^{ab} \chi_{ab}, \qquad \atrchb\coloneqq\in^{ab} \chib_{ab}.
\end{equation*} 

In this paper, we also need the following commutation formulas.
\begin{lemma}[\cite{KS:formula}]\label{Lem:commute}
    For any scalar function $f$, the following commutation identities hold
    \begin{align*}
        [\nab_3, \nab]f=&-\chib\c \nab f+(\eta-\zeta)\nab_3 f+\xib\nab_4 f, \\
         [\nab_4, \nab]f=&-\chi\c \nab f+(\etab+\zeta)\nab_4 f+\xi\nab_3 f, \\
         [\nab_3, \nab_4]f=&2(\eta-\etab)\c \nab f-2\o\nab_3 f+2\omb\nab_4 f.
    \end{align*}
\end{lemma}
\begin{proof}
    See the proof of Lemma 2.35 in \cite{KS:formula}.
\end{proof}
\subsection{Frame Transformation}\label{Subsec:frametrans}
A variety of null frames will be employed throughout this paper. To pass through from one frame to the other, we will use the below general change-of-frame formula provided by Klainerman--Szeftel in \cite{KS:Kerr1}.
\begin{lemma}[\cite{KS:Kerr1}]\label{Lem:frame transform}
    Between two null frames $(e_1, e_2, e_3, e_4)$ and $(e'_1, e'_2, e'_3, e'_4)$, there is a general null frame transformation and it can be written as
    \begin{equation}\label{frame transform formula}
        \begin{split}
            &e_4'=\lambda \left(e_4+f^b e_b+\f14|f|^2e_3\right), \\
            &e_a'=\Big(\delta_a^b+\f12 \fb_a f^b\Big)e_b+\f12 \fb_a e_4+(\f12 f_a+\f{1}{8}|f|^2 \fb_a)e_3, \\
            &e_3'=\lambda^{-1}\Big[ \Big(1+\f12 f\cdot \fb+\f{1}{16}|f|^2|\fb|^2\Big)e_3+\Big(\fb^b+\f{1}{4}|\fb|^2 f^b\Big)e_b+\f14 |\fb|^2 e_4 \Big].
        \end{split}
    \end{equation}
    Here one forms $f^b, \fb^b$ and the scalar function $\la$, in short $(f, \fb, \lambda)$, are called the transition coefficients of the frame transformation.
\end{lemma}
\begin{remark}
    Note that typically it is rather complicated to deal with the null frame transformation with general transition coefficients $(f, \fb, \la)$. If either $\fb=0$ or $f=0$, the frame transformation formula \eqref{frame transform formula} can be reduced to the following simpler form
    \begin{equation*}
        e_4'=\lambda \left(e_4+f^b e_b+\f14|f|^2e_3\right), \qquad e_a'=e_a+f_a e_3, \qquad e_3'=\la^{-1} e_3
    \end{equation*}
    or
    \begin{equation*}
        e_4'=\lambda e_4, \qquad e_a'=e_a+\fb_a e_4, \qquad e_3'=\la^{-1} \qty(e_3+\fb^b e_b+\f14|\fb|^2 e_4).
    \end{equation*}
    In practice, we often split the general frame changes into below two-step decomposition:
    \begin{equation}\label{Eqn:nullframechange1}
        (e_{\mu})\xrightarrow{({}^{(1)}f, {}^{(1)}\fb=0, {}^{(1)}\la)} ({}^{(1)}e_{\mu})\xrightarrow{({}^{(2)}f=0, {}^{(2)}\fb, {}^{(2)}\la=1)} (e_{\mu}').
    \end{equation}
    Here we first transform the original null frame to an intermediate null frame $({}^{(1)}e_{\mu})$ so that ${}^{(1)}e_{4}=e_4'$. Then we shift $({}^{(1)}e_{\mu})$ to the target null frame $(e_{\mu}')$. As a result, it is more convenient to determine the transition coefficients in each of these two steps, while keeping one direction of the null pair unchanged.

    Similarly, by fixing ${}^{(1)}e_{3}=e_3'$, we can also employ the other two-step decomposition
    \begin{equation}\label{Eqn:nullframechange2}
        (e_{\mu})\xrightarrow{({}^{(1)}f=0, {}^{(1)}\fb, {}^{(1)}\la)} ({}^{(1)}e_{\mu})\xrightarrow{({}^{(2)}f, {}^{(2)}\fb=0, {}^{(2)}\la=1)} (e_{\mu}').
    \end{equation}
    We note that both \eqref{Eqn:nullframechange1} and \eqref{Eqn:nullframechange2} are frequently used in this paper.
\end{remark}

Under the general null frame transformation \eqref{frame transform formula}, the Ricci coefficients and curvature components adapted to the new null frame $(e'_1, e'_2, e'_3, e'_4)$, denoted with primes, can be expressed in terms of the geometric quantities with regard to the original null frame $(e_1, e_2, e_3, e_4)$ and the transition coefficients $(f, \fb, \la)$  as follows:
\begin{lemma}[\cite{KS:Kerr1}]\label{Lem:transformation formula}
   Given a frame transformation between two null frames $(e_1, e_2, e_3, e_4)$ and $(e'_1, e'_2, e'_3, e'_4)$ with the transition coefficients $(f, \fb, \lambda)$, the following transformation formulas for Ricci coefficients $\Ga',\Ga$ and curvature components $R',R$ hold true:
\begin{itemize}
\item The transformation formula for $\xi$ is given by 
\begin{equation*}
\begin{split}
\la^{-2}\xi' &= \xi +\frac{1}{2}\la^{-1}\nab_4'f+\frac{1}{4}(\trch f -\atrch\dual f)+\om f +\err(\xi,\xi'),\\
\err(\xi,\xi') &= \frac{1}{2}f\c\hch+\frac{1}{4}|f|^2\eta+\frac{1}{2}(f\c \zeta)\,f -\frac{1}{4}|f|^2\etab \\
&+ \la^{-2}\left( \frac{1}{2}(f\c\xi')\,\fb+ \frac{1}{2}(f\c\fb)\,\xi'   \right) +\lot.
\end{split}
\end{equation*}
\item The transformation formula for $\xib$ is given by 
\begin{equation*}
\begin{split}
\la^2\xib' =& \xib + \frac{1}{2}\la\nab_3'\fb +    \omb\,\fb + \frac{1}{4}\trchb\,\fb - \frac{1}{4}\atrchb\dual\fb +\err(\xib, \xib'),\\
\err(\xib, \xib') =&   \frac{1}{2}\fb\c\hchb - \frac{1}{2}(\fb\c\zeta)\fb +  \frac 1 4 |\fb|^2\etab  -\frac 1 4 |\fb|^2\eta'+\lot.
       \end{split}
\end{equation*}
\item The transformation formulas for $\chi $ are  given by 
\begin{equation*}
\begin{split}
\la^{-1}\trch' =& \trch  +  \div'f + f\c\eta + f\c\zeta+\err(\trch,\trch')\\
\err(\trch,\trch') =& \fb\c\xi+\frac{1}{4}\fb\c\left(f\trch -\dual f\atrch\right) +\om (f\c\fb)  -\omb |f|^2 \\
& -\frac{1}{4}|f|^2\trchb -  \frac 1 4 ( f\c\fb) \la^{-1}\trch' +\frac 1 4  (\fb\wedge f) \la^{-1}\atrch'+\lot, 
\end{split}
\end{equation*}
\begin{equation*}
\begin{split}
\la^{-1}\atrch' &= \atrch  +  \curl'f + f\wedge\eta + f\wedge\zeta +\err(\atrch,\atrch'),\\
\err(\atrch,\atrch') &= \fb\wedge\xi+\frac{1}{4}\left(\fb\wedge f\trch +(f\c\fb)\atrch\right) +\om f\wedge\fb   \\
& -\frac{1}{4}|f|^2\atrchb -  \frac 1 4 ( f\c\fb) \la^{-1}\atrch' +\frac 1 4   \la^{-1}(f\wedge \fb)\trch'+\lot,
\end{split}
\end{equation*}
\begin{equation*}
\begin{split}
\la^{-1}\hch' &= \hch  +  \nab'\hot f + f\hot\eta + f\hot\zeta+\err(\hch,\hch'),\\
\err(\hch,\hch') &=\fb\hot\xi+\frac{1}{4}\fb\hot\left(f\trch -\dual f\atrch\right) +\om f\hot\fb  -\omb f\hot f\\
& -\frac{1}{4}|f|^2\atrchb  +\frac 1 4  (f\hot\fb) \la^{-1}\trch' +\frac 1 4  (\dual f\hot\fb) \la^{-1}\atrch'\\
& +\frac 1 2  \fb\hot (f\c\la^{-1}\hch')+\lot.
\end{split}
\end{equation*}
\item The transformation formulas for $\chib $ are given by 
\begin{equation*}
\begin{split}
\la\trchb' &= \trchb +\div'\fb +\fb\c\etab  -  \fb\c\zeta +\err(\trchb, \trchb'),\\
\err(\trchb, \trchb') &= \frac{1}{2}(f\c\fb)\trchb+f\c\xib -|\fb|^2\om + (f\c\fb)\omb   -\frac 1 4 |\fb|^2\la^{-1}\trch'+\lot,
\end{split}
\end{equation*}
\begin{equation*}
\begin{split}
\la\atrchb' =& \atrchb +\curl'\fb +\fb\wedge\etab  -  \zeta\wedge\fb+\err(\atrchb, \atrchb'),\\
\err(\atrchb, \atrchb') =& \frac{1}{2}(f\c\fb)\atrchb+f\wedge\xib  + (f\wedge\fb)\omb   -\frac 1 4 |\fb|^2\la^{-1}\atrch'+\lot,
\end{split}
\end{equation*}
\begin{equation*}
\begin{split}
\la\hchb' &= \hchb +\nab'\hot\fb +\fb\hot\etab  -  \fb\hot\zeta +\err(\hchb, \hchb'),\\
\err(\hchb, \hchb') &= \frac{1}{2}(f\hot\fb)\trchb  +f\hot\xib -(\fb\hot\fb)\om + (f\hot\fb)\omb   -\frac 1 4 |\fb|^2\la^{-1}\hch'+\lot.
\end{split}
\end{equation*}
\item  The transformation formula for $\zeta$ is given by 
\begin{equation*}
\begin{split}
\zeta' &= \zeta -\nab'(\log\la)  -\frac{1}{4}\trchb f +\frac{1}{4}\atrchb \dual f +\om\fb -\omb f +\frac{1}{4}\fb\trch\\
&+\frac{1}{4}\dual\fb\atrch+\err(\zeta, \zeta'),\\
\err(\zeta, \zeta') &= -\frac{1}{2}\hchb\c f + \frac{1}{2}(f\c\zeta)\fb -  \frac{1}{2}(f\c\etab)\fb +\frac{1}{4}\fb(f\c\eta) + \frac{1}{4}\fb(f\c\zeta)  \\
& + \frac{1}{4}\dual\fb(f\wedge\eta) + \frac{1}{4}\dual\fb(f\wedge\zeta) +  \frac{1}{4}\fb\div'f  + \frac{1}{4}\dual\fb \curl'f +\frac{1}{2}\la^{-1}\fb\c\hch' \\
&  -\frac{1}{16}(f\c\fb)\fb\la^{-1}\trch' +\frac{1}{16}  (\fb\wedge f) \fb\la^{-1}\atrch'  -  \frac{1}{16}\dual\fb ( f\c\fb) \la^{-1}\atrch'\\
& +\frac{1}{16}\dual\fb \la^{-1}(f\wedge \fb)\trch' +\lot.
\end{split}
\end{equation*}

\item   The transformation formula for $\eta$ is given by 
\begin{equation*}
\begin{split}
\eta' &= \eta +\frac{1}{2}\la \nab_3'f  +\frac{1}{4}\fb\trch -\frac{1}{4}\dual\fb\atrch -\omb\, f +\err(\eta, \eta'),\\
\err(\eta, \eta') &= \frac{1}{2}(f\c\fb)\eta +\frac{1}{2}\fb\c\hch
+\frac{1}{2}f(\fb\c\zeta)  -  (\fb\c f)\eta'+ \frac{1}{2}\fb (f\c\eta') +\lot.
\end{split}
\end{equation*}
\item   The transformation formula for $\etab$ is given by 
\begin{equation*}
\begin{split}
\etab' &= \etab +\frac{1}{2}\la^{-1}\nab_4'\fb +\frac{1}{4}\trchb f - \frac{1}{4}\atrchb\dual f -\om\fb +\err(\etab, \etab'),\\
\err(\etab, \etab') &=  \frac{1}{2}f\c\hchb + \frac{1}{2}(f\c\etab)\fb-\frac 1 4  (f\c\zeta)\fb  -\frac 1 4 |\fb|^2\la^{-2}\xi'+\lot.
\end{split}
\end{equation*}

\item   The transformation formula for $\om$ is given by
\begin{equation*}
\begin{split}
\la^{-1}\om' &=  \om -\frac{1}{2}\la^{-1}e_4'(\log\la)+\frac{1}{2}f\c(\zeta-\etab) +\err(\om, \om'),\\
\err(\om, \om') &=   -\frac{1}{4}|f|^2\omb - \frac{1}{8}\trchb |f|^2+\frac{1}{2}\la^{-2}\fb\c\xi' +\lot.
\end{split}
\end{equation*}
\item   The transformation formula for $\omb$ is given by
\begin{equation*}
\begin{split}
\la\omb' =& \omb+\frac{1}{2}\la e_3'(\log\la)  -\frac{1}{2}\fb\c\zeta -\frac{1}{2}\fb\c\eta +\err(\omb,\omb'),\\
\err(\omb,\omb') =& f\c\fb\,\omb-\frac{1}{4} |\fb|^2\om  +\frac{1}{2}f\c\xib + \frac{1}{8}(f\c\fb)\trchb + \frac{1}{8}(\fb\wedge f)\atrchb \\
&-\frac{1}{8}|\fb|^2\trch  -\frac{1}{4}\la \fb\c\nab_3'f    +\frac{1}{2}  (\fb\c f)(\fb\c\eta')- \frac{1}{4}|\fb|^2 (f\c\eta')+\lot.
\end{split}
\end{equation*}
\end{itemize}
Here $\lot$ denotes expressions of the form
\begin{equation*}
         \lot=O\qty((f,\fb)^3)\Ga+O\qty((f,\fb)^3)\Ga
\end{equation*}
and it contains no derivatives of $f$, $\fb$, $\Ga$.
\vspace{2mm}

Meanwhile, the curvature components transform as follows:
\begin{itemize}
\item The transformation formulas for $\a, \ab $  are  given by
\begin{equation*}
\begin{split}
\la^{-2} \a'&=\a +\err(\a, \a'),\\
\err(\a, \a')&=  \big(  f\hot \b  -\dual f \hot \dual\b \big)+ \left( f\hot f-\frac 1 2  \dual f \hot   \dual f \right) \rho
+  \frac 3  2 \big(  f \hot  \dual  f\big) \si +\lot,
\end{split}
\end{equation*}
\begin{equation*}
\begin{split}
\la^2\ab'&=\ab +\err(\a, \a'),\\
\err(\ab, \ab')&=  -\big(  \fb \hot \bb  -\dual \fb \hot \dual  \bb \big)+ \big( \fb \hot \fb-\frac 1 2  \dual \fb \hot   \dual \fb \big) \rho
+  \frac 3  2 \big(  \fb \hot  \dual  \fb\big) \si +\lot.
\end{split}
\end{equation*}
\item   The transformation formulas for $\b , \bb $  are  given by
   \begin{equation*}
  \begin{split}
\la^{-1}   \b'&=\b +\frac 3 2\big(  f \rho+\dual  f  \si\big)+\err(\b, \b'), \\
  \err(\b, \b')&= \frac 1 2 \a\c\fb+\lot,
  \end{split}
  \end{equation*}
  \begin{equation*}
  \begin{split}
  \la\bb'&=\bb -\frac 3 2\big(  \fb \rho+\dual  \fb  \si\big)+\err(\bb, \bb'), \\
  \err(\bb, \bb')&= -\frac 1 2  \ab\c f +\lot.
  \end{split}
  \end{equation*}
  \item The transformation formulas for $\rho$ and $\si $  are  given by
  \begin{equation*}
  \begin{split}
 \rho' &= \rho +\err(\rho, \rho'),\\
\err(\rho, \rho') &= \fb\c\b - f\c\bb +\frac{3}{2}\rho(f\c\fb) -\frac{3}{2}\si (f\wedge\fb) +\lot,
   \end{split}
  \end{equation*}
  \begin{equation*}
  \begin{split}
  \si' &= \si +\err(\si, \si'),\\
\err(\si, \si') &= -\fb\c\dual\b - f\c\dual\bb +\frac{3}{2}\si(f\c\fb) +\frac{3}{2}\rho (f\wedge\fb) +\lot.
   \end{split}
  \end{equation*}  
\end{itemize}
Here $\lot$ denotes expressions of the form
\begin{equation*}
         \lot=O\qty((f,\fb)^3)(\rho, \si) +O\qty((f,\fb)^2)(\a,\b,\ab, \b)
\end{equation*}
and it contains no derivatives of $f$, $\fb$, $\a$, $\b$, $(\rho, \si)$, $\bb$, and $\ab$. 
\end{lemma}
\begin{proof}
 See Proposition 3.3 in \cite{KS:Kerr1}.
 \end{proof}
In the case where either $\fb=0$ or $f=0$, in below we provide more detailed transformation formulas for several geometric quantities.
 \begin{lemma}\label{Lem:frame transform1}
    If $\fb=0$, the following transformation formulas hold true:
\begin{align*}
       \la^{-1}\trch'=&\trch+f\c(\eta+\zeta)+\div f-\f12 f^a f^b\chibh_{ab}-|\fb|^2\omega+\f12 f\c\nab_3 f-\f14 |f|^2(f\c\xib), \\
       \la\trchb'=&\trchb+f\c\xib.
   \end{align*}
   Similarly, if $f=0$, we have
   \begin{align*}
       \la^{-1}\trch'=&\trch+\fb\c\xi, \qquad \la^{-1}\chih'=\chih+\fb\hot\xi,  \qquad\la^{-1}\atrch'=\atrch+\fb\wedge\xi,\\
       \la\trchb'=&\trchb+\fb\c(\etab-\zeta)+\div \fb-\f12\fb^a\fb^b\chih_{ab}-|\fb|^2\omega+\f12 \fb\c\nab_4 \fb-\f14 |\fb|^2(\fb\c\xi),\\
       \etab'=&\etab+\f12\nab_4\fb+\f14|\fb|^2\xi-\fb\o-\f12\fb(\fb\c\xi), \\ \zeta'=&\zeta-\nab'(\log\la)+\fb\o+\f14\fb\trch+\f14\dual\fb\atrch+\f12\fb\c\chih+\f12\fb(\fb\c\xi)\\
       \o'=&\la(\o-\f12\nab_4(\log\la)+\f12 \fb\c\xi),\\
       \la^{-2}\a'=&\a, \qquad\qquad\qquad \la^{-1}\b'=\b+\fb\c\a\\
\rho'=&\rho+\fb\c\b+\f14\fb^a\fb^b\a_{ab}, \qquad \si'=\si-\fb\c\dual\b-\f14\fb^a\fb^b\,\dual\a_{ab}.
   \end{align*}
\end{lemma}
\begin{proof}
    A straightforward verification works.
\end{proof}

\subsection{Principal Geodesic Structure}\label{Subsec:PG structure}
Klainerman--Szeftel's proof of nonlinear Kerr stability with small angular momentum \cite{KS:main} crucially builds upon the geometric formalism of the so-called principal geodesic (PG) structures. There is a non-integrable null frame associated with the PG structure, which most resembles the principal frame of Kerr. The explicit definitions are stated as below:
\begin{definition}\label{Def:outPG}
    The outgoing principal geodesic (PG) structure consists of a pair of null vectors $\{e_3, e_4 \}$ and a scalar function $r$ satisfying
    \begin{equation}\label{outPG cond 1}
        \bfD_{e_4} e_4=0, \quad e_4(r)=1, \quad \nab (r)=0. 
    \end{equation}
    Here $\nab$ denotes the covariant derivative of the horizontal structure $\{e_3, e_4 \}^{\perp}$. Additionally, we can define the corresponding outgoing PG coordinates $(u, \theta, \varphi)$ such that
    \begin{equation}\label{outPG cond 2}
        e_4(u)=e_4(\theta)=e_4(\varphi)=0.
    \end{equation}
\end{definition}
\begin{remark}\label{Rmk:outPG}
   Note that within the outgoing PG structure as defined above, in view of $\bfD_{e_4}e_4=0$, we have
    \begin{equation*}
        \o=\xi=0.
    \end{equation*}
    Furthermore, using \Cref{Lem:commute} and \eqref{outPG cond 1}, we derive
    \begin{equation*}
        \etab+\zeta=[e_4, \nab](r)=e_4(\nab(r))-\nab(e_4(r))=0.
    \end{equation*}
\end{remark}
Interchanging the roles of $e_3$ and $e_4$, we define the incoming PG structure as
\begin{definition}\label{Def:inPG}
    The incoming principle geodesic (PG) structure consists of a pair of null vectors $\{e_3, e_4 \}$ and a scalar function $r$ satisfying
    \begin{equation}\label{PG cond 1}
        \bfD_{e_3} e_3=0, \quad e_3(r)=-1, \quad \nab (r)=0. 
    \end{equation}
     Here $\nab$ again represents the covariant derivative of the horizontal structure $\{e_3, e_4 \}^{\perp}$. Additionally, we have the corresponding incoming PG coordinates $(\ub, \theta, \varphi)$, which satisfies
    \begin{equation}\label{PG cond 2}
        e_3(\ub)=e_3(\theta)=e_3(\varphi)=0.
    \end{equation}
\end{definition}
\begin{remark}
Similar to \Cref{Rmk:outPG}, for the incoming PG structure, we have
    \begin{equation*}
        \omb=\xib=\eta-\zeta=0.
    \end{equation*}
\end{remark}
\vspace{2mm}

    We now list the explicit form of the canonical incoming PG structure in the exact Kerr spacetime. Consider the Kerr metric in Boyer-Lindquist coordinates 
\begin{equation*}
    \bfg_{a, m}=-\f{|q|^2\Delta}{\Si^2}dt^2+\f{\Si^2 \sin^2 \theta}{|q|^2}\Big(d\phi-\f{2amr}{\Si^2} dt \Big)^2+\f{|q|^2}{\Delta} dr^2+|q|^2 d\theta^2,
\end{equation*}
where 
\begin{equation*}
    q\coloneqq r+ia\cos \theta, \qquad  \Delta\coloneqq r^2-2mr+a^2, \qquad \Si^2\coloneqq(r^2+a^2)^2-a^2 \sin^2 \theta \Delta.
\end{equation*}
The incoming principal null pair $\{e_3, e_4 \}$ is chosen as
\begin{equation}\label{Eqn:cano PG e3 e4}
    e_3=\f{r^2+a^2}{\Delta}\pr_t-\pr_r+\f{a}{\De} \pr_{\phi}, \qquad e_4= \f{r^2+a^2}{|q|^2}\pr_t+\f{\De}{|q|^2}\pr_r+\f{a}{|q|^2} \pr_{\phi}.
\end{equation}
And we pick an orthonormal basis of $\{e_3, e_4 \}^{\perp}$ in the form of
\begin{equation}\label{Eqn:cano PG ea}
    e_1= \f{1}{|q|}\pr_{\th}, \qquad e_2=\f{a\sin \th}{|q|}\pr_t+\f{1}{|q|\sin \th}\pr_{\phi}.
\end{equation}

We further define functions $(\ub, \varphi)$ by
\begin{equation*}
    \ub\coloneqq t+f(r) \quad \text{with} \quad f'(r)=\f{r^2+a^2}{\De}, \qquad \varphi\coloneqq\phi+h(r) \quad \text{with} \quad h'(r)=\f{a}{\De}. 
\end{equation*}
It can be readily verified that  $\{e_1, e_2, e_3, e_4 \}$ and $(\ub, r, \th, \varphi)$ together satisfy the conditions \eqref{PG cond 1} and \eqref{PG cond 2} in \Cref{Def:inPG}.

We also remark that, with coordinates $(\ub, r, \th, \varphi)$, the Kerr metric $\bfg_{a, m}$ can be converted into
\begin{equation}\label{Kerr metric:in EF}
    \bfg_{a, m}=-(1-\f{2mr}{|q|^2})d\ub^2+2dr d\ub-2a\sin^2 \th dr d\varphi-\f{4mra\sin^2 \th}{|q|^2}d\ub d\varphi+|q|^2 d\th^2+\f{\Si^2\sin^2\th}{|q|^2}d\varphi^2.
\end{equation}
This is known as the Kerr metric in the Eddington--Finkelstein (EF) coordinates. 

Furthermore, a direct calculation yields
\begin{equation}\label{Kerr coor derivative}
    \begin{aligned}
        e_4(\th)=&0, \ \ \ e_4(\varphi)=\f{2a}{|q|^2},  & \quad  e_4(r)&=\f{\De}{|q|^2}, &\quad e_4(\ub)=&\f{2(r^2+a^2)}{|q|^2}, \\ 
        \nab (\ub)=&(0, \f{a\sin \th}{|q|}), &\quad \nab(\th)&=(\f{1}{|q|}, 0), &\quad \nab (\varphi)=&(0, \f{1}{|q|\sin \th}), \\
        (\nab_1 e_2)_1=&0,  &\quad (\nab_2 e_1)_2&=\f{r^2+a^2}{|q|^3}\cot \th.
    \end{aligned}
\end{equation}
We can also compute the precise values of the corresponding geometric quantities as below:
 \begin{equation}\label{Eqn for Ricci Cur}
     \begin{aligned}
         \trch=&\f{2r \De}{|q|^4}, & \atrch=&\f{2a\cos \th \De}{|q|^4}, &  \eta=&\zeta=\f{a\sin \th}{|q|^3}(-a\cos \th, r), \\
         \trchb=&-\f{2r}{|q|^2}, & \atrchb=&-\f{2a\cos \th}{|q|^2}, &\etab=&-\f{a\sin \th}{|q|^3}(a\cos \th, r), \\
         \omega=&-\f12\pr_r(\f{\De}{|q|^2}),  &
         \rho=&-\f{2mr(r^2-3a^2\cos^2 \th)}{|q|^6},
         & \sigma=&\f{2ma\cos\th(3r^2-a^2\cos^2 \th)}{|q|^6},\\
          \omegab,\chih&, \chibh, \xi,\xib=0, & \a, \b, &\ab, \bb=0.
     \end{aligned}
 \end{equation}

\section{Characteristic Initial Data}\label{Sec:Char Data}
\subsection{Admissible Kerr Black Hole Formation Initial Data}\label{Subsec:admiKerr}
 In this paper, we consider the characteristic initial data along intersected outgoing and incoming null hypersurfaces. Specifically, we employ the double coordinates $(u, v , \th^1, \th^1)$ with the spacetime metric
\begin{equation*}
    \bfg=-2\O^2(du\otimes dv+dv \otimes du)+g_{ab}(d\theta^a-b^a dv)\otimes (d\theta^B-b^b dv).
\end{equation*}
Denote $H_u$ and $\Hb_{v}$ to be the level sets of the optical functions $u$ and $v$, respectively, and let $S_{u, v}$ be the intersection of $H_u$ and $\Hb_{v}$. Now we prescribe the initial data along $H_{u_0}=\qty{u=u_0, \ u_0\le v<\infty}$ with $-u_0\gg 1$ as follows:
\begin{itemize}
    \item We put the Minkowskian initial data along $H_{u_0}\cap\qty{u_0\le v\le 0}$. Then this implies that the data along $\Hb_{0}=\qty{v=0, \ u\in[u_0, 0)}$ is also Minkowskian.
    \item Along $H_{u_0}\cap\qty{v\ge0}$, within the short-pulse region, namely, in $H_{u_0}\cap \qty{0\le v\le \de}$ with $\de=A^{-1}>0$ and $A>0$ being sufficiently large, we define the associated null frame by
\begin{equation}\label{Eqn:null frame pul}
    e_1, e_2\in TS_{u, v}, \qquad e_3=-2\O\bfD v=\O^{-1}\pr_u, \qquad e_4=-2\O \bfD u=\O^{-1}(\pr_{v}+b^a\pr_{\th^a}).
\end{equation}
    We assign the isotropic scale-critical short-pulse data by letting\footnote{In this subsection and \Cref{initial layer}, all estimates are understood to also hold true for higher derivatives $(|u|\nab)^{i}$ up to $N$-th order. Here $\nab$ represents the covariant derivative on $S_{u, v}$ and $N$ is a large integer.}\textsuperscript{,}\footnote{The requirement \eqref{Cond:int chihsq} can be achieved following the same approach addressed in \cite{An:AH}, with $\chih$ sophisticatedly prescribed in $H_{u_{0}}\cap \{ 0\le v \le \de\}$.}
    \begin{equation}\label{Cond:int chihsq}
        \int_0^{\de} |u_0|^2|\chih|^2 dv=4m_0+O(\epsilon_1).
    \end{equation}
    Here $4m_0=1$ and $\epsilon_1>0$ is a small constant satisfying $\de A^{\f12} \ll \epsilon_1\ll 1$. This implies that, along $S_{u_0, \de}$, it holds 
    \begin{equation}\label{Est:trch rho ID}
    |\trch-\f{2}{|u_0|}+\f{4m_0}{|u_0|^2}|+|\omega|\lesssim \f{\delta A^{1/2}}{|u_0|^2}, \qquad |\rho-\f12\chih\c\chibh+\f{2m_0}{|u_0|^3}|\lesssim \f{\delta A^{1/2}}{|u_0|^3}.
    \end{equation}
    For the remaining Ricci coefficients and curvature components, we prescribe the same data as in \cite{A-L}, i.e., they obey the below $L^{\infty}$ bounds\footnote{Note that within the double-null foliation, we have $\zeta=\f12(\eta-\etab)$ and $\xi=\xib=0$.}
    \begin{equation}\label{An-Luk estimate ID}
\begin{aligned}
|\a|\lesssim& \f{\delta^{-1}A^{\f{1}{2}}}{|u_0|}, &\qquad |\b|\lesssim& \f{ A^{\f{1}{2}}}{|u_0|^2}, &\qquad |\rho|\lesssim& \f{\delta A}{|u_0|^3}, \\ 
|\sigma+\f12 \chibh\wedge \chih|\lesssim& \f{\de A^{\f12}}{|u_0|^3},  &\qquad|\bb|\lesssim& \f{\delta^2 A^{\f{3}{2}}}{|u_0|^4}, &\qquad |\ab|\lesssim& \f{\delta^3 A^2}{|u_0|^5},  \\
|\chih|\lesssim& \f{A^{\f{1}{2}}}{|u_0|^2}, &\qquad |\chibh|\lesssim& \f{\delta A^{\f{1}{2}}}{|u_0|^2}, &\qquad |\trch|\lesssim& \f{1}{|u_0|}, \qquad\quad |\eta|\lesssim \f{\delta A^{\f{1}{2}}}{|u_0|^2}, \\
	|\etab|\lesssim& \f{\delta A^{\f{1}{2}}}{|u_0|^2}, &\qquad |\omega|\lesssim& \f{1}{|u_0|}, &\qquad
	|\omegab|\lesssim& \f{\delta A^{\f{1}{2}}}{|u_0|^2}, \qquad |\trchb+\f{2}{|u_0|}|\lesssim \f{\delta A^{\f{1}{2}}}{|u_0|^2}.
\end{aligned}
\end{equation}
Furthermore, while keeping \eqref{Cond:int chihsq} unchanged, we require 
\begin{equation}\label{Eqn:improved chih a ID}
    \de^{-1}|\chih|+|\a|\lesssim \f{\de A^{\f12}}{|u_0|} \qquad \text{on} \quad S_{u_0, \de}.
\end{equation}
\item After the short pulse, along $H_{u_0}\cap \qty{v\ge \de}$ we prescribe converging-to-Kerr data with decay rates such that
\begin{equation}\label{Est:out ID}
    \mathbb{E}^N_{0,out}\lesssim \epsilon_1,
\end{equation}
where relative to the new null frame
\begin{equation*}
     (e_{out})_a=e_a, \qquad  (e_{out})_3=\O_{\SS} e_3, \qquad (e_{out})_4=\O_{\SS}^{-1} e_4,
\end{equation*}
the initial outgoing energy norm $\mathbb{E}^N_{0,out}$ is defined as
\begin{equation}\label{Eqn:out ID1}
\begin{split}
\mathbb{E}^N_{0,out}\coloneqq& \sup\limits_{v\in[\de,\infty)}\|(r\nab)^{\le N}(r^{\f{7+\de_B}{2}} \a_{out}, r^{\f{7+\de_B}{2}}\beta_{out}, r^3(\rho_{out}-\rho_{\SS}), r^3 \sigma_{out}, r^2\bb_{out}, r\ab_{out})\|_{L^\infty(S_{u_0, v})}\\
&+\sup\limits_{v\in[\de,\infty)}\|(r\nab)^{\le N}\Big(r^2 \chih_{out}, r\chibh_{out}, r^2(\trch_{out}-\trch_{\SS}), r^2(\trchb_{out}-\trchb_{\SS}), \\ & r^2 (\eta_{out}, \etab_{out}, \zeta_{out}),  r^3(\omega_{out}-\omega_{\SS}), r^{-2}(g-r^2\mathring{g})\Big)\|_{L^\infty(S_{u_0, v})}
\end{split}
\end{equation}
with $r$ being the area radius of $S_{u, v}$, $0<\de_B\ll 1$ being a small constant, $\mathring{g}$ denoting the standard metric on the round sphere $\ms$ and
\begin{equation}\label{Eqn:linearizeSchw}
   \O_{\SS}\coloneqq \sqrt{1-\f{2m_0}{r}}, \qquad \trch_{\SS}=-\trchb_{\SS}\coloneqq\f{2}{r}\O_{\SS},\qquad  \o_{\SS}\coloneqq -\f{m_0}{2r^2}\O_{\SS}^{-1},\qquad  \rho_{\SS}\coloneqq -\f{2m_0}{r^3}.
\end{equation}
\end{itemize}
\begin{remark}
   Notice that $\omegab$ does not appear in $\mathbb{E}^N_{0,out}$. In \Cref{Subsec:out initial layer} we will fix a choice of gauge from a distant sphere $S_{u_0, v_{*}}$ to ensure that $\omb$ converges to its Schwarzschild value appropriately via a last slice argument.
\end{remark}

In view of aforementioned null frames and the estimates of geometric quantities, we thus define
\begin{definition}\label{Def:admiKerr}
The characteristic initial data along $H_{u_0}$ is called a admissible Kerr formation initial data provided that all above estimates in  \eqref{Est:trch rho ID}, \eqref{An-Luk estimate ID}, \eqref{Eqn:improved chih a ID} and \eqref{Est:out ID} hold true.
\end{definition}
\begin{remark}
 We also provide smooth examples of these admissible Kerr black hole formation initial data in \Cref{Apx:example}. Notably, in this construction, the initial data at later stage can be freely prescribed.
\end{remark}

\subsection{Initial Data Layer}\label{initial layer}
In this section, we solve the initial data layers in the neighborhood of the admissible Kerr formation initial data prescribed along $H_{u_0}\cup \Hb_{0}$ as defined in \Cref{Subsec:admiKerr}.
\vspace{2mm}

Recall that regarding the admissible Kerr formation initial data, in the short-pulse region $\{u_0\leq u\leq -v A/4, \ 0\leq v\leq \de\}$, we prescribe the initial data which are consistent with \cite{A-L}. Note that in this paper we have $\delta=A^{-1}$. Applying the results in \cite{A-L} by An-Luk, we deduce the existence of solutions to Einstein vacuum equations in the short-pulse region and we have the below estimates:
\begin{equation}\label{An-Luk estimate}
\begin{aligned}
|\a|\lesssim& \f{\delta^{-1}A^{\f{1}{2}}}{|u|}, &\qquad |\b|\lesssim& \f{ A^{\f{1}{2}}}{|u|^2}, &\qquad |\rho|\lesssim& \f{\delta A}{|u|^3}, \\ 
|\sigma+\f12 \chibh\wedge \chih|\lesssim& \f{\de A^{\f12}}{|u|^3},  &\qquad|\bb|\lesssim& \f{\delta^2 A^{\f{3}{2}}}{|u|^4}, &\qquad |\ab|\lesssim& \f{\delta^3 A^2}{|u|^5},  \\
|\chih|\lesssim& \f{A^{\f{1}{2}}}{|u|^2}, &\qquad |\chibh|\lesssim& \f{\delta A^{\f{1}{2}}}{|u|^2}, &\qquad |\trch|\lesssim& \f{1}{|u|}, \qquad\quad |\eta|\lesssim \f{\delta A^{\f{1}{2}}}{|u|^2}, \\
	|\etab|\lesssim& \f{\delta A^{\f{1}{2}}}{|u|^2}, &\qquad |\omega|\lesssim& \f{1}{|u|}, &\qquad
	|\omegab|\lesssim& \f{\delta A^{\f{1}{2}}}{|u|^2}, \qquad |\trchb+\f{2}{|u|}|\lesssim \f{\delta A^{\f{1}{2}}}{|u|^2}.
\end{aligned}
\end{equation}
Moreover, proceeding similarly as in the proof of Lemma 3 in \cite{L-Y}, we can obtain more precise estimates for $\trch, \omega$ and $\rho$ along $S_{u_0, \de}$:
\begin{equation*}
    |\trch-\f{2}{|u|}+\f{4m_0}{|u|^2}|+|\omega+\f{m_0}{2|u|^2}-\f{m_0}{2|u_0|^2}|\lesssim \f{\delta A^{1/2}}{|u|^2}, \qquad |\rho-\f12\chih\c\chibh+\f{2m_0}{|u|^3}|\lesssim \f{\delta A^{1/2}}{|u|^3}.
\end{equation*}
Here $4m_0=1$.

Then in view of the improved bounds for $\chih$ and $\a$ on $S_{u_0, \de}$ as in \eqref{Eqn:improved chih a ID}, utilizing
\begin{align*}
\nab_3\chih+\f12\trch \chih=&\nab\hot\eta+2\omb\chih-\f12\trch\chibh+\eta\hot\eta,   
\end{align*}
and hyperbolic estimates in \eqref{An-Luk estimate}, we can propagate the estimates for $\chih$ along $e_3$ direction and obtain the improved bound
\begin{equation*}
\begin{split}
|\chih|\lesssim \f{\delta A^{1/2}}{|u|} \qquad \textrm{for any} \quad u_0\leq u\leq -\delta A  \quad \text{and}\quad  v=\de.
\end{split}
\end{equation*}
Combining with \eqref{An-Luk estimate} and the Codazzi equation
\begin{equation*}
    \div \chih=\f12 \nab \trch-\f12(\eta-\etab)\c (\chih-\f12 \trch)-\b,
\end{equation*}
this further implies better controls of $\si$ and $\b$ along $\Hb_{\delta} \cap \qty{u_0 \leq u \leq -\delta A }$:
\begin{align*}
    |\si|\lesssim&|\sigma+\f12 \chibh\wedge \chih|+|\chibh|\c|\chih| \lesssim \f{\de A^{\f12}}{|u|^3}, \\
    |\b|\lesssim& |\div\chih|+|\nab \trch|+(|\eta|+|\etab|)\c\qty(|\chih|+|\trch|)\lesssim \f{\de A^{\f12}}{|u|^3}.
\end{align*}
Therefore, employing the null Bianchi equation for $\a$
\begin{equation*}
    \nab_3 \a+\f12 \trch\a=\nab\hot\b+4\omegab \a-3(\chih\rho+\dual\chih\sigma)+(\zeta+4\eta)\hot\b,
\end{equation*}
by virtue of the improved estimates for $\chih, \b$ and \eqref{Eqn:improved chih a ID} we arrive at
\begin{equation*}
|\a|\lesssim \f{\delta A^{1/2}}{|u|} \qquad \textrm{along} \quad \Hb_{\delta} \cap \qty{u_0 \leq u \leq -\delta A }.
\end{equation*}

Consequently, in light of all the estimations above, along the incoming null hypersurface $\Hb_{\delta} \cap \qty{u_0 \leq u \leq -\delta A }$, our data satisfy the following condition: 
\begin{equation}\label{Est:E in N}
\mathbb{E}^{N-3}_{0, in}(\de) \lesssim \epsilon_1,
\end{equation}
where 
\begin{equation*}
\begin{split}
\mathbb{E}^N_{0, in}(v)=&\sum_{i=0}^{N}\sup\limits_{u\in[u_0, -\delta A]}\|(r\nab)^i ( \a, \beta, \rho+\f{2m_0}{|u|^3},  \sigma, \bb, \ab\|_{L^{\infty}(S_{u, v})}+\sum_{i=0}^{N}\sup\limits_{u\in[u_0, -\delta A]}\|(r\nab)^i(\chih, \\&\chibh, \trch-\f{2}{|u|}(1-\f{2m_0}{|u|}), \trchb+\f{2}{|u|}, \eta, \etab, \zeta, \omega+\f{m_0}{2|u|^2}-\f{m_0}{2|u_0|^2}, \omb)\|_{L^{\infty}(S_{u, v})}.
\end{split}
\end{equation*}

\subsubsection{Incoming Initial Layer}
Combining the initial energy bounds as in \eqref{Est:out ID}, we can establish the existence of EVEs in the region $\{ u_0 \le u\le -\de A, \ \de \le v\le \de+d\}$ by utilizing a standard local result.\footnote{See for example \cite{Luk}.} Furthermore, the solution is sufficiently close to the Schwarzschild spacetime with mass $m_0$ in the following manner:
\begin{proposition}\label{Prop:Est incoming initial layer}
    Assume $\mathbb{E}^{N}_{0,out}+\mathbb{E}^{N-3}_{0, in}(\de) \lesssim \epsilon_1$ with $N\geq 20$ and with sufficiently small $\epsilon_1>0$. Then there exists a parameter $d$ small enough (independent of $\epsilon_1, \de, A$), such that the solution of Einstein vacuum equations exists in the region $\{u_0\leq u\leq -\de A ,\  v_0 \leq v\le v_0+d \}$, with the detailed hyperbolic estimates satisfying
    \begin{equation*}
	\begin{split}
		&\max\limits_{0\leq i\leq N-5}\sup\limits_{u, v} \|(r\nab)^i (\chih, \chibh, \trch-\f{2}{r}(1-\f{2m_0}{r}), \trchb+\f{2}{r}, \eta, \etab, \zeta, \omega+\f{m_0}{r^2}, \omb)\|_{L^{\infty}(S_{u, v})} \lesssim \epsilon_1, \\
		&\max\limits_{0\leq i\leq N-6}\sup\limits_{u, v} \|(r\nab)^i ( \a, \beta, \rho+\f{2m_0}{r^3},  \sigma, \bb, \ab)\|_{L^{\infty}(S_{u, v})} \lesssim \epsilon_1.
	\end{split}
	\end{equation*}
    Here $r$ denotes the area radius of $S_{u, v}$.
\end{proposition}

\subsubsection{Outgoing initial layer}\label{Subsec:out initial layer}
In this section, we construct a solution to EVEs in a small neighborhood of the outgoing initial null cone $H_{u_0}\cap \{ v_0 \le v<\infty\}$ with $v_0\coloneqq \de$, namely, the thin region $\{u_0\leq u\leq u_0+d ,\  v_0 \leq v<\infty \}$. We call this strip-type region the outgoing initial layer. Note that the outgoing initial layer is no longer a bounded region in $v$-coordinate, hence we have to demonstrate the semi-global existence of EVEs in this region. This is different from applying the local existence result near the incoming initial null cone $\Hb_{\de}\cap \{u_0 \le u\le -\de A\}$ in the preceding section.
\vspace{2mm}

For the convenience of writing, we use a signature $s_1$ introduced by Klainerman--Rodnianski in \cite{KR:Trapped} to carry out $r$-weighted estimates in this section. For each Ricci coefficient and curvature component, Klainerman--Rodnianski \cite{KR:Trapped} introduced a number $s_1$ associated with it. More precisely, for a geometric quantity $\phi$, they defined
\begin{equation}\label{signature}
s_1(\phi)\coloneqq N_4(\phi)+0.5 \cdot N_a(\phi)+0\cdot N_3(\phi)-1.
\end{equation}
Here $N_3(\phi), N_4(\phi)$ and $N_a(\phi)$ represent the number of times that $e_3, e_4$ and $e_a$ appear in the definition of $\phi$, respectively. Gathering these signatures, a signature table is given as below:\footnote{We do not include $\xi, \xib$ in the table, as $\xi=\xib=0$ in the double null foliation.}
\begin{table}[htb]
\centering
\begin{tabular}{|l|l|l|l|l|l|l|l|l|l|l|l|l|l|}
\hline
    & $\a$ & $\b$  & $\rho$ & $\sigma$ & $\bb$ & $\ab$ & $\chi$ & $\omega$ & $\zeta$ & $\eta$ & $\etab$ & $\chib$ & $\omegab$ \\ 
$s_1$ & 2  & 1.5 & 1    & 1      & 0.5 & 0   & 1    & 1      & 0.5   & 0.5  & 0.5   & 0   & 0 \\\hline   
\end{tabular}
\end{table}

Recalling the definition of admissible Kerr black hole formation initial data stated in \Cref{Def:admiKerr}, we have that along $u=u_0$, the outgoing initial energy norm satisfies
\begin{equation*}
\mathbb{E}^{N}_{0,out} \lesssim \epsilon_1.
\end{equation*}
Our main goal of this section is to prove the below proposition:
\begin{proposition}[Estimates of outgoing initial layer]\label{outgoing initial layer}
	Assume $\mathbb{E}^{N}_{0,out}+\mathbb{E}^{N-3}_{0, in}(\de) \lesssim \epsilon_1$ with $N\geq 20$. Then there exists a parameter $d$ small enough (independent of $\epsilon_1$), such that the solution of Einstein vacuum equations exists in the region $\{u_0\leq u\leq u_0+d ,\  v_0 \leq v<\infty \}$. Within this region, employing a new double-null foliation $(u'_{out}, v_{out})$ and the associated new null frame $\qty{(e_{out})_{\mu}'}$, the following bounds for Ricci coefficients $\Gamma_{out}'$ (except $\chih_{out}', \trch_{out}'$) and for curvature components $R_{out}'$ (except $\a_{out}',\b_{out}'$) are satisfied$\,$\footnote{Here  $\trchb_{out}', \omegab_{out}', \rho_{out}'$ should be considered as the renormalized ones, namely, 
    \begin{equation*}
        \trchb_{out}'+\f{2}{r_{out}'}\sqrt{1-\f{2m_0}{r_{out}'}}, \qquad \omb_{out}'-(1-\f{2m_0}{r_{out}'})^{-\f12}\c \f{m_0}{2r_{out}'^2}, \qquad \rho_{out}'+\f{2m_0}{r_{out}'^3}.
    \end{equation*}
    } 
	\begin{equation*}
	\begin{split}
		&\sup\limits_{u'_{out}, v_{out}} \|(r'_{out})^{2s_1( \Gamma_{out}')+1} (r'_{out}\nab'_{out})^{\le N-5} \Gamma_{out}'\|_{L^{\infty}(S'_{u'_{out}, v_{out}})} \lesssim \epsilon_1,\\
		&\sup\limits_{u'_{out}, v_{out}} \|(r'_{out})^{2s_1( R_{out}')+1} (r'_{out}\nab'_{out})^{\le N-6} R_{out}'\|_{L^{\infty}(S'_{u'_{out}, v_{out}})} \lesssim \epsilon_1.
	\end{split}
	\end{equation*}
	For $\chih_{out}', \trch_{out}'$ and $\a_{out}', \b_{out}'$, there hold
	\begin{equation*}
	\begin{split}
	&\sup\limits_{u'_{out}, v_{out}} \| (r'_{out}\nab'_{out})^{\le N-5} \Big((r'_{out})^2\chih_{out}', (r'_{out})^2\qty(\trch_{out}'-\f{2}{r'_{out}}\sqrt{1-\f{2m_0}{r'_{out}}}),\\&(r'_{out})^2\qty(\trchb_{out}'+\f{2}{r'_{out}}\sqrt{1-\f{2m_0}{r'_{out}}})\Big)\|_{L^{\infty}(S'_{u'_{out}, v_{out}})} \lesssim \epsilon_1, \\
	&\sup\limits_{u'_{out}, v_{out}} \| (r'_{out})^{\frac{7+\de_{B}}{2}}(r'_{out}\nab'_{out})^{\le N-6} (\a_{out}', \b_{out}') \|_{L^{\infty}(S'_{u'_{out}, v_{out}})} \lesssim \epsilon_1.
	\end{split}
	\end{equation*}
 Moreover, we have the below estimates for the metric coefficients $\O'_{out}$ and $(b'_{out})^{A}$
\begin{equation}
   \sup\limits_{u'_{out}, v_{out}} \| (r'_{out}\nab'_{out})^{\le N-5}( r'_{out}\log \Big((1-\f{2m_0}{r'_{out}})^{-\f12}\O'_{out}\Big), r'^2(b'_{out})^{a} )\|_{L^{\infty}(S'_{u'_{out}, v_{out}})}\lesssim \epsilon_1.
\end{equation}
And the transition coefficients between $( e_{\mu})$ to $\qty((e'_{out})_{\mu})$ obey
\begin{equation}\label{Est:transition out}
    \sup\limits_{(u,v)\in[u_0, u_0+d]\times [v_0,v_0+d]}
    |(r'_{out}\nab'_{out})^{\le N-5} \Big(f_{out}, \fb_{out}, \log(\O_{\SS}\la_{out})\Big)|\lesssim\epsilon_1.
\end{equation}
 Here $S'_{u'_{out}, v_{out}}$ denotes the intersection of level sets of $u'_{out}$ and level sets of $v_{out}$, with $r'_{out}$ being its area radius and $\nab'_{out}$ being its induced covariant derivative.
\end{proposition}
\begin{remark}
    The upper bound of the initial energy $\epsilon_1$ is not necessarily required to be small in this proposition. We only need the smallness of the characteristic length $d$ in the proof.
\end{remark}
	 Inspired by \cite{L-Z,Shen:Minkext}, we conduct the method of $r^p$-weighted estimates and the approach of the last slice argument. For notational simplicity, we omit the subscripts ${}_{out}$ in all objects (i.e., coordinates, null frames, geometric quantities) throughout this subsection. 
 
 For each $\phi\not\in \{\a', \b', \chih', \trch',\trchb'\}$, we set $p(\phi)\coloneqq 2s_1(\phi)+1$, where $s_1(\phi)$ denotes the signature of $\phi$ as in \eqref{signature}. Then we assign
	\begin{equation*}
	p(\a',\b')=\f{7+\de_B}{2}, \qquad p(\chih', \trch', \trchb')=2.
	\end{equation*}

For any given $v_{\ast}>v_0$, with the given initial data along $H_{u_0}\cup \Hb_{v_0}$, we consider a thin region $D=[u_0, u_0+d]\times[v_0, v_{*}]$. Denoting $\Hb_{v_{*}}$ to be the last slice of our constructed solution in this subsection, we proceed to construct a new outgoing optical function $u'$ as follows:
\begin{itemize}
    \item We identify a canonical foliation along $\Hb_{v_*}$ as introduced in \cite{KNI:book}, which consists of two functions $u'_{*}$ and $\O'$ defined on $\Hb_{v_*}$ satisfying the following system of equations
\begin{equation}\label{last slice equation for log Omega}
\begin{aligned}
\Delta'\log \O'=&\f{1}{2}\qty[\div' \etab'+\f{1}{2}\qty(\chih'\c\chibh'-\ov{\chih'\c\chibh'})-(\rho-\ov{\rho})], \qquad \ov{\log \O'}=\f12\log(1-\f{2m_0}{r'}), \\
-\bfD v'(u_*')=&(2\O')^{-1}, \qquad u_*'\big|_{H_{u_0}\cap\Hb_{v_*}}=u_0.
\end{aligned}
\end{equation}
 Here all the geometric quantities and differential operators are defined with respect to the level sets of $u'_{*}$ along $\Hb_{v_*}$, denoted by $S'_{u'_{*}, v_{*}}$, with $r'$ being its area radius and $\overline{f}$ representing the average of $f$ on $S'_{u'_{*}, v_{*}}$.
\item Then we solve the eikonal equation in $D$:
\begin{equation*}
    \bfg^{\a\b}\pr_{\a} u'\pr_{\b}u'=0 \qquad  \text{with} \quad u'\big|_{\Hb_{v_*}}=u'_{*}.
\end{equation*}
The associated double null foliation are then fixed by
\begin{equation}\label{Eqn:null frame out'}
    e'_1, e'_2\in TS'_{u', v}, \qquad e'_3=-2\O'\bfD v=\O'^{-1}\pr_{u'}, \qquad e'_4=-2\O' \bfD u'=\O'^{-1}(\pr_{v}+b'^a\pr_{\th'^a}).
\end{equation}
Here $(\th'^1, \th'^2)$ is a local chart on $S'_{u', v}$ satisfying $e'_3(\th'^a)=0$.
\end{itemize}
Based on the new double null foliation $(u', v')$ in $D$, we make the bootstrap assumption\footnote{Here $\tilde{\Gamma}'_p$ corresponds to the renormalized Ricci coefficient.}
\begin{equation}\label{BA:gamma}
    \max\limits_{D}\|(r'\nab')^{\le N-5} \qty(r'^{p} \tilde{\Gamma}'_p, (r'\nab')^{\le 1}r'\log \qty((1-\f{2m_0}{r'})^{-\f12}\O'))\|_{L^{\infty}(S'_{u' ,v})}\le \epsilon \qquad \text{with} \quad \epsilon=d^{-\f14}\epsilon_1.
\end{equation}
We also denote $\overline{f}$ to be the average of $f$ on $S'_{u', v}$ and
$\check{f}\coloneqq f-\overline{f}$ for any scalar function $f$, while for any tensor field $\phi$, we set $\check{\phi}=\phi$.
\vspace{2mm}

To close the bootstrap argument, we first control the fluxes of curvature components. Consider the pair $(\check{R}'_1, \check{R}'_2)\in\{(\a',\b'), \qty(\b', (\check{\rho}', \sigma')), \qty((\check{\rho}', \sigma'), \bb'), (\bb', \ab') \}$ and it satisfies the following schematic null Bianchi equations
\begin{equation}\label{Eqn:bianchi no a}
\begin{split}
&\nab'_3 \check{R}'_{p_1}=\mathcal{D'} \check{R}'_{p_2}+N_1, \\
&\nab'_4 \check{R}'_{p_2}+\frac{\max(p_2, 2s_1(R'_{p_2})+1)}{2}\trch' \check{R}'_{p_2}=\mathcal{D'}^\ast \check{R}'_{p_1}+N_2.
\end{split}
\end{equation}
Here we denote $\mathcal{D'}, \mathcal{D'}^\ast$ to be the differential operator on $S'_{u', v}$,  the $L^2(S'_{u', v})$ adjoint operator of $\mathcal{D}'$, respectively. And the nonlinear terms $N_1, N_2$ take the form
\begin{align*}
N_1=\check{\Gamma}'_{p_3} R'_{p_4}+\Gamma'_{p_3'} \check{R}'_{p_4'}, \qquad
N_2=\check{\Gamma}'_{p_5} R'_{p_6}+\Gamma'_{p_5'} \check{R}'_{p_6'}
\end{align*}
with $p_1\le p_2+1$, $p_1\leq \min(p_3+p_4,p'_3+p'_4)$, $p_2+3\leq \min(p_5+p_6, p'_5+p'_6)$.\footnote{Note that $N_2$ does not involve $\trch'$.}

Multiplying by $r^{\prime p_2}$ on both sides, the equations in \eqref{Eqn:bianchi no a} can be transferred into
\begin{equation*}
\begin{split}
&\nab'_3 (r^{\prime p_2}\check{R}'_{p_1})=\mathcal{D'}( r^{\prime p_2} \check{R}'_{p_2})+r^{\prime p_2}N_1+\trchb'\c r^{\prime p_2} \check{R}'_{p_1}, \\
&\nab'_4 (r^{\prime p_2} \check{R}'_{p_2})+\f{\max(0, 2s_1(R'_{p_2})+1)-p_2)}{2}=\mathcal{D'}^\ast (r^{\prime p_2}\check{R}'_{p_1})+r^{\prime p_2}N_2-\frac{p_2}{2}\O^{-1}(\O'\trch'-\overline{\O'\trch'})r^{\prime p_2} \check{R}'_{p_2}.
\end{split}
\end{equation*}
Note the additional nonlinear terms can be collected into $r^{\prime p_2}N_1$ and $r^{\prime p_2}N_2$, respectively.

Employing the integration by parts on $S'_{u',v '}$, we obtain the energy estimates for curvature components in the spacetime region $V=\qty{(u', v)\in D}$:\footnote{Here $H'_{u'}$ denotes the constant $u'$ hypersurface.}
\begin{equation}\label{Est:enenry bianchi pair}
\begin{split}
\|r^{\prime p_2-1} R'_{p_1}\|^2_{L^2(H'_{u'})}+\|r^{\prime p_2-1} R'_{p_2}\|^2_{L^2(\Hb_{v})}\lesssim& \|r^{\prime p_2-1} R'_{p_1}\|^2_{L^2(H'_{u_0})}+\|r^{\prime p_2-1} R'_{p_2}\|^2_{L^2(\Hb'_{v_0})}\\&+\int_{V} r^{\prime 2p_2-2}\qty(|N_1 \check{R}'_{p_1}|+|N_2\check{R}'_{p_2}|).
\end{split}
\end{equation}
The first two terms can be controlled by $\mathbb{E}^{N-3}_{0,out}+\mathbb{E}^{N-3}_{0,int}(\de)$. Using $p_2+1=p_1\le \min(p_3+p_4,p'_3+p'_4)$, the first nonlinear term can be bounded by
\begin{equation*}
\begin{split}
\int_{V} r^{\prime 2p_2-2}|N_1 \check{R}'_{p_1}| 
\lesssim& \qty(\int_{u_0}^{u_0+d}  \|r^{\prime p_1-1} R'_{p_1}\|^2_{L^2(H'_{u'})})^{1/2}\c  \Bigg(\sup\limits_{D}\|r^{\prime p_3}\check{\Gamma}'_{p_3}\|_{L^\infty(S'_{u',v})}\Big(\int_{v_0}^v r^{\prime -2}\| r^{\prime p_4-1}R'_{p_4}\|^2_{L^2(\Hb_{v})}\Big)^{1/2}\\&+\sup\limits_{D}\|r^{\prime p_3'-1}\Gamma'_{p_3'}\|_{L^\infty(S'_{u',v})}\Big(\int_{u_0}^{u_0+d} \| r^{\prime p_4'-1}\check{R}'_{p_4'}\|^2_{L^2(H'_{u'})}\Big)^{1/2}\Bigg) \\
\lesssim& d^{1/2}\sup\limits_{u'}  \|r^{\prime p_2-1} R'_{p_1}\|_{L^2(H'_{u'})}\c \qty(\sup\limits_{v}  \|r^{\prime p_4-1} \check{R}'_{p_4}\|_{L^2(\Hb_{v})}+\sup\limits_{u'}\|r^{\prime p'_4-1} \check{R}'_{p'_4}\|_{L^2(H'_{u'})}+\epsilon).
\end{split}
\end{equation*}

For the second nonlinear term, the part $\check{\Gamma}'_{p_5} {R}'_{p_6}$ can be handled similarly to the estimation of $\check{\Gamma}'_{p_3} R'_{p_4}$ as above. For the remaining part $\Gamma'_{p'_5} \check{R}'_{p'_6}$, observing that
\begin{equation*}
    |r'^{p'_5-1}\Gamma'_{p_5'}|\le |r'^{p'_5-1}\tilde{\Gamma}'_{p_5'}|+1 
    \lesssim \epsilon r'^{-1}+1\lesssim 1,
\end{equation*}
we derive
\begin{align*}
&\int_{V} r^{\prime 2p_2-2}|\Gamma'_{p_5'}| \c|\check{R}'_{p_6'}\check{R}'_{p_2}| \\
\lesssim& \qty(\int_{v}^{v} r^{\prime -2} \|r^{\prime p_2-1} R'_{p_2}\|^2_{L^2(\Hb_{v})})^{1/2}\c\qty(\int_{v_0}^v \qty(r'^{p'_5-2}\Gamma'_{p_5'})^2\| r^{p_6'-1}R'_{p_6'}\|^2_{L^2(\Hb_{v})})^{1/2}\\
\lesssim& \epsilon\c \qty(\int_{v}^{v} r^{\prime -2} \|r^{\prime p_2-1} R'_{p_2}\|^2_{L^2(\Hb_{v})})^{1/2}\c \qty(\int_{v_0}^v r^{\prime -2}\| r^{\prime p_2'-1}R'_{p_2'}\|^2_{L^2(\Hb_{v})})^{1/2}.
\end{align*}
Here in both last step we apply the bootstrap assumption. 
\vspace{2mm}

Hence, all the terms on the right of \eqref{Est:enenry bianchi pair} can be controlled by $\epsilon_1$ owing to the smallness of $d, \epsilon$ and the fact $d^{\f12}\epsilon=d^{\f14}\epsilon_1\ll \epsilon_1$. The energy estimates for higher order derivatives of $R'$ can be derived in a similar manner by commuting \eqref{Eqn:bianchi no a} with $(r'\nab')^{\le N-3}$ and by utilizing the product estimate and the Sobolev embedding to deal with nonlinear terms. As a result, regarding the null Bianchi equations as transport equations, applying the estimates of fluxes of $R'$, the corresponding $L^2(S'_{u', v})$ bounds readily follow with one $r'\nab'$ derivative loss.
\begin{remark}
  It is worth mentioning that as the 2-spheres $S'_{u', v_0}$ along the initial incoming hypersurface $\Hb_{v_0}$ does not coincide with the original 2-sphere foliation $S_{u, v_0}$ along $\Hb_{v_0}$, we also need to derive estimates of curvature components for the initial data along $\Hb_{v_0}$ associated with the null frame $(e_{\mu}')$. This can be achieved by a initialization procedure via deriving appropriate controls of transition coefficients between the original null frame $(e_{\mu})$ and $(e'_{\mu})$ as in \eqref{Est:transition out}. See for example the detailed steps in \cite{KNI:book,L-Z,Shen:Minkext}. We omit the proof in the present paper.
\end{remark}

We proceed to derive the weighted estimates for Ricci coefficients. For $\check{\Gamma}'\in \{\chibh',  \trchb'-\overline{\trchb'}, \etab', \omega'-\overline{\omega'} \}$, we employ the null structure equations along $e'_3$ direction and they can be schematically written as
\begin{equation*}
\nab'_3 (r'^p\check{\Gamma}'_p)=r'^p \underline{R}'_{p'}+r'^p \check{\Gamma}'_{p_1} \Gamma'_{p_2},
\end{equation*}
where $\underline{R}'\in \{\rho'-\overline{\rho'}, \sigma, \bb', \ab' \}$ and $p\leq \min(p', p_1+p_2)$. By applying Gr\"{o}nwall's inequality, we obtain
\begin{equation*}
\|r^{\prime p-1} \check{\Gamma}'_p\|_{L^2(S'_{u', v})}\lesssim \|r^{\prime p-1} \check{\Gamma}'_p\|_{L^2(S'_{u_0, v})}+\int_{u_0}^{u_0+d} \|r^{\prime p-1} \underline{R}'_{p'}\|_{L^2(S'_{u', v})}du'+\int_{u_0}^{u_0+d} \|r^{\prime p-1} \check{\Gamma}'_{p_1} \Gamma'_{p_2}\|_{L^2(S'_{u', v})}du'.
\end{equation*}
On the right, the first term can be controlled by $\mathbb{E}^{N}_{0,out} $ and the second term can be bounded by the flux of curvature components derived previously. The last nonlinear term can be estimated using bootstrap assumptions \eqref{BA:gamma} and the smallness of $d$. Commuting with $(\nab')^{\le N-3}$, repeating this process, we can then derive the estimates for higher order derivatives of $\Ga'$. 

For the remaining Ricci coefficients $\trch',\chih',\eta'$ and $\omegab'$, we appeal to the corresponding null structure equations along $e'_4$ direction, i.e.,
\begin{equation}\label{e4 eqn for trch and chih}
    \nab_4'\trch'+\f12(\trch')^2=-|\chih'|^2-2\o' \trch', \qquad \nab_4'\chih'+\trch'\chih'=-2\o \chih'-\a',
\end{equation}
\begin{align}\label{e4 eqn for eta and omegab}
&\nab'_4 \eta'=-\chi'\cdot (\eta'-\etab')-\beta',\\
&\nab'_4 \omegab'=2\omega' \omegab'+\f{3}{4} |\eta'-\etab'|^2-\f{1}{4} (\eta'-\etab')\cdot (\eta'+\etab')-\f{1}{8}|\eta'+\etab'|^2+\f{1}{2}\rho'. \label{e4 eqn for eta and omegab2}
\end{align}

To obtain the desired decay estimates for above geometric quantities, we need to integrate above equations backwards from the last slice $\Hb_{v_*}$. We start with the estimates of $\trch',\chih',\eta',\omb'$ along $\Hb_{v_*}$.  Introduce the mass aspect function
\begin{equation*}
\mu'\coloneqq-\div' \eta'+\f{1}{2}\chih' \cdot \chibh'-\rho'.
\end{equation*}
In view of \eqref{last slice equation for log Omega} and the fact $\eta'+\etab'=2\nab'\log\O'$, applying the Gauss equation on $S'_{u', v}$, we get 
\begin{equation}\label{Cond:check mu}
    \check{\mu}'=0 \qquad \text{along} \quad \Hb'_{v_\ast}.
\end{equation}
Further applying the Hodge system
\begin{equation*}
\begin{split}
&\div' \eta'=-\check{\mu}'+\f{1}{2}(\chih'\cdot \chibh'-\overline{\chih'\cdot \chibh'})-\check{\rho}', \\
&\curl' \eta'=\sigma'-\f{1}{2}\chih' \wedge \chibh',
\end{split}
\end{equation*}
together with estimations of $\chibh',\check{\rho}', \si'$ and \eqref{Cond:check mu}, we then bound $\eta'$ via
\begin{equation}\label{Est:eta med}
    \|(r'\nab')^{\le 1} r'\eta'\|_{L^2(S'_{u', v_*})}\lesssim \epsilon_1+\epsilon_1 \|r'\chih'\|_{L^2(S'_{u', v_*})}.
\end{equation}
Recalling the transport equation of $\chih'$ along $e_3'$, we have
\begin{align*}
\nab'_3\chih'+(\f12\trchb'-2\omb')\chih'=&\nab'\hot\eta'-\f12\trch'\chibh+\eta'\hot\eta'.
\end{align*}
In light of \eqref{BA:gamma} and the condition \eqref{Cond:check mu}, by utilizing Gr\"{o}nwall's inequality and the smallness of $d$, along $\Hb_{v_*}$ we then infer
\begin{align*}
    \|r'\chih'\|_{L^2(S'_{u', v_*})}\lesssim \epsilon_1+d\|(r'\nab') r'\eta'\|_{L^2(S'_{u', v_*})}.
\end{align*}
Incorporating with \eqref{Est:eta med} and noting that $d\ll1$, we hence deduce
\begin{equation*}
     \|r'\chih'\|_{L^2(S'_{u', v_*})}+ \|(r'\nab')^{\le 1} r'\eta'\|_{L^2(S'_{u', v_*})}\lesssim \epsilon_1.
\end{equation*}
This together with the below equation for $\trch'$
\begin{equation*}
    \nab'_3\trch'+(\f12\trchb'-2\omb')\trchb'=2|\eta'|^2-2\mu'
\end{equation*}
yields the desired estimate of $\wc{\trch'}$.

Via the elliptic equation for $\log\O'$ in \eqref{last slice equation for log Omega}, together with estimates we have already obtained for $\chih', \chibh', \etab'$ and $\check{\rho}$ along $\Hb_{v_*}$, we then derive the desired bound of $(r'\nab')^{\le 1}r'\log \qty((1-\f{2m_0}{r'})^{-\f12}\O')$. The weighted estimate for $\omegab$ follows from an analogous argument. Only, additionally, we need to commute the first equation in \eqref{last slice equation for log Omega} with $\nab'_3$ to obtain the sufficient decaying estimates for $\omegab'=-\f{1}{2}\nab'_3 \log \O'$ along the last slice $\Hb_{v_*}$. 

Next we proceed to control $\trch',\chih',\eta',\omb'$ in the spacetime region $D$. Integrating \eqref{e4 eqn for trch and chih} from $v$ to $v_*$, using the derived estimates of $\o',\a'$ in $D$, we hence derive the estimation of $\trch',\chih'$. Inserting this into \eqref{e4 eqn for eta and omegab}, in view of estimates of $\b'$, we thus obtain the desired weighted estimates for $\eta'$. The estimate of $\omb'$ then follows from the backward integration of \eqref{e4 eqn for eta and omegab2} and obtained estimates for $\o',\eta',\etab',\rho'$. By commuting \eqref{e4 eqn for trch and chih}, \eqref{e4 eqn for eta and omegab}, \eqref{e4 eqn for eta and omegab2} with $(r'\nab')^{\le N-3}$ and repeating the above procedure, we control their high order derivatives as well.

Therefore, employing 
\begin{equation*}
    \nab'_3 \log \O'=-2\omegab', \qquad \f{\pr b'^a}{\pr u'}=-4\O'^2\zeta'^a,
\end{equation*}
along with utilizing the estimates for $\omegab',\zeta'=\f12(\eta'-\etab')$ and the initial data for $\O', b'^A(=0)$ on $H_{u_0}=H'_{u_0}$, we derive the estimations of metric coefficients $\log\O'$ and $b'^A$.

Finally, in view of all above estimates, together with initial conditions along $H_{u_0}\cup \Hb_{v_0}$ and the established estimates along the last slice $\Hb_{v_*}$, we therefore deduce the asymptotic behaviors for the averages of $\log\O'$, $\trch', \trchb', \o', \omb'$ and $\rho'$ as stated. This finishes the proof of this proposition.

\subsection{Adapted Principal Geodesic Foliation in Initial Data Layer}\label{Subsec:initial layer}
In this section, we construct adapted principal geodesic foliations, which are defined in \cite{KS:main} within the outgoing and incoming initial layers: 
\begin{equation*}
    \MM_0=\Int\MM_0\cup \Ext\MM_0\coloneqq\{ u_0 \le u \le -\de A, \ v_0\le v\le v_0+d\}\cup \{u_0 \le u \le u_0+d, \ v_0\le v<\infty \}.
\end{equation*}
Recall that $\Int\MM_0$ is foliated by the original double null coordinates $(u, v, \th^1, \th^2)$ associated with the null frame $\qty{\pul e_{\mu}}$ defined in \eqref{Eqn:null frame pul}, while $\Ext\MM_0$ is equipped with a different double null coordinate system $(u_{out}', v_{out}, \th_{out}'^1, \th_{out}'^2)$ with the adapted null frame $\qty{(e'_{out})_{\mu}}$ given in \eqref{Eqn:null frame out'}.

\subsubsection{Initial Outgoing PG Foliation}\label{Subsusbsec:initial outgoing PG layer}
We first define the outgoing initial data layer $\Ext\LL_0$ associated with the outgoing PG structures as follows:
\begin{itemize}
    \item We set
    \begin{equation*}
        u_{\LL_0}=2(u_{out}'-u_0), \qquad L_0=-2\bfD u_{out}'=-2\bfg^{\mu\nu}\pr_{\mu}u_{out}' \pr_{\nu}.
    \end{equation*}
    Since $u$ is optical, it is easy to verify that
    \begin{equation*}
        L_0(u_{\LL_0})=2L_0(u_{out}')=0.
    \end{equation*}
    \item We define the affine parameter $\Ext r_{\LL_0}$ with respect to $L_0$ and $\Ext r_{\LL_0}$ is initialized along $\Hb_{v_0}$:
    \begin{equation*}
        L_0(\Ext r_{\LL_0})=1, \qquad \Ext r_{\LL_0}=r'_{out} \quad \text{on} \quad S'_{u_{out}', v_0}
    \end{equation*}
    with  $r'_{out}$ being the area radius of $S'_{u_{out}', v_{out}}$.
    \item We denote the outgoing null frame $\qty(\Ext(e_0)_{\mu})$, which satisfies
\begin{equation*}
    \Ext(e_0)_4=L_0, \qquad \Ext(e_0)_a(\Ext u_{\LL_0})=\Ext(e_0)_a(r_{\LL_0})=0.
\end{equation*}
\item We let
\begin{equation*}
   \Ext\LL_0=\{ 0 \le u_{\LL_0} \le 2d, \ \Ext r_{\LL_0}\ge r_0-d\}. 
\end{equation*}
\end{itemize}
\vspace{3mm}

Next we define the transition coefficients between $\qty((e'_{out})_{\mu})$ and $\qty(\Ext(e_0)_\mu)$ as $(\Ext f', \Ext\fb', \Ext\la')$. Since
\begin{equation*}
    \O_{out}'^{-1} (e'_{out})_{4}=-2\bfD u'_{out}=\Ext(e_0)_{4}=\Ext \la'\qty((e'_{out})_{4}+\Ext f'_b \, (e'_{out})_{b}+\f14|\Ext f'|^2 (e'_{out})_{3}),
\end{equation*}
we obtain
\begin{equation*}
    \Ext f'=0, \qquad  \Ext \la'=\O'^{-1}_{out}.
\end{equation*}

We then proceed to estimate $\Ext r_{\LL_0}$. In view of \Cref{outgoing initial layer}, it holds
\begin{equation*}
   (e'_{out})_{4}(\Ext r_{\LL_0}-r'_{out})=\O'_{out}\Ext (e_0)_4(\Ext r_{\LL_0})- (e'_{out})_{4}(r'_{out}) =\O_{out}'-\f{r'_{out}}{2}\O_{out}'^{-1}\c \overline{\O'_{out}\trch'_{out}}=O(\epsilon_1 r'^{-2}_{out}).
\end{equation*}
Integrating $(e'_{out})_{4}(\Ext r_{\LL_0}-r'_{out})$ from $v_0$ to $v$ yields
\begin{equation}\label{Est:compare rLL0 r'}
|\Ext r_{\LL_0}-r'_{out}|\le |\Ext r_{\LL_0}-r'_{out}|\Big|_{S'_{u', v_0}}+O(\epsilon_1r'^{-1}_{out})\le O(\epsilon_1 r'^{-1}_{out}).
\end{equation}
Hence, $\Ext r_{\LL_0}$ and $r'_{out}$ are comparable.
\vspace{2mm}

To evaluate $\Ext \fb'$, we appeal to the equation
\begin{equation*}
   0=\Ext (e_0)_a(\Ext r_{\LL_0})=(e'_{out})_{a}(\Ext r_{\LL_0}) +\f12 \Ext \fb'_a (e'_{out})_{4}(\Ext r_{\LL_0}).
\end{equation*}
Noting that $\O_{out}'^{-1}(e'_{out})_{4}(\Ext r_{\LL_0})=\Ext (e_0)_4(\Ext r_{\LL_0})=1$, this implies
\begin{equation*}
    \Ext\fb'=-2\O_{out}'^{-1} \nab'_{out}(\Ext r_{\LL_0}).
\end{equation*}

Via commuting $(e'_{out})_{4}(\Ext r_{\LL_0})=\O'_{out}$ with $\nab'_{out}$ and applying \Cref{Lem:commute}, we derive the equation for $ \nab'_{out}(\Ext r_{\LL_0})$ and it reads
\begin{equation}\label{transport eqn for nab r}
\begin{aligned}
    (e'_{out})_{4}\Big( \nab'_{out}(\Ext r_{\LL_0})\Big)=&\nab_{out}'(\O'_{out})-\chi'_{out} \c \nab_{out}'(\Ext r_{\LL_0})+(\etab'_{out}+\zeta'_{out})\O \\
    =&2(\etab'_{out}+\zeta'_{out})\O'_{out}-\chi'_{out} \c \nab_{out}'(\Ext r_{\LL_0}).
\end{aligned}
\end{equation}
Here we use the identity
\begin{equation*}
   \etab'_{out}=-\zeta'_{out}+\nab'_{out}(\log \O'_{out}).
\end{equation*}

Meanwhile, we also consider
\begin{equation}\label{Eqn:e4 r etab+zeta}
\begin{aligned}
    (e'_{out})_{4}\Big(\Ext r_{\LL_0}(\etab'_{out}+\zeta'_{out})\Big)=&\O'_{out}(\etab'_{out}+\zeta'_{out})+\Ext r_{\LL_0} (e'_{out})_{4}(\nab'_{out}\qty(\log \O'_{out}))\\
    =&\O(\etab'_{out}+\zeta'_{out})+\Ext r_{\LL_0}\l[(e'_{out})_{4}, \nab'_{out}](\log \O'_{out})+\nab'_{out}((e'_{out})_{4}(\log \O'_{out})) \r \\
    =&\O'_{out}(\etab'_{out}+\zeta'_{out})-\chi'_{out}\c \Ext r_{\LL_0}(\etab'_{out}+\zeta'_{out})\\&-2\om'_{out} \Ext r_{\LL_0}(\etab'_{out}+\zeta'_{out}) 
    -2\Ext r_{\LL_0}\nab_{out}'(\om'_{out}),
\end{aligned}
\end{equation}
where we use the fact that
\begin{equation*}
    (e'_{out})_{4}(\log \O'_{out})=-2\om'_{out}.
\end{equation*}
Combining \eqref{transport eqn for nab r} and \eqref{Eqn:e4 r etab+zeta}, we can eliminate the term $\O(\etab'_{out}+\zeta'_{out})$ and obtain
\begin{equation}\label{Eqn e4 combine r eta zeta}
    \begin{aligned}
    (e'_{out})_{4}\Big( \nab'_{out}(\Ext r_{\LL_0})-2\Ext r_{\LL_0}(\etab'_{out}+\zeta'_{out})\Big)=&-\chi'_{out}\c \Big( \nab'_{out}(\Ext r_{\LL_0})-2\Ext r_{\LL_0}(\etab'_{out}+\zeta'_{out})\Big)\\
    &+4\om'_{out}\Ext r_{\LL_0}(\etab'_{out}+\zeta'_{out})+4\Ext r_{\LL_0}\nab_{out}' (\om'_{out}). \\[2mm]
\end{aligned}
\end{equation}

Multiplying \eqref{Eqn e4 combine r eta zeta} by $r'_{out}$ and utilizing established hyperbolic estimates within $\Ext\MM_0$ by \Cref{outgoing initial layer}, we get
\begin{equation*}
   |(e'_{out})_{4}\Big( \nab'_{out}(\Ext r_{\LL_0})-2\Ext r_{\LL_0}(\etab'_{out}+\zeta'_{out})\Big)|\lesssim \f{\epsilon_1}{r_{out}'^2}+\f{\epsilon_1}{r'^2_{out}}\abs{r\nab'_{out}(\Ext r_{\LL_0})
   }. 
\end{equation*}
Note that along $\Hb'_{v_0}$, the affine parameter $\Ext r_{\LL_0}$ coincides with $r_{out}'$. Thus, it holds
\begin{equation*}
    \nab'_{out}(\Ext r_{\LL_0})=\nab'_{out}(r_{out}')=0 \qquad \text{on} \quad S'_{u_{out}', v_0}.
\end{equation*}
Employing Gr\"onwall's inequality, we hence arrive at
\begin{equation*}
    | \nab'_{out}(\Ext r_{\LL_0})-2\Ext r_{\LL_0}(\etab'_{out}+\zeta'_{out})
    |\lesssim  \f{\epsilon_1 }{r_{out}'}.
\end{equation*}
Together with estimates from \Cref{outgoing initial layer}, this further implies
\begin{equation*}
    |\Ext \fb'|\lesssim | \nab'_{out}(\Ext r_{\LL_0})|\lesssim \f{\epsilon_1 }{r_{out}'}.
\end{equation*}

Regarding the estimates for higher order derivatives of $\Ext \fb'$, we can commute \eqref{Eqn e4 combine r eta zeta} with $(e'_{out})_{3}, r'_{out}\nab'_{out}$ and proceed similarly as above. Together, we conclude
\begin{equation}\label{Est for Ext f'}
   \sup\limits_{\Ext\LL_0}|\qty((e'_{out})_{3}, r'_{out}(e'_{out})_{4}, r'_{out}\nab'_{out})^{\le N-5} \Big(\Ext f', \Ext \fb', \log (\Ext\la' \O'_{out})\Big)|\lesssim \f{\epsilon_1 }{r_{out}'}. 
\end{equation}

Applying \Cref{Lem:transformation formula}, in view of hyperbolic estimates presented in \Cref{outgoing initial layer}, along with \eqref{Est:compare rLL0 r'} and \eqref{Est for Ext f'}, we then verify that relative to the outgoing PG structure within $\Ext \LL_0$, our spacetime is sufficiently close to the Kerr solution $Kerr(a_0=0, m_0)$ in the following sense:
\begin{equation}\label{Est:out}
\begin{aligned}
   \sup\limits_{\Ext\LL_0}|&\Ext\mathfrak{d}_0^{\le N-6}\Big(\Ext r_{\LL_0}^{\f{7+\de_B}{2}} (\Ext\a_{0},\Ext\beta_{0}), \Ext r_{\LL_0}^2(\Ext\Ga_0)_g, \Ext r_{\LL_0}(\Ext\Ga_{0})_b \Big) |\lesssim \epsilon_1,
\end{aligned}
\end{equation}
where $\mathfrak{d}\coloneqq \qty{\nab_3, r\nab_4,r\nab}$ and\footnote{Recall that we have $\etab=-\zeta, \, \om=\xi=0$ within the outgoing PG structure.}
\begin{align*}
    \Ext\Ga_g\coloneqq&\qty{\trch-\f{2}{r}, \chih, \zeta, \trchb+\f{2}{r}(1-\f{2m_0}{r}), r(\rho+\f{2m_0}{r^3}), r\si,r\b, r\a},\\
    \Ext\Ga_{b,1}\coloneqq&\qty{\eta,\chibh, \omb-\f{m_0}{r^2}, \xib, r\bb,r\ab}.
\end{align*}
\begin{remark}
 The renormalization procedure in \eqref{Eqn e4 combine r eta zeta} is needed to avoid the log-loss for the term $\O (\etab'_{out}+\zeta'_{out})$ on the right of \eqref{transport eqn for nab r}. This trick is inspired by Corollary 2.2.5 in \cite{KS:main}.
\end{remark}

\subsubsection{Initial Incoming PG Foliation}\label{Subsubsec:initial incoming PG layer}
Next we turn to construct the incoming PG structure.
\begin{itemize}
\item We choose the incoming null vector $\Lb_0$, which  satisfies
    \begin{equation*}
        \bfD_{\Lb_0}\Lb_0=0 \qquad \text{with} \quad \Lb_0=(1-\f{2m_0}{r_0})^{-1}(\Ext e_0)_3\eqqcolon\la_0^{-1}(\Ext e_0)_3 \quad \text{along} \quad \Ext r_{\LL_0}=r_0.
    \end{equation*}
    Then we solve functions $\ub_{\LL_0}, \Int r_{\LL_0}$ via using
    \begin{equation*}
        \Lb_{0}(\ub_{\LL_0})=0, \qquad \Lb_{0}(\Int r_{\LL_0})=-1.
    \end{equation*}
    Here $\ub_{\LL_0}, \Int r_{\LL_0}$ are initialized as
    \begin{equation*}
        \ub_{\LL_0}=u_{\LL_0}, \quad \Int r_{\LL_0}=\Ext r_{\LL_0} \qquad \text{along} \quad \Ext r_{\LL_0}=r_0.
    \end{equation*}
    \item We then consider the null frame $(\Int(e_0)_1, \Int(e_0)_2, \Int(e_0)_3, \Int(e_0)_4)$, which obeys
\begin{equation*}
    \Int(e_0)_3=\Lb_0, \qquad \Int(e_0)_a(\ub_{\LL_0})=\Int(e_0)_a(\Int r_{\LL_0})=0.
\end{equation*}
\item We define 
\begin{equation*}
   \Int\LL_0=\{ 0 \le \ub_{\LL_0} \le d, \ \ 2m_0(1-2\de_{\HH})\le\Int r_{\LL_0}\le r_0+d\} \qquad \text{with} \quad   0<\de_{\HH}\ll 1.
\end{equation*}
\end{itemize}

We proceed to set the transition coefficients between $(\pul e_{\mu})$ and $\qty(\Int(e_0)_{\mu})$ to be $( \Int f', \Int \fb', \Int \la')$.
For estimating these coefficients, we split the frame transformation from $(\pul e_{\mu})$ to $\qty(\Int(e_0)_{\mu})$ into two steps:\begin{itemize}
    \item 
 First we choose a null frame $( e'_{1} ,e'_{2} ,e'_3 ,e'_{4})$ satisfying
\begin{equation*}
    e'_3=\Int(e_0)_3 \quad \text{and} \quad   {}^{(1)}f=0.
\end{equation*}
Here we define $( {}^{(1)}f,{}^{(1)}\fb,{}^{(1)}\lambda )$ to be the transition coefficients between $(\pul e_{\mu})$ and $( e'_{\mu})$. 
\item Then we denote the transition coefficients from $( e'_{\mu} )$ to $\qty( \Int(e_0)_{\mu})$ to be $({}^{(2)}f,{}^{(2)}\fb,{}^{(2)}\lambda)=({}^{(2)}f,0, 1)$. 
\end{itemize}
Consequently, we have the following relation
\begin{equation*}
    ( \Int f', \Int \fb', \Int \la')=({}^{(2)}f,{}^{(2)}\fb,{}^{(2)}\lambda)\circ( {}^{(1)}f,{}^{(1)}\fb,{}^{(1)}\lambda ).
\end{equation*}

Noting that $\bfD_{\Int(e_0)_3}\Int(e_0)_3=0$, relative to the null frame $( e'_{1} ,e'_{2} ,e'_3 ,e'_{4})$, we hence have\footnote{For simplicity, here the geometric quantities without prime refer to the ones with respect to the null frame $(\pul e_{\mu})$, and quantities corresponding to frames $(e'_{\mu})$ and $\qty(\Int(e_0)_{\mu})$ are denoted by tildes and double tildes, respectively.}
\begin{equation*}
    \xib'=\omegab'=0.
\end{equation*}
Combining with \Cref{Lem:transformation formula}, we thus obtain equations for $({}^{(1)}\fb, {}^{(1)}\la)$ along $e'_3$ direction:
\begin{equation}\label{Eqn transport fb la e3}
    \begin{split}
   \nab'_3({}^{(1)}\fb)=&-2({}^{(1)}\la)^{-1}\l \xib  +    \omb\,{}^{(1)}\fb + \frac{1}{4}\trchb\,{}^{(1)}\fb - \frac{1}{4}\atrchb\dual{}^{(1)}\fb +\err(\xib, \xib')\r,\\
      \nab'_3(\log{}^{(1)}\la)=&-2({}^{(1)}\la)^{-1}\l \omb-\frac{1}{2}{}^{(1)}\fb\c\zeta -\frac{1}{2}{}^{(1)}\fb\c\eta +\err(\omb,\omb')\r, 
    \end{split}
\end{equation}
where $\err(\xib, \xib'), \err(\omb,\omb')$ are error terms in the form of $\Big( ({}^{(1)}\la)^{-1}\Big)^{\le 2} \sum_{i=1}^3 ({}^{(1)}\fb)^{i}\, \Gamma.$

Recall that $e'_3=\Int (e_0)_3$ is initialized on $\Ext r_{\LL_0}=r_0$ and it satisfies
\begin{equation*}
    e'_3=\Int (e_0)_3=\Lb_0=\la_0^{-1}(\Ext e_0)_3=\la_0^{-1} (\Ext\la'')^{-1}\Big(\pul e_3+\Ext \fb''_b \pul e^a+\f14 |\Ext\fb''|^2 \, \pul e_4 \Big).
\end{equation*}
Here we have
\begin{equation*}
    (\Ext f'', \Ext \fb'', \Ext\la'')=(\Ext f', \Ext \fb', \Ext\la')\circ (f_{out}, \fb_{out}, \la_{out}) 
\end{equation*}
with $(f_{out}, \fb_{out}, \la_{out}) $ being the transition coefficients from $(\pul e_{\mu})$ to $\qty((e'_{out})_{\mu})$.

Comparing with
\begin{equation*}
    e'_3=({}^{(1)}\la)^{-1} \Big(\pul e_3+\Int \fb'_b \pul e^a+\f14 |\Int\fb'|^2 \, \pul e_4 \Big),
\end{equation*}
we then obtain the initial conditions of $({}^{(1)}\fb, {}^{(1)}\la)$, i.e.,
\begin{equation*}
    {}^{(1)}\fb=\Ext\fb'', \qquad {}^{(1)}\la=\la_0 \Ext\la'' \qquad \text{along} \quad \Ext r_{\LL_0}=r_0.
\end{equation*}

To solve above equations, we assume first that 
\begin{equation*}
    |{}^{(1)}\fb|+|\log {}^{(1)}\la-\log(\la_0 \Ext\la''\big|_{\{\Ext r_{\LL_0}=r_0\}})|\le 1.
\end{equation*}
In view of hyperbolic estimates in $\MM_0$, the terms on the right of \eqref{Eqn transport fb la e3} can be written as
\begin{equation*}
    O(\epsilon_1)+O({}^{(1)}\fb).
\end{equation*}
Therefore, integrating \eqref{Eqn transport fb la e3} along $e'_3$ and employing Gr\"{o}nwall's inequality as well as inequality \eqref{Est for Ext f'}, we now get an improved bound
\begin{align*}
    &|({}^{(1)}\fb|+|\log {}^{(1)}\la-\log(\la_0 \Ext\la''\big|_{\{\Ext r_{\LL_0}=r_0\}})|\lesssim |\Ext\fb'|_{\{\Ext r_{\LL_0}=r_0\}}+\epsilon_1  \lesssim \epsilon_1.
\end{align*}

In the same manner, first commuting \eqref{Eqn transport fb la e3} with $(\nab'_3, r\nab')$, then employing \eqref{Est:transition out} and \eqref{Est:compare rLL0 r'}, we derive the higher order estimates of $({}^{(1)}\fb, \log {}^{(1)}\la)$ and they satisfy
\begin{equation}\label{Est for (1) fb la}
    |\mathfrak{d}^{\le N-5}  {}^{(1)}\fb|\lesssim\epsilon_1, \qquad|\mathfrak{d}^{\le N-5}\log {}^{(1)}\la)|\lesssim 1\qquad \text{in} \quad \Int\LL_0. \\[2mm]
\end{equation}

Next we proceed to estimate $^{(2)}f$. Since $\Int(e_0)_3(\Int r_{\LL_0})=-1, \Int(e_0)_a(\Int r_{\LL_0})=0$, we then act
\begin{equation*}
   [\Int(e_0)_3, \Int(e_0)_a]=-\chi''_{ab} \Int(e_0)^b+\xib''_a \Int(e_0)_4+(\eta''-\zeta'')_a \Int(e_0)_3
\end{equation*}
on $\Int r_{\LL_0}$ and by virtue of $\xib''=0$ we derive that
\begin{equation*}
    \eta''-\zeta''=0.
\end{equation*}
Utilizing \Cref{Lem:transformation formula}, we now deduce the equation of ${}^{(2)}f$:\footnote{Note that ${}^{(2)}\fb=0$ and ${}^{(2)}\la=1$.}
\begin{equation}\label{Eqn transport f e3}
    \begin{split}
        \Int(\nab_0)_3 ({}^{(2)}f)=&2\l -\eta'+\zeta'-\frac{1}{4}\trchb' \, {}^{(2)}f +\frac{1}{4}\atrchb' \dual \, {}^{(2)}f +\sum_{i=1}^3 ({}^{(2)}f)^{i}\Gamma\r.
    \end{split}
\end{equation}
Thanks to 
\begin{align*}
    0=\Int(e_0)_a(\Int r_{\LL_0})=&e'_a(\Int r_{\LL_0})+\f12 {}^{(2)}f_a e'_3(\Int r_{\LL_0})\\
    =&(\pul e_a+\f12 {}^{(1)}\fb_a \pul e_4)(\Int r_{\LL_0})-\f12 {}^{(2)}f_a,
\end{align*}
we obtain
\begin{equation}\label{Eqn for (2)f a}
    {}^{(2)}f_a=2(\pul e_a+\f12 {}^{(1)}\fb_a \pul e_4)(\Int r_{\LL_0}).
\end{equation}
Recalling that $\Int r_{\LL_0}=\Ext r_{\LL_0}$ along $\Ext r_{\LL_0}=r_0$, we also obtain
\begin{equation*}
    0=\Ext(e_0)_a(\Ext r_{\LL_0})= \Ext(e_0)_a(\Int r_{\LL_0})=(\pul e_a+\f12\Ext \fb'_a) (\Int r_{\LL_0}) \qquad \text{on} \quad \Ext r_{\LL_0}=r_0.
\end{equation*}
Inserting this into \eqref{Eqn for (2)f a} and noting that ${}^{(1)}\fb=\Ext\fb'$ along $\Ext r_{\LL_0}=r_0$, we then derive the initial data of ${}^{(2)}f$, i.e.,
\begin{equation*}
    {}^{(2)}f_a=({}^{(1)}\fb_a-\Ext\fb'_a) e_4(\Int r_{\LL_0})=0 \qquad \text{on} \quad \Ext r_{\LL_0}=r_0.
\end{equation*}
Hence, proceeding similarly as in the derivation of estimates for $({}^{(1)}\fb, \log {}^{(1)}\la)$, we infer from \eqref{Eqn transport f e3} that
\begin{equation}\label{Est for (2) f}
    |\mathfrak{d}^{\le N-5}  ({}^{(2)}f)|\lesssim \epsilon_0\qquad \text{in} \quad \Int\LL_0.
\end{equation}
Finally, due to the decomposition of the null frame transformation
\begin{equation*}
    ( \Int f', \Int \fb', \Int \la')=({}^{(2)}f,{}^{(2)}\fb=0,{}^{(2)}\lambda=1 ) \circ ( {}^{(1)}f=0,{}^{(1)}\fb,{}^{(1)}\lambda ),
\end{equation*}
combining with \eqref{Est for (1) fb la} and \eqref{Est for (2) f}, we arrive at
\begin{equation}\label{Est for Int f'}
  |\mathfrak{d}^{\le N-5}( \Int f', \Int \fb')|\lesssim\epsilon_1, \qquad |\mathfrak{d}^{\le N-5}\log \Int \la'|\lesssim 1\qquad \text{in} \quad \Int\LL_0. \\[2mm]
\end{equation}

As a consequence, according to hyperbolic estimates provided in \Cref{Prop:Est incoming initial layer} in $\Int\MM_0$, applying \Cref{Lem:transformation formula} and \eqref{Est for Int f'}, we verify that, relative to the incoming PG structure within $\Int \LL_0$, our spacetime is sufficiently close to the Kerr solution $Kerr(a_0=0, m_0)$ satisfying
\begin{equation*}
    \sup\limits_{\Int\LL_0} \abs{\Int\mathfrak{d}_0^{\le N-6}\Big((\Int\Ga_0)_{g,1}, (\Int\Ga_0)_{b,1}\Big)}\lesssim \epsilon_1,
\end{equation*}
where $\Int\Ga_{b,1}\coloneqq\qty{\chibh, r\bb, \ab}$ and
\begin{align*} 
    \Int\Ga_{g,1}\coloneqq\qty{\xi, \om+\f{m_0}{r^2}, \trch-\f{2}{r}(1-\f{2m_0}{r}),\chih,\zeta,\etab,\trchb+\f{2}{r}, r(\rho+\f{2m_0}{r^3}),r\si, r\b, r\a}.
\end{align*}
\begin{remark}
    Although we do not have the smallness of $\log\Int\la'$ as shown in \eqref{Est for Int f'}, it is already sufficient for us to establish the desired controls for all curvature components and for most of Ricci coefficients except $\Int\trch_0, \Int\trchb_0, \Int\om_0$. Regarding these exceptional quantities, we first estimate their values on $\Ext r_{\LL_0}=r_0$ using \Cref{Lem:transformation formula}, hyperbolic estimates from \Cref{outgoing initial layer} in $\Ext\LL_0$, along with the initialization conditions on $\Ext r_{\LL_0}=r_0$, i.e.,
    \begin{equation*}
        \Int (e_0)_a=\Ext (e_0)_a,\qquad \Int (e_0)_3=\la_0^{-1} \Ext (e_0)_3, \qquad \Int (e_0)_4=\la_0 \Ext (e_0)_4.
    \end{equation*}
    Then we appeal to the corresponding transport equations along $\Int(e_0)_3$ to obtain the desired asymptotic behaviors for $\Int\trch_0, \Int\trchb_0, \Int\om_0$.
\end{remark}
\begin{remark}
   Here we also demonstrate the benefit of decomposing the frame transformation from $(\pul e_{\mu})$ to $\qty(\Int(e_0)_{\mu})$. If instead we estimate the transition coefficients $( \Int f', \Int \fb', \Int \la')$ directly, then the conditions $\xib''=\omegab''=\eta''-\zeta''=0$ would lead to the following coupled transport equations
   \begin{align*}
        \Int \la' \Int(\nab_0)_3(\Int \fb')=& \Ga+\Ga \c \Int \fb'+E_1\qty( \Int f', \Int \fb', \Int \la', \Ga),\\
      \Int \la' \Int(\nab_0)_3(\log \Int\la')=&\Ga+\Ga \c \Int \fb'+E_2\qty( \Int f', \Int \fb', \Int \la', \Ga),\\
        \Int \la'\Int(\nab_0)_3 (\Int f')= & \Ga+\Int(\nab_0)(\log \Int\la')+\Ga \c (\Int f', \Int \fb')\\&+E_3 \qty(\Int(\nab_0)^{\le 1} (\Int f'), \Int \fb', \Int \la', \Ga),
   \end{align*}
   where $E_1, E_2, E_3$ are nonlinear terms involving $\Int f', \Int \fb'$ and $\nab^{\prime \le 1} \Int f'$. Note that there would be a loss of derivative for the transport equation of $\Int f'$ and it is a major obstruction when attempting to control $( \Int f', \Int \fb', \Int \la')$ simultaneously. While employing the above decomposition of change of frame, we avoid this derivative loss due to the absence of the term $\Int(\nab_0)(\log \Int\la')$ on the right of the last equation.
\end{remark}

\subsubsection{Associated Scalars and 1-Forms in \texorpdfstring{$\LL_0$}{}}
 We also introduce the below scalars and $1$-forms adapted to the initial data layer $\LL_0=\Ext\LL_0\cup \Int \LL_0$.
\begin{itemize}
    \item In $\Ext \LL_0$, we define angular coordinates $(\Ext\th_{\LL_0}, \Ext \varphi_{\LL_0})$ and they obey
    \begin{align*}
    \nab'_{out}(\Ext\th_{\LL_0})=&(\f{1}{\Ext r_{\LL_0}},0), \quad  \nab'_{out}(\Ext\th_{\LL_0})=(0, \f{1}{\Ext r_{\LL_0}\sin\qty(\Ext\th_{\LL_0})})\qquad \text{on} \quad S'_{u_0, v_0}, \\
    \Ext (e_0)_3(\Ext\th_{\LL_0})=&  \Ext (e_0)_3(\Ext\varphi_{\LL_0})=0 \qquad \text{along} \quad \Hb'_{v_0} \\
        \Ext (e_0)_4(\Ext\th_{\LL_0})=&  \Ext (e_0)_4(\Ext\varphi_{\LL_0})=0 \qquad \text{in} \quad \Ext\LL_0.
    \end{align*}
     Then we let
    \begin{align*}
        \Ext J^{(0)}=&\cos (\Ext\th_{\LL_0}), \quad \Ext J^{(+)}=\sin (\Ext\th_{\LL_0})\cos (\Ext\varphi_{\LL_0}), \\
        \Ext J^{(-)}=&\sin (\Ext\th_{\LL_0})\sin (\Ext\varphi_{\LL_0}).
    \end{align*}
    We also define complex $1$-forms $\Ext \JJ, \Ext\JJ_{\pm}$ on $S'_{u_0, v_0}$ as
    \begin{equation*}
        \Ext \JJ=j+i\dual j, \qquad \Ext\JJ_{\pm}=j_{\pm}+i\dual j_{\pm}.
    \end{equation*}
    Here we set
    \begin{align*}
        j\coloneqq&\f{1}{\Ext r_{\LL_0}}\qty(0, \sin\qty(\Ext\th_{\LL_0})), \quad j_+\coloneqq\f{1}{\Ext r_{\LL_0}}\qty(\cos\qty(\Ext\th_{\LL_0})\cos\qty(\Ext\varphi_{\LL_0}), -\sin\qty(\Ext\th_{\LL_0})),\\
        j_-\coloneqq&\f{1}{\Ext r_{\LL_0}}\qty(\cos\qty(\Ext\th_{\LL_0})\sin\qty(\Ext\varphi_{\LL_0}), \cos\qty(\Ext\th_{\LL_0})).
    \end{align*}
    Then we propagate $\Ext \JJ, \Ext\JJ_{\pm}$ in $\Ext\LL_0$ via
    \begin{align*}
        &\Ext(\nab_0)_3 (\Ext r_{\LL_0}\Ext\JJ)= \Ext(\nab_0)_3(\Ext r_{\LL_0}\Ext\JJ_{\pm})=0 \qquad \text{along} \quad \Hb'_{v_0},\\
        &\Ext(\nab_0)(\Ext r_{\LL_0}\Ext\JJ)= \Ext(\nab_0)_4(\Ext r_{\LL_0}\Ext\JJ_{\pm})=0 \qquad \text{in} \quad \Ext\LL_0.
    \end{align*}
    \item In $\Int \LL_0$, we proceed similarly to define angular coordinates $(\Int\th_{\LL_0}, \Int \varphi_{\LL_0})$ by requiring
    \begin{equation*}
        \Int (e_0)_3(\Int\th_{\LL_0})=  \Int (e_0)_3(\Int\varphi_{\LL_0})=0
    \end{equation*}
     with initialization on $\Ext r_{\LL_0}=r_0$ satisfying
     \begin{equation*}
         \Int\th_{\LL_0}=\Ext\th_{\LL_0}, \qquad \Int\varphi_{\LL_0}=\Ext\varphi_{\LL_0}.
     \end{equation*}
     We then set
    \begin{align*}
        \Int J^{(0)}=&\cos (\Int\th_{\LL_0}), \quad \Int J^{(+)}=\sin (\Int\th_{\LL_0})\cos (\Int\varphi_{\LL_0}), \\
        \Int J^{(-)}=&\sin (\Int\th_{\LL_0})\sin (\Int\varphi_{\LL_0})
    \end{align*}
    and define complex $1$-forms $\Int \JJ$, which satisfies 
    \begin{align*}
        &\nab_{\Int (e_0)_3} (\Int r_{\LL_0}\Int \JJ)=\nab_{\Int (e_0)_3} (\Int r_{\LL_0}\Int \JJ_{\pm})=0, \\
        &\Int \JJ=\Ext \JJ, \Int \JJ_{\pm}=\Ext \JJ_{\pm} \qquad \text{on} \quad \Ext r_{\LL_0}=r_0.
    \end{align*}
\end{itemize}

Therefore, following the procedures conducted in Section 5.6 and Sections 6.5--6.6 of \cite{KS:main}, we deduce the below controls for  derivatives of coordinates $(\Ext u_{\LL_0},\Ext r_{\LL_0},\Ext \th_{\LL_0},\Ext \varphi_{\LL_0})$ and for derivatives of $\Ext J^{(0, \pm)}, \Ext\JJ,\Ext\JJ_{\pm}$ in $\Ext\LL_0$:
\begin{equation*}
    \sup\limits_{\Ext\LL_0} |\Ext\mathfrak{d}_0^{\le N-6} \Big(\Ext r_{\LL_0}(\Ext\Ga_0)_{b,2}\Big)|\lesssim \epsilon_1,
\end{equation*}
where $\DD\coloneqq \nab+i\dual\nab$ and
\begin{align*}
    \Ext\Ga_{b,2}\coloneqq&\qty{r^{-1}\qty(e_3(r)+1-\f{2m_0}{r}),\DD J^{(0)}-i\JJ, \DD J^{(\pm)}-\JJ_{\pm}, e_3(J^{(0)}), e_3(J^{(\pm)})}\cup \Big\{r(\ov{\DD} \c\JJ-\f{4i\cos\th}{r^2}), \\&r\DD\hot\JJ,r\qty(\nab_3 \JJ-\f{1}{r}(1-\f{2m_0}{r})\JJ), r(\ov{\DD}\c\JJ_{\pm}+\f{4}{r^2}\JJ^{(\pm)}), r\qty(\nab_3 \JJ_{\pm}-\f{1}{r}(1-\f{2m_0}{r})\JJ_{\pm})\Big\}.
\end{align*}

Then proceeding similarly as in Sections 7.2--7.3 of \cite{KS:main}, we hence obtain the following estimates for derivatives of $(\Int \ub_{\LL_0},\Int r_{\LL_0},\Int \th_{\LL_0},\Int \varphi_{\LL_0})$ and for derivatives of $\Int J^{(0, \pm)}, \Int\JJ,\Ext\JJ_{\pm}$ in $\Int\LL_0$:
\begin{equation*}
    \sup\limits_{\Int\LL_0} |\Int\mathfrak{d}_0^{\le N-6}\Big((\Int\Ga_0)_{g,2},(\Int\Ga_0)_{b,2}\Big)|\lesssim \epsilon_1,
\end{equation*}
where
\begin{align*}
\Int\Ga_{g,2}\coloneqq&\qty{r\qty(e_4(r)-(1-\f{2m_0}{r})), re_4(J^{(0)}), re_4(J^{(\pm)}), r^2 \nab_4\JJ, r^2\nab_4 \JJ_{\pm}}, \\
     \Int\Ga_{b, 2}\coloneqq&\Big\{\DD J^{(0)}-i\JJ, \DD J^{(\pm)}-\JJ_{\pm},r(\ov{\DD} \c\JJ-\f{4i\cos\th}{r^2}), r\DD\hot\JJ,\\&\qty(\nab_4 \JJ+\f{1}{r}(1-\f{2m_0}{r})\JJ), r(\ov{\DD}\c\JJ_{\pm}+\f{4}{r^2}\JJ^{(\pm)}), r\qty(\nab_4 \JJ_{\pm}+\f{1}{r}(1-\f{2m_0}{r})\JJ_{\pm})\Big\}.
\end{align*}

\subsection{Nonlinear Stability of Kerr Spacetime with Small Angular Momentum}
In their celebrated work \cite{KS:main}, Klainerman and Szeftel proved the main theorem of the stability of Kerr black holes with small angular momentum. In this current paper, for the late stage of gravitational collapse, we will employ their main theorems to construct our dynamical Kerr black-hole formation spacetimes. Their main result of \cite{KS:main} can be summarized as follows:
\begin{theorem}[Klainerman--Szeftel \cite{KS:main}]\label{Thm:KS main}
Prescribe initial data set that are perturbations of size $\epsilon_1$ for a Kerr black hole spacetime $Kerr(a_0, m_0)$ with $|a_0|\ll m_0$. Under Einstein vacuum equations \eqref{Eqn:EVE}, arising from these initial data, there always exists a globally hyperbolic development $\mathcal{M}$ that has a complete future null infinity $\II^+$ and $\MM$ converges in its causal past $J^-(\II^+)$ to another nearby Kerr solution $Kerr(a_{\infty},m_{\infty})$ with $|\ainft|\ll \minft$, such that
 \begin{equation*}
     |\ainft-a_0|+|\minft-m_0|\lesssim \epsilon_0\coloneqq \epsilon_1^{\f12}. 
 \end{equation*} 
    Furthermore, in the interior region ${}^{(int)}\mathcal{M}=\mathcal{M}\cap\{r_{\mathcal{H}} \le r\leq r_0\}$ where $r_{\mathcal{H}}=(m_0+\sqrt{m_0^2-a_0^2})(1-\delta_{\mathcal{H}})$ with $r_0\gg m_0$ and $0<\delta_{\mathcal{H}}\ll 1$, there hold
    \begin{enumerate}
        \item the interior region $\intM$ is equipped with an incoming PG null structure $(e_1, e_2, e_3, e_4)$ and  an associated incoming PG coordinate system $(\ub, r, \theta, \varphi)$.
        \item  there exist an integer $k_{small}\ge 3$ and a small constants $\delta_{dec}>0$ independent of $\epsilon_0$ such that, for $0\le k\le k_{small}$ the following estimates for Ricci coefficients $\Gamma$ and curvature components $R$ are satisfied
 \begin{equation*}
	|\mathfrak{d}^k (\Gamma-\Gamma_0, R-R_0)|\lesssim \fepub \qquad \quad \text{in} \quad \intM
	\end{equation*}
 with $\mathfrak{d}\coloneqq\{e_3, re_4, r\nab \}$ and $\Gamma_0, R_0$ represent the corresponding values in the exact Kerr spacetime $Kerr(a_{\infty},m_{\infty})$ provided in \eqref{Eqn for Ricci Cur}.
 \item For $0\le k\le k_{small}$, the metric components in $(\ub, r, \theta, \varphi)$ coordinates obey 
 \begin{equation}\label{Est:kerrmetric}
     |\mathfrak{d}^k (\bfg-\bfg_{a_{\infty}, m_{\infty}})|\lesssim \fepub  \qquad \quad \text{in} \quad \intM,
 \end{equation}
 where $\bfg_{a_{\infty}, m_{\infty}}$ denotes the exact Kerr metric expressed in $(\ub, r, \th, \varphi)$ coordinates in \eqref{Kerr metric:in EF}.
    \end{enumerate}
\end{theorem}
In \Cref{Subsec:initial layer} we construct initial data layers for the admissible Kerr black hole formation initial after the short pulse and derive the desired hyperbolic estimates in $\LL_0=\Ext\LL_0\cup\Int\LL_0$, which validates the initial conditions of \Cref{Thm:KS main} with $a_0=0$. Thus, we can apply \Cref{Thm:KS main} and get the existence of converging-to-Kerr spacetimes and the associated hyperbolic bounds.

\section{Construction of Incoming Null Hypersurface}\label{Sec:Construct null cones}
In this section, we construct incoming null hypersurfaces and the associated incoming geodesic foliation within the interior region $\intM$. To ensure that the apparent horizon smoothly connects to the counterpart solved based on the double null foliation in the short-pulse region $\PP=\{u_0\le u\le -\de A, \  0\le v \le \de \}$, we design a new incoming geodesic foliation in a transition region, that is compatible with foliations in both $\PP$ and $\intM$. We also examine the precise leading-order behavior of the null expansions adapted to our constructed incoming geodesic foliation. This plays a significant role in deriving the a priori estimate for the equation of MOTS in \Cref{Sec:Existence}.

\subsection{Incoming Geodesic Foliation in \texorpdfstring{$\intM$}{}}\label{Subsec:incoming geodesic foliation}
Recall that $\intM$ is equipped with incoming principal geodesic (PG) structures. They are associated with adapted PG coordinates $(\ub, r, \theta, \varphi)$ for $\ub \ge 1, \ 2m_0(1-\de_{\HH})\eqqcolon r_{\mathcal{H}}\le r\le r_0$,\footnote{Here $0<\de_{\HH}\ll 1$, $r_0\gg 1$, $\th\in[0 ,\pi]$ and $\varphi\in[0, 2\pi]$.} and a null frame $(\Int e_1, \Int e_2, \Int e_3, \Int e_4)$ satisfying 
\begin{equation}\label{In PG cond}
    \begin{aligned}
    \bfD_{\Int e_3} \Int e_3=0, \quad \Int e_3 (r)=&-1, \quad \Int e_1 (r)=\Int e_2 (r)=0, \\
    \Int e_3 (\ub)=\Int e_3 (\theta)=&\Int e_3 (\varphi)=0.
\end{aligned}
\end{equation}

For the remaining coordinate derivatives, together with the exact Kerr values provided in \eqref{Kerr coor derivative}, by \Cref{Thm:KS main} we have the following asymptotic behaviors
\begin{equation}\label{Est for coor derivative}
    \begin{aligned}
        \Int e_4(\th)=&O(\fepub), &\quad  \Int e_4(\varphi)=&\f{2\ainft}{|q|^2}+O(\fepub),  \\
     \Int e_4(r)=&\f{\De}{|q|^2}+O(\fepub),  
        &\quad \Int e_4(\ub)=&\f{2(r^2+\ainft^2)}{|q|^2},   \\
        \Int \nab (\ub)=&(0, \f{\ainft\sin \th}{|q|})+O(\fepub), &\quad \Int \nab(\th)=&(\f{1}{|q|}, 0)+O(\fepub), \\
        \Int \nab (\varphi)=&(0, \f{1}{|q|\sin \th})+O(\fepub),  &\quad(\Int\nab_1 \Int e_2)_1=&O(\fepub),\\
       (\Int\nab_2 \Int e_1)_2=&\f{r^2+\ainft^2}{|q|^3}\cot \th+O(\fepub).
    \end{aligned}
\end{equation}
 Here $q=r+i\ainft \cos \th$ and $\De=r^2-2\minft r+\ainft^2$. We also summarize the hyperbolic estimates in \Cref{Thm:KS main} that will be used as below:
 \begin{equation}\label{Est for Ricci Cur}
     \begin{aligned}
         \trch=&\f{2r \De}{|q|^4}+O(\fepub), &\quad ^{(a)}\trch=&\f{2\ainft\cos \th \De}{|q|^4}+O(\fepub), \\
         \trchb=&-\f{2r}{|q|^2}+O(\fepub), &\quad ^{(a)}\trchb=&-\f{2\ainft\cos \th}{|q|^2}+O(\fepub), \\
         \eta=&\zeta=\f{\ainft\sin \th}{|q|^3}(-\ainft\cos \th, r)+O(\fepub), &\quad \etab=&-\f{\ainft\sin \th}{|q|^3}(\ainft\cos \th, r)+O(\fepub), \\
         \omega=&-\f12\pr_r(\f{\De}{|q|^2})+O(\fepub), &\quad \omegab=&\xib=0, \\
         \rho=&-\f{2\minft r(r^2-3\ainft^2\cos^2 \th)}{|q|^6}+O(\fepub),
         &\quad \sigma=&\f{2\minft \ainft\cos\th(3r^2-\ainft^2\cos^2 \th)}{|q|^6}+O(\fepub), \\
          \chih, \chibh&, \xi,\a, \b, \ab, \bb=O(\fepub).
     \end{aligned}
 \end{equation}
It consequently follows from \eqref{In PG cond} and \eqref{Est for coor derivative} that
\begin{equation*}
    \bfg^{\a \b} \pr_{\a} \ub \pr_{\b}\ub=|\nab \ub|^2=\f{a_{\infty}^2 \sin^2 \th}{|q|^2}+O(\fepub),
\end{equation*}
which suggests that in \cite{KS:main} the level set of $\ub$ is generally a timelike hypersurface.

In order to solve the MOTS along incoming null hypersurfaces, we now aim to solve the eikonal equation
\begin{equation}\label{incoming eikonal eqn}
    \bfg^{\a \b} \pr_{\a} \tub \pr_{\b}\tub=0.
\end{equation}
 
\begin{figure}[ht]
    \centering

\tikzset{every picture/.style={line width=0.75pt}} %set default line width to 0.75pt        

\begin{tikzpicture}[x=0.75pt,y=0.75pt,yscale=-1,xscale=1]
%uncomment if require: \path (0,415); %set diagram left start at 0, and has height of 415

%Straight Lines [id:da8079092738221392] 
\draw [color={rgb, 255:red, 0; green, 0; blue, 0 }  ,draw opacity=1 ] [dash pattern={on 0.84pt off 2.51pt}]  (392.93,60.85) -- (467.93,134.25) ;
%Curve Lines [id:da676152194537932] 
\draw [color={rgb, 255:red, 0; green, 0; blue, 0 }  ,draw opacity=1 ]   (281.93,100.05) .. controls (323.66,91.52) and (344.33,83.88) .. (392.21,61.2) ;
\draw [shift={(392.93,60.85)}, rotate = 334.63] [color={rgb, 255:red, 0; green, 0; blue, 0 }  ,draw opacity=1 ][line width=0.75]      (0, 0) circle [x radius= 2.01, y radius= 2.01]   ;
%Straight Lines [id:da06987422313122216] 
\draw    (392.6,61.12) -- (396.2,243.87) ;
%Straight Lines [id:da672226814546376] 
\draw  [dash pattern={on 0.84pt off 2.51pt}]  (318.37,91.97) -- (394.66,169.34) ;
\draw [shift={(394.66,169.34)}, rotate = 45.4] [color={rgb, 255:red, 0; green, 0; blue, 0 }  ][fill={rgb, 255:red, 0; green, 0; blue, 0 }  ][line width=0.75]      (0, 0) circle [x radius= 1.34, y radius= 1.34]   ;
%Curve Lines [id:da1359319093978204] 
\draw    (281.93,100.05) .. controls (349.87,168.87) and (377.2,207.21) .. (396.2,243.87) ;
%Curve Lines [id:da20560498799934346] 
\draw    (396.2,243.87) .. controls (405.2,217.87) and (435.53,169.87) .. (467.93,134.25) ;

% Text Node
\draw (326.74,155.8) node [anchor=north west][inner sep=0.75pt]  [font=\scriptsize,rotate=-50.3]  {$\underline{u} =1$};
% Text Node
\draw (391.16,43.4) node [anchor=north west][inner sep=0.75pt]  [font=\scriptsize]  {$i^{+}$};
% Text Node
\draw (438.76,89.8) node [anchor=north west][inner sep=0.75pt]  [font=\scriptsize]  {$\mathscr{I}^{+}$};
% Text Node
\draw (327.26,71.5) node [anchor=north west][inner sep=0.75pt]  [font=\scriptsize]  {$\mathcal{A}$};
% Text Node
\draw (396.09,142.73) node [anchor=north west][inner sep=0.75pt]  [font=\scriptsize]  {$\mathcal{T}$};
% Text Node
\draw (351.16,138.4) node [anchor=north west][inner sep=0.75pt]  [font=\scriptsize]  {$\widetilde{\underline{H}}_{\widetilde{\underline{u}}}$};
% Text Node
\draw (355.36,102.27) node [anchor=north west][inner sep=0.75pt]  [font=\scriptsize]  {$^{( int)}\mathcal{M}$};
% Text Node
\draw (399.86,104.27) node [anchor=north west][inner sep=0.75pt]  [font=\scriptsize]  {$^{( ext)}\mathcal{M}$};

\end{tikzpicture}

    \caption{Construction of Incoming Null Cones within $\intM$}
    \label{fig:constructnullcone}
\end{figure}
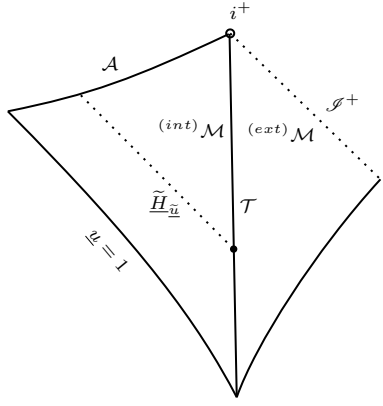

As shown in \Cref{fig:constructnullcone}, since the exterior boundary of $\intM$ is $\TT=\{r=r_0 \}$, we initialize the (incoming) optical function $\tub$ along $\TT=\{r=r_0 \}$, such that 
\begin{equation*}
    \tub=\tub_0(\ub, r_0, \th).
\end{equation*}
 Here $\tub_0=\tub_0 (\ub, r, \theta)$ is the (incoming) optical function of the exact Kerr family $Kerr(\ainft, \minft)$ constructed in \Cref{Appendix:coordinate in Mint}. The function $\tub_0$ is known as the Pretorius--Israel coordinate function as first derived in \cite{P-I}. Since our constructed spacetimes are perturbed Kerr solutions, we expect that $\tub$ and $\tub_0$ differ by lower order terms of size $\fepub$. This is indeed the case and it is achieved by linearizing \eqref{incoming eikonal eqn} around the $\tub_0$ and by utilizing the detailed geometric information of $\tub_0$ that describes the Kerr geometry.

Applying \Cref{Appendix:coordinate in Mint}, we have that the explicit expression of $\tub_0$ reads
\begin{equation*}
    \tub_0=\ub-f(r)+\rs(\rthe)=\ub+\fs(\rthe) \qquad \text{with} \quad f'(r)=\f{r^2+\ainft^2}{\De}.
\end{equation*}
Here $\fs(\rthe)$ is a bounded function and $\rs(\rthe)$ satisfies
\begin{equation}\label{Est for D rs}
    \begin{aligned}
   \pr_{r} \rs=&\f{\sqrt{(r^2+\ainft^2)^2-\ainft^2\sin^2 \ths \De}}{\De}\eqqcolon\f{\Sis}{\Delta}, \\
    \pr_{\th} \rs=&-\sgn(r-r_{+,\infty})|\ainft|\sqrt{\sin^2 \ths-\sin^2 \th}\eqqcolon\bs,
\end{aligned}
\end{equation}
with $\ths=\ths(\rthe)\in [0, \pi]$ being a smooth function of $r, \th$ and $r_{+,\infty}\coloneqq m_{\infty}+\sqrt{\minft^2-\ainft^2}$.  Employing \eqref{In PG cond}, \eqref{Est for coor derivative} and \eqref{Est for D rs}, we obtain
\begin{equation}\label{Est for Dtub0}
    \begin{aligned}
        e_3(\tub_0)=&e_3(\ub-f(r)+\rs) 
        =\f{r^2+\ainft^2-\sqrt{(r^2+\ainft^2)^2-\ainft^2 \sin^2 \ths \De}}{\De}+O(\fepub), \\
        e_4(\tub_0)=&e_4(\ub-f(r)+\rs)=\f{r^2+\ainft^2+\sqrt{(r^2+\ainft^2)^2-\ainft^2 \sin^2 \ths \De}}{|q|^2}+O(\fepub), \\
        \nab(\tub_0)=&\nab(\ub-f(r)+\rs)=\f{1}{|q|}(-\sgn(r-r_{+,\infty})|\ainft|\sqrt{\sin^2 \ths-\sin^2 \th}, \ainft\sin \th)+O(\fepub).
    \end{aligned}
\end{equation}
Together with \eqref{In PG cond}, \eqref{Est for coor derivative}, this yields
\begin{equation}\label{Est:inner product}
    \begin{aligned}
        \bfD \tub_0 \cdot \bfD \ub=&\f{\ainft^2}{|q|^2}(\sin^2 \th-\f{(r^2+\ainft^2)\sin^2 \ths}{r^2+a^2+\Sis} )+O(\fepub), &\quad \bfD \tub_0 \cdot \bfD r=&\f{\Sis}{|q|^2}+O(\fepub), \\
   \bfD \tub_0 \cdot \bfD \th=&-\f{|\ainft|\sgn(r-r_{+,\infty})}{|q|^2}\sqrt{\sin^2\ths-\sin^2\th}+O(\fepub), &\quad
    \bfD \tub_0 \cdot \bfD \varphi=&\f{\ainft}{|q|^2}(1-\f{\ainft^2\sin^2\ths}{r^2+\ainft^2+\Sis}),\\
    \bfD r \cdot \bfD r=& \f{\De}{|q|^2}+O(\fepub), &\quad \bfD \tub_0 \cdot \bfD \ub_0=&O(\fepub).
    \end{aligned}
\end{equation}
We note that in \Cref{Appendix:coordinate in Mint}, we provide precise definitions and detailed properties of $\ths(\rthe), \rs(\rthe)$ and $ f_{*}(\rthe)$. 
\vspace{2mm}

For notational simplicity, we will drop the prefixes $\Int$ of the null frame $(\Int e_1, \Int e_2, \Int e_3, \Int e_4)$ and the subscripts $_\infty$ for $\ainft, \minft$ if there is no danger of confusion. Denote $\nab$ as the covariant derivative of $\text{span}\{e_1, e_2 \}$. Inspired by Section 8.2.2 as in \cite{KS:main}, we then establish the existence and asymptotics of $\tub$ as below.
\begin{proposition}\label{Lem:est for tub}
    The eikonal equation \eqref{incoming eikonal eqn} admits a unique solution $\tub$ in $\intM$ with $\tub=\tub_0$ along $\TT=\{r=r_0 \}$. Moreover, for $0\le k\le k_{small}$ the following estimate holds
    \begin{equation*}
        |\mathfrak{d}^k(\tub-\tub_0)|\lesssim \fepub \qquad \text{in} \quad \intM,
    \end{equation*}
    where $\mathfrak{d}\coloneqq\{e_3, re_4, r\nab \}$.
\end{proposition}
\begin{proof}
   Letting $\tub=\tub_0+\hb$, then $\hb= 0$ along $\TT$. And the eikonal equation \eqref{incoming eikonal eqn} takes the form of
   \begin{align*}
       0=\bfg^{\a \b} \pr_{\a} (\tub_0+\hb) \pr_{\b}(\tub_0+\hb)=&-e_3(\tub_0+\hb)e_4(\tub_0+\hb)+|\nab(\tub_0+\hb)|^2 \\
       =&-e_3(\tub_0)e_4(\hb)-e_4(\tub_0)e_3(\hb)+2\nab \tub_0\cdot \nab h\\&+\bfg^{\a \b} \pr_{\a} \hb \pr_{\b}\hb+\bfg^{\a \b} \pr_{\a} \tub_0\pr_{\b}\tub_0.
   \end{align*}
   Note that the above form of eikonal equation is equivalent to
   \begin{equation}\label{nonlinear transpoar eqn}
       (2\bfD \tub_0+\bfD \hb)\cdot \bfD\hb=-\bfD \tub_0\cdot \bfD \tub_0.
   \end{equation}
   Using the fact that $\hb\equiv 0$ along $r=r_0$, we can write $\bfD\hb|_{\TT}=\mu \bfD r$ with a scalar function $\mu$. Plugging this into \eqref{nonlinear transpoar eqn}, we derive 
   \begin{equation}\label{quadratic eqn for mu}
       \l (\bfD r\cdot \bfD r)\mu^2+ 2(\bfD \tub_0 \cdot \bfD r)\mu+\bfD \tub_0\cdot \bfD \tub_0 \r\Big|_{\TT}=0.
   \end{equation}
   According to \eqref{Est:inner product}, we have
   \begin{equation*}
       (\bfD \tub_0 \cdot \bfD r)^2-(\bfD r\cdot \bfD r)(\bfD \tub_0\cdot \bfD \tub_0)=\f{\Si^{*2}}{|q|^4}+O(\fepub)\f{\De}{|q|^2}>0.
   \end{equation*}
This ensures that the quadratic equation  \eqref{quadratic eqn for mu} processes two real roots
   \begin{equation*}
       \mu_{\pm}=\f{-\bfD \tub_0\cdot \bfD r\pm\sqrt{(\bfD \tub_0 \cdot \bfD r)^2-(\bfD r\cdot \bfD r)(\bfD \tub_0\cdot \bfD \tub_0)}}{\bfD r\cdot \bfD r}.
   \end{equation*}
   Choosing $\mu=\mu_+$ so that the corresponding optical function is incoming, we then deduce the initial data for $\bfD\hb$ along $\TT$, namely, $ \bfD\hb|_{\TT}=\mu_+ \bfD r$. Notice that it can be inferred from \eqref{Est:inner product} that
   \begin{equation}\label{Est smallness of mu+}
       \mu_+=\f{-\f{\Sis}{|q|^2}+O(\fepub)+\sqrt{\f{\Si^{*2}}{|q|^4}+O(\fepub)}}{\f{\De}{|q|^2}+O(\fepub)}=O(\fepub). \\[2mm]
   \end{equation}
   
   For $x=(\ub, r, \th, \varphi)\in \intM$ and $p=(p_{\ub}, p_{r}, p_{\th}, p_{\varphi})$, we define 
   \begin{equation*}
       F(x, p)=\bfg^{\a \b} (\pr_{\a} \tub_0+p_{\a}) (\pr_{\b}\tub_0+p_{\b}).
   \end{equation*}
With the estimates of $\tub_0$ in \eqref{Est for Dtub0}, we get
\begin{equation*}
    F(x, p)=\bfg^{\a \b} \pr_{\a} \tub_0 \pr_{\b} \tub_0+2\bfg^{\a \b} \pr_{\a} \tub_0 \, p_{\b}+\bfg^{\a \b} p_{\a}p_{\b}=O(1)(p\c p+p)+O(\fepub).
\end{equation*}
   
   We also notice that The eikonal equation \eqref{incoming eikonal eqn} can be transferred to the form
   \begin{equation*}
       F(x, D\hb)=0 \qquad \text{with} \quad  x=(\ub, r,\th,\varphi), \ D\hb\coloneqq(\pr_{\ub} \hb, \pr_{r} \hb, \pr_{\th} \hb, \pr_{\varphi} \hb).
   \end{equation*}
   Based on the theory of first-order nonlinear PDE, we further convert the above equation to the below Hamiltonian system for the the characteristic curve $x=x(s)$ with $p=\f{dx(s)}{ds}$:
   \begin{equation}\label{Eqn: Character for hb}
      \left\{ \begin{aligned}
           &\f{d}{ds} p_{\a}=D_{x^{\a}}F(x, p), \\
           &\f{d}{ds} x^{\a}=-D_{p_{\a}} F(x, p)=-2\bfg^{\a \b}(\pr_{\b} \tub_0+p_{\b}), \\
       \end{aligned} \right.
   \end{equation}
with the initial conditions
\begin{equation*}
     x(0)=(\ub, r_0, \th, \varphi), \quad p(0)=\mu_+ Dr(\ub, r_0, \th, \varphi).
\end{equation*}
Once \eqref{Eqn: Character for hb} can be solved, integrating along the characteristic curve $x=x(s)$, the solution to \eqref{incoming eikonal eqn} is obtained through
\begin{equation*}
    \hb(x(s))=\hb(x(0))-\int_{0}^{s} D_{p_{\a}} F(x(s'), p(s')) p_{\a} (s') ds'=-2\int_{0}^{s} \bfg^{\a \b}p_{\a} (\pr_{\b} \tub_0+p_{\b}) ds'.
\end{equation*}
Also, the following relation holds
\begin{equation*}
    D\hb(x(s))=p(x(s)). \\[2mm]
\end{equation*}

We then proceed to solve \eqref{Eqn: Character for hb}. Given any point $(\ub, r_0, \th, \varphi)\in \TT$, we make a bootstrap assumption as below:
\begin{equation}\label{BA for p}
    |p|=\sqrt{p_{\ub}^2+p_{r}^2+p_{\th}^2+p_{\varphi}^2}\le \epsilon\coloneqq\epsilon_0^{\f{2}{3}} \qquad \text{for all} \quad 0\le s\le s_0.
\end{equation}
Here $s_0>0$ is a constant independent of $\epsilon_0$. 

Consider the second equation of \eqref{Eqn: Character for hb}. Employing the estimates for metric components from \Cref{Thm:KS main}, together with \eqref{Est:inner product} and noting that $|a|\lesssim 1$, we obtain
\begin{equation}\label{Est for pr s x}
    \begin{aligned}
    \f{d}{ds} r(s)=&-2\bfg^{r\b}(\pr_{\b} \tub_0+p_{\b})=-2\bfD \tub_0 \cdot \bfD r+O(\epsilon)=-\f{2\Sis}{|q|^2}+O(\epsilon), \\
    \f{d}{ds} x^{\a}(s)=&-2\bfg^{\a\b}(\pr_{\b} \tub_0+p_{\b})=-2\bfD \tub_0 \cdot \bfD x^{\a}+O(\epsilon)=O(a)+O(\epsilon)=O(1) \quad \text{with} \ \a\in \{\ub, \th, \varphi \}. 
\end{aligned}
\end{equation}
Hence, there exist constants $C>c>0$ independent of $\epsilon_0, \epsilon$ and $\ub$, so that
\begin{align*}
    r_0-Cs\le r(s)\le r_0-cs, \qquad |\ub(s)-\ub|+|\th(s)-\th|+|\varphi(s)-\varphi|\lesssim s.
\end{align*}
Since $(\ub(s_0), r(s_0), \th(s_0), \varphi(s_0))$ still belongs to $\intM=\{ r_{\mathcal{H}} \le r\le r_0  \}$, it consequently gives
\begin{equation*}
   r_{\mathcal{H}}\le r(s_0)\le r_0-cs.
\end{equation*}
This leads to the desired control of $s_0$, i.e.,
\begin{equation}\label{Est for s0}
    s_0\le c^{-1}(r_0-r_{\mathcal{H}})\lesssim r_0.
\end{equation}

Now, in view of estimates for metric components as in \Cref{Thm:KS main}, together with \eqref{Est for Dtub0}, we derive\footnote{When the spacetime metric degenerates near $\th=0, \f{\pi}{2}$, we can use the other regular coordinate system $(x^1, x^2)=(\sin\th\cos\varphi, \sin\th\sin\varphi)$ on $S_{\ub, r}$ for $\th\in [0, \f{\pi}{3}]\cup [\f{2\pi}{3}, \pi]$. And the similar estimates as in \eqref{Est for D x F} still hold with $(\th, \varphi)$ replaced by $(x^1, x^2)$.}
\begin{equation}\label{Est for D x F}
    |D_{x^{\a}}F|\lesssim |p|^2+|p|+\fepub \lesssim (1+\epsilon)|p|+\epsilon_0.
\end{equation}
This gives
\begin{equation*}
    \f{d}{ds}|p|\lesssim (1+\epsilon)|p|+\epsilon_0.
\end{equation*}
Applying  Gr\"{o}nwall's inequality, we thus deduce
\begin{equation*}
    |p(s)|\le e^{Cs}\big((1+\epsilon)|p_0|+ \epsilon_0 \big)=e^{Cs}\big((1+\epsilon)|\mu_+ Dr(\ub, r_0, \th, \varphi)|+ \epsilon_0 \big)\lesssim e^{Cs_0}\epsilon_0.
\end{equation*}
The bootstrap assumption \eqref{BA for p} is hence improved due to the uniform bound for $s_0$ in \eqref{Est for s0}.

Therefore, we conclude that the characteristic equation \eqref{Eqn: Character for hb} admits a unique solution in $[0, s_0]$ satisfying
\begin{equation}\label{Est for p, x}
\begin{aligned}
     |p(s)|\lesssim \fepub, \quad r_0-Cs\le r(s)\le& r_0-cs, \\
     |\ub(s)-\ub|+|\th(s)-\th|+|\varphi(s)-\varphi|\lesssim& s_0.
\end{aligned}
\end{equation}
The corresponding solution to $F(x, D\hb)=0$ now obeys
\begin{align*}
    |\hb(x(s))|=|2\int_{0}^{s} \bfg^{\a \b}p_{\a} (\pr_{\b} \tub_0+p_{\b}) ds'|\lesssim& \fepub, \qquad
    |D\hb(x(s))|=|p(s)|\lesssim \fepub. 
\end{align*}

Define the flow map $\Phi: (s, y)\to x(s, y)$ with $y=(r,\th,\varphi)$. Denoting $D$ to be the derivatives in variables $(\ub, r, \th, \varphi)$ and using $\nab$ to represent the gradient with respect to $(s, y)$, we next prove that $\Phi$ is a diffeomorphism and $\Phi$ satisfies the following property with $0\le k\le k_{small}-1$:
\begin{equation}\label{Esf for Phi p}
\begin{aligned}
    |D\Phi^{-1}|+\max\limits_{1\le k\le k_{small}-1}|\nab^k \Phi|\lesssim 1, \qquad
    |\nab^{k} p|\lesssim \fepub.
\end{aligned}
\end{equation}

We start with the estimates for the first derivatives of $\Phi$ and $p$. For $\pr_s \Phi$ and $\pr_s p$, it follows from \eqref{Eqn: Character for hb}, \eqref{Est for D x F}, \eqref{Est for pr s x}, \eqref{Est for p, x}, that
\begin{equation}\label{Est for pr x pr s}
 \begin{aligned}
           &|\pr_s p|=|D_{x}F|\lesssim |p|^2+|p|+\fepub\lesssim \fepub, \\
           &|\pr_s x^r|\sim 1, \qquad |\pr_s x^{\a}| \lesssim 1 \quad \text{with} \ \a\in \{\ub, \th, \varphi \}.\\
       \end{aligned}
   \end{equation}
Regarding derivatives of $p(s, y)$ and $x(s, y)$ with respect to $y=(\ub, \th, \varphi)$, we aim to establish
 \begin{equation}\label{Est for der of x,p in y}
     \begin{aligned}
     |\f{\pr x^{\a}}{\pr y^{\b}}-\de^{\a}_{\b}|\lesssim s_0, \qquad |\pr_y p|\lesssim \fepub.
     \end{aligned}
 \end{equation}
 Differentiating \eqref{Eqn: Character for hb} with respect to $y^{\b}$, we obtain a system of equations
 \begin{equation}\label{Eqn for pr y x,p}
\left\{ \begin{aligned}
           &\f{d}{ds} \f{\pr p_{\a}}{\pr y^{\b}}=D_{x^{\gamma}}D_{x^{\a}}F \f{\pr x^{\gamma}}{\pr y^{\b}}+D_{p_{\gamma}}D_{x^{\a}}F \f{\pr p_{\gamma}}{\pr y^{\b}}, \\
           &\f{d}{ds} \f{\pr x^{\a}}{\pr y^{\b}}=-D_{x^{\gamma}}D_{p_{\a}} F \f{\pr x^{\gamma}}{\pr y^{\b}}-D_{p_{\gamma}}D_{p_{\a}} F \f{\pr p_{\gamma}}{\pr y^{\b}}, \\
       \end{aligned} \right.
   \end{equation}
with the initial conditions
\begin{equation*}
     \f{\pr x^{\a}}{\pr y^{\b}}(0)=\de^{\a}_{\b}, \quad \f{\pr p}{\pr y}(0)=\pr_{y} (\mu_+ Dr)(\ub, r_0, \th, \varphi)=O(\fepub).
\end{equation*}
Equipped with above equations, we proceed to impose the bootstrap assumption $|\pr_y p|\le 1$. Then from the second equation of \eqref{Eqn for pr y x,p} and the fact that $|D_{x}D_x F|,|D_p D_x F|\lesssim 1$, we deduce
\begin{equation}\label{Est:prs pry x}
  | \f{d}{ds} \pr_y x|\lesssim |\pr_y x|+|\pr_y p|\lesssim |\pr_y x|+1.
\end{equation}
Hence, by  Gr\"{o}nwall's inequality, we get
\begin{equation}\label{Est for pr y x bounded}
    |\pr_y x|\lesssim 1.
\end{equation}
Plugging this into the first equation in \eqref{Eqn for pr y x,p}, we infer
\begin{equation*}
    |\f{d}{ds} \pr_y p|\lesssim |\pr_y p|+\fepub |\pr_y x|\lesssim |\pr_y p|+\fepub,
\end{equation*}
where we use the fact that 
\begin{equation*}
   |D_{x^{\gamma}}D_{x^{\a}}F|\lesssim |p|^2+|p|+\fepub\lesssim \fepub.
\end{equation*}
Employing Gr\"{o}nwall's inequality again, we then derive the improved estimate
\begin{equation}\label{Est for improved pr y p}
    |\pr_y p|\lesssim |\pr_y p(0)|+\fepub\lesssim\fepub
\end{equation}
provided that $\epsilon_0$ is sufficiently small. 

To control $\pr_y x$, we insert \eqref{Est for pr y x bounded} into \eqref{Est:prs pry x} and deduce
\begin{equation*}
     | \f{d}{ds} \pr_y x|\lesssim 1.
\end{equation*}
This immediately yields our desired estimates
\begin{align*}
    |\f{\pr x^{\a}}{\pr y^{\b}}-\de^{\a}_{\b}|=&|\f{\pr x^{\a}}{\pr y^{\b}}-\f{\pr x^{\a}}{\pr y^{\b}}(0)|\lesssim s_0.
\end{align*}
\vspace{2mm}

Recalling the boundedness of $\f{\pr x^{\a}}{\pr s}$ in \eqref{Est for pr x pr s} and viewing $\f{\pr x^{\a}}{\pr s}$, $\f{\pr x^{\a}}{\pr y^{\b}}$ as entries of the Jacobian matrix $\nab\Phi$, we obtain
\begin{equation*}
    \nab\Phi= \f{\pr x}{\pr (s, y)}=
  \begin{pmatrix}
    \pr_s x^r &  &  &    \\
    * & 1 &  &    \\
    * &   & 1 &  \\
    * &   &  &  1\\
  \end{pmatrix}+O(s_0).
\end{equation*}
From this form, we can see that $\nab\Phi$ is invertible if choosing $s_0\ll 1$ and it holds $|D\Phi^{-1}|\lesssim 1$.

Then following from the inverse function theorem, there exists a positive constant $\de$ independent of $\epsilon_0, \ub$, such that  
\begin{equation*}
    \intM\cap \{ \ r_0-\de\le r\le r_0 \}\subset \text{Im}(\Phi).
\end{equation*}

The estimates for higher order derivatives of $\Phi$ and $p$ can be derived in the same fashion by differentiating equations \eqref{Eqn: Character for hb} for multiple times and by conducting the same inductive argument. Hence, we prove that \eqref{Esf for Phi p} holds true in $\intM\cap \{ \ r_0-\de\le r\le r_0 \}$, i.e.,
\begin{equation*}
\begin{aligned}
    |D\Phi^{-1}|+\max\limits_{1\le k\le k_{small}-1}|\nab^k \Phi|\lesssim& 1, \qquad
    |\nab^{k} p|\lesssim \fepub.
\end{aligned}
\end{equation*}
Combining with the equality
\begin{equation*}
    D\hb(x(s))=p(x(s)),
\end{equation*}
we thus conclude that for $0\le k\le k_{small}$, we have
    \begin{equation}\label{Est for hb first interval}
        |D^k \hb|\lesssim \fepub \qquad \text{in} \quad \intM\cap \{ r_0-\de\le r\le r_0 \}.
    \end{equation} 

Repeating the above process, we can then solve  $F(x, D\hb)=0$ with
\begin{equation*}
  \quad \hb=\hb(\ub, r_0-(l-1)\de, \th, \varphi), \quad D\hb=D\hb(\ub, r_0-(l-1)\de, \th, \varphi) \qquad \text{along} \quad  \{r= r_0-(l-1)\de\}
\end{equation*}
 for $l\in \mathbb{N}$ and we also establish the desired estimates \eqref{Est for hb first interval}  in the region $\intM\cap \{ r_0-l\de\le r\le  r_0-(l-1)\de \}$ until it exceeds the inner boundary of $\intM$. This completes the proof of this proposition.   
\end{proof}
\begin{remark}
    The above proof does not utilize the condition $|a/m|\ll 1$ and is valid for the full subextremal Kerr solutions with $|a/m|<1$.
\end{remark}
\begin{remark}\label{Rmk: stable eikonal eqn}
   Following the proof of \Cref{Lem:est for tub}, we can see that the solutions to the eikonal equation \eqref{incoming eikonal eqn} are stable subject to tiny perturbations of initial data along $r=r_0$. Specifically, if ${}^{(pert)}\tub$ solves \eqref{incoming eikonal eqn} with ${}^{(pert)}\tub={}^{(pert)}\tub_0$ along $r=r_0$ and 
   \begin{equation*}
       \sup\limits_{\intM}|\mathfrak{d}^{\le k_{small}}({}^{(pert)}\tub_0-\tub_0)|\ll 1,
   \end{equation*} 
   then
   \begin{equation*}
      \sup\limits_{\intM} |\mathfrak{d}^{\le k_{small}}({}^{(pert)}\tub-\tub)|\lesssim  \sup\limits_{\intM} |\mathfrak{d}^{\le k_{small}}({}^{(pert)}\tub_0-\tub_0)|.
   \end{equation*}
\end{remark}
Consequently, by \Cref{Lem:est for tub} and \eqref{Est for Dtub0}, we also obtain the estimations for derivatives of $\tub$:
\begin{equation}\label{Est for Dtub}
    \begin{aligned}
        e_3(\tub)=&e_3(\tub_0)+O(\fepub) 
        =\f{r^2+a^2-\sqrt{(r^2+a^2)^2-a^2 \sin^2 \ths \De}}{\De}+O(\fepub), \\
        e_4(\tub)=&e_4(\tub_0)+O(\fepub)=\f{r^2+a^2+\sqrt{(r^2+a^2)^2-a^2 \sin^2 \ths \De}}{|q|^2}+O(\fepub), \\
        \nab(\tub)=&\nab(\tub_0)+O(\fepub)=\f{1}{|q|}(-\sgn(r-r_{+})|a|\sqrt{\sin^2 \ths-\sin^2 \th}, a\sin \th)+O(\fepub). \\[2mm]
    \end{aligned}
\end{equation}

 We proceed to define an incoming geodesic foliation of $\intM$ adapted to the incoming optical function $\tub$ as follows: 
\begin{itemize}
     \item   Set $\te_3=-\bfD \tub$ and define
     \begin{equation}\label{Def:t r}
         \t{r}(r,\th)\coloneqq e^{\kappa \rs(r,\th)}+r_+ \qquad \text{with} \quad \kappa=\f{r_+-r_-}{2mr_+}.
     \end{equation}
     Here $r_{\pm}=m\pm\sqrt{m^2-a^2}$. Recalling that $\rs(r, \th)=f(r)+\fs(r, \th)$ and
     \begin{equation*}
         f(r)=\int \f{r^2+a^2}{\De} dr=r+\f{2m}{r_+-r_-}\qty(r_+\log|r-r_+|-r_-\log|r-r_-|),
     \end{equation*}
     we can thus express $\t{r}=\t{r}(r,\th)$ as
     \begin{equation*}
         \t{r}=e^{\kappa (r+\fs(r,\th))}(r-r_-)^{-\f{r_-}{r_+}}(r-r_+)+r_+\eqqcolon g(r,\th)(r-r_+)+r_+,
     \end{equation*}
     where $g(r, \th)\sim 1$ is a smooth function for $r, \th\in [r_{\HH}, \infty)\times [0, \pi]$.
     
     \item Denote $\tHb_{\tub}$ to be the level set of $\tub$ and define $\tS_{\tub, \t{r}}$ as the level set of $\t{r}$ along $\tHb_{\tub}$. We choose an orthonormal frame $\{\te_1, \te_2 \}$ of $\tS_{\tub, \t{r}}$ and let $\te_4$ be the unique outgoing null vectorfield that is orthogonal to $\tS_{\tub, \t{r}}$ and is normalized by $\bfg(\te_3, \te_4)=-2$. 

    \item  To define angular coordinates $(\t{\th}^1, \t{\th}^2)$ on each $\tS_{\tub, \t{r}}$, we first pick a coordinate system $(\t{\th}^1, \t{\th}^2)$ on $\tS_{1, r_+}$. Then we
 propagate this coordinate system to $\TT_+\coloneqq\qty{\t{r}=r_+}=\qty{r=r_+}$, such that
 \begin{equation}\label{Eqn for tth along r=r0}
     (-e_3(r)e_4+e_4(r) e_3)(\t{\th}^a)=0 \qquad \text{with} \quad a=1,2.
 \end{equation}
We further transport this coordinate system from $\TT_+=\{\t{r}=r_+\}$ along the integral curve of $\te_3$ under the requirement
\begin{equation*}
    \te_3(\t{\th}^a)=0 \qquad \text{with} \quad a=1,2.
\end{equation*}
 \end{itemize}

In order to evaluate the transition coefficients for the change of frame from $(e_1, e_2, e_3, e_4)$ to $(\te_1, \te_2, \te_3, \te_4)$, proceeding similarly as in \Cref{Subsubsec:initial incoming PG layer},  we decompose this frame transformation into two steps as below:
\begin{itemize}
    \item The transition coefficients from  $(e_{\mu})$ to $( {}^{(1)}e_{\mu})$ are given by $({}^{(1)}f,{}^{(1)}\fb,{}^{(1)}\lambda)$. Here ${}^{(1)}e_{3}=\te_3$ and ${}^{(1)}f=0$.
    \item The transition coefficients from  $( {}^{(1)}e_{\mu})$ to $( \te_{\mu})$ are given by $({}^{(2)}f,{}^{(2)}\fb,{}^{(2)}\lambda)$. Here ${}^{(2)}\fb=0$ and ${}^{(2)}\la=1$.
\end{itemize}
Denoting the transition coefficients from $(e_{\mu})$ to $(\te_{\mu})$ by $(f,\fb, \la)$, we have the following relation
\begin{equation}\label{Eqn:f fb la}
    ( f, \fb, \la)=({}^{(2)}f,{}^{(2)}\fb=0,{}^{(2)}\lambda=1)\circ( {}^{(1)}f=0,{}^{(1)}\fb,{}^{(1)}\lambda ).
\end{equation}

It suffices to control $( {}^{(1)}f,{}^{(1)}\fb,{}^{(1)}\lambda ) $ and $({}^{(2)}f,{}^{(2)}\fb,{}^{(2)}\lambda)$.  Recalling the frame transformation formula \eqref{frame transform formula}, there holds
    \begin{equation}\label{Eqn:trans1}
        ^{(1)}e_3=^{(1)}\lambda^{-1}\qty(e_3+{}^{(1)}\fb^b e_b+\f14 |^{(1)}\fb|^2 e_4 ),  \qquad \te_a={}^{(1)}e_a+\f{1}{2} {}^{(2)}f_a {}^{(1)}e_3.
    \end{equation}
Comparing the first equality with 
\begin{equation*}
    ^{(1)}e_3=\te_3=-\bfD\tub=\f12 e_4(\tub)e_3\f12 e_3(\tub)e_4-e^a(\tub)e_a,
\end{equation*}
we derive
\begin{equation*}
    {}^{(1)}\lambda=\f{2}{e_4(\tub)}, \qquad {}^{(1)}\fb=-\f{2\nab \tub}{e_4(\tub)}.
\end{equation*}
Combining with \eqref{Est for Dtub}, this further renders
\begin{equation}\label{Est:la fb 1}
       {}^{(1)}\fb=-\f{2|q|}{r^2+a^2+\Sis}(|a|\bs, a\sin\th)+O(\fepub),  \quad   {}^{(1)}\lambda=\f{2|q|^2}{r^2+a^2+\Sis}+O(\fepub).
\end{equation}
Here $\Sis=\sqrt{(r^2+a^2)^2-a^2\De\sin^2\ths}$ and $\bs=-\sgn(r-r_+)\sqrt{\sin^2\ths-\sin^2\th}$.

Next, since $\te_1, \te_2 \in T \tS_{\tub, \t{r}}$, from \eqref{Eqn:trans1} we obtain\footnote{Note that it automatically holds $\te_a(\tub)={}^{(1)}e_a(\tub)+\f12{}^{(2)}f_a {}^{(1)}e_3(\tub)=0$.}
\begin{equation*}
    0=\te_a(\t{r})={}^{(1)}e_a(\t{r})+\f{1}{2} {}^{(2)}f_a {}^{(1)}e_3(\t{r}),
\end{equation*}
from which we deduce that 
\begin{equation}\label{Eqn:f 2}
     {}^{(2)}f=-\f{2{}^{(1)}\nab \t{r}}{{}^{(1)}e_3(\t{r})}.
\end{equation}
Notice that by virtue of \eqref{In PG cond}, \eqref{Est for coor derivative} and \eqref{Est for D rs}, the derivatives of $\t{r}$ obey the bounds
\begin{equation}\label{Est:der t r}
    \begin{aligned}
        e_3(\t{r})=&\kappa e^{\kappa \rs}e_3(\rs)=-\kappa e^{\kappa \rs}\c \f{\Sis}{\De}, \\
        e_4(\t{r})=&\kappa e^{\kappa \rs} e_4(\rs)=\kappa e^{\kappa \rs}\qty(\f{\Sis}{|q|^2}+O(\fepub)), \\
        \nab(\t{r})=&\kappa e^{\kappa \rs} \nab(\rs)=\kappa e^{\kappa \rs}\qty(\f{|a|}{|q|}(\bs,0)+O(\fepub)).
    \end{aligned}
\end{equation} 

As a consequence, together with \eqref{Est for Dtub}, we derive
\begin{equation}\label{Est for te4 tub,r}
\begin{aligned}
    \te_4(\tub)=&\bfg(\te_4, \bfD \tub)=-\bfg(\te_4, \te_3)=2,  \\
    \te_3(\t{r})=&-\bfg(\bfD \tub, \bfD \t{r})=\f12 e_3(\tub)e_4(\t{r})+\f12 e_4(\tub)e_3(\t{r})-\nab(\tub)\c\nab(\t{r})\\
     =&- \f{\kappa e^{\kappa \rs}}{\De |q|^2}\qty((\Sis)^2+a^2\bs^2\De+O(\fepub))
     =-\f{\kappa g}{|q|^2(r-r_-)} \qty(\Si^2+O(\fepub))\sim -1.
\end{aligned}
\end{equation}
Here we utilize the fact that $e^{\kappa \rs}=g(r,\th)(r-r_+)$ with $g\sim 1$. 

In view of
\begin{equation*}
    -\te_3(\t{r})\te_4(\t{r})=\bfg(\bfD \t{r}, \bfD \t{r})=-e_3(\t{r})e_4(\t{r})+|\nab \t{r}|^2,
\end{equation*}
we then infer that
\begin{align}\label{Est for te3 r}
    \te_4(\t{r})=\f{e_3(\t{r})e_4(\t{r})-|\nab \t{r}|^2}{\te_3(\t{r})}=
    \kappa (\t{r}-r_+)+O(\fepub).
\end{align}
Hence, plugging \eqref{Est for Dtub}, \eqref{Est for Dtub}, \eqref{Est:der t r} and \eqref{Est for te4 tub,r} into \eqref{Eqn:f 2}, we arrive at
\begin{equation}\label{Est:f 2}
    \begin{aligned}
     {}^{(2)}f=-\f{2}{\te_3(\t{r})}\qty(\nab\t{r}+\f12 {}^{(1)}\fb e_4(\t{r}))=
     \f{2|q|\De }{\Si^2(r^2+a^2+\Sis)} (|a|\bs(r^2+a^2), -a\sin\th \Sis)+O(\fepub).
    \end{aligned}
\end{equation}

Collecting estimates for transition coefficients as in \eqref{Est:la fb 1} and \eqref{Est:f 2} and utilizing the decomposition relation \eqref{Eqn:f fb la}, we conclude
\begin{equation}\label{Est: f fb la}
    |\mathfrak{d}^{\le k_{small}-1} (f, \fb)|\lesssim |a|+\fepub, \quad |\mathfrak{d}^{\le k_{small}-1} \log \la|\lesssim 1 \qquad \text{with} \quad \mathfrak{d}=\qty{\nab_3,r\nab_4, r\nab}.
\end{equation}
 
Therefore, with the assistance of estimations for $(f, \fb, \lambda)$, we derive the below asymptotic behaviors of geometric quantities associated with the new null frame $(\te_1, \te_2, \te_3, \te_4)$.
\begin{proposition}\label{Lem:hyper est in incom geo frame}
    Relative to the null frame $(\te_1, \te_2, \te_3, \te_4)$, the Ricci coefficients $\t{\Gamma}$ and curvature components $\t{R}$ obey
 \begin{equation}\label{Est for Ga R new frame}
	|\mathfrak{d}^{\le k_{small}-2} (\t{\Gamma}, \t{R})|\lesssim 1.  \quad \intM
	\end{equation}
 Moreover, letting $\qty{(\te_\mu)_0}$ with $(\te_3)_0=-\bfD\tub_0$ be the null frame adapted to 2-spheres $\tS_{\tub_0, \t{r}}\coloneqq\qty{\tub_0'=\tub_0} \cap\qty{\t{r}'=\t{r}}$ in the exact Kerr spacetime $Kerr(a, m)$ and denoting the corresponding null expansions by $\tr{\t{\chi}}_0$ and $\tr{\t{\chib}}_0$, in our considered spacetimes we have the leading order estimates
 \begin{equation}\label{Est for trch new frame}
    \begin{aligned}
             \tr{\t{\chi}}=\tr{\t{\chi}}_0+O(\fepub), \qquad  \tr{\t{\chib}}=\tr{\t{\chib}}_0+O(\fepub). 
    \end{aligned}
    \end{equation}
\end{proposition}
\begin{remark}
    In \Cref{Appendix:Null expansion}, we further show that
    \begin{align*}
        \tr{\t{\chi}}_0=F\c(\t{r}-r_+) \quad \text{with} \quad F\sim1 \qquad \text{and} \qquad \tr{\t{\chib}}_0<0.
    \end{align*}
    Notably, this proof is applicable to all subextremal Kerr solutions with $|a/m|< 1$.
\end{remark}
\begin{proof}[Proof of \Cref{Lem:hyper est in incom geo frame}]
  From \eqref{Est: f fb la}, we have the bounds of the transition coefficients $(f, \fb, \la)$ up to $(k_{small}-1)$-th order. Plugging these into the transformation formulas provided in \Cref{Lem:transformation formula}, we then obtain \eqref{Est for Ga R new frame}. We proceed to prove \eqref{Est for trch new frame}. Denote
   \begin{align*}
       {}^{(1)}\fb_0=& -\f{2|q|}{r^2+a^2+\Sis}(|a|\bs, a\sin\th),  \qquad   {}^{(1)}\lambda_0=\f{2|q|^2}{r^2+a^2+\Sis}, \\  {}^{(2)}f_0=&
     \f{2|q|\De }{\Si^2(r^2+a^2+\Sis)} (|a|\bs(r^2+a^2), -a\sin\th \Sis)
   \end{align*}
   and define
   \begin{equation*}
       (f_0, \fb_0, \la_0)=({}^{(2)}f_0,{}^{(2)}\fb_0=0, {}^{(2)}\lambda_0=1)\circ ({}^{(1)}f_0=0,{}^{(1)}\fb_0, {}^{(1)}\lambda_0).
   \end{equation*}
   Here $(f_0, \fb_0, \la_0)$ corresponds to the transition coefficients from the canonical (incoming) PG null frame $\qty{(e_{\mu})_0}$ introduced in \eqref{Eqn:cano PG e3 e4} and \eqref{Eqn:cano PG ea} to the adapted geodesic null frame $\qty{(\te_\mu)_0}$ in the exact Kerr spacetime $Kerr(a, m)$.

   According to \eqref{Est:la fb 1}, \eqref{Est:f 2} and \eqref{Eqn:f fb la}, we have\footnote{The estimates for tensors are understood in the sense of components.} 
   \begin{equation*}
       |\mathfrak{d}^{\le k_{small}-1} (f-f_0, \fb-\fb_0, \log \f{\la}{\la_0})|\lesssim \fepub.
   \end{equation*}
   The desired asymptotics \eqref{Est for trch new frame} thus follow from the transformation formulas as provided in \Cref{Lem:transformation formula}.
\end{proof}
We also need the bounds of $\te_4(\tth^a)$ and $\tnab(\tth^a)$ in $\intM$. To this end, we first commute \eqref{Eqn for tth along r=r0} with $\nab$. Utilizing the commutation formulas in \Cref{Lem:commute}, along $r=r_+$ we derive that
\begin{equation}\label{Eqn:eb ttha}
\begin{aligned}
  \Big(-e_3(r)e_4+e_4(r) e_3\Big)(e_b(\t{\th}^a))=&[-e_3(r)e_4+e_4(r) e_3, e_b](\tth^a) \\
  =&\l-e_3(r)[e_4, e_b]+e_b(e_3(r)) e_4+e_4(r)[e_3, e_b]-e_b(e_4(r))e_3 \r (\tth^a) \\
  =&\Big(e_3(r)\chi_{b}^{\ c}-e_4(r) \chib_{b}^{\ c}\Big) e_c(\tth^a)+ (\eta+\etab)_b \Big(-e_3(r)e_4+e_4(r) e_3\Big)(\t{\th}^a) \\
  =&\Big(e_3(r)\chi_{b}^{\ c}-e_4(r) \chib_{b}^{\ c}\Big) e_c(\tth^a).
\end{aligned}
\end{equation}
Applying hyperbolic estimates as in \eqref{In PG cond}, \eqref{Est for coor derivative} and \eqref{Est for Ricci Cur}, we further compute the coefficients in front of $e_c(\tth^a)$ and it obeys
\begin{align*}
    e_3(r)\chi_{b}^{\ c}-e_4(r) \chib_{b}^{\ c}=&-\qty(\chih_{b}^{\ c}+\f12 \trch \de_{b}^{\ c}+\f12 \atrch\in_{b}^{\ c})-e_4(r)\qty(\chibh_{b}^{\ c}+\f12 \trchb \de_{b}^{\ c}+\f12 \atrchb\in_{b}^{\ c})\\
    =&\f12\qty(-\f{2r\De}{|q|^4}-\f{\De}{|q|^2}\c\f{-2r}{|q|^2})\de_{b}^{\ c}+\f12\qty(-\f{2a\cos\th\De}{|q|^4}-\f{\De}{|q|^2}\c\f{-2a\cos\th}{|q|^2})\in_{b}^{\ c}+O(\fepub)\\
    =&O(\fepub).
\end{align*}
In view of the fact that 
\begin{equation*}
    \Big(-e_3(r)e_4+e_4(r) e_3\Big)(\ub)=e_4(\ub)=\f{2(r^2+a^2)}{|q|^2}+O(\fepub)\sim 1
\end{equation*}
and the integrability of $\ub^{-1-\de_{dec}}$,\footnote{This is the only place where this integrability property is used.} back to \eqref{Eqn:eb ttha}, by Gr\"{o}nwall's inequality we obtain
\begin{equation*}
    |\nab(\tth^a)|\lesssim 1 \qquad \text{along} \quad r=r_+.
\end{equation*}

Then we proceed to estimate $e_3(\tth^a)$ and $e_4(\tth^a)$ along $\TT_+=\qty{r=r_+}$. Employing \eqref{Eqn for tth along r=r0}, the fact $-\bfD\tub(\tth^a)=\te_3(\tth^a)=0$, together with $e_3(r)=-1$, we derive the following system of equations
\begin{equation}\label{Eqn for etth}
    \left\{\begin{aligned}
    e_4(r)e_3(\tth^a)+e_4(\tth^a)=&0, \\
        e_4(\tub)e_3(\tth^a)+e_3(\tub) e_4(\tth^a)=&2e^b(\tub)e_b(\tth^a). 
    \end{aligned}\right.
\end{equation}
Owing to the derived estimates of the transition coefficients $(f, \fb, \la)$ as in \eqref{Est for Dtub}, employing $e_4(r)=\f{\De}{|q|^2}+O(\fepub)$, a direct calculation hence yields
\begin{equation*}
\begin{aligned}
    \det  \begin{pmatrix}
    e_4(r) &  1  \\
     e_4(\tub) & e_3(\tub)     
  \end{pmatrix}
  =e_4(r)e_3(\tub)-e_4(\tub)
  =-\f{2\Sis}{|q|^2}+O(\fepub)\sim -1.
\end{aligned}
\end{equation*}
Back to \eqref{Eqn for etth}, we hence arrive at
\begin{equation}\label{Est:e tth r+}
    |e_3(\tth^a)|+|e_4(\tth^a)|\lesssim |\nab(\tth^a)|\lesssim 1 \qquad \text{along} \quad r=r_+. 
\end{equation}
Together with \Cref{Lem:frame transform}, \eqref{Est for Dtub}, \eqref{Est: f fb la}, this further implies
\begin{equation*}
    |\te_3(\tth^a)|+|\te_4(\tth^a)|+|\tnab(\tth^a)|\lesssim 1 \qquad \text{along} \quad r=r_+. 
\end{equation*}

Next, commuting $\te_3(\tth^a)=0$ with $\te_4$ and $\tnab$ respectively and noting that $\t{\omb}=\t{\xib}=0$ as $\te_3$ is geodesic, we then deduce
\begin{align}
    \te_3(\te_4(\tth^a))=&[\te_3, \te_4](\tth^a)=-2\t{\omega} \te_3(\tth^a)+2\t{\omegab} \te_4(\tth^a)+2(\t{\eta}-\t{\etab})\cdot \tnab (\tth^a)
    =2(\t{\eta}-\t{\etab})\cdot \tnab (\tth^a), \nonumber\\
    \te_3(\te_b(\tth^a))=&[\te_3, \te_b](\tth^a)=-\t{\chib}_{b}^{\ c}\te_c(\tth^a)+\t{\xib}_b \te_4(\tth^a)+(\t{\eta}_b-\t{\zeta}_b)\te_3(\tth^a)=-\t{\chib}_{b}^{\ c}\te_c(\tth^a). \label{Eqn:eb ttha1}
\end{align}
It then follows from \Cref{Lem:hyper est in incom geo frame} and Gr\"{o}nwall's inequality again that\footnote{Since $\te_3(r)\sim -1$ and $r\in [r_{\HH}, r_0]$ within $\intM$, the integral interval is bounded, and there is no issue of integrability compared to the estimates along $r=r_+$.}
\begin{equation*}
    |\te_4(\tth^a)|+|\tnab(\tth^a)|\lesssim 1 \qquad \text{in} \quad \intM. 
\end{equation*}

For later use, we also estimate the inverse matrix of $(\te_b(\tth^a))$, denoted by $(\tX_{b}^{\ a})$. We first compare $\te_b(\tth^a)$ and $e_b(\tth^a)$ along $r=r_+$. From \eqref{Est:f 2} we have ${}^{(2)}f=O(\fepub)$ along $r=r_+$. Thus, in light of \Cref{Lem:frame transform}, \eqref{Eqn:f fb la}, \eqref{Est:la fb 1}, \eqref{Est:f 2} and \eqref{Est:e tth r+}, together with \eqref{Eqn for etth} and the fact that $e_4(r)=\f{\De}{|q|^2}+O(\fepub)$, along $r=r_+$ we derive
\begin{equation}\label{Est:te tth e tth}
\begin{aligned}
    \te_b(\tth^a)={}^{(1)}e_b(\tth^a)+\f12 {}^{(2)}f_b {}^{(1)}e_3(\tth^a)=&e_b(\tth^a)+\f12 {}^{(1)}\fb_b e_4(\tth^a)+O(\fepub)\\
    =&e_b(\tth^a)-\f12 {}^{(1)}\fb_b e_4(r)e_3(\tth^a)+O(\fepub)=e_b(\tth^a)+O(\fepub).
\end{aligned}
\end{equation}

Now, employing \eqref{Eqn:eb ttha} and \eqref{Eqn:eb ttha1}, we deduce
\begin{align*}
\Big(-e_3(r)e_4+e_4(r) e_3\Big)\Big(\det \big(e_b(\tth^a)\big) \Big)=&\Big(e_3(r)\tr\chi-e_4(r) \tr\chib\Big) \det \big(e_b(\tth^a)\big)=O(\fepub)\det \big(e_b(\tth^a)\big),\\
    \te_3\Big(\det \big(\te_b(\tth^a)\big) \Big)=&-\tr \t{\chi} \det \big(\te_b(\tth^a)\big).
\end{align*}
This together with \Cref{Lem:hyper est in incom geo frame}, \eqref{Est:te tth e tth} and Gr\"{o}nwall's inequality yields the boundedness of $\log \det \big(\te_b(\tth^a)\big)$, and hence $\tX_{b}^{\ a}$ is also bounded in $\intM$. The estimates for the higher order derivatives of $\te_4(\tth^a), \tnab(\tth^a)$ and $\tX_{b}^{\ a}$ can be derived in the same fashion. Therefore, we conclude
\begin{equation}\label{Esf for ttha tX a b}
      |\mathfrak{d}^{\le k_{small}-2}(\te_4(\tth^a), \tnab(\tth^a), \tX_{b}^{\ a})|\lesssim 1 \qquad \text{in} \quad  \intM.
\end{equation}

\subsection{Adapted Foliation in the Transition Region}
\begin{figure}[ht]
    \centering

\tikzset{every picture/.style={line width=0.75pt}} %set default line width to 0.75pt        

\begin{tikzpicture}[x=0.75pt,y=0.75pt,yscale=-1,xscale=1]
%uncomment if require: \path (0,415); %set diagram left start at 0, and has height of 415

%Shape: L Shape [id:dp14521572527356863] 
\draw  [color={rgb, 255:red, 255; green, 255; blue, 255 }  ,draw opacity=1 ][fill={rgb, 255:red, 155; green, 155; blue, 155 }  ,fill opacity=0.45 ] (237.22,59.75) -- (265.53,32.44) -- (376.79,147.77) -- (443.64,83.28) -- (470.87,111.5) -- (375.7,203.31) -- cycle ;
%Straight Lines [id:da04148539083938896] 
\draw [color={rgb, 255:red, 0; green, 0; blue, 0 }  ,draw opacity=1 ] [dash pattern={on 0.84pt off 2.51pt}]  (372.93,14.81) -- (469.2,110.5) ;
%Curve Lines [id:da3789953131295283] 
\draw [color={rgb, 255:red, 0; green, 0; blue, 0 }  ,draw opacity=1 ]   (261.93,54.01) .. controls (307.64,47.73) and (324.37,37.87) .. (372.21,15.16) ;
\draw [shift={(372.93,14.81)}, rotate = 334.63] [color={rgb, 255:red, 0; green, 0; blue, 0 }  ,draw opacity=1 ][line width=0.75]      (0, 0) circle [x radius= 2.01, y radius= 2.01]   ;
%Straight Lines [id:da8594658011416159] 
\draw    (372.6,15.08) -- (376.2,197.83) ;
%Straight Lines [id:da20745982101594695] 
\draw    (298.37,45.93) -- (374.66,123.3) ;
\draw [shift={(374.66,123.3)}, rotate = 45.4] [color={rgb, 255:red, 0; green, 0; blue, 0 }  ][fill={rgb, 255:red, 0; green, 0; blue, 0 }  ][line width=0.75]      (0, 0) circle [x radius= 1.34, y radius= 1.34]   ;
%Curve Lines [id:da8756276133891708] 
\draw    (261.93,54.01) .. controls (329.87,122.83) and (357.2,161.17) .. (376.2,197.83) ;
%Curve Lines [id:da12288960939135152] 
\draw    (376.2,197.83) .. controls (385.2,171.83) and (415.53,123.83) .. (447.93,88.21) ;
%Straight Lines [id:da9320302873822118] 
\draw [color={rgb, 255:red, 155; green, 155; blue, 155 }  ,draw opacity=1 ]   (265.53,32.44) -- (377.86,149.1) ;
%Straight Lines [id:da7550702648192735] 
\draw [color={rgb, 255:red, 155; green, 155; blue, 155 }  ,draw opacity=1 ]   (237.22,59.75) -- (375.7,203.31) ;
%Straight Lines [id:da3383148817663877] 
\draw [color={rgb, 255:red, 155; green, 155; blue, 155 }  ,draw opacity=1 ]   (469.2,110.5) -- (375.7,203.31) ;
%Straight Lines [id:da8656525677601468] 
\draw [color={rgb, 255:red, 155; green, 155; blue, 155 }  ,draw opacity=1 ]   (443.64,83.28) -- (377.86,149.1) ;
%Straight Lines [id:da36689465807760224] 
\draw  [dash pattern={on 0.84pt off 2.51pt}]  (256.53,42.17) -- (395.03,184.97) ;

% Text Node
\draw (371.16,-2.64) node [anchor=north west][inner sep=0.75pt]  [font=\scriptsize]  {$i^{+}$};
% Text Node
\draw (433.09,43.76) node [anchor=north west][inner sep=0.75pt]  [font=\scriptsize]  {$\mathscr{I}^{+}$};
% Text Node
\draw (307.26,25.46) node [anchor=north west][inner sep=0.75pt]  [font=\scriptsize]  {$\mathcal{A}$};
% Text Node
\draw (376.09,96.69) node [anchor=north west][inner sep=0.75pt]  [font=\scriptsize]  {$\mathcal{T}$};
% Text Node
\draw (349.83,78.36) node [anchor=north west][inner sep=0.75pt]  [font=\scriptsize]  {$\widetilde{\underline{H}}_{\widetilde{\underline{u}}}$};
% Text Node
\draw (327.36,50.23) node [anchor=north west][inner sep=0.75pt]  [font=\scriptsize]  {$^{( int)}\mathcal{M}$};
% Text Node
\draw (375.86,51.23) node [anchor=north west][inner sep=0.75pt]  [font=\scriptsize]  {$^{( ext)}\mathcal{M}$};
% Text Node
\draw (395.03,185.71) node [anchor=north west][inner sep=0.75pt]  [font=\scriptsize]  {$\underline{H}_{v}$};

\end{tikzpicture}

    \caption{Transition Region between $\PP$ and $\MM$}
    \label{fig:Transition}
\end{figure}
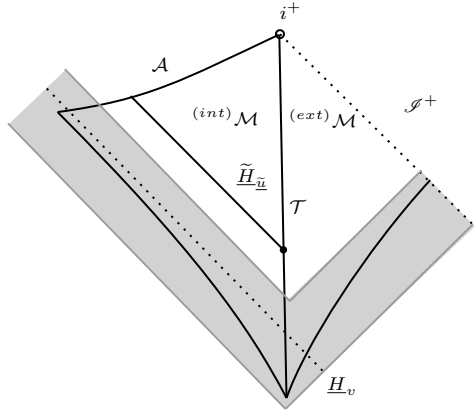

Recall that the short-pulse region $\PP$ treated in \cite{A-L} is equipped with a double foliation and its constant $v$ incoming null hypersurfaces might not coincide with those of the geodesic foliation derived in \Cref{Subsec:incoming geodesic foliation}. This subsection is dedicated to the construction of a smooth incoming optical function within $\Int\MM_0\cap \intM$, that connects the constant $v$ hypersurface in $\PP$ and the constant $\tub$ hypersurface in $\intM$ as depicted in \Cref{fig:Transition}.

Also recall that in \Cref{Sec:Char Data} we have derived two coordinate systems for $\Int\MM_0$, i.e., $(u, v, \th^1, \th^2)$ and $(\ub, r, \th, \varphi)$. Let $(\pul e_{\mu})$ and $(\Int e_{\mu})$ be the corresponding null frames, respectively. Now we consider the transition region $\Int\LL_0\cap \intM$, which is also equipped with the incoming geodesic foliation $(\tub, r, \t{\th}^a)$ associated with the null frame $(\te_{\mu})$. 
Denote the transition coefficients from $(\pul e_{\mu})$ to $(\te_{\mu})$ by $(f''', \fb''', \la''')$. Then we have
\begin{lemma}\label{Lem:est for tran coef1}
    The following estimates hold for the transition coefficients $(f''', \fb''', \la''')$ between null frames $(\pul e_1, \pul e_2, \pul e_3, \pul e_4)$ and $(\te_1, \te_2, \te_3, \te_4)$:
    \begin{equation*}
        |\mathfrak{d}^{\le k_{small}-1}(f''', \fb''')|\lesssim \epsilon_0, \qquad |\mathfrak{d}^{\le k_{small}-1}\log \la'''|\lesssim 1 \qquad \text{in} \quad \Int\LL_0\cap \intM.
    \end{equation*}
\end{lemma}
\begin{proof}
We denote the transition coefficients from  $\qty{\Int (e_0)_{\mu}}$  to $\qty{\Int e_{\mu}}$ by $(f'', \fb'', \la'')$.\footnote{Recall that in \Cref{Subsubsec:initial incoming PG layer} we define the transition coefficients from $\qty{\pul e_{\mu}}$ to 
$\qty{\Int (e_0)_{\mu}}$ by $(\Int f', \Int \fb',  \Int \la')$.} In \Cref{fig:transition coef} below, we can see that
\begin{equation}\label{Eqn:relation f'''}
    ( f''', \fb''', \lambda''' )=( f, \fb, \lambda )\circ ( f'', \fb'', \lambda'' ) \circ ( f', \fb', \lambda' ).
\end{equation}
\begin{figure}[htp]
    \centering
\[\begin{tikzcd}[sep=huge]
	{(\pul e_{\mu})} &&& {(\Int (e_0)_{\mu})} \\
	\\
	{( \te_{\mu})} &&& {(\Int e_{\mu})}
	\arrow["{(\Int f', \Int \fb',  \Int \la')}", from=1-1, to=1-4]
	\arrow["{( f''', \fb''', \lambda''' )}", from=1-1, to=3-1]
	\arrow["{( f'', \fb'', \lambda'' )}", from=1-4, to=3-4]
	\arrow["{( f, \fb, \lambda )}"', from=3-4, to=3-1]
\end{tikzcd}\]
    \caption{Relation of Transition Coefficients between Different Null Frames}
    \label{fig:transition coef}
\end{figure}
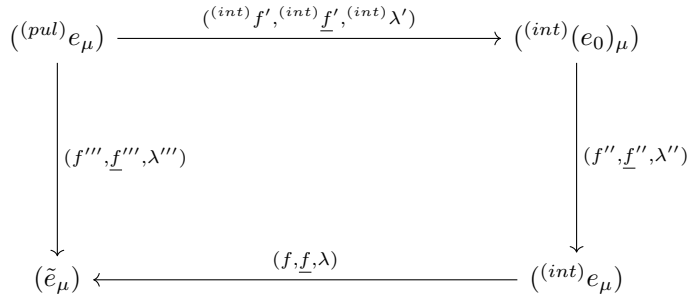

By \eqref{Est for Int f'} in \Cref{Subsubsec:initial incoming PG layer}, we already have
\begin{equation*}
    |\mathfrak{d}^{\le N-5}( \Int f', \Int \fb')|\lesssim\epsilon_1=\epsilon_0^2, \qquad |\mathfrak{d}^{\le N-5}\log \Int \la'|\lesssim 1\qquad \text{in} \quad \Int\LL_0.
\end{equation*}
Meanwhile, following steps of Section 8.3 in  \cite{KS:main}, we deduce\footnote{From (8.3.68) in \cite{KS:main} we have the desired estimates along $\ub=1$. We can then propagate all the estimates from the last slice $\Si_{\ast}\cap \Ext \LL_0$ to $\{r=r_0 \}\cap \LL_0$ and extend the desired estimate from $\{r=r_0 \}\cap \LL_0$ to $\Int\LL_0\cap \intM$.}
\begin{equation*}
     |\mathfrak{d}^{\le k_{small}}(f'', \fb'', \log \la'')|\lesssim \epsilon_0 \qquad \text{in} \quad \Int\LL_0\cap \intM.
\end{equation*}
According to \eqref{Est: f fb la}, we also have
\begin{equation*}
     |\mathfrak{d}^{\le k_{small}-1}(f, \fb)|\lesssim |a| \lesssim \epsilon_0, \qquad |\mathfrak{d}^{\le k_{small}-1}\log \la|\lesssim 1 \qquad \text{in} \quad \Int\LL_0\cap \intM.
\end{equation*}
Combining above estimates with \eqref{Eqn:relation f'''}, we then obtain the control of $( f''', \fb''', \lambda''' )$ with $k_{small}-1\le N-5$
\begin{align*}
     |\mathfrak{d}^{\le k_{small}-1}(f''', \fb''')|\lesssim& |\mathfrak{d}^{\le k_{small}-1}(f'-f, \fb'-\fb)|+|\mathfrak{d}^{\le k_{small}-1}(f'', \fb'')|\lesssim \epsilon_0, \\
     |\mathfrak{d}^{\le k_{small}-1}\log \la'''|\lesssim& |\mathfrak{d}^{\le k_{small}-1}(\log \la', \log \la,  \log \la'')|\lesssim 1.
\end{align*}
as desired.
\end{proof}
\vspace{2mm}

Then we move to construct a family of spacelike 2-surfaces along $\widetilde{\TT}\coloneqq\{\t{r}=\t{r}_0 \}$ that smoothly transit from $v=\text{constant}$ to $\tub=\text{constant}$. Here $\t{r}$ is introduced in \eqref{Def:t r} satisfying
\begin{equation*}
    \t{r}=g(r,\th)\c(r-r_+)+r_+ \qquad \text{with} \quad C^{-1}\le g(r,\th)\le C
\end{equation*}
for some constant $C\ge 1$ and $\t{r}_0\coloneqq C^{-1}r_0$. This gives that $\widetilde{\TT}\subset \Int\MM$.

Note that we have
\begin{align*}
   \te_4 (v)=&\bfg(\te_4, \bfD v)=-\f12 \O^{-1} \bfg(\te_4, \pul e_3)=\O^{-1}\la''', \\
   \te_a (v)=&\bfg(\te_a, \bfD v)=-\f12 \O^{-1} \bfg(\te_a, \pul e_3)=\f12\O^{-1} \fb'''_a, \\
   \te_3 (v)=&\bfg(\te_3, \bfD v)=-\f12 \O^{-1} \bfg(\te_3, \pul e_3)=\f14\O^{-1}(\la''')^{-1} |\fb'''|^2.
\end{align*}
Combining with 
\begin{align*}
    \te_4=\te_4(\tub) \pr_{\tub}+\te_4(\t{r})\pr_{\t{r}}+\te_4(\tth^a)\pr_{\tth^a},\qquad
    \te_3=\te_3(\t{r})\pr_{\t{r}}, \qquad \te_a=\te_a(\tth^b)\pr_{\tth^b},
\end{align*}
we then deduce
\begin{align*}
    \O\te_4(\tub) \pr_{\tub} v=&\O \l \te_4(v)-\f{\te_4(\t{r})}{\te_3(\t{r})}e_3(v)-\te_4(\tth^a)\tX_{a}^{\ b} \te_b(v) \r \\
    =&\la'''-\f{\te_4(\t{r})}{4\te_3(\t{r})}(\la''')^{-1}|\fb'''|^2-\f12\te_4(\tth^a)\tX_{a}^{\ b} \fb_b''',
\end{align*}
where $\tX_{a}^{\ b}$ satisfies
\begin{equation*}
    \tX_{a}^{\ b} \te_b(\tth^c)=\de_a^c.
\end{equation*}
In view of estimates for $(f''', \fb''', \la''')$ in \Cref{Lem:est for tran coef1}, together with \eqref{Est for te4 tub,r}, \eqref{Est for te3 r}, \eqref{Esf for ttha tX a b} and the fact that $\O\sim 1$, we then arrive at
\begin{equation*}
    \pr_{\tub} v\sim  1.
\end{equation*}
This implies that $(v, \t{r}, \tth^1, \tth^2)$ also forms a coordinate system within $\Int\MM_0\cap \intM$. Hence, the optical function $\tub$ can be written as
\begin{equation*}
    \tub=F(v, \t{r}, \tth^1, \tth^2).
\end{equation*}
Applying to the chain rule and \Cref{Lem:est for tran coef1}, we obtain
\begin{equation}\label{Est for pr F}
    \begin{aligned}
    \pr_v F=&(\pr_{\tub} v)^{-1}\sim 1, \\ 
    \pr_{\tth^a} F=&-\pr_v F \pr_{\tth^a} v=-(\pr_{\tub} v)^{-1} \tX_a^{\ b} \te_b(v)=-\f12(\O\pr_{\tub} v)^{-1} \tX_a^{\ b} \fb_b'''=O(\epsilon_0).
\end{aligned}
\end{equation}
This indicates that the oscillation of $F(v, \t{r}, \t{\th}^a)$ for $(\t{\th}^1, \tth^2)\in \mathbb{S^2}$ is of size $\epsilon_0$ .

Next, at each point $(\tub, \t{r}, \tth^1, \tth^2)\in \Int\MM_0\cap \intM$, we proceed to define 
\begin{align*}
    V=&\la v(\tub, \t{r}, \tth^1, \tth^2)+(1-\la)v(\tub, \t{r}, \tth^1_1, \tth^2_1).
\end{align*}
Here $(\t{\th}^1_1, \t{\th}^2_1)$ is a fixed point on $\mathbb{S}^2$ and $\la=\la(\tub)$ is a smooth cut-off function satisfying
\begin{equation*}
    \la(\tub)=\left\{\begin{array}{lll}
         1&  \quad \text{when} \quad \tub\le& F(v_1, \t{r}_0, \tth^1_1, \tth^2_1),\\
         0&  \quad \text{when} \quad \tub \ge& \tub_1. 
    \end{array}\right.
\end{equation*}
For the parameters, we choose $v_1\in (\de, \de+d)$ and $\tub_1>\f32$ such that 
\begin{equation*}
    1<\min\limits_{\t{\th}^1, \tth^2}F(v_1, \t{r}_0, \t{\th}^a) \le \max\limits_{\t{\th}^1, \tth^2}F(v_1, \t{r}_0, \t{\th}^a)<\f{3}{2}<\tub_1<2.
\end{equation*}

Then we set
\begin{equation*}
    \tran \tub_0=F(V, \t{r}, \tth^1_1, \tth^2_1) \qquad \text{along}  \quad \Int\MM_0\cap \intM.
\end{equation*}
The function $\tran \tub_0$ serves as the initial data along $\t{r}=\t{r}_0$ and with it we solve for the new optical function within the transition region $\Int\MM_0\cap \intM$. More specifically, we examine the eikonal equation
\begin{equation}\label{Eikonal Eqn for glo tub}
    \left\{\begin{array}{rll}
         \bfg^{\mu \nu} \pr_{\mu} \tran \tub \pr_{\nu} \tran \tub=&0 \qquad &\text{in} \quad \Int\MM_0\cap \intM,  \\
          \tran \tub=&\tran \tub_0 \qquad &\text{on} \quad \t{r}=\t{r}_0.
    \end{array} \right.
\end{equation}
Mimicking the proof as in \Cref{Lem:est for tub}, in the transition region $\Int\MM_0\cap \intM$ as illustrated in \Cref{fig:transition foliation}, we then obtain the existence and estimates of $\glo \tub$ as stated in below proposition.
\begin{figure}[ht]
    \centering
 \tikzset{every picture/.style={line width=0.75pt}} %set default line width to 0.75pt        

\begin{tikzpicture}[x=0.75pt,y=0.75pt,yscale=-1,xscale=1]
%uncomment if require: \path (0,415); %set diagram left start at 0, and has height of 415

%Shape: Polygon [id:ds6237261309302556] 
\draw  [draw opacity=0][fill={rgb, 255:red, 155; green, 155; blue, 155 }  ,fill opacity=1 ] (374.33,80.63) -- (374.33,80.63) -- (375.52,164.85) -- (267.85,53.84) -- (329.53,35.83) -- cycle ;
%Curve Lines [id:da3789953131295283] 
\draw [color={rgb, 255:red, 0; green, 0; blue, 0 }  ,draw opacity=1 ]   (261.93,54.01) .. controls (304.87,44.83) and (310.87,43.83) .. (351.6,24.75) ;
%Straight Lines [id:da8594658011416159] 
\draw    (373.87,47.83) -- (376.2,197.83) ;
%Straight Lines [id:da20745982101594695] 
\draw    (329.53,35.83) -- (360.73,66.5) ;
\draw [shift={(360.73,66.5)}, rotate = 44.5] [color={rgb, 255:red, 0; green, 0; blue, 0 }  ][fill={rgb, 255:red, 0; green, 0; blue, 0 }  ][line width=0.75]      (0, 0) circle [x radius= 1.34, y radius= 1.34]   ;
%Curve Lines [id:da8756276133891708] 
\draw    (261.93,54.01) .. controls (329.87,122.83) and (357.2,161.17) .. (376.2,197.83) ;
%Curve Lines [id:da12288960939135152] 
\draw    (376.2,197.83) .. controls (385.2,171.83) and (399.8,143.45) .. (432.2,107.83) ;
%Straight Lines [id:da9320302873822118] 
\draw [color={rgb, 255:red, 155; green, 155; blue, 155 }  ,draw opacity=1 ]   (351.6,24.75) -- (452.53,127.83) ;
%Straight Lines [id:da7550702648192735] 
\draw [color={rgb, 255:red, 155; green, 155; blue, 155 }  ,draw opacity=1 ]   (237.22,59.75) -- (375.7,203.31) ;
%Straight Lines [id:da3383148817663877] 
\draw [color={rgb, 255:red, 155; green, 155; blue, 155 }  ,draw opacity=1 ]   (452.53,127.83) -- (375.7,203.31) ;
%Straight Lines [id:da18827894494949793] 
\draw  [dash pattern={on 0.84pt off 2.51pt}]  (266.87,52.17) -- (362.15,151.5) ;
\draw [shift={(362.15,151.5)}, rotate = 46.19] [color={rgb, 255:red, 0; green, 0; blue, 0 }  ][fill={rgb, 255:red, 0; green, 0; blue, 0 }  ][line width=0.75]      (0, 0) circle [x radius= 1.34, y radius= 1.34]   ;
%Straight Lines [id:da5824185761525147] 
\draw    (294.87,46.83) -- (361.53,113.83) ;
\draw [shift={(361.53,113.83)}, rotate = 45.15] [color={rgb, 255:red, 0; green, 0; blue, 0 }  ][fill={rgb, 255:red, 0; green, 0; blue, 0 }  ][line width=0.75]      (0, 0) circle [x radius= 2.01, y radius= 2.01]   ;
%Straight Lines [id:da7050609010360948] 
\draw    (360.2,34.5) -- (360.73,66.5) -- (361.53,113.83) -- (362.15,151.5) -- (362.53,174.17) ;
%Straight Lines [id:da7079336426699215] 
\draw    (361.53,113.83) -- (374.87,127.5) ;
%Straight Lines [id:da3913701128364271] 
\draw    (360.73,66.5) -- (374.33,80.63) ;
%Straight Lines [id:da9413291926985649] 
\draw  [dash pattern={on 0.84pt off 2.51pt}]  (362.15,151.5) -- (395.03,184.97) ;

% Text Node
\draw (303.59,23.79) node [anchor=north west][inner sep=0.75pt]  [font=\scriptsize]  {$\mathcal{A}$};
% Text Node
\draw (362.09,86.36) node [anchor=north west][inner sep=0.75pt]  [font=\scriptsize]  {$\widetilde{\mathcal{T}}$};
% Text Node
\draw (337.49,32.69) node [anchor=north west][inner sep=0.75pt]  [font=\tiny]  {$\widetilde{\underline{H}}_{\widetilde{\underline{u}}_{1}}$};
% Text Node
\draw (326.37,97.71) node [anchor=north west][inner sep=0.75pt]  [font=\scriptsize]  {$\underline{H}_{v_{1}}$};
% Text Node
\draw (284.7,126.04) node [anchor=north west][inner sep=0.75pt]  [font=\scriptsize]  {$\underline{H}_{\delta }$};
% Text Node
\draw (307.46,42.96) node [anchor=north west][inner sep=0.75pt]  [font=\tiny,rotate=-44.27]  {$^{( tran)}\widetilde{\underline{u}} =const$};
% Text Node
\draw (403.7,52.71) node [anchor=north west][inner sep=0.75pt]  [font=\scriptsize]  {$\underline{H}_{\delta +d}$};
% Text Node
\draw (426.03,154.71) node [anchor=north west][inner sep=0.75pt]  [font=\scriptsize]  {$H_{u_{0}}$};

\end{tikzpicture}

    \caption{Transition Region $\Int\MM_0\cap \intM$}
    \label{fig:transition foliation}
\end{figure}
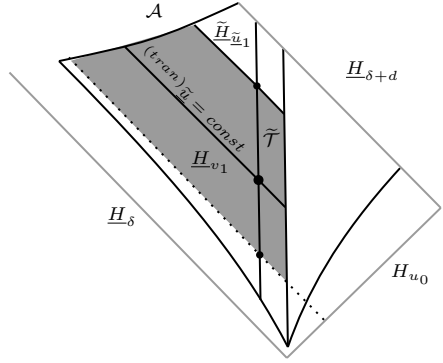
\begin{proposition}\label{Lem:est for glo tub}
    The eikonal equation \eqref{Eikonal Eqn for glo tub} admits a unique solution $\glo \tub$ in $\Int\MM_0\cap \intM$ and for all $0\le k\le k_{small}-2$ it obeys
    \begin{equation}\label{Est for tran tub}
        |\mathfrak{d}^k(\tran\tub-\tub_0)|\lesssim \epsilon_0 \qquad \text{in} \quad \Int\MM_0\cap \intM.
    \end{equation}
    Furthermore, the solution $\tran\tub$ is smooth in $\Int\MM_0\cap \intM$ and satisfies
    \begin{equation}\label{Eqn:transition glo tub}
    \tran \tub=\left\{\begin{array}{cc}
         F(v, \t{r}_0, \tth^1_1, \tth^2_1)&  \quad \text{when} \quad v\le v_1,\\
         \tub&  \quad \text{when} \quad \tub \ge \tub_1. 
    \end{array}\right.
\end{equation}
\end{proposition}
\begin{remark}
    The conclusion \eqref{Eqn:transition glo tub} indicates that the optical function $\tran\tub$ smoothly connects the optical function $F(v, \t{r}_0, \t{\th}^a_1)$ in $\Int \MM_0$ and the optical function $\tub$ in $\intM$.
\end{remark}
\begin{proof}[Proof of \Cref{Lem:est for glo tub}]
   Note that along $\{\t{r}=\t{r}_0 \}\cap \Int\MM_0\cap \intM$, we have $\tub=\tub_0=F(v, \t{r}_0, \tth^1, \tth^2)$ and it also holds
   \begin{align*}
      \tran \tub_0-\tub_0=&F(V, \t{r}_0, \tth^a_1)-F(v, \t{r}_0, \tth^a) \\
      =&\Big(\int_{0}^1 \pr_v F(tV+(1-t)v, \t{r}_0, t\tth^a_1+(1-t)\tth^a)dt \Big)(V-v)\\&+\Big(\int_{0}^1 \pr_{\tth^a} F(tV+(1-t)v, \t{r}_0, t\tth^a_1+(1-t)\tth^a)dt\Big)(\tth^a_1-\tth^a), \\
      V-v=&(1-\la) \Big(v(\tub, \t{r}_0, \tth^a_1)-v(\tub, \t{r}_0, \tth^a) \Big)\\
      =&(1-\la)\Big(\int_{0}^1 \pr_{\tth^a} v(\tub, \t{r}_0, t\tth^a_1+(1-t)\tth^a)dt \Big)(\tth^a_1-\tth^a).
   \end{align*}
Recall that from \eqref{Esf for ttha tX a b} and \Cref{Lem:est for tran coef1} we have $|\tX_a^{\ b}|\lesssim 1$ and $|\fb'''|\lesssim \epsilon_0$. These give
\begin{equation*}
    \pr_{\tth^a} v=\tX_a^{\ b} \fb_b'''=O(\epsilon_0).
\end{equation*}
Combining with \eqref{Est for pr F}, we infer that 
\begin{equation*}
    |\mathfrak{d}^{\le k_{small}-2}(\tran \tub_0-\tub_0)|\lesssim \epsilon_0 \qquad \text{along} \quad \{\t{r}=\t{r}_0 \}\cap \Int\MM_0\cap \intM. 
\end{equation*}
Therefore, by virtue of \Cref{Rmk: stable eikonal eqn}, proceeding similar as in the proof of \Cref{Lem:est for tub}, we obtain the existence and uniqueness of $\tran \tub$ and \eqref{Est for tran tub}.

It remains to prove \eqref{Eqn:transition glo tub}.  Observing that along $\t{r}=\t{r}_0$, we have
\begin{equation*}
    \tran \tub_0=\left\{\begin{array}{cc}
         F(v, \t{r}_0, \tth^1_1, \tth^2_1)&  \quad \text{when} \quad v\le v_1,\\
         \tub_0&  \quad \text{when} \quad \tub \ge \tub_1. 
    \end{array}\right.
\end{equation*}
Also noticing that both $F(v, \t{r}_0, \t{\th}^a_1)$ and $\tub$ are optical functions, by the uniqueness of solutions to the (incoming) eikonal equation, we thus conclude 
\begin{equation*}
    \tran \tub=\left\{\begin{array}{cc}
         F(v, \t{r}_0, \tth^1_1, \tth^2_1)&  \quad \text{when} \quad v\le v_1,\\
         \tub&  \quad \text{when} \quad \tub \ge \tub_1. 
    \end{array}\right.
\end{equation*}
 This completes the proof of \Cref{Lem:est for glo tub}.
\end{proof}
\begin{remark}
  The conditions $\pr_v F, \pr_{\tub} v\sim 1$ and $\pr_{\tth^a} F, \pr_{\tth^a} v=O(\epsilon_0)$ play important roles in the proof of \Cref{Lem:est for glo tub}. We emphasize that we have these properties because we have obtained the precise control of the transition coefficients between the short-pulse double-null frame and the incoming geodesic null frame constructed in \Cref{Subsec:incoming geodesic foliation}, which is established in \Cref{Lem:est for tran coef1}.
\end{remark}
Finally, we construct the global incoming geodesic foliation in $\Int\MM_0\cup \intM$ as follows.  We first smoothly extend $\tran \tub$ to be the global (incoming) optical function $\glo \tub$ in $\PP\cup \Int\MM_0\cup \intM $ such that
\begin{equation*}
    \glo \tub=\left\{\begin{array}{cl}
         ^{(e)}F(v)&  \quad \text{in} \quad \PP\cup (\Int\MM_0 \backslash\intM),\\
         \tran \tub& \quad \text{in} \quad \Int\MM_0\cap \intM,\\
         \tub&  \quad \text{in} \quad \intM\backslash \Int\MM_0. 
    \end{array}\right.
\end{equation*}
Here $^{(e)}F(v)$ represents a smooth extension of the function $F(v)\coloneqq F(v, \t{r}_0, \tth^1_1, \tth^2_1)$ for $0\le v\le v_1$ such that $^{(e)}F(0)\equiv 0$ and $^{(e)}F(v)$ is strictly increasing in $v$. It then follows from \Cref{Lem:est for tub} and \Cref{Lem:est for glo tub} that, for all $0\le k\le k_{small}-2$ we have
\begin{equation}\label{Est for glotub}
     |\mathfrak{d}^k(\glo\tub-\tub_0)|\lesssim \fepub \qquad \text{in} \quad \intM.
\end{equation}

Therefore, by repeating the process conducted in \Cref{Subsec:incoming geodesic foliation}, we are now able to construct an incoming geodesic foliation $(\glo\tub, \, r, \glo\tth^1, \glo\tth^2)$ associated with the null frame $(\glo\te_1, \glo\te_2, \glo\te_3, \glo\te_4)$ within $\intM$. Furthermore, in light of \eqref{Est for glotub}, all the hyperbolic estimates  derived in \Cref{Subsec:incoming geodesic foliation} still hold true with regard to $(\glo\te_{\mu})$. And these estimates will be crucially used in \Cref{Sec:Existence}, \Cref{Sec:physical} and \Cref{Sec:penrose ineq}.

For the sake of simplicity, we will henceforth omit the prefixes $\glo$ for $\glo\tub, \glo\tth^1, \glo\tth^2$ and for $(\glo\te_{\mu})$ whenever there is no risk of confusion.

\subsection{Asymptotics of Null Expansions with Large \texorpdfstring{$a$}{}}\label{Appendix:Null expansion}
This subsection is dedicated to deriving precise estimates of null expansions $\tr\t{\chi}_0$ and $\tr\t{\chib}_0$ for 2-spheres $\tS_{\tub_0, \t{r}}=\qty{\tub_0'=\tub_0} \cap\qty{\t{r}'=\t{r}}$ embedded in the exact Kerr spacetime $Kerr(a, m)$ with large $a$ in the full range $|a|<m$. 

Note that $\tu_0$ and $\tub_0$ represent the outgoing and incoming optical functions in the exact Kerr spacetime $Kerr(a, m)$. As shown in \Cref{Appendix:coordinate in Mint}, in the Boyer-Lindquist coordinates $(t, r, \th,\phi)$ of Kerr metric $\bfg_{a, m}$, they take the forms of 
\begin{equation}\label{Eqn:tu0 tub01}
\begin{aligned}
    \tu_0=&t-\rs(r,\th)=\ub-f(r)-\rs(r,\th),  \\ 
    \tub_0=&t+\rs(r, \th)=\ub-f(r)+\rs(r,\th)
\end{aligned}
\end{equation}
with $f'(r)=\f{r^2+a^2}{\De}$ and $\De=r^2-2mr+a^2$. And the corresponding null frame $\qty{(\te_0)_{\mu}}$ is chosen such that $(\te_0)_3=-\bfD \tub_0$.

Recall that to suppress the singularity of $\rs(r,\th)$ at $r=r_+$, in \Cref{Subsec:incoming geodesic foliation} we introduce a regular coordinate $\t{r}=\t{r}(r,\th)$ replacing $\rs(r,\th)$ as
\begin{equation}\label{Apx:Eqn:t r1}
    \t{r}=e^{\kappa \rs}+r_+=g(r,\th)(r-r_+)+r_+ \qquad \text{with} \quad \kappa=\f{r_+-r_-}{2mr_+}, \ \ g\sim \f{1}{r},
\end{equation}
where $r_{\pm}=m\pm\sqrt{m^2-a^2}$.

In view of \eqref{Eqn:tu0 tub01} and \eqref{Apx:Eqn:t r1}, we have that the level sets of $\tu_0$ and $\tub_0$ intersect at the 2-spheres $\tS_{\tub_0, \t{r}}$. For convenience, we sometimes write $\tS_{\tub_0, \t{r}}=\tS_{\tu_0, \t{r}}$.

The main result of this subsection can be stated as below:
\begin{proposition}\label{Apx:prop trch trchb}
    Let  $r_0>r_+$ be a fixed large constant. Then within the exact subextremal Kerr spacetime $Kerr(a, m)$ with $|a|<m$,  there exist constants $C_{r_0}\ge 1, c_{r_0}>0$ such that for all  $m\le r\le r_0$, the following estimate hold for the null expansions $\tr\t{\chi}_0$ and $\tr\t{\chib}_0$ of 2-spheres $\tS_{\tub_0, \t{r}}$:
    \begin{equation*}
        \tr\t{\chi}_0=F(r,\th)\c(\t{r}-r_+) \quad \text{with} \quad C_{r_0}^{-1}\le F(r,\th)\le C_{r_0} \qquad  \text{and} \qquad\tr\t{\chib}_0\le -c_{r_0}.
    \end{equation*}
\end{proposition}

Let $\qty{(e_{0})_{\mu}}$ be the canonical (incoming) PG null frame introduced in \eqref{Eqn:cano PG e3 e4} and \eqref{Eqn:cano PG ea} in \Cref{Subsec:PG structure}. Using the above explicit expressions of $\tu_0, \tub$, together with \eqref{Apx:eqn:pr rs}, a direct calculation yields
\begin{equation}\label{Apx:Eqn:der tu tub}
    \begin{aligned}
        (e_{0})_3(\tub_0)=&\f{r^2+a^2-\Sis}{\De}, &\qquad (e_{0})_4(\tub_0)=&\f{r^2+a^2+\Sis}{|q|^2}, &\qquad \nab_0(\tub_0)=&\f{1}{|q|}(|a|\bs, a\sin\th), \\
        (e_{0})_3(\tu_0)=&\f{r^2+a^2+\Sis}{\De}, &\qquad (e_{0})_4(\tu_0)=&\f{r^2+a^2-\Sis}{|q|^2}, &\qquad \nab_0(\tu_0)=&\f{1}{|q|}(-|a|\bs, a\sin\th), \\
        (e_{0})_3(\rs)=&-\f{\Sis}{\De}, &\qquad (e_{0})_4(\rs)=&\f{\Sis}{|q|^2}, &\qquad \nab_0(\rs)=&\f{1}{|q|}(|a|\bs, 0).
    \end{aligned}
\end{equation}
Here $q=r+ia\cos\th$, $\Sis=\sqrt{(r^2+a^2)^2-a^2\sin^2\ths\De}$, $\bs=-\sgn(r-r_+)\sqrt{\sin^2\ths-\sin^2\th}$ with $\ths=\ths(r,\th)$ being the smooth function defined in \Cref{Appendix:coordinate in Mint}.
\vspace{2mm}

For the sake of simplicity, in below we omit the subscripts ${}_0$ in $\tu_0, \tub_0$, $\qty{(e_{0})_{\mu}}, \qty{(\te_0)_{\mu}}$ and in the corresponding geometric quantities if there is no danger of confusion. 
\vspace{2mm}

Since  $\tu=t-\rs(r, \th)$ is singular near the event horizon $r=r_+$, we introduce
\begin{equation*}
    \Tu\coloneqq-e^{-\f12\kappa \tu}=-e^{\kappa(\rs-\f12\tub)}=-e^{-\f12\kappa \tub} g(r,\th)(r-r_+).
\end{equation*}
With the assistance of $\Tu$, we conduct a frame transformation and define a new null frame
\begin{equation*}
    \te_3'=\la^{-1}e_3, \qquad \te_a'=e_a+\f12 f_a e_3, \qquad \te_4'=\la\qty(e_4+f^b e_b+\f14 |f|^2 e_3)=-\bfD\Tu.
\end{equation*}
With the assumption
\begin{equation*}
    \la\qty(e_4+f^b e_b+\f14 |f|^2 e_3)=-\bfD\Tu=\f12 e_3(\Tu)e_4+\f12e_4(\Tu)e_3-e^a e_a(\Tu),
\end{equation*}
by applying \eqref{Apx:Eqn:der tu tub}, we infer that
\begin{equation}\label{Apx:Eqn la f}
\begin{aligned}
\la=&\f12 e_3(\Tu)=\f12 \Tu'(\tu) e_3(\tu)=\f{\kappa}{4} e^{-\f{\kappa}{2}\tub} g(r,\th)\c\f{r^2+a^2+\Sis}{r-r_-},   \\
    f=&-\f{2\nab(\Tu)}{e_3(\Tu)}=-\f{2\nab(\tu)}{e_3(\tu)}=-\f{2\De}{|q|(r^2+a^2+\Sis)}(-|a|\bs, a\sin\th).
\end{aligned}
\end{equation}

To explore the connection between $\tr\t{\chi}$ adapted to the frame $(\te_{\mu})$ and the new null expansion $\tr\t{\chi}'$ adapted to the null frame $(\te_{\mu}')$, we proceed to perform the below change of frame
\begin{equation*}
      \te^{''}_3=\la'^{-1}\qty(\te'_3+\fb'^{b} \te'_b+\f14 |\fb'|^2 \te'_4), \qquad \te_a^{''}=\te'_a+\f12 \fb'_a \te'_4, \qquad \te_4^{''}=\la'\te_4',
\end{equation*}
where
\begin{equation*}
    \la'\coloneqq -\f{2}{\bfg(\bfD\tub, \bfD\Tu)}, \qquad \fb'\coloneqq -\f{2\nab'(\t{r})}{\te_4'(\t{r})}.
\end{equation*}
Note that by virtue of \eqref{Apx:Eqn:der tu tub} and \eqref{Apx:Eqn la f}, there hold
\begin{align}
\bfg(\bfD\tub, \bfD\Tu)=&\Tu'(\tu)\qty(-\f12 e_3(\tu)e_4(\tub)-\f12 e_4(\tu)e_3(\tub)+\nab\tu\c\nab\tub)=-\f{\kappa e^{-\f12\kappa\tub} g\Si^2}{(r-r_-)|q|^2}, \nonumber\\
\tnab'(\t{r})=&\nab_a(\t{r})+\f12 f e_3(\t{r})=\f{\kappa e^{\kappa\rs}}{|q|(r^2+a^2+\Sis)}\qty(|a|\bs(r^2+a^2), a\sin\th \Sis), \nonumber\\
    \te'_4(\t{r})=&\la\qty(e_4(\t{r})+f^b e_b(\t{r})+\f14 |f|^2 e_3(\t{r}))=\kappa e^{\kappa\rs}\c\f{\kappa e^{-\f12\kappa \tub}g}{2(r-r_-)}\c \f{\Si}{|q|^2}.\label{Apx:eqn la fb prime}
\end{align}
Here $\Si=\sqrt{(r^2+a^2)^2-a^2\De\sin^2\th}$.

Hence we obtain
\begin{equation}\label{Apx:eqn la prime}
    \la'=\f{2e^{\f12\kappa \tub}|q|^2(r-r_-)}{\kappa g\Si^2}>0.
\end{equation}
and both $\la', \fb'$ are well-defined.

It can be check readily that
\begin{equation*}
    \te_a^{''}(\tu)=\te_a^{''}(\Tu)=0, \quad \te_a^{''}(\t{r})=0, \qquad \bfg(\te_4^{''}, \te_3)=\la'\bfg(\bfD\Tu, \bfD\tub)=-2.
\end{equation*}
Recalling that $(\te_{\mu})$ is adapted to the 2-sphere $\tS_{\tub, \t{r}}$, this implies that 
\begin{equation*}
    \spn\qty{\te_1'', \te_2''}=\spn\qty{\te_1, \te_2}=T\tS_{\tub, \t{r}} \qquad \text{and} \qquad  \te_{\mu}^{''}=\te_{\mu} \quad \text{for} \quad \mu=3,4.
\end{equation*}
Utilizing \Cref{Lem:frame transform1}, owing to the fact that $\te_4'$ is geodesic, we hence obtain
\begin{align}\label{Apx:eqn:relation trchi trchi prime}
\la'^{-1}\t{\chi}=\t{\chi}'+\fb'\c\t{\xi}'=\t{\chi}'.
\end{align}
\begin{remark}
    In fact, in the above process we employ the two-step decomposition for the change of frame
    \begin{equation*}
        (e_{\mu})\xrightarrow{(f,\fb=0,\la)}(\te'_{\mu})\xrightarrow{(f'=0,\fb',\la')}(\te''_{\mu}=\te_{\mu}).
    \end{equation*}
    Since $\te_4'$ is geodesic, in view of \eqref{Apx:eqn:relation trchi trchi prime}, we have a relatively simple expression for the outgoing null expansion, that is, $\tr\t{\chi}=\la' \tr\t{\chi}'$ in the second step. And we only need to compute $\tr\t{\chi}'$ in terms of $(f,\fb=0,\la)$ in the first null frame transformation from $(e_{\mu})$ to $(\te'_{\mu})$.
\end{remark}

Now we turn to the proof of \Cref{Apx:prop trch trchb}. The key idea is to analyze the behavior of $\tr\t{\chi}'$ both in the far-away region $r\gg r_+$ and near $r=r_+$, and utilize the monotonicity of $\tr\t{\chi}'$ along $e_4'$ direction. We first derive an important formula for $\tr\t{\chi}'$, when $r$ is large enough.
\begin{lemma}\label{Apx:lem: asymp trch}
    The following asymptotic behavior for $\tr\t{\chi}'$ holds for $r\gg r_+$:
    \begin{equation*}
        \tr\t{\chi}'=\f{2\la r\De}{|q|^4}\qty(1+O(\f{a}{r})).
    \end{equation*}
\end{lemma}
\begin{proof}
    Applying \Cref{Lem:frame transform1} with $\fb=0$ and using expressions in \eqref{Eqn for Ricci Cur}, we obtain
    \begin{equation}\label{new trch express}
       \la^{-1}\tr{\t{\chi}}'=\trch+f\c(\eta+\zeta)+\div f+\f12 f^a \nab_3 f_a,
   \end{equation}
   where
   \begin{equation*}
   \begin{aligned}
           \trch=\f{2r\De}{|q|^4}, \qquad   \eta=\zeta=\f{a\sin \th}{|q|^3}(-a\cos \th, r).
   \end{aligned}
   \end{equation*}
   According to \eqref{Apx:Eqn la f}, there holds
   \begin{equation*}
       |(r\nab)^{\le 1}f|\lesssim \f{|a|\De}{r^3}, \qquad |\nab_3 f|\lesssim \f{|a|}{r^2}.
   \end{equation*}
  Inserting above two estimates into \eqref{new trch express}, the desired estimate thus follows.
\end{proof}
To track the behavior of $\tr\t{\chi}'$ near the event horizon $r=r_+$, we also compute the limit of $\tr\t{\chi}'/\De$ as $r\to r_+$.
\begin{lemma}\label{Apx:lem trch near}
   In the exact Kerr spacetime $Kerr(a, m)$ with $|a/m|<1$, the below limit equality holds
    \begin{equation*}
        \lim\limits_{r\to r_+}\f{\la^{-1}|q|^2\tr\t{\chi}'}{ \De}=\f{1}{r_+^2+a^2}\qty(2r_+-\f{(r_+-m)a^2\sin^2\th}{r_+^2+a^2}).
    \end{equation*}
\end{lemma}
\begin{proof}
    Applying the calculation in \Cref{Apx:lem: asymp trch}, along with \eqref{Kerr coor derivative}, \eqref{Eqn for Ricci Cur}, \eqref{Apx:Eqn la f}, we infer
    \begin{align*}
        f\c(\eta+\zeta)=&\f{-4\De a^2\sin^2\th}{|q|^4(r^2+a^2+\Sis)}\qty(r+|a|\bs\cot\th),\\
        \div f=&e^a(f_a)-(\nab_1 e_1)_2 f_2-(\nab_2 e_2)_1 f_1\\
        =&\f{1}{|q|}\pr_{\th}\qty(\f{2\De|a|\bs}{|q|(r^2+a^2+\Sis)})+\f{r^2+a^2}{|q|^3}\cot\th \c \f{2\De|a|\bs}{|q|(r^2+a^2+\Sis)}\\
        =&\f{2|a|\De}{|q|}\qty(\Big(\pr_{\th}\big(\f{1}{|q|(r^2+a^2+\Sis)}\big)+\f{(r^2+a^2)\cot\th}{|q|^3(r^2+a^2+\Sis)}\Big)\bs+\f{\pr_{\th} \bs}{|q|(r^2+a^2+\Sis)}),\\
        \f12 f^a \nab_3 f_a=&\f14\nab_3(|f|^2)=-\pr_r\qty(\f{a^2 \sin^2\ths}{|q|^2(r^2+a^2+\Sis)^2})\De^2-\f{4(r-m)a^2 \sin^2\ths\De}{|q|^2(r^2+a^2+\Sis)^2}.
    \end{align*}
As a result, in light of \eqref{new trch express}, we deduce
\begin{align*}
    \f{\la^{-1}|q|^4\tr\t{\chi}'}{\De}=&2r-\f{4 a^2\sin^2\th}{r^2+a^2+\Sis}\qty(r+|a|\bs\cot\th)+2|q|^3|a|\Big(\pr_{\th}\big(\f{1}{|q|(r^2+a^2+\Sis)}\big)+\f{(r^2+a^2)\cot\th}{|q|^3(r^2+a^2+\Sis)}\Big)\c \bs\\
    &+\f{2|a|\c|q|^2 }{r^2+a^2+\Sis}\c \pr_{\th} \bs-\f{4(r-m)a^2 |q|^2\sin^2\ths}{(r^2+a^2+\Sis)^2}-\pr_r\qty(\f{a^2 \sin^2\ths}{|q|^2(r^2+a^2+\Sis)^2})\c|q|^4\De.
\end{align*}

Note that from \Cref{Appendix:coordinate in Mint} we have $\ths(r,\th)=\th$ along $r=r_+$, and hence we obtain
\begin{equation*}
    \bs=-\sgn(r-r_+)\sqrt{\sin^2\ths-\sin^2\th}\to 0 \qquad \text{as}\quad r\to r_+.
\end{equation*}
On the other hand, incorporating with \Cref{Apx:tla limit}, we can compute $\pr_{\th}\bs$ and get\footnote{Here we use $\t{\la}$ to denote $\sin^2\ths$.}
\begin{equation*}
   \lim\limits_{r\to r_+} \pr_{\th}\bs=\lim\limits_{r\to r_+}-\sgn(r-r_+)\c\f{\pr_{\th}\t{\la}-\sin(2\th)}{2\sqrt{\sin^2\ths-\sin^2\th}}=0 \qquad \text{for all} \quad \th\in(0, \pi/2)\cup (\pi/2, \pi).
\end{equation*}

Combining all above equalities and noting that $\Sis\to r_+^2+a^2$ as $r\to r_+$, for any $\th\in(0, \pi/2)\cup (\pi/2, \pi)$ we arrive at
\begin{equation}\label{Apx:eqn:lim}
\begin{aligned}
    \lim\limits_{r\to r_+}\f{\la^{-1}|q|^4\tr\t{\chi}'}{ \De}=&2r_+-\f{2 a^2r_+\sin^2\th}{r_+^2+a^2}-\f{(r_+-m)a^2\sin^2\th|q|^2}{(r_+^2+a^2)^2}\\
    =&\f{|q|^2}{r_+^2+a^2}\qty(2r_+-\f{(r_+-m)a^2\sin^2\th}{r_+^2+a^2}).
    \end{aligned}
\end{equation}

For the case $\th\in \qty{0, \pi/2, \pi}$, due to our construction of $\tub$ and $\Tu$, the fraction $\la^{-1}|q|^4\tr\t{\chi}'\De^{-1}$ is regular near the event horizon $r=r_+$ for all $\th\in[0, \pi]$. Therefore, by sending $\th\to \qty{0, \pi/2, \pi}$ in \eqref{Apx:eqn:lim}, we also get the desired stated equality.
\end{proof}
\vspace{2mm}

Both \Cref{Apx:lem: asymp trch} and \Cref{Apx:lem trch near} allow us to establish the asymptotic formulas for  $\tr\t{\chi}'$ and $\tr\t{\chi}$.
\begin{proposition}\label{Apx:prop trch}
    For any fixed $r_0\gg r_+$, there exists a constant $C_{r_0}\ge 1$ such that for all $m\le r\le r_0$, it holds
    \begin{equation*}
        \tr\t{\chi}=\la' \tr\t{\chi}'=F(r,\th)\c(\t{r}-r_+) \qquad \text{with} \quad C_{r_0}^{-1}\le F(r,\th)\le C_{r_0}.
    \end{equation*}
\end{proposition}
\begin{proof}
From the definition of $\t{r}$ in \eqref{Apx:Eqn:t r1}, we have
\begin{equation*}
    \t{r}-r_+=g(r,\th)(r-r_+)\sim \f{r-r_+}{r}.
\end{equation*}
    Thus, it suffices to prove that
    \begin{equation*}
         \tr\t{\chi}=G(r,\th)\c(r-r_+) \qquad \text{with} \quad C_{r_0}^{-1}\le G\coloneqq gF\le C_{r_0} \quad \text{for all} \quad m\le r\le r_0.
    \end{equation*}
    
    In light of \eqref{Apx:eqn:relation trchi trchi prime}, \eqref{Apx:Eqn la f} and \eqref{Apx:eqn la prime}, it holds
    \begin{equation}\label{Apx:eqn:trch 1}
        \tr\t{\chi}=\la' \tr\t{\chi}'=\la'\la \c (\la^{-1}\tr\t{\chi}')=\f{|q|^2(r^2+a^2+\Sis)}{2\Si^2}\c (\la^{-1}\tr\t{\chi}').
    \end{equation}
    Based on the calculation of $\la^{-1}\tr\t{\chi}'$ in the proof of \Cref{Apx:lem trch near}, we can write
    \begin{equation*}
        \la^{-1}\tr\t{\chi}'=H(r,\th)\c(r-r_+).
    \end{equation*}
    Back to \eqref{Apx:eqn:trch 1}, we then obtain
    \begin{equation*}
        \tr\t{\chi}=\f{|q|^2(r^2+a^2+\Sis)}{2\Si^2}\c H(r,\th)\c(r-r_+)\eqqcolon G(r,\th)\c(r-r_+).
    \end{equation*}
    Since $ \tr\t{\chi}$ is smooth in $\qty{r\ge m}$ and the set $\qty{m\le r\le r_0, \th\in[0,\pi]}$ is compact, for the remaining it is sufficient to show that $F(r,\th)>0$ for all $r\ge m$. To this end, we consider the following three scenarios as portrayed in \Cref{fig:proofnullexpansion}.
    \begin{figure}[htb]
        \centering
\tikzset{every picture/.style={line width=0.75pt}} %set default line width to 0.75pt        

\begin{tikzpicture}[x=0.75pt,y=0.75pt,yscale=-0.9,xscale=0.9]

%Straight Lines [id:da6038387219290717] 
\draw    (221.85,143.09) -- (350.49,100.14) ;
%Straight Lines [id:da7547266640889024] 
\draw    (349.78,100.85) -- (264.9,185.73) ;
\draw [shift={(350.49,100.14)}, rotate = 135] [color={rgb, 255:red, 0; green, 0; blue, 0 }  ][line width=0.75]      (0, 0) circle [x radius= 2.01, y radius= 2.01]   ;
%Straight Lines [id:da6250856622740689] 
\draw    (249.8,76.95) -- (350.49,100.14) ;
%Straight Lines [id:da4240969725975927] 
\draw    (350.49,100.14) -- (350.67,239.25) ;
%Straight Lines [id:da13438721720964775] 
\draw    (255.75,109.19) -- (221.85,143.09) ;
\draw [shift={(255.75,109.19)}, rotate = 135] [color={rgb, 255:red, 0; green, 0; blue, 0 }  ][fill={rgb, 255:red, 0; green, 0; blue, 0 }  ][line width=0.75]      (0, 0) circle [x radius= 1.34, y radius= 1.34]   ;
%Straight Lines [id:da8587008878134385] 
\draw    (350.49,180.14) -- (316.59,214.04) ;
\draw [shift={(316.59,214.04)}, rotate = 0] [color={rgb, 255:red, 0; green, 0; blue, 0 }  ][fill={rgb, 255:red, 0; green, 0; blue, 0 }  ][line width=0.75]      (0, 0) circle [x radius= 1.34, y radius= 1.34]   ;

% Text Node
\draw (354.16,86.86) node [anchor=north west][inner sep=0.75pt]  [font=\scriptsize]  {$i^{+}$};
% Text Node
\draw (282.38,148.14) node [anchor=north west][inner sep=0.75pt]  [font=\scriptsize,rotate=-315]  {$r=r_{+}$};
% Text Node
\draw (280.04,108.45) node [anchor=north west][inner sep=0.75pt]  [font=\scriptsize,rotate=-342.52]  {$r=r_{1}$};
% Text Node
\draw (293.92,72.82) node [anchor=north west][inner sep=0.75pt]  [font=\scriptsize,rotate=-13.29]  {$r=m$};
% Text Node
\draw (356.3,169.4) node [anchor=north west][inner sep=0.75pt]  [font=\scriptsize]  {$r=r_{0}$};
% Text Node
\draw (215.38,126.14) node [anchor=north west][inner sep=0.75pt]  [font=\scriptsize,rotate=-315]  {$r< r_{+}$};
% Text Node
\draw (311.91,195.53) node [anchor=north west][inner sep=0.75pt]  [font=\scriptsize,rotate=-315]  {$r >r_{+}$};
% Text Node
\draw (250.15,90.97) node [anchor=north west][inner sep=0.75pt]  [font=\scriptsize]  {$\tilde{S}_{\tilde{u} ,\tilde{r}}$};
% Text Node
\draw (296.15,208.97) node [anchor=north west][inner sep=0.75pt]  [font=\scriptsize]  {$\tilde{S}_{\tilde{u} ,\tilde{r}}$};

\end{tikzpicture}

        \caption{Picture for Proof of \Cref{Apx:prop trch}}
        \label{fig:proofnullexpansion}
    \end{figure}
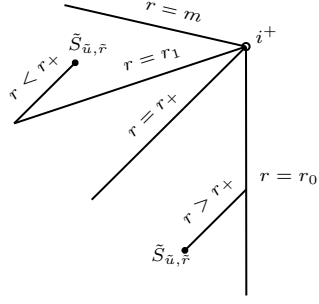
    
\begin{enumerate}
    \item  \textbf{The case $r>r_+$:} Noting that $\te_4'$ is geodesically outgoing null, by the Raychaudhuri equation, we have
    \begin{equation}\label{Apx:eqn:Ray}
        \te_4'(\tr\t{\chi}')=-\f12(\tr\t{\chi}')^2-|\t{\chih}'|^2.
    \end{equation}
Recall that from \eqref{Apx:eqn la fb prime} we have
\begin{equation*}
    \te_4'(\t{r})=\kappa g(r,\th)(r-r_+)\c\f{\kappa e^{-\f12\kappa \tub}g}{2(r-r_-)}\c \f{\Si}{|q|^2}>0.
\end{equation*}
Together with \eqref{Apx:eqn:Ray}, we infer that $\tr\t{\chi}'\big|_{\tS_{\tub, \t{r}}}$ is non-increasing in $\t{r}$, namely,
\begin{equation*}
    \tr\t{\chi}'\big|_{\tS_{\tu, \t{r}}}\ge \tr\t{\chi}'\big|_{\tS_{\tu, \t{r}_1}} \qquad \text{for any} \quad r_+<\t{r}<\t{r}_1.
\end{equation*}
We then fix $\t{r}$ and pick $\t{r}_1$ sufficiently large. By employing \Cref{Apx:lem: asymp trch}, we deduce that, for any $r=r(\t{r},\th)>r_+$, we have\footnote{Since $\pr_r \t{r}=\kappa e^{\kappa \rs}\pr_r \rs=\f{\kappa g \Sis}{r-r_-}>0$, we can find $r=r(\t{r},\th)$ so that $\t{r}\qty(r(\t{r},\th), \th)=\t{r}$ for any $\t{r}$ and $\th$.}
\begin{equation*}
    \tr\t{\chi}'\big|_{\tS_{\tu, \t{r}}}\ge \tr\t{\chi}'\big|_{\tS_{\tu, \t{r}_2}}=\f{2\la r\De}{|q|^4}\qty(1+O(\f{a}{\t{r}_1}))\ge c_1\qty(r(\t{r}_1,\th)-r_+)\ge c_2(\t{r}_1-r_+)>0.\footnote{Here we use the fact that $\t{r}$ and $r$ are comparable when $r\gg r_+$.}
\end{equation*}
Combining with \eqref{Apx:eqn la prime}, this thus implies
\begin{equation*}
    G(r,\th)=\f{\la'\tr\t{\chi}'\big|_{\tS_{\tu, \t{r}}}}{r-r_+}>0 \qquad \text{for all} \quad r>r_+.
\end{equation*}

\item \textbf{The case $r=r_+$:} Applying \Cref{Apx:lem trch near} and \eqref{Apx:eqn:trch 1}, we directly derive
\begin{equation}\label{Apx:eqn:F r+}
\begin{aligned}
    F(r_+,\th)=\lim\limits_{r\to r_+}\f{\tr\t{\chi}\big|_{\tS_{\tub, \t{r}}}}{r-r_+}=&\lim\limits_{r\to r_+}\f{(r^2+a^2+\Sis)(r-r_-)}{2\Si^2}\c\lim\limits_{r\to r_+}\f{\la^{-1}|q|^2\tr\t{\chi}'}{\De}\\
    =&\f{r_+-r_-}{(r_+^2+a^2)^2}\qty(2r_+-\f{(r_+-m)a^2\sin^2\th}{r_+^2+a^2})\ge \f{(r_+-r_-)(r_++m)}{(r_+^2+a^2)^2}\ge \f{\sqrt{m^2-a^2}}{4m^3}>0,
\end{aligned}
\end{equation}
where we utilize the fact that $r_+=m+\sqrt{m^2-a^2}\in [m, 2m]$.

    \item \textbf{The case $m\le r<r_+$:} Proceeding similarly as in the proof of Case (1) and observing that
    \begin{equation*}
         \te_4'(\t{r})=\kappa g(r,\th)(r-r_+)\c\f{\kappa e^{-\f12\kappa \tub}g}{2(r-r_-)}\c \f{\Si}{|q|^2}<0,
    \end{equation*}
    we can now get that
    \begin{equation*}
    \tr\t{\chi}'\big|_{\tS_{\tu, \t{r}}}\le \tr\t{\chi}'\big|_{\tS_{\tu, \t{r}_1}} \qquad \text{for any} \quad \t{r}<\t{r}_1<r_+.
\end{equation*}
By taking $\t{r}_1$ sufficiently close to $r_+$, together with \eqref{Apx:eqn:F r+}, we infer\footnote{As $\t{r}-r_+=e^{\kappa \rs}=-\Tu e^{\f12\kappa \tub}$, we can let $\tub\to-\infty$ to make $\t{r}$ arbitrarily close to $(r_+)^-$.}
\begin{equation*}
     \tr\t{\chi}'\big|_{\tS_{\tu, \t{r}}}\le\tr\t{\chi}'\big|_{\tS_{\tu, \t{r}_1}}=(\la')^{-1}F\qty(r(\t{r}_1,\th), \th)\c\qty(r(\t{r}_1,\th)-r_+)\le \f12(\la')^{-1}F(r_+, \th)\c\qty(r(\t{r}_1,\th)-r_+)<0,
\end{equation*}
from which we conclude
\begin{equation*}
    G(r,\th)=\f{\la'\tr\t{\chi}'\big|_{\tS_{\tu, \t{r}}}}{r-r_+}>0 \qquad \text{for all} \quad m\le r<r_+.
\end{equation*}
\end{enumerate}
This completes the proof of this proposition.
\end{proof}

Similarly, we can also derive the desired property for the incoming null expansion $\tr\t{\chib}$, while the proof is simpler.
\begin{proposition}\label{Apx:prop trchb}
     For any fixed $r_0\gg r_+$, there exists a constant $c_{r_0}>0$ such that for all $m\le r\le r_0$, it holds
    \begin{equation*}
        \tr\t{\chib}\le -c_{r_0}.
    \end{equation*}
\end{proposition}
\begin{proof}
    For this proposition, it is sufficient to prove that $\tr\t{\chib}<0$. Following the two-step frame decomposition conducted in \Cref{Subsec:incoming geodesic foliation} and applying the proof of \Cref{Lem:hyper est in incom geo frame}, we can now find an intermediate null frame $({}^{(1)}e_{\mu})$ such that
    \begin{itemize}
        \item The transition coefficients from  $(e_{\mu})$ to $( {}^{(1)}e_{\mu})$ are given by 
        \begin{equation}\label{Apx:est fb la 1}
            {}^{(1)}f=0, \qquad {}^{(1)}\fb=-\f{2|q|}{r^2+a^2+\Sis}(|a|\bs, a\sin\th),  \qquad   {}^{(1)}\lambda=\f{2|q|^2}{r^2+a^2+\Sis}.
        \end{equation}
        Note that we have ${}^{(1)}e_{3}=\te_3$.
    \item The transition coefficients from  $( {}^{(1)}e_{\mu})$ to $( \te_{\mu})$ satisfy
    \begin{equation*}
      {}^{(2)}f=
     \f{2|q|\De }{\Si^2(r^2+a^2+\Sis)} (|a|\bs(r^2+a^2), -a\sin\th \Sis)\qquad   {}^{(2)}\fb=0,\qquad {}^{(2)}\la=1.
    \end{equation*}
    \end{itemize}
    
    Notice that by \Cref{Lem:frame transform1} and the fact that ${}^{(1)}e_{3}=\te_3$ is geodesic, we have $\tr\t{\chib}=\tr{}^{(1)}\chib$. Consequently, with the transition coefficients $({}^{(1)}f=0,{}^{(1)}\fb,{}^{(1)}\lambda)$, we conduct \Cref{Lem:frame transform1} again, and along with \eqref{Eqn for Ricci Cur} we get
    \begin{equation}\label{Apx:eqn t trchb}
        ({}^{(1)}\lambda)^{-1}\tr\t{\chib}=({}^{(1)}\lambda)^{-1}\tr{}^{(1)}\chib=\trchb+{}^{(1)}\fb\c(\etab-\zeta)+\div {}^{(1)}\fb-|{}^{(1)}\fb|^2\omega+\f12 {}^{(1)}\fb\c\nab_4 {}^{(1)}\fb.
    \end{equation}
    Here 
    \begin{equation*}
        \trchb=-\f{2}{r}+O(r^{-2}), \qquad \etab,\zeta=O(r^{-2}), \qquad \omega=O(r^{-2}).
    \end{equation*}
    By virtue of \eqref{Apx:est fb la 1}, we also obtain
    \begin{equation*}
        |(r\nab, r\nab_4)^{\le 1}({}^{(1)}\fb)|\lesssim \f{1}{r}, \qquad {}^{(1)}\lambda\sim 1.
    \end{equation*}
    Substituting these back to \eqref{Apx:eqn t trchb}, we hence deduce
    \begin{equation*}
        ({}^{(1)}\lambda)^{-1}\tr\t{\chib}=-\f{2}{r}+O(r^{-2}).
    \end{equation*}
    In particular, for $\t{r}\gg r_+$ we infer
    \begin{equation*}
({}^{(1)}\lambda)^{-1}\tr\t{\chib}\big|_{\tS_{\tub, \t{r}}}=-\f{2}{r(\t{r},\th)}+O(r^{-2})\le -\f{1}{r(\t{r},\th)}<0.
    \end{equation*}
    The desired inequality in the statement thus follows by applying the Raychaudhuri equation along $\te_3$ direction.
\end{proof}

Therefore, combining \Cref{Apx:prop trch} and \Cref{Apx:prop trchb}, we conclude \Cref{Apx:prop trch trchb}.
\vspace{2mm}

\section{Existence and Uniqueness of MOTS}\label{Sec:Existence}
The aim of this section is to solve the MOTS along each incoming null hypersurface $\tHb_{\tub}$ constructed in \Cref{Sec:Construct null cones}. Note that the MOTS along each $\Hb_{v}$ within the short-pulse region $\qty{0< v\le \de}=\qty{0 \le \tub\le F(\de)}$ has already been identified in \cite{An:AH}, and the method there extends to the transition region $F(\de)\le \tub\le F(v_1)$. Consequently, we obtain an apparent horizon originating from $O$ with coordinates $u=0, v=0$, which is sliced by the unique MOTS $M_{\tub}$ along each $\tHb_{\tub}$ for all $0< \tub\le F(v_1)$. We proceed to solve the MOTS along each $\tHb_{\tub}$ with $\tub\ge F(v_1)$.

\subsection{Derivation of Elliptic Equation}
Recall that in \Cref{Subsec:incoming geodesic foliation}, we derive a new incoming geodesic foliation $(\tub, \t{r}, \t{\th}^1, \t{\th}^2)$ in $\intM$ with $\tub$ being the constructed incoming optical function and with $(\te_1, \te_2, \te_3, \te_4)$ being the associated null frame obeying
 \begin{equation*}
     \te_3=-\bfD \tub, \qquad \bfD_{\te_3} \te_3=0, \qquad \te_3(\tth^1)=\te_3(\tth^2)=0. 
 \end{equation*}
 
We now construct the MOTS $M_{\tub}\coloneqq\{(\tub, \t{r}, \tth^1, \tth^2): \t{r}=R(\tub, \thth) \}$ along each $\tHb_{\tub}$ with $R(\tub, \tth^1, \tth^2)$ being an unknown function. At each point of $M_{\tub}$, the outgoing null expansion vanishes and the incoming null expansion stays negative. It can be easily checked that adapted to the MOTS $M_{\tub}$
\begin{equation}\label{new frames}
e_3'=\te_3, \qquad e_a '=\te_a+f \te_a (R) \te_3, \qquad e_4'=\te_4+2f\te^a(R) \te_a+f^2|\tnab R|^2 \te_3, 
\end{equation}
form a null frame, where $f=[\te_3(\t{r})]^{-1}$ and $\nab$ represents the induced covariant derivative on $S_{\ub, r}$. 

Note that by \eqref{Est for te4 tub,r}, we have $\te_3(\t{r})\sim -1$. Hence, $f=[\te_3(\t{r})]^{-1}$ is well-defined and is comparable with $-1$. From $\te_3(\tub)=\te_3(\tth^a)=0$, it also holds $\te_3(R)=0$, which gives $e'_a(R)=\te_a(R)$. 

In view of \eqref{new frames} and $\bfD_{\te_3}\te_3=0$, employing \Cref{Lem:frame transform1}, we infer that $\trchb'=\tr\t{\chib}$. Thus, applying \eqref{Est for trch new frame} in \Cref{Lem:hyper est in incom geo frame} and using \Cref{Apx:prop trch trchb}, we obtain
\begin{equation*}
    \trchb'\le -c+O(\tfepub)=-\f{c}{2}<0 \qquad \text{with} \quad c>0 \quad \text{being a constant}. \\[2mm]
\end{equation*}

Following the derivation in \cite{An-He}, we proceed to state the expression of the outgoing null expansion for the MOTS $M_{\tub}$.
\begin{lemma}[An-He \cite{An-He}]
Employing the null frame $(e'_1, e'_2, e'_3, e'_4)$ adapted to the MOTS $M_{\tub}$, we have
\begin{equation}\label{deformation formula}
	\trch'=\tr\t{\chi}+2f\tDe R+2(\tnab f+f(\t{\eta}+\t{\zeta}))\cdot \tnab R-4f^2\t{\chibh}_{bc}\tnab^b R\tnab^c R+(2f\te_3(f)-f^2\\tr\t{\chib}-4\t{\omegab} f^2)|\tnab R|^2.
	\end{equation}
 Here $\tDe$ represents the Laplace--Beltrami operator on $\tS_{\ub, \t{r}}$.
\end{lemma}

Therefore, $\trch'\equiv 0$ along the MOTS $M_{\tub}$ is equivalent to the below elliptic equation for $R$, i.e.,
\begin{equation}\label{MOTS main equation}
   \begin{split}
    0=L(R, \tub)\coloneqq(\f{1}{2} f^{-1} \trch')|_{\tS_{\tub, R}}=
&\tDe_{\tS_{\tub, R}} R+\big(f^{-1}\tnab f+(\t{\eta}+\t{\zeta})\big)\cdot \tnab R-2f\t{\chibh}_{ac}\tnab^a R \tnab^c R\\&+(\te_3(f)-\f{1}{2}f\tr\t{\chib}-2\t{\omegab} f)|\tnab R|^2+\f{1}{2}f^{-1}\tr\t{\chi},
\end{split}
\end{equation}
where $\tS_{\tub, R}\coloneqq\{(\tub, \t{r}, \tth^1, \tth^2): r=R(\tub, \tth^1, \tth^2) \}$ and $\tDe_{\tS_{\tub, R}} R\coloneqq(\tDe R)\big|_{\tS_{\tub, R}}$. 

Utilizing \Cref{Lem:hyper est in incom geo frame}, after the transition region $\qty{F(\de)\le\tub\le F(v_1)}$ we can transfer \eqref{MOTS main equation} into the following form of a quasilinear elliptic equation. 
\begin{proposition}\label{MOTS main equation new prop}
    For any $\tub\ge F(v_1)$, the equation for MOTS \eqref{MOTS main equation} can be written as the below quasilinear elliptic equation on $\mathbb{S}^2$:
    \begin{equation}\label{MOTS main equation new}
        a^{ij}(\tub, \omega, R,  DR)D_{ij} R+b(\tub, \omega, R, DR)+c(\tub, \omega, R)(R-r_{+})=0,
    \end{equation}
    where $\omega\in\mathbb{S}^2$,\footnote{In this section, as the the Ricci coefficient $\omega=-\f14\bfg(\bfD_{4} e_3, e_4)$ does not appear in the equation of MOTS, we use $\omega$ to represent angular variables in a local chart on $\mathbb{S}^2$ if there is no danger of confusion.} and $a^{ij}(\tub,\omega, \t{r},  p), b(\tub, \omega, \t{r},  p), c(\tub, \omega, \t{r})$ are smooth functions satisfying below properties:
    \begin{enumerate}
        \item $a^{ij}(\tub, \omega, \t{r},  p)=a^{ji}(\tub, \omega, \t{r},  p)$;
        \item There exist constants $\mu\ge \nu>0$ and $C_0>c_0>0$ independent of $\epsilon_0, \tub, \t{r}, p$, such that there hold
        \begin{align}
            \nu |\xi|^2 \le& a^{ij}(\tub, \omega, \t{r},  p) \xi_i \xi_j \le \mu |\xi|^2 \qquad \text{for any} \quad \xi\in \mathbb{R}^2, \label{property for a} \\ 
            |b(\tub, \omega, \t{r},  p)|\lesssim&  |p|^{2}+|p|+\tfepub, \qquad -C_0\le c(\tub, \omega, \t{r})\le -c_0. \label{property for b}
        \end{align}
        \item The derivatives of $b$ satisfy the bound
        \begin{equation}\label{property for pr b}
            |\pr_{\t{r}}b|+|\pr_{\tub} b|\lesssim  |p|^{2}+|p|+\tfepub.
        \end{equation}
    \end{enumerate}
\end{proposition}
\begin{remark}
    In fact, to study the MOTS $M_{\tub}$ along an incoming null hypersurface $\tHb_{\tub}$, we have $a^{ij}(\omega, \t{r},  p)=a^{ij}(\omega, \t{r})$.\footnote{For notational simplicity, we will drop $\tub$ in $(\tub, \o)$ $(\tub, \o, \t{r})$ and $(\tub, \o, \t{r}, p)$ whenever there is no danger of confusion.} In this current paper, we only utilize the uniform ellipticity of $a^{ij}(\omega, \t{r}, p)$ to derive the a priori estimates for the solution $R(\o)$ to the equation of MOTS \eqref{MOTS main equation new}. This generality enables us to extend our elliptic techniques in this paper to solving MOTS along slightly timelike or slightly spacelike hypersurfaces in future works, where $a^{ij}(\omega, \t{r},  p)$ also depends on $p=DR$.
\end{remark}
\begin{proof}[Proof of \Cref{MOTS main equation new prop}]
Notice that in local coordinates $\omega=(\tth^1, \tth^2)$, the Laplace--Beltrami operator on $\tS_{\tub, \t{r}}$ takes the form of
\begin{equation}\label{Express L-B operator}
	\begin{split}
	\tDe_{\tS_{\tub, \t{r}}} f=&\f{1}{\sqrt{\det \t{g}}}\f{\partial}{\partial \tth^i} (\sqrt{\det \t{g}}\,\t{g}^{\tth^i \tth^l}\f{\partial f}{\partial \tth^l})\\
	=&\t{g}^{\tth^i\tth^j}\f{\pr^2 f}{\pr\tth^i\pr\tth^j}+\f{\pr}{\pr\tth^i}(\t{g}^{\tth^i\tth^j}) \f{\pr f}{\pr\tth^j}+\f12 \t{g}^{\tth^k\tth^j} \t{g}^{\tth^i\tth^l} \f{\pr}{\pr\tth^i}(\t{g}_{\tth^j\tth^k}) \f{\pr f}{\pr\tth^l}.
	\end{split}
	\end{equation}
	Here $i,j,k,l=1,2$ and $\t{g}^{\tth^k \tth^j}$ denotes the inverse matrix of $(\t{g}_{\tth^k \tth^j})$ with $\t{g}$ being the induced metric on $\tS_{\tub, \t{r}}$. Then we denote 
 \begin{equation*}
     a^{ij}(\tub, \omega, \t{r}, p)=\t{g}^{\th^i \th^j}(\tub, \t{r}, \tth^1, \tth^2), \qquad D_{i}=\f{\pr}{\pr \tth^i}.
 \end{equation*}
 Recalling that $\te_i=\te_i(\tth^{j})\pr_{\tth^j}$, we hence get
\begin{equation*}
    \de_{ik}=\te_i(\tth^{j})\te_k(\tth^l) \t{g}_{\tth^j \tth^{l}}.
\end{equation*}
 This yields
 \begin{equation*}
     a^{ij}=\t{g}^{\th^i \th^j}=\sum_{k=1}^2\te_k(\tth^i)\te_k(\tth^j).
 \end{equation*}
As a result, for any $\xi\in \mathbb{R}^2$, it holds that
\begin{equation*}
    a^{ij}\xi_i \xi_j=|\sum_{k=1}^{2} \te_k(\tth^i) \xi_i|^2=|\t{E}\xi |^2,
\end{equation*}
where we set $\t{E}=(\te_i(\tth^j))$. By the definition of $\tX_i^j$, we also have $\t{E}^{-1}=(\tX_i^j)$. Due to the boundedness of $\te_{i}(\tth^j)$ and $\tX_i^j$ as in \eqref{Esf for ttha tX a b}, we also infer that
\begin{equation*}
  C^{-1}| \xi|^2 \le \|\t{E}^{-1}\|^{-2} |\xi |^2  \le a^{ij}\xi_i \xi_j \le \|\t{E}\|^2 | \xi |^2\le C|\xi |^2
\end{equation*}
for some constant $C\ge 1$ independent of $\epsilon_0, \tub, \t{r}$. Here the norm of metric is defined as
\begin{equation*}
    \| (m_{ij}) \|\coloneqq(\sum_{i, j=1}^{2} m_{ij}^2)^{\f12}.
\end{equation*}
Note that by setting $\mu=C$ and $\nu=C^{-1}$, we then satisfy the desired bounds for $a^{ij}$ in \eqref{property for a}.
\vspace{2mm}
 
Next we proceed to select $c(\omega, \t{r})$. Applying \Cref{Lem:hyper est in incom geo frame}, \Cref{Apx:prop trch trchb} and using that $\ub\sim \tub$ along $\tHb_{\tub}$, we obtain
\begin{equation*}
     \tr{\widetilde{\chi}}= F\c(\t{r}-r_+)+O(\tfepub) \qquad \text{with} \quad F\sim1.
 \end{equation*}
 Since $f\sim -1$, we then derive  
 \begin{equation*}
     c(\omega, \t{r})= f^{-1}F\sim -1.
 \end{equation*}

 Finally, we collect the rest of the terms into $b(\omega, \t{r},  p)$, which takes the form of\,\footnote{Here $\c$ denotes any type of contraction for the vector $p$ on $T\mathbb{S}^2$, not necessarily the inner product.}
 \begin{equation*}
     b(\omega, \t{r},  p)=\t{\Ga}(p\c p+p)+O(\tfepub).
 \end{equation*}
We then have that the conditions \eqref{property for b} and \eqref{property for pr b} follow readily from the hyperbolic estimates in \Cref{Lem:hyper est in incom geo frame}.
\end{proof}
Due to the uniform ellipticity of \eqref{MOTS main equation new}, it can be checked that the matrix $\Big(a^{ij}(\omega, \t{r}, p)\Big)$ satisfies the so-called Campanato condition introduced in \cite{Cam1, Cam2}.
\begin{lemma}[Campanato condition \cite{Cam1, Cam2}]\label{Lem:Campanato}
    For $A=\Big(a^{ij}(\omega, \t{r}, p)\Big)$ defined in \Cref{MOTS main equation new prop}, there exist a positive function $a=a(\omega, \t{r}, p) \in [\mu^{-1}, \nu^{-1}]$ such that 
    \begin{equation*}
        |\tr X-a\tr(AX)|\le (1-\f{\nu}{\mu})\|X\| \qquad \text{for any $2\times 2$ real matrix} \ X. 
    \end{equation*}
\end{lemma}
\begin{proof}
    Since $A=(a^{ij})$ is symmetric, there exists an orthogonal matrix $P$ such that 
    \begin{equation*}
        P^{T}AP=\left(\begin{array}{cc}
          \la   &  \\
             &  \La
        \end{array}\right),
    \end{equation*}
    where $\la(\omega, \t{r}, p), \La(\omega, \t{r}, p)$ are two eigenvalues of $A$ satisfying
    \begin{equation*}
        \nu \le \la(\omega, \t{r}, p)\le \La(\omega, \t{r}, p)\le \mu.
    \end{equation*}
    Denote $Y=P^T X P$ and choose $a( \omega, r, p)= [\La(\omega, \t{r}, p)]^{-1}\in [\mu^{-1}, \nu^{-1}]$. We then obtain
    \begin{align*}
        \tr X-a\tr(AX)=\tr Y-a \tr\l\left(\begin{array}{cc}
          \la   &  \\
             &  \La
        \end{array}\right) Y\r=y_{11}+y_{22}-\La^{-1}(\la y_{11}+\La y_{22})=(1-\f{\la}{\La})y_{11}.
    \end{align*}
    This implies 
    \begin{equation*}
        |\tr X-a\tr(AX)|\le (1-\f{\la}{\La})|y_{11}|\le (1-\f{\nu}{\mu})\|Y\|=(1-\f{\nu}{\mu})\|X\|
    \end{equation*}
 as stated.
\end{proof}

We proceed to start with the a priori $C^0$ estimate for solutions to the equation of MOTS \eqref{MOTS main equation new}. As a direct implication of the maximum (minimum) principal on $\mathbb{S}^2$, we derive 
\begin{proposition}\label{C0 est}
    Let $R$ be a solution to \eqref{MOTS main equation new}. Then we have
    \begin{equation*}
        |R-r_+|\lesssim \tfepub \qquad \text{with} \quad r_+=m+\sqrt{m^2-a^2}.
    \end{equation*}
\end{proposition}
\begin{proof}
    Denote   $R(\omega_0)\coloneqq\max\limits_{\omega\in \mathbb{S}^2} R(\omega)$. Then at $\omega=\omega_0$ we have $D R(\omega_0)=0$ and  $a^{ij}(\omega, R,  DR)D_{ij} R(\omega_0)\le0$. Plugging these into \eqref{MOTS main equation new}, we deduce that
   \begin{equation*}
       0=a^{ij}(\omega, R,  DR)D_{ij} R+b(\omega, R, DR)+c(\omega, R)(R-r_+)\le c(\omega, R)(R(\omega_0)-r_+)+O(\tfepub).
   \end{equation*}
   Using the fact that $c(\o, \t{r})\sim -1$, this consequently implies
   \begin{equation*}
       \max\limits_{\omega\in \mathbb{S}^2} R(\omega)=R(\omega_0)\le r_+ +O(\tfepub).
   \end{equation*}
   Similarly, we also get
       \begin{equation*}
       \min\limits_{\omega\in \mathbb{S}^2} R(\omega)\ge r_+ -O(\tfepub).
   \end{equation*}
   The desired estimate thus follows.
\end{proof}
\subsection{A Leray--Schauder Fixed Point Argument}
To prove the existence of solutions to equation \eqref{MOTS main equation new}, we appeal to Leray--Schauder fixed point theorem in \cite{G-T} as stated below.
\begin{theorem}\label{L-S fixed point thm}
    Let $M>0$ and $\mathcal{B}$ be a Banach space. Suppose $T$ is a compact mapping of a closed ball $\overline{B}_M\coloneqq\{x\in \mathcal{B}: \ \|x\|_{\mathcal{B}} \le M  \}$ into $\mathcal{B}$. Further assume that for all  $x\in \overline{B}_M$ and $\sigma\in [0, 1]$ satisfying $x=\sigma Tx$, it holds
    \begin{equation}\label{condition for L-S fixed point thm}
        \|x\|_{\mathcal{B}}< M.
    \end{equation}
  Then $T$ has a fixed point.
\end{theorem}
\begin{proof}
    We remark that the proof of this conclusion is summarized as Theorem 11.3 in \cite{G-T} on page 280--281. 
\end{proof}
With the aid of \Cref{L-S fixed point thm}, once the a priori $C^{1, \a}$ estimate\footnote{Since the curvature component $\a_{ab}=\bfR(e_a, e_4, e_b, e_4)$ also does not appear in the equation of MOTS, throughout this section we use $\a$ to denote the H\"older exponent if there is no danger of confusion.} is obtained, we can employ the below proposition to prove the existence of solution to the equation of MOTS.
\begin{proposition}\label{Prop:general elliptic existence}
    Consider the quasilinear elliptic equation on $\mathbb{S}^2$ in the following general form 
    \begin{equation}\label{Eqn general elliptic}
        a^{ij}(\omega, R,  DR)D_{ij} R+b(\omega, R, DR)+c(\omega, R)(R-r_+)=0,
    \end{equation}
     where $\omega\in\mathbb{S}^2$ and $a^{ij}(\omega, \t{r},  p), b(\omega, \t{r},  p), c(\omega, \t{r})$ are smooth functions satisfying below properties:
    \begin{enumerate}
        \item $a^{ij}(\omega, \t{r},  p)=a^{ji}(\omega, \t{r},  p)$;
        \item There exist constants $\mu\ge \nu>0$ and $C_0>c_0>0$ independent of $\t{r}, p$ such that
        \begin{align*}
            \nu |\xi|^2 \le& a^{ij}(\omega, \t{r},  p) \xi_i \xi_j \le \mu |\xi|^2 \qquad \text{for any} \quad \xi\in \mathbb{R}^2, \\
            \qquad -C_0\le& c(\omega, \t{r})\le -c_0.
        \end{align*}
    \end{enumerate}
    Assume that for any $\sigma\in [0, 1]$ and any solution $R$ to 
    \begin{equation*}
         a^{ij}(\omega, R,  DR)D_{ij} R+\sigma b(\omega, R, DR)+c(\omega, R)(R-r_+)=0,
    \end{equation*}
     it holds that
\begin{equation}\label{General a priori est}
    \| R-r_+ \|_{C^{1, \a}}< M
\end{equation}
with $\a \in (0, 1)$ and $M>0$ being constants independent of $\sigma, R$. Then the elliptic equation \eqref{Eqn general elliptic} admits a $C^{1, \alpha}$ solution.
\end{proposition}
\begin{proof}
    We first write \eqref{Eqn general elliptic} as
    \begin{equation*}
        Q (R)\coloneqq a^{ij}(\omega, R,  DR)D_{ij} R+b(\omega, R, DR)+c(\omega, R)(R-r_+)=0.
    \end{equation*}
    Given any $S\in C^{1, \a}$, we define the operator $T: C^{1, \a}\to C^{1, \a}$ by requiring $R=T(S-r_+)+r_+$ to be the unique solution in $C^{2, \a}$ to the linear equation
    \begin{equation}\label{def for mapping T}
        a^{ij}(\omega, S,  DS)D_{ij} R+b(\omega, S, DS)+c(\omega, S)(R-r_+)=0.
    \end{equation}
   The solvability and solution's uniqueness of this equation follows from the above conditions imposed on $a^{ij}(\omega, \t{r},  p), b(\omega, \t{r},  p)$ and $c(\omega, \t{r})$, as well as using the linear theory of elliptic equations. Also notice that, for the operator $T$, the equation $R-r_+=\sigma T(R-r_+)$ is equivalent to the form
    \begin{equation*}
        Q_{\sigma} (R)\coloneqq a^{ij}(\omega, R,  DR)D_{ij} R+\sigma b(\omega, R, DR)+c(\omega, R)(R-r_+)=0.
    \end{equation*}
    Thus, in view of \eqref{General a priori est}, we obtain that for all $R \in C^{1, \alpha}$ and $\sigma\in [0, 1]$ satisfying $R-r_+=\sigma T(R-r_+)$, there must hold
    \begin{equation*}
        \|R-r_+\|_{C^{1,\a}} < M,
    \end{equation*}
    which is exactly \eqref{condition for L-S fixed point thm}.
    
    To employ \Cref{L-S fixed point thm}, it remains to verify the compactness of the mapping $T$.  From linear theory we known that $T$ maps $\overline{B}_{M}\coloneqq\{R-r_+\in C^{1, \a}: \ \|R-r_+\|_{C^{1, \a}}\le M \}$ to a bounded set in $C^{2, \a}$, and by Arzel\`a–Ascoli theorem $T(\overline{B}_M)$ is precompact in $C^{2}$ (and also in $C^{1, \a}$). To prove the continuity of $T$, we consider a sequence $\{S_n-r_+\}$ that converges to some $S-r_+$ in $C^{1, \a}$.  Since $T$ is precompact, for every subsequence we can extract a sub-subsequence $\{T(S_{n_k}-r_+) \}$ with its limit $R-r_+\in C^2$. Plugging in \eqref{def for mapping T} and letting $k\to \infty$, we deduce
    \begin{align*}
        &a^{ij}(\omega, S,  DS)D_{ij} R+b(\omega, S, DS)+c(\omega, S)(R-r_+)\\
        =&\lim\limits_{k\to \infty}\l a^{ij}(\omega, S_{n_k},  DS_{n_k})D_{ij} T(S_{n_k}-r_+) +b(\omega, S_{n_k}, DS_{n_k})+c(\omega, S_{n_k})T(S_{n_k}-r_+) \r=0.
    \end{align*}
    This gives $R-r_+=T(S-r_+)$. Therefore, we must have $T(S_n-r_+)\to T(S-r_+)$ as $n\to \infty$.\footnote{Suppose $\{T(S_n-r_+)\}$ dose not converge to $T(S-r_+)$. Then there exist $\epsilon_0>0$ and a subsequence $\{S_{n_k}-r_+\}$ such that $ \|T(S_n-r_+)-T(S-r_+)\|_{C^{1,\a}}\ge \epsilon_0$ for all $k\in \mathbb{N}$. Letting $\{T(S_{n_{k_l}}-r_+) \}$ be its convergent sub-subsequence, we hence obtain $T(S_{n_l}-r_+)\to T(S-r_+)$ as $l\to \infty$, which leads to a contradiction.} This completes the proof of this proposition. 
\end{proof}
\begin{remark}
  In practical applications of \Cref{Prop:general elliptic existence}, it is usually sufficient to verify the condition \eqref{General a priori est} with $\sigma=1$. For $\si\in[0, 1)$, the verification is typically the same.
\end{remark}
\subsection{Gradient Estimate with Quasiconformal Method}\label{Subsec:gradient est}
In this section, for the linear elliptic equation
\begin{equation}\label{Eqn:linear ellip 2d}
    Lu=au_{xx}+2bu_{xy}+cu_{yy}=f,
\end{equation}
we aim to establish the $C^{1, \a}$ estimate that demands only $L^{\infty}$ bounds of the coefficients $a, b,c$ without any additional regularity assumptions on $a, b,c$. This is in contrast with the classical $W^{2, p}$ estimate and the Schauder estimate, which require the modulus continuity or the H\"older continuity of the coefficients.

To proceed, we generalize the notion of $(K, K')$-quasiconformal maping in a two-dimensional domain $\Omega$. 
\begin{definition}\label{Def:quasiconformal}
Given a continuously differentiable mapping $w: \Omega \to \mathbb{R}^2$ denoted by $w(z)=(p(z), q(z))$ with $z=(x,y)\in \Omega$, we say that $w$ is generalized $(K, K'(z))$-quasiconformal in $\Omega$, if there exist a constant $K\ge 1$ and a function $K'=K'(z)\ge 0$, such that it holds
\begin{equation}\label{K, K' quasi ineq}
    p_x^2+p_y^2+q_x^2+q_y^2\le 2K(p_x q_y-p_y q_x)+K'(z).
\end{equation}
\end{definition}
We also define the Dirichlet integral $\mathcal{D}(r; z)$ to be
\begin{equation*}
    \mathcal{D}(r; z)\coloneqq\int_{B_r(z)} |Dw|^2 dx dy=\int_{B_r(z)} (|w_x|^2+|w_y|^2) dx dy,
\end{equation*}
where $B_r(z)$ is the disk centering at $z$ with radius $r$. For a generalized quasiconformal mapping $w=(p, q)$, in this paper we prove the following lemma, which is a generalization of Lemma 12.1 in \cite{G-T}.
\begin{lemma}\label{generalized K K' quasiconformal mapping}
    Let $w=(p, q)$ be generalized $(K, K'(z))$-quasiconformal in a disk $B_R=B_R(z_0)$ satisfying \eqref{K, K' quasi ineq} with $K>1$ and $K'(z)\ge0$. Denote $\alpha=K-(K^2-1)^{\f12}\in (0, 1)$. Suppose $K'\in L^{\beta}$ for some $\beta>\f{1}{1-\a}$ and $|p|\le M$ in $B_R$, then for all $r\le R/2$, the following estimate holds
    \begin{equation*}
        \mathcal{D}(r; z_0)\le C(\f{r}{R})^{2\alpha}, 
    \end{equation*}
    where $C=C_1(K, \b)(M^2+\|K'\|_{L^{\b}(B_R)}R^{2-\f{2}{\b}})$.
\end{lemma}
\begin{remark}
The proof of Lemma 12.1 in \cite{G-T} relies on the $L^{\infty}$ bound of $K'$. In this paper, we extend that lemma with weaker assumption for $K'\in L^{\b}$ with some $\b>\f{1}{1-\a}$. Note that only the weaker assumption can be verified in this paper.
\end{remark}
\begin{proof}
    Here we generalize the proof strategy of Lemma 12.1 in \cite{G-T}, with more delicate and more complicated treatments for the integrals involving $K'$. In below writing, we omit $z_0$ if there is no risk of confusion. 
    
    To achieve our goal, we first establish the bound for the Dirichlet integral in $B_{R/2}$ and we aim to prove
    \begin{equation}\label{bound for mD(R/2)}
        \mD(R/2)\le \f{8\pi}{\log 2} M^2 K^2+\|K'\|_{L^p\beta} (\pi R^2)^{1-\f{1}{\beta}}.
    \end{equation}
    Using \eqref{K, K' quasi ineq}, in any concentric disk $B_r\subset B_R$, we have
    \begin{equation}\label{ineq for mD 1}
        \begin{split}
             \mD(r)=\int_{B_r} |Dw|^2 dx dy\le& 2K\int_{B_r} \f{\partial(p, q)}{\partial(x, y)} dx dy+\int_{B_r} K' dxdy \\
             =& 2K \int_{C_r} p \f{\partial q}{\partial s} ds+\int_{B_r} K' dxdy.
        \end{split}
    \end{equation}
    Here the third equality is a consequence of Green's formula and $s$ denotes the arc length along the circle $C_r=\partial B_r$ in the counter-clockwise direction. Noting that $\mD'(r)=\int_{C_r} |Dw|^2 ds$ and $|\f{\partial q}{\partial s}|=|Dq \cdot (-\sin \theta, \cos \theta)|\le |Dq|$, we then bound the first integral on the right of \eqref{ineq for mD 1} by
    \begin{align*}
        \int_{C_r} p \f{\partial q}{\partial s} ds \le (\int_{C_r} p^2 ds)^{\f12}(\int_{C_r} |Dq|^2)^{\f12}\le  M(2\pi r \mD'(r))^{\f12}.
    \end{align*}
    Since $K'\in L^{\b}$, we also derive
    \begin{equation}
        \int_{B_r} K' dxdy \le \|K'\|_{L^{\b}} (\pi r^2)^{1-\f{1}{\b}} 
        \le \|K'\|_{L^\b} (\pi R^2)^{1-\f{1}{\b}}\eqqcolon k_1.
    \end{equation}
    Substituting above two estimates into \eqref{ineq for mD 1}, we now obtain
    \begin{equation}\label{ineq for mD 2}
        \mD(r)-k_1 \le 2KM(2\pi r \mD'(r))^{\f12}.
    \end{equation}
    
    If $\mD(R/2)\le k_1$, we finish the desired estimate. Otherwise, we can find some $r_0<R/2$ such that $\mD(r_0)>k_1$ and hence $\mD(r)>k_1$ for all $r_0\le r\le R$. Taking the square of \eqref{ineq for mD 2}, we derive
    \begin{equation}\label{ineq for mD 3}
        [\mD(r)-k_1]^2 \le k_2 r \mD'(r) \quad \text{for any} \ r\in (r_0, R) \qquad \text{with} \quad k_2=8\pi M^2 K^2.
    \end{equation}
   We then integrate the differential inequality \eqref{ineq for mD 3} from $R/2$ to $R$ and deduce
    \begin{equation*}
        \f{1}{\mD(R/2)-k_1}\ge \int_{R/2}^{R} \f{\mD'(r) dr}{[\mD(r)-k_1]^2}\ge \f{1}{k_2}\int_{R/2}^{R} \f{dr}{r}=\f{1}{k_2} \log 2,
    \end{equation*}
    which yields
    \begin{equation*}
        \mD(R/2)\le \f{k_2}{\log 2}+k_1=\f{8\pi}{\log 2} M^2 K^2+\|K'\|_{L^\beta} (\pi R^2)^{1-\f{1}{\beta}}.
    \end{equation*}
    This is the desired \eqref{bound for mD(R/2)}.
    
    We then proceed to prove the growth estimate for $\mD(r)$ when $r\le R/2$. Observe that
    \begin{equation*}
        p_x q_y-p_y q_x \le \f{\a}{2}p_x^2+\f{1}{2\a} q_y^2+\f{\a}{2}q_x^2+\f{1}{2\a} p_y^2=\f{\a}{2}|w_x|^2+\f{1}{2\a} |w_y|^2.
    \end{equation*}
    Here $\alpha>0$ is a parameter and will be determined later. Plugging this into \eqref{K, K' quasi ineq}, we derive
    \begin{equation*}
        |w_x|^2+|w_y|^2 \le 2K (\f{\a}{2}|w_x|^2+\f{1}{2\a} |w_y|^2 )+K'(z).
    \end{equation*}
    Choosing $\a=K-(K^2-1)^{\f12}\in (0, 1)$, or equivalently $K=\f{1+\a^2}{2\a}$, it can be inferred that
    \begin{equation*}
        |w_x|^2\le \f{1}{\a^2}|w_y|^2+\f{2K'(z)}{1-\a^2},
    \end{equation*}
    which gives
    \begin{equation}\label{ineq for w x}
        |w_x|^2=\f{1}{1+\a^2}(|w_x|^2+\a^2 |w_x|^2)\le \f{1}{1+\a^2}\Big(|Dw|^2+\f{2\a^2 K'(z)}{1-\a^2} \Big).
    \end{equation}
    Noticing that \eqref{K, K' quasi ineq} is invariant under the rotation, we can then replace $(w_x, w_y)$ by $(w_s, w_r)$ in the above argument, where $w_s\coloneqq Dw \cdot (-\sin \theta, \cos \theta)=\f{\pr w}{\pr s}$ and  $w_r\coloneqq Dw \cdot (\cos \theta, \sin \theta)$. We observe that the inequality \eqref{ineq for w x} still remains valid for $w_s$, i.e.,
    \begin{equation}\label{ineq for w s}
        |w_s|^2\le \f{1}{1+\a^2}\Big(|Dw|^2+\f{2\a^2 K'(z)}{1-\a^2} \Big).
    \end{equation}
    Denote $\overline{p}(r)\coloneqq\f{1}{2\pi r} \int_{C_r} p ds$. With the aid of \eqref{ineq for w s}, we then derive a more precise estimation for $\int_{C_r} pq_s ds$ on the right of \eqref{ineq for mD 1}:
    \begin{align*}
        \int_{C_r} p q_s ds =\int_{C_r} (p-\overline{p})q_s ds \le& \f12 \int_{C_r} \Big( \f{(p-\overline{p})^2}{r}+rq_s^2 \Big) ds \\ \le& \f{r}{2}\int_{C_r} (|p_s|^2+|q_s|^2) ds=\f{r}{2}\int_{C_r} |w_s|^2 ds \\
        \le& \f{r}{2(1+\a^2)} \int_{C_r} |Dw|^2 ds+\f{\a^2 r}{1-\a^4} \int_{C_r} K' ds.
    \end{align*}
    Here in the second line we use the Wirtinger inequality
    \begin{equation*}
        \int_{0}^{2\pi} [p(r, \theta)-\overline{p}]^2 d\theta\le \int_{0}^{2\pi} p_{\theta}^2 d\theta,
    \end{equation*}
    that is,
    \begin{equation*}
    \int_{C_r} (p-\overline{p})^2 ds \le r^2 \int_{C_r} p_s^2 ds.
    \end{equation*}
 Inserting \eqref{ineq for w s} into \eqref{ineq for mD 1} and utilizing the relation $\mD'(r)=\int_{C_r} |Dw|^2 ds$ again, we now obtain the following inequality
 \begin{equation}\label{Est:D(r)}
 \begin{aligned}
     \mD(r)\le& \f{Kr}{1+\a^2} \mD'(r)+\f{2K\a^2}{1-\a^4}\int_{C_r} K' ds+\int_{B_r} K' dxdy \\
     =&\f{r}{2\a} \Big( \mD'(r)+\f{2\a^2}{1-\a^2} \int_{C_r} K' ds\Big)+ \int_{B_r} K' dxdy.
 \end{aligned}
 \end{equation} 
 
 To proceed, we set
 \begin{equation*}
     \widetilde{\mD}(r)\coloneqq\mD(r)+\f{2\a^2}{1-\a^2}\int_{B_r} K' dxdy.
 \end{equation*}
Hence the above inequality \eqref{Est:D(r)} can be rewritten as 
\begin{equation*}
    \widetilde{\mD}(r)\le \f{r}{2\a}\widetilde{\mD}'(r)+\f{1+\a^2}{1-a^2} \int_{B_r} K' dxdy \le \f{r}{2\a}\widetilde{\mD}'(r)+ \f{1+\a^2}{1-a^2}\|K'\|_{L^\beta} (\pi r^2)^{1-\f{1}{\beta}}.
\end{equation*}
Denote $k=\f{1+\a^2}{1-a^2}\|K'\|_{L^\beta} \pi^{1-\f{1}{\beta}}$. It then follows
\begin{equation*}
    -\f{d}{dr}(r^{-2\a} \widetilde{\mD}(r)) \le 2\a k r^{1-\f{2}{\b}-2\a}.
\end{equation*}
Note that from the condition $\b>\f{1}{1-\a}$, we get $\f{1}{\b}+\a<1$. By integration from $r$ to $R/2$, we arrive at
\begin{equation*}
    \widetilde{\mD}(r)\le \Big(\widetilde{\mD}(R/2)+\f{\a k}{1-\f{1}{\b}-\a} (R/2)^{2-\f{2}{\beta}}\Big)(\f{r}{R/2})^{2\a}.
\end{equation*}
Combining the estimate for $\mD(R/2)$ in \eqref{bound for mD(R/2)}, we thus conclude
\begin{equation*}
    D(r)\le\widetilde{\mD}(r) \le C_1(K, \beta)(M^2+\|K'\|_{L^{\b}}R^{2-\f{2}{\b}})(\f{r}{R})^{2\a}.
\end{equation*}
as desired.
\end{proof}
We will also need the following type of Morrey's inequality:
\begin{lemma}\label{Morrey est}
    Let $\Omega'\subset \subset \Omega\subset\mathbb{R}^2$. Suppose that, for some $\alpha\in (0, 1)$ and $R_0>0$, we have $w\in H^{1}_{loc}(\Omega)$ satisfies 
    \begin{equation*}
        \int_{B_r(z)} |Dw|^2 dz \le M^2 r^{2\alpha} \quad \text{for any} \ B_r(z)\subset \Omega  \ \text{with} \ 0<r\le R_0 \ \text{and} \ z\in \Omega'.
    \end{equation*}
    Then $w\in C^{\alpha}(\Omega')$ and it obeys
    \begin{equation*}
        \|w\|_{C^{0, \alpha}(\Omega')}\le C(\alpha)(M+d^{-\a}\|w\|_{L^2}(\O)), \quad d=\text{dist}(\Omega', \partial\Omega)>0.
    \end{equation*}
\end{lemma}
   For the proof of this lemma, we refer to the verifications of Theorem 3.1 and Corollary 3.2 in Chapter 3 of \cite{H-L:ellipticpdetextbook} on page 48--50.
\vspace{2mm}

In the context of H\"{o}lder regularity, for a function $u\in C^{1, \alpha}(\Omega)$ we introduce the following interior norms and seminorms:
\begin{equation}\label{def interior norms}
\begin{split}
 &[u]^*_{1}\coloneqq\sup\limits_{z\in \Omega} d_z |Du(z) |,\\
    &[u]^*_{1,\a}\coloneqq\sup\limits_{z_1, z_2\in \Omega} d_{z_1, z_2}^{1+\alpha}  \f{|Du(z_1)-Du(z_1)|}{|z_2-z_1|^{\a}}, \\
    &|u|^*_{1, \a}\coloneqq\sup\limits_{\Omega} |u|+[u]^*_{1}+[u]^*_{1,\a},
    \end{split}
\end{equation}
where $d_z\coloneqq\text{dist}(z, \partial \Omega)$ and $d_{z_1, z_2}\coloneqq\min(d_{z_1}, d_{z_2})$. 

Using these definitions, we now present an interpolation inequality from Lemma 6.32 in \cite{G-T}:
\begin{lemma}\label{interplation ineq}
    Let $u\in C^{1, \alpha}(\Omega)$ and $\epsilon>0$. With $|u|_0\coloneqq\sup\limits_{\Omega} |u|$, then there holds
    \begin{equation*}
        [u]^*_1\le \epsilon [u]^*_{1, \a}+C(\epsilon) |u|_0.
    \end{equation*}
\end{lemma}

Finally, incorporating \Cref{generalized K K' quasiconformal mapping} with \Cref{Morrey est} and \Cref{interplation ineq}, we establish the desired interior $C^{1, \alpha}$ estimate for linear elliptic equations \eqref{Eqn:linear ellip 2d}.
\begin{lemma}\label{Appendix Holder est for linear eqn in two variables}
   In a bounded region $\O$ in $\mathbb{R}^2$, let $u$ be a $C^2(\Omega)$ solution of the equation
    \begin{equation*}
        Lu=au_{xx}+2bu_{xy}+cu_{yy}=f.
    \end{equation*}
    Denote $\la=\la(z), \La=\La(z)$ as the eigenvalues of the associated coefficient matrix, satisfying
    \begin{equation*}
        \lambda(z) (\xi^2+\eta^2)\le a\xi^2+2b\xi \eta+c\eta^2\le \Lambda(z)  (\xi^2+\eta^2) \quad \text{for any} \ (\xi, \eta)\in \mathbb{R}^2. 
    \end{equation*}
    Assume that the operator $L$ is uniformly elliptic, i.e.,  $\f{\Lambda}{\lambda} \le \gamma$ for some constant $\gamma \ge 1$. Further suppose that $\f{f}{\lambda}\in L^{2\beta}$ with $\beta>\f{1}{1-\a}$ and $\a=1+\gamma-(\gamma^2+2\gamma)^{\f12}\in (0, 1)$. Then we have
    \begin{equation*}
         |u|^*_{1, \a}\le C(\gamma, \b, \Omega) (|u|_0+\|f/\lambda\|_{L^{2\b}}).
    \end{equation*}
\end{lemma}
\begin{proof}
  Setting $p=u_x$ and $q=u_y$, we first rewrite $Lu=f$ as the system
  \begin{equation}
      ap_x+2b p_y+cq_y=f, \quad p_y=q_x.
  \end{equation}
  Multiplying  the first equation by $p_x$, we obtain
  \begin{equation*}
      \lambda (p_x^2+p_y^2) \le ap_x^2+2b p_xp_y+cp_y^2= cJ+fp_x,
  \end{equation*}
  where $J\coloneqq q_xp_y-q_yp_x$. Similarly, it also holds
  \begin{equation*}
      \lambda (q_x^2+q_y^2)\le aJ+fq_y.
  \end{equation*}
  Combining these inequalities and noting that $ \f{a+c}{\lambda}=1+\f{\Lambda}{\lambda} \in [2, 1+\gamma]$, with $\mu\coloneqq\f{|f|}{\lambda}$ we deduce
  \begin{align*}
      |Dp|^2+|Dq|^2\le \f{a+c}{\lambda}J+\f{f}{\lambda} (p_x+q_y) 
      =&  \f{a+c}{\lambda}[J+\f{\lambda}{a+c}\f{f}{\lambda}  (p_x+q_y)] 
      \\
      \le& (1+\gamma) J+\f{1+\gamma}{2}\mu (|p_x|+|q_y|)
  \end{align*}
 On the right, inserting the inequality
  \begin{equation*}
      (1+\gamma)\mu (|p_x|+|q_y|)\le (p_x^2+p_y^2)+\f14 (1+\gamma)^2 \mu^2,
  \end{equation*}
  we obtain
\begin{equation}
    |Dp|^2+|Dq|^2\le 2(1+\gamma)J+\f{(1+\gamma)^2\mu^2}{4}.
\end{equation}

Therefore, according to \Cref{Def:quasiconformal}, we have that $w=(p, -q)$ defines a generalized $(K, K'(z))$-quasiconformal mapping in $B_1(z_0)$ with
\begin{equation*}
    K=1+\gamma>1, \quad K'(z)=\f{(1+\gamma)^2\mu^2}{4}\ge 0.
\end{equation*}
Since $\mu=\f{|f|}{\lambda}\in L^{2\beta}(B_1(z_0))$, we also have $K'\in L^{\beta}(B_1(z_0))$ and $\|K'\|_{L^{\beta}(B_1(z_0))}=\f{(1+\gamma)^2}{4}\|\f{f}{\lambda}\|^2_{L^{2\beta}(B_1(z_0))}$. 

For any $z_1, z_2 \in \Omega$, $z_1\neq z_2$, denote $d\coloneqq\f12 \min (d_{z_1}, d_{z_2})$. Let $\Omega'\coloneqq\{z\in \Omega: \  \text{dist}(z, \partial \Omega)>d \}$ and $\Omega''\coloneqq\{z\in \Omega': \ \text{dist}(z, \partial \Omega')>d \}$. Notice that we have $z_1, z_2 \in \overline{\Omega}''$. Now we apply  \Cref{generalized K K' quasiconformal mapping} in $B_d(z)$ with center $z\in \Omega''$. For all $ r\le d/2$ we then deduce that
\begin{align*}
    \int_{B_r(z)} |Dw|^2 dz\le& C_1(\gamma, \b) (\|p\|^2_{C^0(B_{d}(z))}+\|K'\|_{L^\beta(B_{d}(z))}d^{2-\f{2}{\b}})(\f{r}{d})^{2\alpha} \\
    \le& C_2(\gamma, \b)(\|w\|^2_{C^0(\Omega')}+\|\mu\|^2_{L^{2\beta}(\Omega')}d^{2-\f{2}{\b}})(\f{r}{d})^{2\alpha}.
\end{align*}
Hence, employing \Cref{Morrey est} with $(\Omega, \Omega')=(\Omega', \Omega'')$ and $R_0=d/2$, we obtain
\begin{align*}
    |w|_{C^{0, \alpha}(\Omega'')}\le& C(\gamma, \b)(\|w\|_{C^0(\Omega')}+\|\mu\|_{L^{2\beta}(\Omega')}d^{1-\f{1}{\b}}+\|w\|_{L^2(\Omega)})d^{-\a} \\
    \le& C(\gamma, \b, \Omega)(\|w\|_{C^0(\Omega')}+\|\mu\|_{L^{2\beta}(\Omega')}d^{1-\f{1}{\b}})d^{-\a}.
\end{align*}
As an immediate implication, the following estimate holds
\begin{equation*}
    d^{\a} \f{|Du(z_1)-Du(z_1)|}{|z_2-z_1|^{\a}} \le C (\sup\limits_{\Omega'} |Du|+d^{1-\f{1}{\b}}\|\mu\|_{L^{2\beta}(\Omega')}).
\end{equation*}
For any $z_1, z_2\in \Omega$, this then yields
\begin{align*}
    d_{z_1, z_2}^{1+\alpha}  \f{|Du(z_1)-Du(z_1)|}{|z_2-z_1|^{\a}} \le& C (\sup\limits_{\Omega'} d_z |Du(z)|+d^{2-\f{1}{\b}}\|\mu\|_{L^{2\beta}(\Omega)}) \\
    \le& C(\sup\limits_{\Omega} d_z |Du(z)|+\|\mu\|_{L^{2\beta}(\Omega)}).
\end{align*}
 Utilizing the notions of interior norms and seminorms in \eqref{def interior norms}, we can write it as
 \begin{equation}
      [u]^*_{1,\a}\le C([u]^*_{1}+\|f/\lambda\|_{L^{2\b}}).
 \end{equation}
 Applying the interpolation inequality from \Cref{interplation ineq}, we deduce
 \begin{equation*}
     [u]^*_{1,\a}\le C(\epsilon [u]^*_{1,\a}+C_1|u|_0+\|f/\lambda\|_{L^{2\b}}).
 \end{equation*}
 Choosing $C\epsilon=\f12$, we hence arrive at
 \begin{equation*}
      [u]^*_{1,\a}\le C(|u|_0+\|f/\lambda\|_{L^{2\b}})
 \end{equation*}
 with constant $C=C(\gamma, \b, \Omega)>0$.  Combining with \Cref{interplation ineq}, we conclude
 \begin{equation*}
     |u|^*_{1, \a}\le C(\gamma, \b, \Omega) (|u|_0+\|f/\lambda\|_{L^{2\b}})
 \end{equation*}
as stated.
 \end{proof}
\subsection{A Priori \texorpdfstring{$C^{1,\alpha}$}{} Estimate}\label{Subsec:C1aest}
In this subsection, we derive the a priori $C^{1,\a}$ estimate for the solution $R$ to the MOTS equation \eqref{MOTS main equation new}
\begin{equation}\label{MOTS main equation new2}
     a^{ij}(\omega, R,  DR)D_{ij} R+b(\omega, R, DR)+c(\omega, R)(R-r_+)=0.
\end{equation}
To achieve so, we conduct an iteration scheme. This process is inspired by \cite{W} and \cite{Soft}. In light of \Cref{MOTS main equation new prop}, we observe that $a^{ij}(\omega, \t{r},  p), b(\omega, \t{r},  p), c(\omega, \t{r})$ are smooth functions satisfying 
\begin{enumerate}
   \item \label{Cond1} $a^{ij}(\omega, \t{r},  p)=a^{ji}(\omega, \t{r},  p)$;
        \item \label{Cond2} There exist constants $\mu\ge \nu>0$, $C_0>c_0>0$ independent of $\epsilon_0, \tub, \t{r}, p$ and a constant $N>0$, such that
        \begin{align}
            \nu |\xi|^2 \le& a^{ij}(\omega, \t{r},  p) \xi_i \xi_j \le \mu |\xi|^2 \qquad \text{for any} \quad \xi\in \mathbb{R}^2, \nonumber \\ 
            |b(\omega, \t{r},  p)|\lesssim&  |p|^{2}+|p|+N, \qquad -C_0\le c(\omega, \t{r})\le -c_0. \label{property for b new}
        \end{align}
    \end{enumerate}

It is worthy to mention that, in the regime we are considering for the equation of the MOTS, we would have 
\begin{equation}\label{Choice of N}
    N=\tfepub\ll 1.
\end{equation}
However, our elliptic method developed here is more robust and can also be applied to the non-perturbative regime. Hence, we keep and allow $N$ to be large throughout this subsection. Furthermore, for tracing the dependence of estimates on $N$, we remark that all the inequalities involving $\lesssim$ are independent of $N$.
\vspace{2mm}

Denote $\mathscr{L}(S)R\coloneqq a^{ij}(\omega, S,  DS) D_{ij} R+c(\omega, S)(R-r_+)$. We first set 
$$\gamma\coloneqq\f{\mu}{\nu} \qquad \text{and} \qquad \a\coloneqq1+\gamma-(\gamma^2+2\gamma)^{\f12}\in (0, 1).$$ 
Given $S\in C^{1, \alpha}$, we consider the following one-parameter family of elliptic equations 
\begin{equation}\label{eqn for R sigma}
    \mathscr{L}(S)R+b(\omega, R, DR)-(1-\sigma)b(\omega, r_+, 0)=0 \qquad \text{with} \quad \sigma\in[0, 1].
\end{equation}
When $\sigma=0$, it is straightforward to verify that $R\equiv r_+$ satisfies the above equation. On the other hand, when $\sigma=1$, we encounter the equation
\begin{equation}\label{eqn for R 1}
     \mathscr{L}(S)R+b(\omega, R, DR)=0.
\end{equation} 
Define $R_{\sigma}$ to be the solution of \eqref{eqn for R sigma} on $\mathbb{S}^2$ with $\sigma\in [0, 1]$.  Since $\mathscr{L}(S)$ is uniformly elliptic and $c(\omega, \t{r})\le -c_0$, we obtain
\begin{equation}\label{C0 est for R sigma}
    |R_{\sigma}-r_+|\lesssim N \qquad \text{for all} \quad \sigma\in [0, 1].
\end{equation}

Next, we present a key proposition, which allows us to bound the $C^{1, \alpha}$ norm for the difference of $R_{\sigma_1}$ and $R_{\sigma_2}$, provided that $|\sigma_1-\sigma_2|$ is sufficiently small.
\begin{proposition}\label{R1-R2 C 1 apha}
     Let $S\in C^{1, \alpha}$ with $\a=\a(\mu, \nu)\in (0, 1)$ and assume that $\partial_{\t{r}} b(\omega, \t{r},  p)\le 0$. For solutions $R_{\sigma_i}$ solving \eqref{eqn for R sigma} with $\sigma=\sigma_i$ ($i=1, 2$), there exists a constant $\delta\sim N^{-1}>0$ and it holds
\begin{equation*}
    \|R_{\sigma_1}-R_{\sigma_2}\|_{C^{1, \alpha}}\lesssim \|DR_{\sigma_1}\|^2_{C^0}+\|DR_{\sigma_1}\|_{C^0}+N \qquad  \text{if} \quad  |\sigma_1-\sigma_2|\le \delta.
\end{equation*}
\end{proposition}
\begin{proof}
    Using the definitions of $R_{\sigma_1}$ and $R_{\sigma_2}$, we get
    \begin{align*}
        \mathscr{L}(S)(R_{\sigma_1}-R_{\sigma_2})+b(\omega, R_{\sigma_1}, DR_{\sigma_1})-b(\omega, R_{\sigma_2}, DR_{\sigma_2})+(\sigma_1-\sigma_2)b(\omega, r_+, 0)=0.
    \end{align*}
    Notice that
    \begin{align*}
        &b(\omega, R_{\sigma_1}, DR_{\sigma_1})-b(\omega, R_{\sigma_2}, DR_{\sigma_2})\\=&\int_{0}^1 \partial_{\t{r}} b (\omega, tR_{\sigma_1}+(1-t)R_{\sigma_2}, tDR_{\sigma_1}+(1-t)DR_{\sigma_2}) dt (R_{\sigma_1}-R_{\sigma_2})\\
        &+\sum_{i=1}^2\int_{0}^1 \partial_{p_i} b (\omega, tR_{\sigma_1}+(1-t)R_{\sigma_2}, tDR_{\sigma_1}+(1-t)DR_{\sigma_2}) dt D_i(R_{\sigma_1}-R_{\sigma_2})\\
        \eqqcolon&-C (R_{\sigma_1}-R_{\sigma_2})+B^i D_i(R_{\sigma_1}-R_{\sigma_2}).
    \end{align*}
    Thus, we can rewrite the above equation for $R_{\sigma_1}-R_{\sigma_2}$ as
    \begin{equation*}
        \mathscr{L}(S)(R_{\sigma_1}-R_{\sigma_2})+B^i D_i(R_{\sigma_1}-R_{\sigma_2})-C (R_{\sigma_1}-R_{\sigma_2})+(\sigma_1-\sigma_2)b(\omega, r_+, 0)=0,
    \end{equation*}
    where 
    \begin{equation*}
       C=-\int_{0}^1 \partial_{\t{r}} b (\omega, tR_{\sigma_1}+(1-t)R_{\sigma_2}, tDR_{\sigma_1}+(1-t)DR_{\sigma_2}) dt \ge 0.
    \end{equation*}
    Applying the maximum principle for the above equation, we obtain
    \begin{equation}\label{L infty est for R1-R2}
        |R_{\sigma_1}-R_{\sigma_2}|\lesssim  |\sigma_1-\sigma_2| |b(\omega, r_+, 0)| \lesssim |\sigma_1-\sigma_2|N.
    \end{equation}
    
    On the other hand, utilizing \eqref{property for b} and \eqref{C0 est for R sigma}, we also obtain the below estimation for the difference of $b(\omega, R_{\sigma_1}, DR_{\sigma_1})$ and $b(\omega, R_{\sigma_2}, DR_{\sigma_2})$:
    \begin{align*}
        &|b(\omega, R_{\sigma_1}, DR_{\sigma_1})-b(\omega, R_{\sigma_2}, DR_{\sigma_2})|\\\lesssim& |DR_{\sigma_1}|^2+|DR_{\sigma_1}| +(|D(R_{\sigma_1}-R_{\sigma_2})|^2+(|DR_{\sigma_1}|^2)+|D(R_{\sigma_1}-R_{\sigma_2})|)+|DR_{\sigma_1}|+N \\
        \lesssim& |DR_{\sigma_1}|^2+|DR_{\sigma_1}|+|D(R_{\sigma_1}-R_{\sigma_2})|^2+|D(R_{\sigma_1}-R_{\sigma_2})|+N.
    \end{align*}
    Combining with the fact that $|b(\omega, r_+, 0)|\lesssim N$ from \eqref{property for b new}, we deduce
    \begin{equation}\label{L infty bound for L(S)(R1-R2)}
        |\mathscr{L}(S)(R_{\sigma_1}-R_{\sigma_2})|\lesssim |DR_{\sigma_1}|^2+|DR_{\sigma_1}|+|D(R_{\sigma_1}-R_{\sigma_2})|^2+|D(R_{\sigma_1}-R_{\sigma_2})|+N.
    \end{equation}
    Taking the $L^2$ norm on both sides, applying \eqref{L infty est for R1-R2} we then derive
 \begin{equation}\label{L2 est for L(S)(R1-R2)}
     \begin{split}
          &\|\mathscr{L}(S)(R_{\sigma_1}-R_{\sigma_2})\|_{L^2}\\
          \lesssim& \|DR_{\sigma_1}\|_{L^4}^2+\|DR_{\sigma_1}\|_{L^2}+\|D(R_{\sigma_1}-R_{\sigma_2})\|_{L^4}^2+\|D(R_{\sigma_1}-R_{\sigma_2})\|_{L^2}+N  \\
\lesssim& F(\|DR_{\sigma_1}\|_{C^0})+\|R_{\sigma_1}-R_{\sigma_2}\|_{L^\infty} \|R_{\sigma_1}-R_{\sigma_2}\|_{H^2}+\|R_{\sigma_1}-R_{\sigma_2}\|_{L^\infty}^{\f12} \|R_{\sigma_1}-R_{\sigma_2}\|_{H^2}^{\f12}+N \\
\lesssim& F(\|DR_{\sigma_1}\|_{C^0})+|\sigma_1-\sigma_2|N \cdot\|R_{\sigma_1}-R_{\sigma_2}\|_{H^2}+\l |\sigma_1-\sigma_2|N \cdot\|R_{\sigma_1}-R_{\sigma_2}\|_{H^2}\r^{\f12}+N.
     \end{split}
 \end{equation}
    Here $F(x)\coloneqq x^2+x$ and in the first inequality we also use the interpolation inequality on $\mathbb{S^2}$, i.e.,
    \begin{equation*}
        \|DR\|^2_{L^4}\lesssim \|R\|_{L^\infty} \|R\|_{H^2} \quad \text{for any} \ R\in H^2.
    \end{equation*}
    
    To estimate $ \|R_{\sigma_1}-R_{\sigma_2}\|_{H^2}$, we appeal to using the Campanato condition verified in \Cref{Lem:Campanato}. With the selection of $X=(D_{ji}R)$, we get
\begin{equation*}
    \left|\De (R_{\sigma_1}-R_{\sigma_2})-a\l a^{ij}D_{ij} (R_{\sigma_1}-R_{\sigma_2})\r \right|\le (1-\gamma^{-1})\|D^2 (R_{\sigma_1}-R_{\sigma_2} )\|,
\end{equation*}
where $D^2=(D_{ij})$ and $\De=D_{11}+D_{22}$. We also have $a=a(\omega, \t{r}, p)\in [\mu^{-1}, \nu^{-1}]$ and $\gamma=\f{\mu}{\nu}\ge 1$. Taking the $L^2$ norm on both sides, we deduce
\begin{align*}
    \|\De (R_{\sigma_1}-R_{\sigma_2})\|_{L^2}\le&  \left\|\De (R_{\sigma_1}-R_{\sigma_2})-a\l a^{ij}D_{ij} (R_{\sigma_1}-R_{\sigma_2})\r \right\|_{L^2}+\left\|a\l a^{ij}D_{ij} (R_{\sigma_1}-R_{\sigma_2})\r \right\|_{L^2} \\
    \le& (1-\gamma^{-1})\|D^2 (R_{\sigma_1}-R_{\sigma_2})\|_{L^2}+\nu^{-1} \|a^{ij}D_{ij} (R_{\sigma_1}-R_{\sigma_2})\|_{L^2}.
\end{align*}
By Miranda--Talenti inequality on $\mathbb{S}^2$ proved in \Cref{Apx:M-T ineq on S2} in Appendix, we then obtain
\begin{equation}\label{Ineq:MT}
    \|D^2 (R_{\sigma_1}-R_{\sigma_2})\|_{L^2}\le \|\De (R_{\sigma_1}-R_{\sigma_2})\|_{L^2},
\end{equation}
It thus renders
\begin{align*}
    \|D^2 (R_{\sigma_1}-R_{\sigma_2})\|_{L^2}\le& \gamma\nu^{-1}\|a^{ij}D_{ij} (R_{\sigma_1}-R_{\sigma_2})\|_{L^2} \\
    \le& \gamma\nu^{-1}\l\|\mathscr{L}(S) (R_{\sigma_1}-R_{\sigma_2})\|_{L^2}+C_0\|R_{\sigma_1}-R_{\sigma_2}\|_{L^2} \r.
\end{align*}
This thus implies the regularity estimate for $R_{\sigma_1}-R_{\sigma_2}$:
    \begin{equation}\label{Est:H2 R1-R2}
        \|R_{\sigma_1}-R_{\sigma_2}\|_{H^2}\lesssim  \|\mathscr{L}(S)(R_{\sigma_1}-R_{\sigma_2})\|_{L^2}+\|R_{\sigma_1}-R_{\sigma_2}\|_{L^2} \lesssim \|\mathscr{L}(S)(R_{\sigma_1}-R_{\sigma_2})\|_{L^2}+N.
    \end{equation}
    Plugging the above inequality back to \eqref{L2 est for L(S)(R1-R2)}, we hence arrive at
    \begin{align*}
         \|\mathscr{L}(S)(R_{\sigma_1}-R_{\sigma_2})\|_{L^2}\le& C\Big( F(\|DR_{\sigma_1}\|_{C^0})+|\sigma_1-\sigma_2|N \cdot(\|\mathscr{L}(S)(R_{\sigma_1}-R_{\sigma_2})\|_{L^2}+N)\\&+\big( |\sigma_1-\sigma_2|N \cdot(\|\mathscr{L}(S)(R_{\sigma_1}-R_{\sigma_2})\|_{L^2}+N)\big)^{\f12}+N \Big)\\
         \le& C\Big( F(\|DR_{\sigma_1}\|_{C^0})+N|\sigma_1-\sigma_2|\cdot\|\mathscr{L}(S)(R_{\sigma_1}-R_{\sigma_2})\|_{L^2}+N\Big).
    \end{align*}
    We also note that the term $\|\mathscr{L}(S)(R_{\sigma_1}-R_{\sigma_2})\|_{L^2}$ on the right can be absorbed to the left if 
    \begin{equation}\label{Cond:si1-si2}
        |\sigma_1-\sigma_2|\le \delta\coloneqq\min\{\f12C^{-1}N^{-1}, 1\}.
    \end{equation}
   With this observation, under \eqref{Cond:si1-si2} we then get
    \begin{equation*}
         \|\mathscr{L}(S)(R_{\sigma_1}-R_{\sigma_2})\|_{L^2}\lesssim F(\|DR_{\sigma_1}\|_{C^0})+N.
    \end{equation*}
      Applying \eqref{Est:H2 R1-R2}, we thus derive the $H^2$ bound for $R_{\sigma_1}-R_{\sigma_2}$ and it states
    \begin{equation*}
         \|R_{\sigma_1}-R_{\sigma_2}\|_{H^2} \lesssim F(\|DR_{\sigma_1}\|_{C^0})+N. \\[2mm]
    \end{equation*}
    
    Now we turn to estimate the $L^p$ norm of $\mathscr{L}(S)(R_{\sigma_1}-R_{\sigma_2})$ for $2\le p<\infty$. In view of Sobolev embedding on $\mathbb{S}^2$, we have
    \begin{equation*}
        \|R\|_{L^p} \lesssim_p \|R\|_{H^1}.
    \end{equation*}
    Employing \eqref{L infty bound for L(S)(R1-R2)}, we then obtain
    \begin{align*}
         \|\mathscr{L}(S)(R_{\sigma_1}-R_{\sigma_2})\|_{L^p}\lesssim& \|DR_{\sigma_1}\|_{L^{2p}}^2+\|DR_{\sigma_1}\|_{L^{p}}+\|D(R_{\sigma_1}-R_{\sigma_2})\|_{L^{2p}}^2+\|D(R_{\sigma_1}-R_{\sigma_2})\|_{L^{p}}+N  \\
\lesssim&_p F(\|DR_{\sigma_1}\|_{C^0})+ \|R_{\sigma_1}-R_{\sigma_2}\|_{H^2}^2+\|R_{\sigma_1}-R_{\sigma_2}\|_{H^2}+N \\
\lesssim&  F(\|DR_{\sigma_1}\|_{C^0})+N.
    \end{align*}
   This along with \eqref{L infty est for R1-R2} gives
    \begin{equation*}
       \|a^{ij}D_{ij}(R_{\sigma_1}-R_{\sigma_2})\|_{L^p}\le \|\mathscr{L}(S)(R_{\sigma_1}-R_{\sigma_2})\|_{L^p}+C_0\|R_{\sigma_1}-R_{\sigma_2} \|_{L^p}\lesssim F(\|DR_{\sigma_1}\|_{C^0})+N.
    \end{equation*}
    Further applying \Cref{Appendix Holder est for linear eqn in two variables} by choosing $p>\f{1}{1-\a}$, we therefore derive the desired H\"{o}lder estimate
    \begin{equation*}
        \|R_{\sigma_1}-R_{\sigma_2}\|_{C^{1, \a}}\lesssim \|R_{\sigma_1}-R_{\sigma_2}\|_{C^{0}}+\|R_{\sigma_1}-R_{\sigma_2}\|_{L^p}\lesssim F(\|DR_{\sigma_1}\|_{C^0})+N.
    \end{equation*}
    This completes the proof of this proposition.
\end{proof}
\begin{remark}
    In the derivations of the $H^2$ or $C^{1,\a}$ estimates for $R_{\sigma_1}-R_{\sigma_2}$, we remark that we cannot directly employ the classical $W^{2, p}$ estimate. This is because the classical $W^{2, p}$ estimate demands the control of the modulus of continuity for the second-order coefficients $a^{ij}(\omega, S, DS)$, which also depends on $S, DS$. To overcome this difficulty, we employ the following two ideas:
    \begin{itemize}
        \item Regarding the $H^2$ bound of $R_{\sigma_1}-R_{\sigma_2}$, we appeal to verifying the Campanato condition and utilizing the Miranda--Talenti inequality on $\mathbb{S}^2$ in \eqref{Ineq:MT}. 
        \item To control $\|R_{\sigma_1}-R_{\sigma_2}\|_{C^{1, \a}}$, we apply \Cref{Appendix Holder est for linear eqn in two variables} via quasi-conformal method. The conditions to apply this method are less restrictive and it only requires the boundedness of the coefficients and the uniform ellipticity of the elliptic operator $\mathscr{L}(S)$.
    \end{itemize}
\end{remark}
We are now ready to derive the $C^{1, \alpha}$ estimates for $R_{\sigma}$ as below.
\begin{proposition}\label{R sigma C 1 alpha}
   Assume that $R$ is a $C^2$ solution to \eqref{eqn for R 1}. Under the same assumptions of \Cref{R1-R2 C 1 apha}, 
   the following inequality holds 
    \begin{equation*}
         \|R-r_+\|_{C^{1, \alpha}}\lesssim \max(N^{2^{CN}}, N)
    \end{equation*}
    with the constant $C\ge 1$ being independent of $N$.
\end{proposition}
\begin{proof}
    Let $\de\sim N^{-1}$ be the same constant as in \Cref{R1-R2 C 1 apha}. Assume first that $R_{\delta}$ satisfies \eqref{eqn for R sigma} with $\sigma=\delta$. Picking $R_0\equiv r_+$, applying \Cref{R1-R2 C 1 apha} (with $(\sigma_1, \sigma_2)=(0, \de)$) and \eqref{C0 est for R sigma}, we then infer the below a priori estimate for $R_{\de}$:
    \begin{equation*}
        \|R_{\delta}-r_+\|_{C^{1, \alpha}}\le C\l \|DR_{0}\|_{C^{1, \alpha}}+\|DR_{0}\|_{C^0}^2+\|DR_{0}\|_{C^0}+N\r=CN.
    \end{equation*}
   Thus, by \Cref{Prop:general elliptic existence} with the selection of $M=2CN$, we obtain the existence of the solution to \eqref{eqn for R sigma} with $\sigma=\delta$, namely, $R=R_{\de}$.
   
   Repeating this process, we then show that the equation \eqref{eqn for R sigma} with $\sigma=n\de$ admits a solution $R_{n\de}\in C^{1, \a}$ for $n=2, \dots , \lfloor\f{1}{\delta}\rfloor$, and the following estimate holds
    \begin{equation*}
        \|R_{n\delta}-r_+\|_{C^{1, \alpha}}\lesssim \max(N^{2^{n-1}}, N) \quad \text{for all} \ n=0, 1, \dots , \lfloor\f{1}{\delta}\rfloor.
    \end{equation*}
    
    Note that
    \begin{equation*}
        1-\sigma_{\lfloor\f{1}{\delta}\rfloor}=1-\lfloor\f{1}{\delta}\rfloor\de \in [0, 1].
    \end{equation*}
    Employing \Cref{R1-R2 C 1 apha} again for the pair $(\sigma_1, \sigma_2)=(\lfloor\f{1}{\delta}\rfloor\sigma, 1)$, we thus deduce
    \begin{equation*}
        \|R-r_+\|_{C^{1, \alpha}}\lesssim\|R-R_{\sigma_{\lfloor\f{1}{\delta}\rfloor}}\|_{C^{1, \alpha}}+  \|DR_{\sigma_{\lfloor\f{1}{\delta}\rfloor}}\|_{C^{1, \alpha}}^2+\|DR_{\sigma_{\lfloor\f{1}{\delta}\rfloor}}\|_{C^0}+N \lesssim \max(N^{2^{\lfloor\f{1}{\delta}\rfloor}}, N).
    \end{equation*}
    This completes the proof of this proposition.
\end{proof}
\vspace{2mm}

We proceed to conclude the desired $C^{1, \a}$ estimate for the solution to \eqref{MOTS main equation new2}.
\begin{proposition}\label{MOTS eqn C 1, alpha}
    Assume that $R$ is a $C^2$ solution to \eqref{MOTS main equation new2}. Then the following estimate holds
    \begin{equation*}
        \|R-r_+\|_{C^{1, \alpha}} \lesssim \max(N^{2^{CN}}, N).
    \end{equation*}
\end{proposition}
\begin{proof}
    Recall that from \eqref{property for b new} we have
    \begin{equation}\label{property for b new1}
        |b(\omega, R, DR)|\lesssim |DR|^2+|DR|+N.
    \end{equation}
    Set
    \begin{equation*}
        \widetilde{b}(\omega, \t{r}, p)\coloneqq\f{b(\omega, R, DR)}{|DR|^2+|DR|+N}\cdot (|p|^2+|p|+N).
    \end{equation*}
    For $\widetilde{b}(\omega, \t{r}, p)$, using \eqref{property for b new1}, we also get
    \begin{equation*}
         |\widetilde{b}(\omega, \t{r}, p)|\lesssim  |p|^{2}+|p|+N.
    \end{equation*}
    Additionally, since $\widetilde{b}(\omega, \t{r}, p)$ is independent of $\t{r}$, it holds
    \begin{equation*}
        \partial_{\t{r}} \widetilde{b}(\omega, \t{r}, p)=0.
    \end{equation*}
     
     Now we consider the elliptic equation in the form of
    \begin{equation*}
        \mathscr{L}(R)\widetilde{R}+\widetilde{b}(\omega, \widetilde{R}, D\widetilde{R})=0.
    \end{equation*}
    In view of
    \begin{equation*}
        \widetilde{b}(\omega, R, DR)=\f{b(\omega, R, DR)}{|DR|^2+|DR|+N}\cdot (|DR|^2+|DR|+N)=b(\omega, R, DR),
    \end{equation*}
    it can be checked directly that $\tR=R$ solves the above elliptic equation. Employing \Cref{R sigma C 1 alpha} by choosing $S=R$, we hence deduce
    \begin{equation*}
        \|R-r_+\|_{C^{1, \alpha}} \lesssim \max(N^{2^{CN}}, N).
    \end{equation*}
    This then completes the proof of this proposition.
\end{proof}
Consequently, by the standard Schauder estimates, we further get
\begin{corollary}\label{MOTS eqn C 2 alpha}
     Assume that $R$ is a $C^2$ solution to \eqref{MOTS main equation new2}. Then the following estimate holds
    \begin{equation*}
        \|R-r_+\|_{C^{2, \alpha}} \lesssim \max(N^{2^{CN}}, N).
    \end{equation*}
\end{corollary}
\vspace{2mm}

Although in above we only derive the crucial a priori $C^{1, \a}$ estimate for solutions to the equation
\begin{equation*}
    Q_{\sigma}(R)=a^{ij}(\omega, R,  DR)D_{ij} R+\sigma b(\omega, R, DR)+c(\omega, R)(R-r_+)=0
\end{equation*} 
with $\sigma=1$, the preceding argument applies equally to all $\sigma\in[0, 1]$. Therefore, utilizing \Cref{Prop:general elliptic existence} via Leray--Schauder fixed point argument and employing the Schauder estimate for the difference quotient, we thus obtain the existence of the solution to the equation of MOTS \eqref{MOTS main equation new2}.
\begin{proposition}\label{MOTS eqn C infty}
    The equation \eqref{MOTS main equation new2} admits a $C^{2, \a}(\mathbb{S}^2)$ solution $R=R(\omega)$ satisfying
    \begin{equation*}
        \|R-r_+\|_{C^{2, \alpha}} \les \max(N^{2^{CN}}, N).
    \end{equation*}
     Moreover, we can improve its regularity to $R\in C^{\infty}(\mathbb{S}^2)$. 
\end{proposition}
In particular, in the perturbative Kerr setting, with $N=\tfepub\ll 1$, we deduce
\begin{theorem}\label{Thm:MOTS exis}
    For any fixed $\tub\ge F(v_1)$, in perturbative Kerr regime with $\epsilon_0>0$ being sufficiently small, the equation for MOTS \eqref{MOTS main equation} admits a $C^{\infty}(\mathbb{S}^2)$ solution $R=R(\omega)$. Furthermore, there holds that 
    \begin{equation*}
        \|R-r_+\|_{C^{2, \alpha}} \les \f{\epsilon_0}{\tub^{1+\de_{dec}}}.
    \end{equation*}
\end{theorem}
\begin{remark}
    The proof of \Cref{Thm:MOTS exis} does not grant the uniqueness of solutions for the equation of MOTS \eqref{MOTS main equation}. The proof of uniqueness of MOTS is via a different approach and it is presented in the next subsection.
\end{remark}
\subsection{Uniqueness of MOTS}
The goal of this section is to prove that the MOTS solved in \Cref{Thm:MOTS exis} is unique along each incoming null hypersurface $\tHb_{\tub}$ within $\intM$. 

Recall that the interior region $\intM$ is equipped with the incoming geodesic foliation $(\tub, \t{r}, \t{\th}^1, \t{\th}^2)$ and the associated null frame $(\te_1, \te_2, \te_3, \te_4)$. As derived in \Cref{MOTS main equation new prop}, the equation of MOTS reads
\begin{equation}\label{MOTS operator}
       L(R, \tub)\coloneqq a^{ij}(\tub, \omega, R,  DR)D_{ij} R+b(\tub, \omega, R, DR)+c(\tub, \omega, R)(R-r_+)=0.
    \end{equation}
A direct computation yields
\begin{equation}\label{Eqn:pr R L}
\begin{aligned}
    \pr_R L(R, \tub)[W]=&a^{ij} D_{ij}W+\l\pr_{p^i} a^{jk} D_{jk}R+\pr_{p^i} b\r \cdot D_i W\\&+\l \pr_{\t{r}} a^{ij} D_{ij}R+\pr_{\t{r}} b+\pr_{\t{r}} c\c(R-r_+)+c  \r W
\end{aligned}
\end{equation}
and
\begin{equation}\label{Eqn:pr tub L}
    \pr_{\tub}L(R, \tub)=\pr_{\tub}a^{ij} D_{ij} R+\pr_{\tub}b+\pr_{\tub} c\c(R-r_+).
\end{equation}
We can then use the structure of \eqref{Eqn:pr R L} to obtain the desired uniqueness for the solution to \eqref{MOTS main equation}.
\begin{proposition}\label{Prop:unique MOTS}
   For any fixed $\tub\ge F(v_1)$, suppose that $R_1, R_2\in C^2$ both solve equation \eqref{MOTS main equation}. Then it must hold $R_1=R_2$ at each point of $\ms$.
\end{proposition}
\begin{proof}
    According to \Cref{MOTS eqn C 2 alpha}, with $N=\tfepub$ we have
    \begin{equation}\label{Est for R1 R2}
        \|R_i-r_+\|_{C^{2, \alpha}} \lesssim \tfepub \qquad \text{for} \quad i=1, 2.
    \end{equation}
    We then consider the elliptic operator
    \begin{align*}
       \LL[R_1-R_2]=L(R_1)-L(R_2)=&\int_{0}^1 \pr_R L(tR_1+(1-t)R_2) dt[R_1-R_2] \\
        \coloneqq&A^{ij}(R_1, R_2)D_{ij}(R_1-R_2)+B^{i}(R_1, R_2)D_i (R_1-R_2)\\&+C(R_1, R_2)(R_1-R_2),
    \end{align*}
    where 
    \begin{align*}
        A^{ij}(R_1, R_2)=&\int_{0}^1 a^{ij}(\omega, R(t), DR(t))dt, \\
        B^{i}(R_1, R_2)=&\int_{0}^1 \l\pr_{p^i} a^{jk} D_{jk}R+\pr_{p^i} b\r(\omega, R(t), DR(t)) dt, \\
        C(R_1, R_2)=&\int_{0}^1 \l \pr_{\t{r}} a^{ij} D_{ij}R+\pr_{\t{r}} b+\pr_{\t{r}} c\c(R-r_+)+c\r(\omega, R(t), DR(t)) dt
    \end{align*}
    with $R(t)=tR_1+(1-t)R_2$ for $t\in[0, 1]$.

    In view of estimates for $a^{ij}(\omega, \t{r}, p), b^i(\omega, \t{r}, p), c(\omega, \t{r})$ in \Cref{MOTS main equation new prop}, together with \eqref{Est for R1 R2}, we infer that
    \begin{equation*}
        A^{ij}(R_1, R_2)\xi_i \xi_j\ge \nu|\xi|^2 \qquad \text{for all} \quad \xi\in\mathbb{R}^2
    \end{equation*}
    and if $\epsilon_0$ is small enough it also holds
    \begin{equation*}
        C(R_1, R_2)\le -c_0+O(\tfepub) \le -\f{c_0}{2}<0.
    \end{equation*}
    
    Applying the maximum principle on $\mathbb{S}^2$ for the elliptic equation $\LL[R_1-R_2]=0$, we then prove that $ R_1\equiv R_2$ on $\ms$ as stated.
\end{proof}

\section{Physical Properties of Apparent Horizon}\label{Sec:physical} 
In \Cref{Sec:Existence}, with $\tub\ge F(v_1)$ we prove that, along each incoming null hypersurface $\tHb_{\tub}$, there exists a unique MOTS $M_{\tub}=\{(\tub, \t{r}, \tth^1, \tth^2): \, \t{r}=R(\tub, \tth^1, \tth^2), (\tth^1, \tth^2)\in \mathbb{S}^2 \}$. Together with the apparent horizon constructed in the short-pulse and transition region $\qty{0\le \tub \le F(v_1)}$, collectively, these MOTSs form an apparent horizon $\AH\coloneqq\cup_{\tub>0} M_{\tub}$ emerging from the spacetime point $O=(u=0, v=0)$. This section is devoted to establishing the asymptotics and the physical properties of the apparent horizon $\AH$. We will prove that our solved apparent horizon $\AH$ must be smooth and (locally) achronal. This further enables us to derive the area increasing law (the second law of black hole thermodynamics) along $\AH$.

\subsection{Smoothness and Asymptotics of Apparent Horizon}
Note that in \Cref{Thm:MOTS exis} we have already shown that $R(\tub, \tth^1, \tth^2)$, the unique solution to \eqref{MOTS main equation new}, depends smoothly on $(\tth^1, \tth^2)\in \mathbb{S}^2$. We proceed to prove that $R(\tub, \tth^1, \tth^2)$ is also smooth in $\tub$. To this end, in analogous to Proposition 3.8 in \cite{An-He}, we consider the operator
\begin{equation*}
     L(R, \tub)=a^{ij}(\tub, \omega, R,  DR)D_{ij} R+b(\tub, \omega, R, DR)+c(\tub, \omega, R)(R-r_+).
\end{equation*}
Proceeding the same as in the proof of \Cref{Prop:unique MOTS}, by virtue of \eqref{Eqn:pr R L}, \eqref{Eqn:pr tub L}, the boundedness of $a^{ij}(\omega, \t{r}, p), b^i(\omega, \t{r}, p), c(\omega, \t{r})$ in \Cref{MOTS main equation new prop} and the estimates of $R$ in \Cref{Thm:MOTS exis}, we deduce that
\begin{equation}\label{Eqn:prL}
\begin{aligned}
     \pr_R L(R(\tub), \tub)[W]=&a^{ij} D_{ij}W+\l\pr_{p^i} a^{jk} D_{jk}R+\pr_{p^i} b\r \cdot D_i W+\l c+O(\tfepub)  \r W, \\
     |\pr_{\tub}L(R(\tub), \tub)|\lesssim& |\pr_{\tub} b(\tub, R(\tub), DR(\tub))|+\tfepub\lesssim \tfepub.
\end{aligned}
\end{equation}
Note that when the perturbation is small, i.e., $\epsilon_0\ll 1$, the zeroth-order coefficient of $\pr_R L(R(\tub), \tub)$ satisfies
\begin{equation*}
    c+O(\tfepub)\le -c_0+O(\tfepub)\le -\f{c_0}{2},
\end{equation*}
where $c_0$ is a positive constant. Hence, the elliptic operator $\pr_R L$ is invertible at $R=R(\tub)$. Since $L(R, \tub)$ is smooth in $\tub$ and $R$, by the implicit function theorem, we hence prove that $R=R(\tub, \tth^1, \tth^2)$ is smooth in $\tub, \tth^1, \tth^2$ with $\tub\ge F(v_1)$ and $(\tth^1, \tth^2)\in \mathbb{S}^2$.

Combining with the smoothness of the initial apparent horizon solved in $\qty{0\le\tub\le F(v_1)}$ and utilizing the uniqueness of MOTS along each $\tHb_{\tub}$,\footnote{Notice that the uniqueness of MOTS on $\tHb_{\tub}$ is independent of the choice of coordinates along $\tHb_{\tub}$.} we obtain a globally smooth apparent horizon $\AH=\cup_{\tub>0} M_{\tub}$.
\vspace{3mm}

We turn to prove $\AH$ is asymptotically null. As our current focus is on the late-time behavior of $\AH$, we restrict our attention to the region where $\tub\ge F(v_1)$. We start with deriving the estimate for $\pr_{\tub} R$. Differentiating $L(R(\tub), \tub)=0$ in $\tub$, it gives
\begin{equation*}
    \pr_R L(R(\tub), \tub)[\pr_{\tub} R]+\pr_{\tub}L(R(\tub), \tub)=0.
\end{equation*}
Employing the structures in \eqref{Eqn:prL} and applying maximum principle for $\partial_{\ub} R$ and $-\partial_{\ub} R$ respectively, we then deduce
\begin{equation}\label{Est for pr tub R}
    \max\limits_{\mathbb{S}^2} |\pr_{\tub} R(\tub, \tth^1, \tth^2)|\le 2c_0^{-1} \max\limits_{\mathbb{S}^2} |\pr_{\tub}L(R(\tub), \tub)|\lesssim \tfepub.
\end{equation}
To show the asymptotic degeneracy of induced metric on $\AH=\{\t{r}=R(\tub, \tth^1, \tth^2) \}$, we further consider its tangential vector 
\begin{equation}\label{Eqn tangent vec AH}
    X= e'_4-\f{e'_4(\t{r}-R)}{e'_3(\t{r}-R)}e'_3 ,
\end{equation}
where 
\begin{equation*}
    e_3'=\te_3, \quad e_a '=\te_a+f \te_a (R) \te_3\in TM_{\tub}, \quad e_4'=\te_4+2f\te^a(R) \te_a+f^2|\tnab R|^2 \te_3 \qquad \text{with} \quad f=[\te_3(\t{r})]^{-1}
\end{equation*}
form the null frame adapted to the MOTS $M_{\tub}$. We remark that $X\in T\AH$ is normal to all the MOTSs $M_{\tub}$.

Recalling the estimates \eqref{Est for te4 tub,r}, \eqref{Est for te3 r}, \eqref{Esf for ttha tX a b} in \Cref{Sec:Construct null cones}, we have
\begin{align*}
    \te_4(\tub)=2, \qquad |\te_4(\t{r})|\lesssim |\t{r}-r_+|+\tfepub, \qquad f=[\te_3(\t{r})]^{-1}\sim -1, \qquad |\te_4(\tth^a), \te_a(\tth^b)|\lesssim 1.
\end{align*}

Together with \Cref{Thm:MOTS exis}, by using the inequality \eqref{Est for pr tub R} and noting that $\pr_{\t{r}} R=0$, we then derive the below estimates for $\te_4(R)$ and $\te_a(R)$:
\begin{align*}
    \te_4(R)=&\te_4(\tub)\pr_{\tub}R+\te_4(\t{r})\pr_{\t{r}} R+\te_4(\tth^a)\pr_{\tth^a}R=O(\tfepub), \\
    \te_a(R)=&\te_a(\tth^b)\pr_{\tth^b} R=O(\tfepub).
\end{align*}
Inserting above into \eqref{Eqn tangent vec AH} and applying \Cref{Thm:MOTS exis} again,  we infer that
\begin{align*}
    X(\tub)=&e_4'(\tub)=\te_4(\tub)=2, \qquad e'_3(\t{r}-R)=\te_3(\t{r})\sim -1, \\
    e'_4(\t{r}-R)=&\te_4(\t{r}-R)-f|\tnab R|^2=
    O(1)(R-r_+)+O(\tfepub)=O(\tfepub).
\end{align*}
As a consequence, it implies
\begin{equation*}
    \bfg(X, X)=\f{4e'_4(\t{r}-R)}{e'_3(\t{r}-R)}=O(\tfepub)\to 0 \qquad \text{as} \quad \tub\to \infty.
\end{equation*}
With $\tub$ being the affine parameter of $X$ along $\AH$, this indicates that the vector $X$ represents an asymptotically degenerate tangential direction of $\AH$.  
\vspace{2mm}

Here we also demonstrate that establish that the our constructed apparent horizon $\AH$ eventually converges to the event horizon $\HH^+$. Denote $\Si_{\tub}$ as the intersection of the event horizon $\HH^+$ and the incoming null hypersurface $\tHb_{\tub}$. Let $\Si_{\tub}$ be associated with the coordinates $\{\t{r}=R_{\HH^+}(\tub, \tth^1, \tth^2)  \}$ along $\tHb_{\tub}$. In \cite{KS:main} Klainerman--Szeftel showed that the event horizon $\HH^+=\qty{\t{r}=R_{\HH^+}(\tub, \tth^1,\tth^2)}$ is close to $r_{+,\infty}$ and it satisfies
\begin{equation*}
     r_{+, \infty}(1 +\f{\sqrt{\epsilon_0}}{\tub^{1+\de_{dec}}})\ge R_{\HH^+}(\tub, \tth^1, \tth^2) \ge r_{+, \infty}(1-\f{\sqrt{\epsilon_0}}{\tub^{1+\de_{dec}}}) \qquad \text{for all} \quad \tub\ge F(v_1), \ (\tth^1, \tth^2) \in \mathbb{S}^2,
\end{equation*}
where $r_{+, \infty}=m_{\infty}+\sqrt{m_{\infty}^2-a_{\infty}^2}$.

Together with our obtained estimate for $R=R(\tub, \tth^1, \tth^2)$ in \Cref{Thm:MOTS exis}, we hence infer
\begin{equation*}
   |R(\tub, \tth^1, \tth^2)-R_{\HH^+}(\tub, \tth^1, \tth^2)|\lesssim
    \f{\sqrt{\epsilon_0}}{\tub^{1+\de_{dec}}}+\tfepub\lesssim \f{\sqrt{\epsilon_0}}{\tub^{1+\de_{dec}}}\to 0 \qquad \text{as} \quad \tub\to\infty.
\end{equation*}

To sum up, we hence prove
\begin{proposition}\label{Prop:AH smooth}
     The apparent horizon $\mathcal{AH}=\cup_{\tub>0} M_{\tub}$ solved in our setting emerges from a spacetime point $O$, is a globally smooth, asymptotically null and asymptotes to the event horizon $\HH^+$ as $\tub$ tends to $\infty$.
\end{proposition}
\subsection{Achronality of Apparent Horizon}\label{Subsec:AH achronal}
We further prove the achronality of $\AH$ in this subsection. Recall that in \cite{An-He} we provide a criteria, called the null comparison principle and it guarantees the local achronality of the apparent horizon. Its precise definition is as below:
\begin{definition}\label{comparison principle}
    Along each incoming null hypersurface $\tHb_{\tub}$,  the null comparison principle with respect to the MOTS $M_{\tub}=\{ \t{r}=R(\tub, \tth^1, \tth^2)\}$ says that for any smooth 2-surface $\Sigma'=\{r=R'(\tth^1, \tth^2) \}$ lying in $\tHb_{\tub}$ and being close to $M_{\tub}$, if its outgoing null expansion satisfies $\tr \chi'\le 0$ pointwise, then we have $R'(\tth^1, \tth^2) \le R(\tub, \tth^1, \tth^2)$ for all $(\tth^1, \tth^2)\in \mathbb{S}^2$ and if $\tr \chi'\ge 0$ pointwise then it yields $R'(\tth^1, \tth^2) \ge R(\tub, \tth^1, \tth^2)$ on $\mathbb{S}^2$.
\end{definition}
In \cite{An-He} we proved
\begin{theorem}[An-He \cite{An-He}]\label{AH piecewise spacelike or null}
	Along the apparent horizon $\mathcal{AH}=\cup_{\tub} M_{\tub}$, assuming that the null comparison principle holds true for the MOTS $M_{\tub}$ along $\tHb_{\tub}$, then when restricted to $M_{\tub}$, $\mathcal{AH}$ is either spacelike everywhere or (outgoing) null everywhere. That is, the tangent vector of $\AH$ that is normal to $M_{\tub}$ must be either spacelike everywhere or (outgoing) null everywhere on $M_{\tub}$.
\end{theorem}
For the later portion of our constructed apparent horizon $\AH\cap\qty{\tub\ge F(v_1)}=\{\t{r}=R(\tub, \tth^1, \tth^2), \ \tub\ge F(v_1) \}$, we proceed to verify the null comparison principle with respect to each MOTS $M_{\tub}$ along $\AH$.
\begin{proposition}\label{Prop:verify NCP1}
    For every $\tub\ge F(v_1)$, the null comparison principle holds true for each MOTS $M_{\tub}$ along $\AH$.
\end{proposition}
\begin{proof}
    We conduct in the same fashion as for the proof of \Cref{Prop:unique MOTS}. As in \Cref{fig:NCP figure}, along $\tHb_{\tub}$,  let $\Sigma'=\{r=R'(\tth^1, \tth^2) \}$ be a spacelike $2$-surface close to the MOTS $M_{\tub}$ so that $\|\tR-R\|_{C^2(\mathbb{S}^2)}\ll 1$. We denote $\tr \chi'|_{\Sigma'}$ as the outgoing null expansion with respect to $\Sigma'$ and assume that $\tr \chi'|_{\Sigma'} \leq  0$ pointwise.
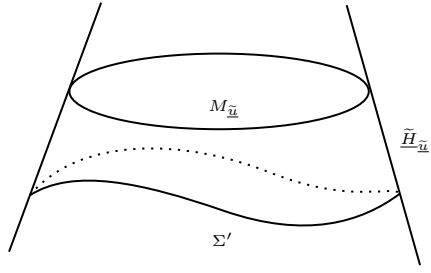
\begin{figure}[htp]
    \centering

\tikzset{every picture/.style={line width=0.75pt}} 

\begin{tikzpicture}[x=0.75pt,y=0.75pt,yscale=-1,xscale=1]

\draw   (230.56,89.18) .. controls (230.56,78.68) and (264.01,70.16) .. (305.28,70.16) .. controls (346.55,70.16) and (380,78.68) .. (380,89.18) .. controls (380,99.68) and (346.55,108.2) .. (305.28,108.2) .. controls (264.01,108.2) and (230.56,99.68) .. (230.56,89.18) -- cycle ;

\draw    (246.4,44.56) -- (200.44,169.8) ;

\draw    (368.96,45.8) -- (405.6,176.56) ;

\draw    (210.91,141.37) .. controls (232.65,127.6) and (266.57,134.97) .. (307.71,149.26) .. controls (348.86,163.54) and (376.3,155.03) .. (395.43,140.69) ;

\draw  [dash pattern={on 0.84pt off 2.51pt}]  (210.91,141.37) .. controls (235.77,111.48) and (288,112.97) .. (330.29,129.83) .. controls (372.57,146.69) and (392.53,135.76) .. (395.43,140.69) ;

\draw (298.45,91.86) node [anchor=north west][inner sep=0.75pt]  [font=\tiny]  {$M_{\tub}$};

\draw (300.34,158.56) node [anchor=north west][inner sep=0.75pt]  [font=\tiny]  {$\Si'$};

\draw (394.27,106.49) node [anchor=north west][inner sep=0.75pt]  [font=\tiny]  {$\tHb_{\tub}$};

\end{tikzpicture}

    \caption{Verification of Null Comparison Principle for MOTS $M_{\tub}$ along $\tHb_{\tub}$}
    \label{fig:NCP figure}
\end{figure}

 From \eqref{MOTS main equation} and the proof of \Cref{MOTS main equation new prop}, it holds
    \begin{equation*}
        (\f{1}{2} f^{-1} \trch')|_{\tS_{\tub, R'}}=a^{ij}(\omega, R',  DR')D_{ij} R'+b(\omega, R', DR')+c(\omega, R')(R'-r_+)=L(R', \tub).
    \end{equation*}
    Since $f<0$, we deduce 
    \begin{equation*}
        L(R', \tub)\ge L(R, \tub)=0,
    \end{equation*}
    which implies
   \begin{align*}
       0\le \LL[R'-R]=&L(R')-L(R)=\int_{0}^1 \pr_R L(tR_1+(1-t)R_2) dt[R'-R] \\
        =&A^{ij}(R', R)D_{ij}(R'-R)+B^{i}(R', R)D_i (R'-R)+C(R', R)(R'-R).
    \end{align*}
    Here the coefficients of $\LL$ are given by 
    \begin{align*}
        A^{ij}(R', R)=&\int_{0}^1 a^{ij}(\omega, R(t), DR(t))dt, \\
        B^{i}(R', R)=&\int_{0}^1 \l\pr_{p^i} a^{jk} D_{jk}R(t)+\pr_{p^i} b\r(\omega, R(t), DR(t)) dt, \\
        C(R', R)=&\int_{0}^1 \l \pr_{\t{r}} a^{ij} D_{ij}R(t)+\pr_{\t{r}} b+\pr_{\t{r}} c\c(R(t)-r_+)+c  \r(\omega, R(t), DR(t)) dt
    \end{align*}
    with $R(t)=tR'+(1-t)R$ for $t\in[0, 1]$.

    In view of \Cref{MOTS eqn C infty} and the fact that $\|\tR-R\|_{C^2(\mathbb{S}^2)}\ll 1$, we get
    \begin{equation*}
        \|R(t)-r_+\|_{C^{2}} \ll 1 \qquad \text{for any} \quad t\in[0, 1].
    \end{equation*}
    Combining with estimates of $a^{ij}(\omega, \t{r}, p), b^i(\omega, \t{r}, p), c(\omega, \t{r})$ in \Cref{MOTS main equation new prop}, there then hold
    \begin{align*}
        &A^{ij}(R', R)\xi_i \xi_j\ge \nu|\xi|^2 \qquad \text{for all} \quad \xi\in\mathbb{R}^2, \\
        &C(R', R)\le -c_0+O(\tfepub) \le -\f{c_0}{2}<0.
    \end{align*}
    Thus, after applying the maximum principle on $\mathbb{S}^2$, we conclude
    \begin{equation*}
        R'(\tth^1, \tth^2)\le R(\tub, \tth^1, \tth^2) \qquad \text{for all} \quad (\tth^1, \tth^2)\in \mathbb{S}^2.
    \end{equation*}
    
    The case $\tr \chi'|_{\Sigma'}\geq  0$ can be dealt with in the same manner. For both cases, we have hence verified the null comparison principle. This completes the proof of \Cref{Prop:verify NCP1}.
\end{proof}
Proceeding similarly, we can also validate the null comparison principle for the MOTS along the initial part of the apparent horizon $\AH\cap\{ 0<\tub\le F(v_1)\}$. Employing \Cref{AH piecewise spacelike or null}, we then obtain
\begin{proposition}\label{Prop:verify NCP}
    The entire apparent horizon $\AH=\cup_{\tub>0} M_{\tub}$ is locally achronal and it is either piecewise spacelike or piecewise null.
\end{proposition}

\subsection{Area Increasing Law of MOTS}\label{Subsec:area law}
 Define $A_M(\tub)$ to be the area of the MOTS $M_{\tub}$. Once $\AH$ has been proved to be either piecewise spacelike or piecewise (outgoing) null as in \Cref{Prop:verify NCP}, a direct utilization of Proposition 5.4 in \cite{An-He} gives
 \begin{proposition}\label{Prop:area law}
    Along $\AH=\cup_{\tub>0} M_{\tub}$, the area of the MOTS $A_M(\tub)$ is non-decreasing as a function of $\tub$ for all $\ub>0$. More specifically, along $\AH$ we have 
\begin{equation*}
    \f{dA_M}{d\tub}=0 \, \, \text{in the null piece}  \quad \text{and} \quad \f{dA_M}{d\tub}>0 \,\, \text{in the spacelike piece}.
\end{equation*}
\end{proposition}

\section{Penrose Inequality in Perturbed Kerr Regime}\label{Sec:penrose ineq}
\subsection{Dynamical Penrose Inequality and Spacetime Penrose Inequality}
In this section, we demonstrate a new approach to connect the area of the MOTS $M_{\tub}$ to the final mass and final angular momentum at timelike infinity. This approach was proposed by Penrose in 1973 and is rigorously realized in this paper. In particular, in this section, in perturbed Kerr regime we prove the desired Penrose inequality without requiring the time symmetric assumption.

To achieve our goal, we start with the precise calculation for the area of the 2-sphere $\tS_{\tub, r_{+, \infty}}$ in our incoming geodesic foliation $(\tub, \t{r}, \tth^1, \tth^2)$ constructed in \Cref{Sec:Construct null cones}. Here $r_{+, \infty}\coloneqq m_{\infty}+\sqrt{m^2_{\infty}-a_{\infty}^2}$ with $m_{\infty}$ and $a_{\infty}$ being the final mass and the final angular momentum of the perturbed Kerr spacetime.
\begin{lemma}\label{Lem:area of tS}
    The area of $\tS_{\tub, r_{+, \infty}}$ obeys the following asymptotic estimate:
    \begin{equation*}
        \textrm{Area}\,(\tS_{\tub, r_{+, \infty}})=4\pi(r_{+, \infty}^2+a_{\infty}^2)+O(\sqrt{\tfepub}).
    \end{equation*}
\end{lemma}
\begin{proof}
    For notational simplicity, we use $r_+, m, a$ to represent $r_{+,\infty}, m_{\infty}, a_{\infty}$ throughout this proof if there is no danger of confusion. In $(\ub, r, \th, \varphi)$ coordinates adapted to the PG structure within $\intM$, we have
    \begin{equation*}
        e_{\mu}=e_{\mu}(\ub)\pr_{\ub}+e_{\mu}(r) \pr_r+e_{\mu}(\th)\pr_{\th}+e_{\mu}(\varphi)\pr_{\varphi} \qquad \text{with} \quad \mu=1, 2, 3, 4.
    \end{equation*}
    By virtue of \eqref{In PG cond} and \eqref{Est for coor derivative}, we can express the corresponding coordinate derivatives via using the null frame $\{e_1, e_2, e_3, e_4 \}$ as follows
   \begin{equation}
    \label{Eqn coor derv}
    \begin{aligned}
        \pr_{\ub}=&\f12\l e_4+\f{\De}{|q|^2}e_3-\f{2a\sin \th}{|q|}e_2 \r+\lot, \qquad \pr_{r}=-e_3, \\
    \pr_{\th}=&|q|e_1+\lot, \qquad
    \pr_{\varphi}=-\f{a^2\sin^2 \th}{2}\l  e_4+\f{\De}{|q|^2}e_3-\f{2(r^2+a^2)}{a|q|\sin \th}e_2 \r+\lot,
    \end{aligned}
    \end{equation}
    where $q=r+ia\cos \th$, $\De=r^2+a^2-2mr$ and $\lot$ denotes expressions of the terms satisfying
    \begin{equation*}
        \lot=O(\fepub)\c(e_1, e_2, e_3, e_4).
    \end{equation*}
    
    Next we proceed to write $\tS_{\tub, r_{+}}$ as a graph over $(\th, \varphi)\in\mathbb{S}^2$ in $(\ub, r, \th, \varphi)$ coordinates. For any fixed $\tub$, by \eqref{Est for Dtub} and \eqref{Eqn coor derv}, we have
    \begin{align*}
        \pr_{\ub} \tub=&\f12\l e_4(\tub)+\f{\De}{|q|^2}e_3(\tub)-\f{2a\sin \th}{|q|}e_2(\tub) \r+O(\tfepub) \\
        =&\f12\l \f{r^2+a^2+\Sis}{|q|^2}+ \f{\De}{|q|^2}\c\f{r^2+a^2-\Sis}{\De}-\f{2a\sin \th}{|q|}\c \f{a\sin \th}{|q|} \r+O(\tfepub)\\
       =&1+O(\tfepub), 
       \end{align*}
       \begin{align*}
        \pr_{\th} \tub=&|q|e_1(\tub)+O(\fepub)=|a|\bs+O(\fepub),\\
        \pr_{\varphi} \tub=&-\f{a^2\sin^2 \th}{2}\l  e_4(\tub)+\f{\De}{|q|^2}e_3(\tub)-\f{2(r^2+a^2)}{a|q|\sin \th}e_2(\tub) \r+O(\tfepub)\\
        =&-\f{a^2\sin^2 \th}{2}\l \f{r^2+a^2+\Sis}{|q|^2}+ \f{\De}{|q|^2}\c\f{r^2+a^2-\Sis}{\De}- \f{2(r^2+a^2)}{a|q|\sin \th}\c \f{a\sin \th}{|q|}\r+O(\tfepub) \\
        =&O(\tfepub).
    \end{align*}
    Here $\ths=\ths(\rthe)\in [\th, \pi]$ is defined in \Cref{Apx:existence of ths} as in \Cref{Appendix:coordinate in Mint} and
    \begin{equation*}
        \Sis=\sqrt{(r^2+a^2)^2-a^2\sin^2 \ths \De}, \qquad \bs=\sgn(r-r_+)\sqrt{\sin^2 \ths-\sin^2 \th}.
    \end{equation*}
    
    Applying the implicit function theorem for $\tub=\tub(\ub, r, \th, \varphi)$ with respect to the parameter $\ub$, we can show that there exists a function $\Ub=\Ub\,(\tub, r, \th, \varphi)$ such that
    \begin{equation*}
        \ub=\Ub\,(\tub, r, \th, \varphi).
    \end{equation*}
   Differentiating the identity
    \begin{equation*}
        \tub=\tub(\Ub\,(\tub, r, \th, \varphi), r, \th, \varphi)
    \end{equation*}
    with respect to $\th, \varphi$, we derive
    \begin{equation}\label{Est:der Ub}
    \begin{aligned}
       \pr_{\th}\, \Ub=&-(\pr_{\ub} \tub)^{-1}\pr_{\th} \tub=-|a|\bs+O(\tfepub), \\
          \pr_{\varphi}\, \Ub=&-(\pr_{\ub} \tub)^{-1}\pr_{\varphi} \tub=O(\tfepub).
    \end{aligned}
       \end{equation}
    
    We then regard $\tS_{\tub, r_{+}}=\{(\ub=\Ub\,(\tub, r_+, \th, \varphi), r=r_+, \th, \varphi) \}$\footnote{Note that $\t{r}=r_+\Longleftrightarrow r=r_+$ according to the definition of $\t{r}$ in \eqref{Def:t r}.} as a graph over $(\th, \varphi)\in \mathbb{S}^2$ and its induced metric $\t{g}$ in $(\th, \varphi)$ coordinates takes the form of
    \begin{align*}
        \t{g}_{\th \th}=&\bfg(\pr_{\th}\, \Ub \, \pr_{\ub}+\pr_{\th}, \pr_{\th}\, \Ub \, \pr_{\ub}+\pr_{\th})\\
        =&\f{a^2(a^2\sin^2\th-\De)\bs^2}{|q|^2}+|q|^2+O(\tfepub),\\
        \t{g}_{\th \varphi}=&\bfg(\pr_{\th}\, \Ub \, \pr_{\ub}+\pr_{\th}, \pr_{\varphi}\, \Ub \, \pr_{\ub}+\pr_{\varphi}) \\
        =&\f{2mra|a|\sin^2 \th\bs}{|q|^2}+O(\tfepub),\\
        \t{g}_{\varphi \varphi}=&\bfg(\pr_{\varphi}\, \Ub \, \pr_{\ub}+\pr_{\varphi}, \pr_{\varphi}\, \Ub \, \pr_{\ub}+\pr_{\varphi}) \\
        =&\f{\sin^2 \th}{|q|^2}\l (r^2+a^2)-a^2\sin^2 \th \De \r+O(\tfepub)=\f{\sin^2 \th \Si^2}{|q|^2}+O(\tfepub),
    \end{align*}
    where we utilize \eqref{Est:kerrmetric}, \eqref{Est:der Ub} and $\Si=\sqrt{(r^2+a^2)-a^2\sin^2 \th \De}$.

Therefore, the determinant of $\t{g}$ reads
\begin{align*}
    \det (\t{g})=\t{g}_{\th \th} \t{g}_{\varphi \varphi}-( \t{g}_{\th \varphi})^2
    =\l \Si^2-|a|^2(\sin^2\ths-\sin^2\th)\De \r \sin^2\th 
    =(\Sis \sin \th)^2+O(\fepub).
\end{align*}
With this, we now compute the area of $\tS_{\tub, r_{+}}$ and get
\begin{align*}
    \text{Area}\,(\tS_{\tub, r_{+}})=&\int_{0}^{\pi}\int_{0}^{2\pi} \sqrt{ \det (\t{g})} d\th d\varphi \\
    =&\int_{0}^{\pi}\int_{0}^{2\pi} \sqrt{(r^2+a^2)-a^2\sin^2 \ths \De}\Big|_{r=r_{+}} \sin \th d\th d\varphi+O(\sqrt{\tfepub}) \\
    =&\int_{0}^{\pi}\int_{0}^{2\pi} (r_{+, \infty}^2+a^2) \sin \th d\th d\varphi+O(\sqrt{\tfepub}) \\
    =&4\pi(r_{+}^2+a^2)+O(\sqrt{\tfepub}).
\end{align*}
Here we use the fact that $\De\big|_{r_{+}}=0$.
\end{proof}
In view of \Cref{MOTS eqn C infty}, we have accurate control of the location and the shape of the MOTS $M_{\tub}$ and it is sufficiently close to the 2-sphere $\tS_{\tub, r_{+, \infty}}$ with deviations decaying in $\tub$. This enables us to approximate the area of the MOTS $M_{\tub}$ by the area of $\tS_{\tub, r_{+, \infty}}$. 
\begin{lemma}\label{Lem:compare MOTS tS}
    For the area of the MOTS $M_{\tub}$, we have
    \begin{equation*}
        A_M(\tub)=4\pi(r_{+, \infty}^2+a_{\infty}^2)+O(\sqrt{\tfepub}).
    \end{equation*}
\end{lemma}
\begin{proof}
    According to \Cref{Lem:area of tS}, it is sufficient to show that
    \begin{equation}\label{Ineq:compare}
        \left|A_M(\tub)-\text{Area}\,(\tS_{\tub, r_{+, \infty}})\right|\lesssim  \tfepub.
    \end{equation}
    Notice that in $(\tub, \t{r}, \tth^1, \tth^2)$ coordinates, the induced metric on the MOTS $M_{\tub}=\{(\tub, \t{r}, \tth^1, \tth^2): \ \t{r}=R(\tub, \tth^1, \tth^2) \}$ takes the form of
    \begin{equation*}
        g'_{\tth^a \tth^b}=\bfg(\pr_{\tth^a}+\pr_{\tth^a} R \, \pr_{\t{r}}, \pr_{\tth^b}+\pr_{\tth^b} R\, \pr_{\t{r}})=\bfg_{\tth^a \tth^b}\eqqcolon \t{g}_{\tth^a \tth^b}.
    \end{equation*}
    Here we use the fact that $\pr_{\t{r}}=f\te_3 $ and $\pr_{\tth^a}\in T\tS_{\tub, r_{+,\infty}}=\text{span}\{\te_1, \te_2 \}$.

    Consequently, we then compute the area of $M_{\tub}$ as below:
    \begin{align*}
        A_M(\tub)=&\int_{\mathbb{S}^2} \sqrt{\det(g')}(\tub, R(\tub, \tth^1, \tth^2), \tth^1, \tth^2) d\tth^1 d\th^2\\
        =&\int_{\mathbb{S}^2} \sqrt{\det(\t{g})}(\tub, R(\tub, \tth^1, \tth^2), \tth^1, \tth^2) d\tth^1 d\tth^2.
    \end{align*}
    Combining with
    \begin{equation*}
         \text{Area}\,(\tS_{\tub, r_{+, \infty}})=\int_{\mathbb{S}^2} \sqrt{\det(\t{g})}(\tub, r_{+, \infty}, \tth^1, \tth^2) d\tth^1 d\tth^2,
    \end{equation*}
    from the mean value theorem and the a priori estimates of $R=R(\tub, \tth^1, \tth^2)$ provided in \Cref{MOTS eqn C 2 alpha}, we infer that
    \begin{align}\label{Est:AM-AS}
        \left| A_M(\tub)-\text{Area}\,(\tS_{\tub, r_{+, \infty}})\right|\lesssim \sup\limits_{r, \tth^1, \tth^2} \left|\pr_{\t{r}} \l\sqrt{\det(\t{g})}\r\right|\sup\limits_{\tth^1, \tth^2} |R-r_{+, \infty}|\lesssim \sup\limits_{r, \tth^1, \tth^2} \left|\pr_{\t{r}} \l\sqrt{\det(\t{g})}\r\right| \tfepub.
    \end{align}
    
    To control $\pr_{\t{r}} (\sqrt{\det(\t{g})})$, we appeal to the first variational formula
    \begin{equation*}
        \LL_{\te_3} \t{g}=2\t{\chib}.
    \end{equation*}
    In $(\tth^a, \tth^b)$ coordinates, it reads
    \begin{equation*}
        \pr_{\t{r}}(\t{g}_{\tth^a \tth^b})=2f\t{\chib}(\pr_{\tth^a}, \pr_{\tth^b})\eqqcolon 2f \t{\chib}_{\tth^a \tth^b}.
    \end{equation*}
    By taking the trace, then it yields
    \begin{equation*}
        \pr_{\t{r}}\Big(\det(\t{g})\Big)=2f\tr\t{\chib} \det(\t{g}).
    \end{equation*}
    In view of \eqref{Est for te3 r}, \eqref{Esf for ttha tX a b} and \Cref{Lem:hyper est in incom geo frame}, the terms on the right obey the bounds
    \begin{align*}
       | f, \tr\t{\chib}|\lesssim 1, \qquad
       |\det(\t{g})|=|\det(\te_a(\tth^b))|^2\lesssim 1.
    \end{align*}
    Therefore, for $\pr_{\t{r}} (\sqrt{\det(\t{g})})$ we derive
    \begin{equation*}
        \left|\pr_{\t{r}} \l\sqrt{\det(\t{g})}\r\right|=|\sqrt{\det(\t{g})}\c f\tr\t{\chib}|\lesssim 1.
    \end{equation*}
    Back to \eqref{Est:AM-AS}, the desired inequality \eqref{Ineq:compare} thus follows.
\end{proof}
Letting $\tub\to\infty$, we hence conclude from \Cref{Lem:compare MOTS tS} that
\begin{proposition}\label{Prop:limit area MOTS}
    At the timelike infinity, the limit of the area of the MOTS $M_{\tub}$ exists and it obeys
    \begin{equation*}
        \lim\limits_{\tub \to \infty} A_M(\tub)=4\pi (r_{+, \infty}^2+a_{\infty}^2)=8\pi \minft \qty(\minft+\sqrt{\minft^2-\ainft^2}).
    \end{equation*}
\end{proposition}
 Now, we are ready to prove the dynamical Penrose inequality in the perturbative Kerr regime as demonstrated in \Cref{fig:spacetime penrose ineq}. 
\begin{theorem}[Dynamical Version]\label{Thm:Penrose ineq 1}
  In the perturbative regime of all subextremal Kerr black holes, if the corresponding stability estimates as in \cite{KS:main} hold, then the area $A_M(\tub)$ of each MOTS $M_{\tub}$ along the apparent horizon $\AH$ obeys the inequality
    \begin{equation*}
        M_B(\tu)\ge m_{\infty}\ge \sqrt{\f{\minft \qty(\minft+\sqrt{\minft^2-\ainft^2})}{2}} \ge \sqrt{\f{A_M(\tub)}{16\pi}}.
    \end{equation*}
    Here $a_{\infty}$ and $m_{\infty}$ are final angular momentum and final mass of the perturbed Kerr spacetime satisfying $|a_{\infty}/m_{\infty}|< 1$, $r_{+,\infty}=m_{\infty}+\sqrt{m_{\infty}^2-a_{\infty}^2}$ and $M_B(\tu)$ is the Bondi mass along $\II^+$ with $\tu$ being an outgoing optical function.
\end{theorem}
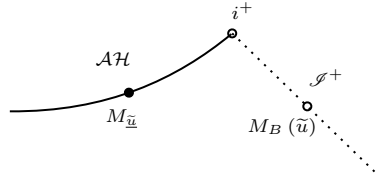
\begin{figure}[htp]
    \centering
\tikzset{every picture/.style={line width=0.75pt}} %set default line width to 0.75pt        

\begin{tikzpicture}[x=0.75pt,y=0.75pt,yscale=-1,xscale=1]
%uncomment if require: \path (0,415); %set diagram left start at 0, and has height of 415

%Straight Lines [id:da04148539083938896] 
\draw [color={rgb, 255:red, 0; green, 0; blue, 0 }  ,draw opacity=1 ] [dash pattern={on 0.84pt off 2.51pt}]  (372.93,14.81) -- (447.93,88.21) ;
\draw [shift={(410.43,51.51)}, rotate = 44.38] [color={rgb, 255:red, 0; green, 0; blue, 0 }  ,draw opacity=1 ][line width=0.75]      (0, 0) circle [x radius= 2.01, y radius= 2.01]   ;
%Curve Lines [id:da3789953131295283] 
\draw [color={rgb, 255:red, 0; green, 0; blue, 0 }  ,draw opacity=1 ]   (261.93,54.01) .. controls (310.56,54.4) and (345.25,37.77) .. (372.52,15.15) ;
\draw [shift={(372.93,14.81)}, rotate = 320.18] [color={rgb, 255:red, 0; green, 0; blue, 0 }  ,draw opacity=1 ][line width=0.75]      (0, 0) circle [x radius= 2.01, y radius= 2.01]   ;
\draw [shift={(321.38,44.62)}, rotate = 339.53] [color={rgb, 255:red, 0; green, 0; blue, 0 }  ,draw opacity=1 ][fill={rgb, 255:red, 0; green, 0; blue, 0 }  ,fill opacity=1 ][line width=0.75]      (0, 0) circle [x radius= 2.01, y radius= 2.01]   ;

% Text Node
\draw (371.16,-2.64) node [anchor=north west][inner sep=0.75pt]  [font=\scriptsize]  {$i^{+}$};
% Text Node
\draw (409.56,31.49) node [anchor=north west][inner sep=0.75pt]  [font=\scriptsize]  {$\mathscr{I}^{+}$};
% Text Node
\draw (303.13,22.59) node [anchor=north west][inner sep=0.75pt]  [font=\scriptsize]  {$\mathcal{AH}$};
% Text Node
\draw (308.13,51.4) node [anchor=north west][inner sep=0.75pt]  [font=\scriptsize]  {$M_{\widetilde{\underline{u}}}$};
% Text Node
\draw (379.13,55.73) node [anchor=north west][inner sep=0.75pt]  [font=\scriptsize]  {$M_{B}\left(\tu\right)$};

\end{tikzpicture}
    \caption{Dynamical Penrose Inequality}
    \label{fig:spacetime penrose ineq}
\end{figure}
\begin{proof}
     By employing the monotonicity of $A_M(\tub)$ proved in \Cref{Prop:area law} and applying \Cref{Prop:limit area MOTS}, we derive
    \begin{equation*}
        \sqrt{\f{A_M(\tub)}{16\pi}}\le \sqrt{\f{4\pi (r_{+, \infty}^2+a_{\infty}^2)}{16\pi}}=\f{\sqrt{r_{+, \infty}^2+a_{\infty}^2}}{2}.
    \end{equation*}
    Note that here we have the below explicit algebraic calculation:
    \begin{equation}\label{Basic ineq 1}
        r_{+, \infty}^2+a_{\infty}^2=(\minft+\sqrt{\minft^2-\ainft^2})^2+\ainft^2=2\minft^2+2\minft\sqrt{\minft^2-\ainft^2}\le 4\minft^2.
    \end{equation}
    This yields
    \begin{equation*}
        m_{\infty}\ge \sqrt{\f{\minft \qty(\minft+\sqrt{\minft^2-\ainft^2})}{2}}=\f{\sqrt{r_{+, \infty}^2+a_{\infty}^2}}{2}.
    \end{equation*}
     
     Meanwhile, in view of \Cref{Prop:Bondi}, along $\II^+$ we have the Bondi mass loss formula and it gives
    \begin{equation*}
        M_B(\tu)\ge m_{\infty}.
    \end{equation*}
Combining these inequalities together, we complete the proof of this theorem.
\end{proof}
\begin{remark}
    This dynamical version of Penrose inequality also holds for our constructed Kerr black-hole formation spacetimes. For all $\tub>0$, our apparent horizon is always achronal and by \Cref{Prop:area law} the area of MOTS is non-decreasing towards the timelike infinity. Note that $\tub=0$ corresponds to the center of gravitational collapse.
\end{remark}
We then move on to establish the spacetime Penrose inequality without imposing time symmetric conditions.  Let $(\Si, g, k)$ be an initial data set, that consists of an asymptotically flat 3-Riemannian manifold $(\Si, g)$ and a symmetric 2-tensor $k$ on $\Si$ satisfying the constraint equations
    \begin{align*}
        R(g)+(\tr_g k)^2-|k|_g^2=&0, \\
        \text{div}_g \, k-d\tr_g k=&0.
    \end{align*}
    Here $R(g)$ is the scalar curvature of $g$. With Kerr metric $\bfg_{a,m}$, we denote $(\Si, g_{a, m}, k_{a, m})$ to be an associated initial data set.
    
    With these notations, we can formulate the spacetime Penrose inequality in perturbative Kerr regime as follows:
\begin{theorem}[Spacetime Version]\label{Thm:Penrose ineq 2}
    With $0\le |a|\ll m$, suppose there exist $N\in\mathbb{N}$ suitable large,\footnote{This coincides with the number of derivatives required in \cite{KS:main}.} $\epsilon_0>0$ sufficiently small and a coordinate system $(x_1, x_2, x_3)$ on $\Si$ outside a compact set $K\subset \Si$, such that\footnote{The existence of such initial data set is guaranteed by \cite{F-S-T}.} with $r=\sqrt{x_1^2+x_2^2+x_3^2}$ and $0<\de\ll 1$ it holds
    \begin{equation}\label{Cond:g-gam}
        |(r\pr)^{\le N}(g-g_{a, m})|\le \epsilon_0 (1+r)^{-\f52-\de}, \quad |(r\pr)^{\le N-1}(k-k_{a, m})|\le \epsilon_0 (1+r)^{-\f72-\de}.
    \end{equation}
Then  $(\Si, g)$ admits ADM mass $m$ and there exists a MOTS $M_0$ along $\Si$ located near $r=r_+\coloneqq m+\sqrt{m^2-a^2}$. Assume $M_0$ is close to the exact Kerr MOTS $r=r_+$ and denote the area of $M_0$ to be $A$. We then have  the following spacetime Penrose inequality
    \begin{equation}\label{Eqn:Penroseineq2}
        m\ge \sqrt{\f{A}{16\pi}}.
    \end{equation}
with equality if and only if the Gauss curvature of the MOTS $M_0$ is pointwise equal to $\f{1}{4m^2}$.
    
   Moreover, assuming the validity of nonlinear Kerr stability for the full subextremal range with $|a/m|<1$, we have both \eqref{Eqn:Penroseineq2} and the associated rigidity also hold in the perturbative regime of this full range.
\end{theorem}
\begin{proof}
 The proof of this theorem is divided into several steps.\footnote{The proof of rigidity is put in \Cref{Subsec:rigidity}.}
\vspace{2mm}
 
 \noindent\textbf{Step 1.} Given $(\Si, g, k)$, we first establish the global existence of solutions to EVEs \eqref{Intro:EVE eqn} arising from this initial data. The stated condition \eqref{Cond:g-gam} of $(\Si,g,k)$ enables the application of the nonlinear Kerr stability. Using  Klainerman--Szeftel's Kerr stability for small angular momentum \cite{KS:main} in region $III$, along with a standard local existence result in region $I$ and stability of Kerr in the external region $II$ by \cite{C-N,Shen:Kerr}, we obtain the hyperbolic future development $(\MM, \bfg)$ of initial data triplet $(\Si, g, k)$ and $\MM$ possesses a complete future null infinity $\II^+$. Furthermore, by \cite{KS:main} and \cite{C-N,Shen:Kerr} we also have that $\MM$ contains a domain of outer communication and it converges (globally) to a nearby Kerr solution $Kerr(a_{\infty}, m_{\infty})$. We draw \Cref{fig:future development ID} below for a schematic demonstration. Here $i^0$ and $i^+$ denote the spacelike infinity and the timelike infinity, $\mathcal{A}$ serves as a spacelike inner boundary of $\MM$ and the existence of regions $I,II,III$ are guaranteed by hyperbolic arguments.
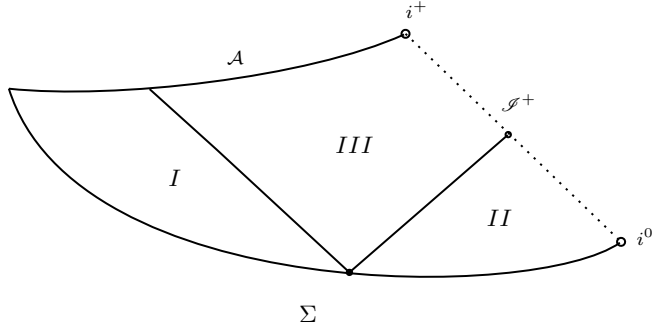
\begin{figure}[htp]
    \centering

\tikzset{every picture/.style={line width=0.75pt}} %set default line width to 0.75pt        

\begin{tikzpicture}[x=0.75pt,y=0.75pt,yscale=-1,xscale=1]

%Curve Lines [id:da599180705506545] 
\draw    (119.6,121.68) .. controls (153.43,229.94) and (382.89,228.73) .. (424.59,199.17) ;
\draw [shift={(425.2,198.72)}, rotate = 323.13] [color={rgb, 255:red, 0; green, 0; blue, 0 }  ][line width=0.75]      (0, 0) circle [x radius= 2.01, y radius= 2.01]   ;
%Straight Lines [id:da9249598285412924] 
\draw  [dash pattern={on 0.84pt off 2.51pt}]  (317.6,94.08) -- (425.2,198.72) ;
%Curve Lines [id:da16323498668352543] 
\draw    (119.6,121.68) .. controls (172.53,127.25) and (269.03,116.19) .. (316.88,94.41) ;
\draw [shift={(317.6,94.08)}, rotate = 335.19] [color={rgb, 255:red, 0; green, 0; blue, 0 }  ][line width=0.75]      (0, 0) circle [x radius= 2.01, y radius= 2.01]   ;
%Straight Lines [id:da2279706504090223] 
\draw    (189.87,121.9) -- (223.34,152.33) -- (289.6,214.08) ;
%Straight Lines [id:da03823923770601123] 
\draw    (289.6,214.08) -- (368.61,144.99) ;
\draw [shift={(368.87,144.77)}, rotate = 318.83] [color={rgb, 255:red, 0; green, 0; blue, 0 }  ][line width=0.75]      (0, 0) circle [x radius= 1.34, y radius= 1.34]   ;
\draw [shift={(289.6,214.08)}, rotate = 318.83] [color={rgb, 255:red, 0; green, 0; blue, 0 }  ][fill={rgb, 255:red, 0; green, 0; blue, 0 }  ][line width=0.75]      (0, 0) circle [x radius= 1.34, y radius= 1.34]   ;

% Text Node
\draw (316.16,76.36) node [anchor=north west][inner sep=0.75pt]  [font=\scriptsize]  {$i^{+}$};
% Text Node
\draw (431.36,190.36) node [anchor=north west][inner sep=0.75pt]  [font=\scriptsize]  {$i^{0}$};
% Text Node
\draw (363.76,122.76) node [anchor=north west][inner sep=0.75pt]  [font=\scriptsize]  {$\mathscr{I}^{+}$};
% Text Node
\draw (226.96,101.36) node [anchor=north west][inner sep=0.75pt]  [font=\scriptsize]  {$\mathcal{A}$};
% Text Node
\draw (263.09,229.56) node [anchor=north west][inner sep=0.75pt]  [font=\small]  {$\Sigma $};
% Text Node
\draw (197.76,160.76) node [anchor=north west][inner sep=0.75pt]  [font=\small]  {$I$};
% Text Node
\draw (281.09,144.43) node [anchor=north west][inner sep=0.75pt]  [font=\small]  {$III$};
% Text Node
\draw (356.76,181.09) node [anchor=north west][inner sep=0.75pt]  [font=\small]  {$II$};

\end{tikzpicture}

    \caption{Future Development of Initial Data Set $(\Si, g, k)$}
    \label{fig:future development ID}
\end{figure}

\noindent\textbf{Step 2.} We then consider the MOTS along $\Si$. Employing the barrier argument as in \cite{A-Met,E}, we infer that there exists a smooth MOTS $M_0$ on $\Si$ near $r=r_+$. As portrayed in \Cref{fig:Penroseineq}, now we can divide $\MM$ into two parts: the interior region $\intM$ and the exterior region $\Ext\MM$. We also set the timelike boundary $\TT$ to be far away from $\mathcal{A}$, so that $\Int\Si\coloneqq\Si\cap\intM$ contains $M_0$.
\vspace{2mm}

\noindent\textbf{Step 3.} Within $\intM$, we proceed to foliate the spacetime by constructing incoming null hypersurfaces in below steps. The construction here is similar to the process conducted in \Cref{Subsec:incoming geodesic foliation}.
\begin{itemize}
    \item In $\intM$ we first solve the eikonal equation
    \begin{equation*}
        \bfg^{\a \b}\pr_{\a} \tub  \pr_{\b} \tub=0.
    \end{equation*}
    Here we initialize the incoming optical function $\tub$ along $\Int\Si\cup \TT$ by requiring $\tub=0$ on the MOTS $M_0$.\footnote{Recall that in \Cref{Subsec:incoming geodesic foliation} we construct Pretorius--Israel type incoming null cones in perturbative Kerr spacetimes. Although the MOTS $M_0$ along $\Si$ may not lie in these null cones, we note that the eikonal equation is stable as illustrated in \Cref{Rmk: stable eikonal eqn} and that in our perturbative regime $M_0$ is assumed to be sufficiently close to the $2$-sphere $r=r_+$, which is the intersection of the Pretorius--Israel type incoming null cone and $\Si$. We can thus perturb these null cones a bit so that $\tub=0$ on $M_0$.}
    \item Set $\te_3=-\bfD \tub$ and denote $\tHb_{\tub}$ to be the level set of $\tub$. We now assign a coordinate system $(\t{r}, \tth^1, \tth^2)$ on $\tHb_{\tub}$ such that 
    \begin{equation*}
          \te_3(\t{r})\sim -1, \qquad \te_3(\tth^1)=\te_3(\tth^2)=0.
    \end{equation*}
    \item We define $\tS_{\tub, \t{r}}$ as the level set of $\t{r}$ along $\tHb_{\tub}$. Choose an orthonormal frame $\{\te_1, \te_2 \}$ of $\tS_{\tub, \t{r}}$ and let $\te_4$ be the unique outgoing null vector that is orthogonal to $\tS_{\tub, \t{r}}$ and satisfies $\bfg(\te_3, \te_4)=-2$. 
\end{itemize}
\vspace{2mm}

 \noindent\textbf{Step 4.} We then apply \Cref{Thm:MOTS exis}, \Cref{Prop:AH smooth}, \Cref{Prop:area law} and \Cref{Prop:limit area MOTS}. These results together give rise to a smooth apparent horizon $\AH$ composed of MOTSs $M_{\tub}$ with $\tub\ge0$ and the area of MOTS $M_{\tub}$, denoted by $A_M(\tub)$, is non-decreasing in $\tub$ and converging to $4\pi(r_{+,\infty}^2+a_{\infty}^2)$. In particular, for the area $A$ of the initial MOTS $M_0$, we obtain
\begin{equation*}
    4\pi(r_{+,\infty}^2+a_{\infty}^2)\ge A_M(0)=A.
\end{equation*}

\begin{figure}[ht]
    \centering
\tikzset{every picture/.style={line width=0.75pt}} %set default line width to 0.75pt        

\begin{tikzpicture}[x=0.75pt,y=0.75pt,yscale=-1,xscale=1]

%Curve Lines [id:da599180705506545] 
\draw    (119.6,121.68) .. controls (153.43,229.94) and (382.89,228.73) .. (424.59,199.17) ;
\draw [shift={(425.2,198.72)}, rotate = 323.13] [color={rgb, 255:red, 0; green, 0; blue, 0 }  ][line width=0.75]      (0, 0) circle [x radius= 2.01, y radius= 2.01]   ;
%Straight Lines [id:da5569341305376619] 
\draw    (214,200.08) -- (317.6,94.08) ;
%Straight Lines [id:da9249598285412924] 
\draw  [dash pattern={on 0.84pt off 2.51pt}]  (317.6,94.08) -- (425.2,198.72) ;
%Curve Lines [id:da16323498668352543] 
\draw    (119.6,121.68) .. controls (172.53,127.25) and (269.03,116.19) .. (316.88,94.41) ;
\draw [shift={(317.6,94.08)}, rotate = 335.19] [color={rgb, 255:red, 0; green, 0; blue, 0 }  ][line width=0.75]      (0, 0) circle [x radius= 2.01, y radius= 2.01]   ;
%Curve Lines [id:da8665236965398172] 
\draw    (162.8,175.28) .. controls (208.4,164.88) and (277.6,124.08) .. (317.6,94.08) ;
\draw [shift={(162.8,175.28)}, rotate = 347.15] [color={rgb, 255:red, 0; green, 0; blue, 0 }  ][fill={rgb, 255:red, 0; green, 0; blue, 0 }  ][line width=0.75]      (0, 0) circle [x radius= 2.68, y radius= 2.68]   ;
%Straight Lines [id:da2514986180680092] 
\draw  [dash pattern={on 0.84pt off 2.51pt}]  (129.14,142.54) -- (162.8,175.28) ;
%Straight Lines [id:da2279706504090223] 
\draw  [dash pattern={on 0.84pt off 2.51pt}]  (190.42,121.66) -- (223.34,152.33) -- (289.6,214.08) ;
%Straight Lines [id:da03823923770601123] 
\draw  [dash pattern={on 0.84pt off 2.51pt}]  (331.6,216.88) -- (383.37,159.13) ;
\draw [shift={(383.6,158.88)}, rotate = 311.88] [color={rgb, 255:red, 0; green, 0; blue, 0 }  ][line width=0.75]      (0, 0) circle [x radius= 1.34, y radius= 1.34]   ;
\draw [shift={(331.6,216.88)}, rotate = 311.88] [color={rgb, 255:red, 0; green, 0; blue, 0 }  ][fill={rgb, 255:red, 0; green, 0; blue, 0 }  ][line width=0.75]      (0, 0) circle [x radius= 1.34, y radius= 1.34]   ;
%Straight Lines [id:da8594658011416159] 
\draw    (317.6,94.08) -- (289.6,214.08) ;
%Straight Lines [id:da27143287102420766] 
\draw  [dash pattern={on 0.84pt off 2.51pt}]  (190.42,121.66) -- (223.34,152.33) ;
\draw [shift={(223.34,152.33)}, rotate = 42.98] [color={rgb, 255:red, 0; green, 0; blue, 0 }  ][fill={rgb, 255:red, 0; green, 0; blue, 0 }  ][line width=0.75]      (0, 0) circle [x radius= 1.34, y radius= 1.34]   ;

% Text Node
\draw (316.16,76.36) node [anchor=north west][inner sep=0.75pt]  [font=\scriptsize]  {$i^{+}$};
% Text Node
\draw (431.36,190.36) node [anchor=north west][inner sep=0.75pt]  [font=\scriptsize]  {$i^{0}$};
% Text Node
\draw (363.76,122.76) node [anchor=north west][inner sep=0.75pt]  [font=\scriptsize]  {$\mathscr{I}^{+}$};
% Text Node
\draw (226.96,101.36) node [anchor=north west][inner sep=0.75pt]  [font=\scriptsize]  {$\mathcal{A}$};
% Text Node
\draw (287.53,123.02) node [anchor=north west][inner sep=0.75pt]  [font=\scriptsize]  {$\mathcal{H}^{+}$};
% Text Node
\draw (230.56,124.96) node [anchor=north west][inner sep=0.75pt]  [font=\scriptsize]  {$\mathcal{AH}$};
% Text Node
\draw (140.16,176.56) node [anchor=north west][inner sep=0.75pt]  [font=\scriptsize]  {$M_{0}$};
% Text Node
\draw (210.16,158.56) node [anchor=north west][inner sep=0.75pt]  [font=\scriptsize]  {$M_{\widetilde{\underline{u}}}$};
% Text Node
\draw (388.16,146.16) node [anchor=north west][inner sep=0.75pt]  [font=\scriptsize]  {$M_{B}\left(\tilde{u}\right)$};
% Text Node
\draw (324.16,220.96) node [anchor=north west][inner sep=0.75pt]  [font=\scriptsize]  {$S_{\tilde{u} ,-\tilde{u}}$};
% Text Node
\draw (311.13,126.67) node [anchor=north west][inner sep=0.75pt]  [font=\scriptsize]  {$\mathcal{T}$};
% Text Node
\draw (235.76,216.56) node [anchor=north west][inner sep=0.75pt]  [font=\small]  {$\Sigma $};
% Text Node
\draw (260.85,156.1) node [anchor=north west][inner sep=0.75pt]  [font=\scriptsize]  {$^{( int)}\mathcal{M}$};
% Text Node
\draw (318.85,155.25) node [anchor=north west][inner sep=0.75pt]  [font=\scriptsize]  {$^{( ext)}\mathcal{M}$};
% Text Node
\draw (366.39,182.45) node [anchor=north west][inner sep=0.75pt]  [font=\scriptsize]  {$\tilde{H}_{\tilde{u}}$};
% Text Node
\draw (183.06,129.78) node [anchor=north west][inner sep=0.75pt]  [font=\scriptsize]  {$\widetilde{\underline{H}}_{\widetilde{\underline{u}}}$};

\end{tikzpicture}

    \caption{Proof of the Dynamical Penrose Inequality}
    \label{fig:Penroseineq}
\end{figure}
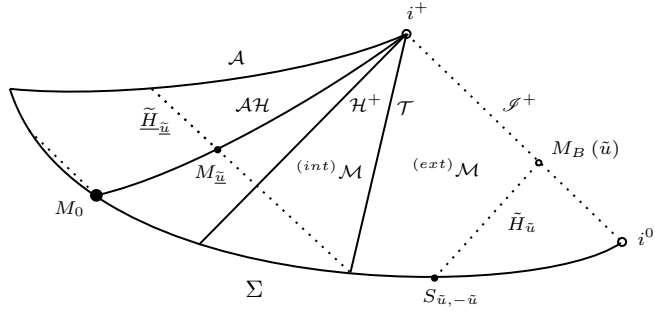

\noindent\textbf{Step 5.} In view of \eqref{Basic ineq 1}, it suffices for us to prove $m\ge m_{\infty}$. 
To achieve this, we follow a similar approach provided in the proof of \Cref{Prop:Bondi}. There we construct a suitable double null foliation in the exterior region $\Ext\MM$. In addition to the steps taken in \Cref{Appendix:property Bondi mass}, we also initialize the incoming optical function $\Ext\tub$ on $\Ext\Si=\Si\cap\Ext\MM$ so that $\tH_{\tu}\cap \Si=\tS_{\tu, -\tu}$. Here $\tH_{\tu}$ is the level set of the solved outgoing optical function $\tu$ and $\tS_{\tu, \Ext\tub}$ denotes the intersection of $\tH_{\tu}$ and the constant $\Ext\tub$ hypersurface. Applying the estimates \eqref{ineq Bondi2} established in the proof of \Cref{Prop:Bondi}, we deduce\footnote{Here $\Ext\t{r}$ is the area radius of $\tS_{\tu, \Ext\tub}$.}
\begin{equation}\label{ineq Bondi}
\begin{aligned}
      |m_H(\tu, \Ext\tub)-M_B(\tu)|\lesssim& \f{1}{\Ext\t{r}}, \\
    \pr_{\tu} M_B(\tu)=-\f{1}{4}\oint_{\mathbb{S}^2} |\t{\Thetab}|^2\le& 0 \qquad \text{with} \qquad \Thetab\coloneqq\lim\limits_{ \Ext\tub\to\infty}\t{r}\t{\chibh}\big|_{\tS_{\tu, \Ext\tub}}.
\end{aligned}
\end{equation}
By virtue of \eqref{ineq Bondi}, together with the fact that
\begin{equation}\label{Eqn:limit -infty mH}
    \lim\limits_{\tu\to-\infty} m_H(\tu, -\tu)=m, 
\end{equation}
we obtain
\begin{equation*}
    |M_B(\tu)-m|\le |M_B(\tu)-m_H(\tu, -\tu)|+|m_H(\tu, -\tu)-m|\lesssim \f{1}{\Ext\t{r}(\tu, -\tu)}+|m_H(\tu, -\tu)-m|.
\end{equation*}
Letting $\tu\to -\infty$, we get $\Ext\t{r}(\tu, -\tu)\to \infty$. Combining with \eqref{Eqn:limit -infty mH}, we then derive
\begin{equation*}
    M_B(-\infty)=m.
\end{equation*}
Therefore, utilizing the Bondi mass loss formula stated in \eqref{ineq Bondi}, we hence arrive at 
\begin{equation*}
    m=M_B(-\infty)\ge M_B(\infty)=m_{\infty}.
\end{equation*}
This completes the proof of the spacetime Penrose inequality in perturbative Kerr regime.
\end{proof}

\begin{remark}
   Both conclusions of \Cref{Thm:Penrose ineq 1} and \Cref{Thm:Penrose ineq 2} remain valid for the full subextremal range of Kerr black holes, i.e., $|a_{\infty}/m_{\infty}|<1$, provided that the corresponding Kerr stability result holds and the associated hyperbolic estimates in \cite{KS:main} are applicable. This is because our construction of incoming null cones in \Cref{Subsec:incoming geodesic foliation}, the elliptic arguments in \Cref{Sec:Existence}, discussions in \Cref{Sec:physical} are essentially independent of the smallness of $a_{\infty}$.
\end{remark}

\subsection{Rigidity of Penrose Inequality}\label{Subsec:rigidity}
By assuming the equality of the spacetime Penrose inequality \eqref{Eqn:Penroseineq2}, utilizing the proof of \eqref{Thm:Penrose ineq 2}, we derive that
\begin{equation*}
    16\pi m^2=16\pi M_B^2(\tu)= 16 \pi m^2_{\infty}= 8\pi m_{\infty}(m_{\infty}+\sqrt{\minft^2-\ainft^2})=A_M(\tub)=A.
\end{equation*}
Here $M_B(\tu)$ denotes the Bondi mass along the outgoing null hypersurface $\tH_{\tu}$ and $A_M(\tub)$ is the area of MOTS $M_{\tub}$.

This immediately implies that $m_{\infty}=m$ and $a_{\infty}=0$. Incorporating with Bondi mass loss formula along $\II^+$ as stated in \Cref{Prop:Bondi} from \Cref{Appendix:property Bondi mass} and the area increasing law of MOTS along $\AH$ established in \Cref{Prop:area law} from \Cref{Subsec:area law}, we further infer
\begin{proposition}\label{Prop:rigidity}
      Under the assumptions of \Cref{Thm:Penrose ineq 2}, if equality holds in \eqref{Eqn:Penroseineq2}, then the spacetime $(\MM, \bfg)$, evolved from initial data set $(\Si, g\, k)$, is non-radiating and the apparent horizon $\AH$ is a non-expanding horizon. Specifically, we have that the Bondi news $\t{\Thetab}\coloneqq\lim\limits_{\tH_{\tu}, \Ext\tub\to \infty}\t{r}\t{\chibh}=0$ along $\II^+$, and the apparent horizon is a null hypersurface.
\end{proposition}
\begin{proof}
    From \Cref{Appendix:property Bondi mass} and \Cref{Prop:area law} we obtain
    \begin{equation*}
        \pr_{\tu}M_B(\tu)=-\f{1}{4}\oint_{\mathbb{S}^2} |\t{\Thetab}|^2.
    \end{equation*}
    Along $\AH$ it also holds
    \begin{equation*}
          \f{dA_M}{d\tub}=0 \, \, \text{in the null piece}  \quad \text{and} \quad \f{dA_M}{d\tub}>0 \,\, \text{in the spacelike piece}.
    \end{equation*}
    As both $M_B(\tu)$ and $A_M(\tub)$ remain constant, we derive that $\t{\Thetab}\equiv 0$ and $\AH$ is null.
\end{proof}

Combining with the hyperbolic estimates in $\MM$, we further deduce that various geometric quantities take their Schwarzschild values along $\AH$.
\begin{proposition}\label{Lem:rigidity}
    Under the same assumptions of \Cref{Prop:rigidity}, relative to the null frame $(e'_{\mu})$ adapted to the MOTS $M_{\tub}$ along $\AH$, we have
    \begin{equation}\label{Eqn:rigidity}
        \chih'=\alpha'=\b'=\si'=0, \qquad \rho'=-\f{1}{4m^2}.
    \end{equation}
\end{proposition}
\begin{proof}
   Recall the following relation between the null frames used in $\intM$:
   \begin{equation*}
       (e_{\mu})\xrightarrow{(f,\fb,\la)}(\te_{\mu})\xrightarrow{(2\t{f}\tnab R, 0, 1)} (e'_{\mu}).
   \end{equation*}
   Here $(e_{\mu})$ is the incoming PG null frame constructed in \cite{KS:main}, $(\te_{\mu})$ denotes the incoming geodesic null frame associated with our Pretorius--Israel type coordinates $(\tub, \t{r}, \tth^1,\tth^2)$ and $\t{f}\coloneqq [\te_3(\t{r})]^{-1}\sim -1$.

   Employing \eqref{Eqn:f fb la}, \eqref{Est:la fb 1} and \eqref{Est:f 2} as established in \Cref{Subsec:incoming geodesic foliation}, along with the fact that $\ainft=0$ for the case of equality, we obtain
\begin{equation*}
    {}^{(1)}\fb, {}^{(2)}f=O(\tfepub), \qquad {}^{(1)}\lambda=1+O(\tfepub)
\end{equation*}
   and thus
   \begin{equation*}
       ( f, \fb, \la)=({}^{(2)}f,{}^{(2)}\fb=0,{}^{(2)}\lambda=1)\circ( {}^{(1)}f=0,{}^{(1)}\fb,{}^{(1)}\lambda )=(0, 0, 1)+O(\tfepub).
   \end{equation*}
   Meanwhile, using the estimate for $R$ obtained in \Cref{Thm:MOTS exis}, we have $2\t{f}\tnab R=O(\tfepub)$. This implies that the transition coefficients between $(e_{\mu})$ and $(e'_{\mu})$ satisfy
   \begin{equation*}
       (f', \fb',\la')=(2\t{f}\tnab R, 0, 1)\circ(f,\fb,\la)=(0, 0, 1)+O(\tfepub).
   \end{equation*}
   Consequently, applying the transformation formulas from \Cref{Lem:transformation formula} and \eqref{Est for Ricci Cur}, noting that along $\AH$ the $r$ coordinate obeys
   \begin{equation*}
       r\big|_{\AH}=[g(r,\th)]^{-1}(R-r_+)+r_+=O(\tfepub)+2m,
   \end{equation*}
   we derive that along $\AH$, all geometric quantities are of size $O(\tfepub)$ except
   \begin{equation}\label{Est:Rigidity}
       \trchb'=-\f{1}{m}+O(\tfepub), \qquad \om'=-\f{1}{4m}+O(\tfepub), \qquad \rho'=-\f{1}{4m^2}+O(\tfepub).
   \end{equation}

   Now we turn to prove \eqref{Eqn:rigidity}. Since $\AH$ is outgoing null and $e_4'\in (TM_{\tub})^{\perp}$, we have $e_4'\in T\AH$ and hence 
   \begin{equation*}
       \xi'_a=\f12\bfg(\bfD_{e_4'} e_4', e_a')=-\f12\bfg( e_4', \bfD_{e_4'}e_a')=-\f12\bfg(e_4', [e_4', e_a']+\bfD_{e_a'}e_4')=0.
   \end{equation*}
   Then inserting the condition that $\trch'=0$ along $\AH$ into the Raychaudhuri equation
   \begin{equation*}
       \nab_4' \trch'+\f12(\trch')^2=-|\chih'|^2-2\om'\trch',
   \end{equation*}
   we deduce that $\chih'=0$ along $\AH$. This further implies
   \begin{equation*}
       \a'=-\nab'_4\chih'-\trch'\chih'-2\om'\chih'=0 \qquad \text{along} \quad \AH.
   \end{equation*}
   To determine $\b'$, we consider its corresponding transport equation
   \begin{equation*}
\nab_4'\b'=\div'\a'-2\trch'\b'-2\om'\b'+\a'\c(2\zeta'+\eta')=-2\om'\b'.
   \end{equation*}
   Here we utilize the derived equality $\trch'=\a'=0$ along $\AH$.

   Introducing the integrating factor\footnote{Note that $e_4'(\tub)=\te_4(\tub)=-\bfg(\te_4, \te_3)=2$.} $\mu=e^{\int 2\om'd\tub}$, we infer that
   \begin{equation}\label{Eqn:rigidity:b'}
       \nab_4'(\mu|\b'|^2)=0.
   \end{equation}
   Also, from \eqref{Est:Rigidity} we have $\mu\sim e^{-\f{\tub}{2m}}$ and thus these imply
    \begin{equation*}
       \mu|\b'|^2=e^{-\f{\tub}{2m}} \c O(\f{\epsilon_0^2}{\tub^{2+2\de_{dec}}})\to 0 \qquad \text{as} \quad \tub\to\infty.
   \end{equation*}
   Combining with \eqref{Eqn:rigidity:b'}, this leads to $\b'=0$ along $\AH$.

   With all these proven vanishing quantities along $\AH$, for $\rho', \si'$ we consider their transport equations along $e_4'$ and get
   \begin{equation*}
       \nab_4'(\rho', \si')=(\div'\b', -\curl'\b')+\f32\trch'(\rho',\si')+(2\eta'+\zeta)'\c (\b', -\dual \b')-\f12\chibh'\c(\a', -\dual\a') =0.
\end{equation*}
Incorporating with \eqref{Est:Rigidity}, we conclude
\begin{equation*}
    \rho'=\lim\limits_{\tub\to\infty}\rho'=-\f{1}{4m^2}, \qquad \si'=\lim\limits_{\tub\to\infty}\si'=0
\end{equation*}
as stated.
\end{proof}

Finally, via employing the Gauss equation on the initial MOTS $M_0$, we further establish a version of rigidity for our spacetime Penrose inequality.
\begin{theorem}[Rigidity]
    The equality of \eqref{Eqn:Penroseineq2} in \Cref{Thm:Penrose ineq 2} holds if and only if the Gauss curvature of the MOTS $M_0$ embedded $\Si$ is pointwise equal to $\f{1}{4m^2}$.
\end{theorem}
\begin{proof}
    The if part is a direct consequence of the Gauss--Bonnet theorem. For the only if part, by virtue of the Gauss equation on $M_0$, together with \Cref{Lem:rigidity} we deduce
    \begin{equation*}
        \bfK'=-\rho'+\f12\chih'\c\chibh'-\f14\trch'\trchb'=\f{1}{4m^2}.
    \end{equation*}
    This completes the proof of this theorem.
\end{proof}

\appendix

\section{Smooth Examples of Admissible Kerr Black Hole Formation Initial Data}\label{Apx:example}
\subsection{Prescription of Characteristic Initial Data}

In this subsection, we aim to construct the smooth examples of admissible characteristic initial data along $u=u_0$ with $-u_0\gg 1$ as introduced in \Cref{Subsec:admiKerr}. Recall that in \Cref{Subsec:admiKerr}, the spacetime metric in the double null coordinates $(u, v, \vartheta=(\th^1, \th^2))$ take the form of
\begin{equation*}
    \bfg=-2\O^2(du\otimes dv+dv \otimes du)+g_{ab}(d\theta^a-b^a dv)\otimes (d\theta^B-b^b dv).
\end{equation*}
The corresponding null frame is given by
\begin{equation}\label{apx:Eqn:null frame pul}
    e_1, e_2\in TS_{u, v}, \qquad e_3=-2\O\bfD v=\O^{-1}\pr_u, \qquad e_4=-2\O \bfD u=\O^{-1}(\pr_{v}+b^a\pr_{\th^a}),
\end{equation}
where $S_{u ,v}$ represents the intersection of the constant $u$ hypersurface $H_{u}$ and the constant $v$ hypersurface $\Hb_{v}$.

Inspired by \cite{Chr:book}, we set the characteristic data along $H_{u_0}$ via a conformal argument. Let $\mathring{g}\big|_{S_{u_0, 0}}$ be the standard metric on the unit round sphere $\mathbb{S}^2$. The induced metric $g|_{S_{u_0, v}}$ can be prescribed within the below conformal class
\begin{equation}\label{hatg}
g\big|_{S_{u_0, v}}=(\phi\big|_{S_{u_0, v}})^2 \hat{g}\big|_{S_{u_0, v}} \qquad \text{with} \quad \det \hat{g}_{ab}=\det \mathring{g}_{ab}.
\end{equation}

As shown in \cite{Chr:book}, the conformal factor $\phi$ then satisfies the following second-order ordinary differential equation\footnote{Compared with \cite{Chr:book} note that in our setting there is an additional term involving the first-order derivative of $\phi$, since our $\O$ is not set to be a constant along $H_{u_0}$.}
\begin{equation}\label{2nd order ODE}
\f{\partial^2 \phi}{\partial v^2}-\f{2\partial (\log \Omega)}{\partial v} \f{\partial \phi}{\partial v}+e\phi=0
\end{equation}
with initial conditions 
\begin{equation*}
\phi|_{S_{u_0, 0}}=|u_0|, \quad \f{\partial \phi}{\partial v}\Big|_{S_{u_0, 0}}=1
\end{equation*}
and 
\begin{equation*}
e\coloneqq\f{1}{8}(\hat{g}^{-1})^{ac}(\hat{g}^{-1})^{bd}\f{\partial \hat{g}_{ab}}{\partial v}\f{\partial \hat{g}_{cd}}{\partial v}.
\end{equation*}

By prescribing a smooth, symmetric, positive-definite 2-tensor $\hat{g}(u_0, v ,\vartheta)$ on $S_{u_0, v}$ and a function $\Omega(u_0, v, \vartheta)$ along $H_{u_0}$, we can find a $v_1>0$ such that $1/2\le \phi<\infty$ for $0\leq v\leq v_1$. Consequently, the conformal metric $g|_{S_{u_0, v}}$ is then fixed and well-defined for $0\leq v\leq v_1$, and we can express $\trch, \chih, \a$ as
\begin{equation}\label{3.1}
\Omega\chih_{ab}=\f{1}{2}\phi^2 \f{\partial \hat{g}_{ab}}{\partial v}, \qquad \Omega \trch=\f{2}{\phi} \f{\partial \phi}{\partial v}, \qquad 
\Omega\a_{ab}=\f{\partial}{\partial v}(\f{1}{2}\phi^2\f{\partial \hat{g}_{ab}}{\partial v}).
\end{equation}
The remaining geometry quantities can further be derived through EVEs schematically. See \cite{Chr:book} for details. 
\vspace{2mm}

Recall that in \cite{A-L}, An-Luk set $|\chih|\sim \f{A^{1/2}}{|u_0|}$ for $0\leq v\leq \delta$ and obtained the existence of EVEs and the hyperbolic estimates in the scale-critical short-pulse region $\{u_0\leq u\leq - Av/4,\  0\leq v \leq \delta\}$. Denote $v_0=\delta$ and set $A=\de^{-1}\gg1$, $m_0=\delta A/4=1/4$. In below we extend their initial data to the outgoing hypersurface $H_{u_0}\cap \{v\ge v_0 \}$ beyond the short-pulse region. 

Firstly we set $\Omega=\O_{\SS}\coloneqq\sqrt{1-\f{2m_0}{r_{\SS}(v)}}$ after the short pulse along $H_{u_0}$, where the background radius $r_{\SS}=r_{\SS}(v)$ satisfies\footnote{Although in \cite{Chr:book} and \cite{A-L}, $\O$ is set to be $1$ along $H_{u_0}\cup \Hb_{0}$, we have the freedom to re-prescribe $\O$ on $H_{u_0}\cup \Hb_{0}$ by a change of variables for the pair of optical functions $(u, v)\to(U(u), V(v))$.
}
\begin{equation}\label{Eqn:pr v rs}
\begin{cases}
\f{dr_{\SS}(v)}{dv}=1-\f{2m_0}{r_{\SS}(v)}, \\
r_{\SS}(0)=|u_0|.
\end{cases} 
\end{equation}
Since $r_{\SS}(v)\geq |u_0|>4m_0$, we have $1/2 \leq \Omega(u_0, v)\leq 1$, and thus $\f{v+|u_0|}{2}\leq r_{\SS}(v) \leq v+|u_0|$.
\vspace{2mm}

With the help of $\Omega_{\SS}$ and $r_{\SS}=r_{\SS}(v)$, the corresponding non-vanishing geometric quantities adapted to the associated Schwarzschild metric can be given as
\begin{equation}\label{background sch value}
\O_{\SS}\text{tr} \chi_{\SS}=-\O_{\SS}\trchb_{\SS}=\f{2}{r_{\SS}}(1-\f{2m_0}{r_{\SS}}), \quad \O_{\SS}\omega_{\SS}=-\O_{\SS}\omegab_{\SS}=-\f{m_0}{2r_{\SS}^2}, \quad \rho_{\SS}=-\f{2m_0}{r_{\SS}^3}.
\end{equation}

We then introduce a weighted norm for any $S_{u_0 ,v}$-tensor $\xi$ with $N\in \mathbb{N}$ and $k\in \mathbb{R}$:
\begin{equation*}
\|\xi\|_{C^N_k(v_1)}=\max\limits_{|i|+j\leq N}\sup\limits_{(v, \vartheta)\in [v_0, v_1]\times \mathbb{S}^2} \|r_{\SS}^{j+k} (\f{\partial}{\partial \vartheta})^{i}(\f{\partial}{\partial v})^{j}\xi(u_0, v, \vartheta)\|. 
\end{equation*}
Here $\|\xi\|=\qty(\sum_{a_i, b_j\in\{1,2\}}|\xi_{a_1\cdots a_p}^{b_1\cdots b_q}|^2)^{1/2}$ is the Euclidean norm in stereographic coordinates $\vartheta=(\theta^1, \theta^2)\in \mathbb{S}^2$, and for a pair of non-negative integers $a=(a_1, a_2)$, we denote
\begin{equation*}
\qty(\f{\partial}{\partial \vartheta})^{a}=\qty(\f{\pr}{\pr\th^1})^{a_1}\qty(\f{\pr}{\pr\th^2})^{a_2} \qquad \text{and} \qquad |a|=a_1+a_2.
\end{equation*}

Denoting $C^N_k(\infty)=C^N_k$, with our definition we immediately obtain that $\|r_{\SS}^{-1}\|_{C^{N}_{1}}$ and $\|\O\|_{C^{N}_0}, \|\O^{-1}\|_{C^{N}_0}$ are bounded for any $N\geq 0$. 
\vspace{2mm}

The following product estimate is helpful in our construction of the desired initial data.
\begin{lemma}\label{Est Product}
For any $k, l\in \mathbb{R}$, it holds
	$$\|fg\|_{C^N_k}\lesssim \|f\|_{C^N_{k-l}}\|g\|_{C^N_l}.$$
\end{lemma}
\begin{proof}
	Utilizing the Leibniz rule, we have
 \begin{equation*}
     r_{\SS}^{j+k}(\prtheta)^{i} \f{\pr^j}{\pr v^j} (fg)=\sum_{i'\le i}\sum_{j'\le j}\binom{i}{i'}\binom{j}{j'}r_{\SS}^{j'+k-l}(\prtheta)^{i'}(\f{\pr}{\pr v})^{j'}f \cdot r_{\SS}^{j-j'+l}(\prtheta)^{i-i'}(\f{\pr}{\pr v})^{j-j'}g.
 \end{equation*}
 This implies
 \begin{equation*}
      |r_{\SS}^{j+k}(\prtheta)^{i} \f{\pr^j}{\pr v^j} (fg)|\lesssim \|f\|_{C^N_{k-l}}\|g\|_{C^N_l}.
 \end{equation*}
 Taking the supremum over $|i|+j\le N$ and over $(v, \vartheta)\in [v_0, \infty)\times \mathbb{S}^2$, the desired estimate thus follows.
\end{proof}

For future use, we also establish the below evolution lemma:
\begin{lemma}\label{gronwall lemma1}
	Set $k\in \mathbb{R}$ and let $f=f(v, \vartheta)$ be a function satisfying the transport equation
	\begin{equation}\label{gronwall lemma eqn}
	\O^{-1}\f{\partial f}{\partial v}+\f{k}{2}\trch_{\SS} f=b.
	\end{equation} 
	Then for any $v\geq v_0$, there holds
	\begin{equation}\label{M=0}
	|r_{\SS}^{k}f(v, \vartheta)|\lesssim |f(v_0, \vartheta)|+\int_{v_0}^{v} r_{\SS}^{k} |b| dv'.
	\end{equation}
	Furthermore, if $\trch \in C^{N-2}_1(v)$ and $b\in C^{M}_{m}(v)$ with $0\leq M\leq N-2$ and $k\neq m-1$, then $f\in C^{M}_{\min(k, m-1)}(v)$  and it obeys
	\begin{equation}\label{C^{N}_k estimate}
	\|f\|_{C^{M}_{\min(k, m-1)}(v)}\lesssim \max\limits_{|i|\leq M}\sup\limits_{\vartheta \in \mathbb{S}^2}|(\f{\partial}{\partial \vartheta})^{i}f(v_0, \vartheta)|+\|b\|_{C^{M}_{m}(v)}.
	\end{equation}
	When $k=m-1$, we have that $f\in C^{M}_{k-1}(v)$ and the following estimate holds
	\begin{equation}\label{C^{N}_k weak estimate}
	\|f\|_{C^{M}_{k-1}(v)}\lesssim \max\limits_{|i|\leq M}\sup\limits_{\vartheta \in \mathbb{S}^2}|(\f{\partial}{\partial \vartheta})^{i}f(v_0, \vartheta)|+\|b\|_{C^{M}_{m}(v)}.
	\end{equation}
\end{lemma}
\begin{proof}
	Multiplying \eqref{gronwall lemma eqn} by $\O r_{\SS}^k$ and applying $\f{d r_{\SS}}{d v}=\f12 r_{\SS}\O \trch_{\SS}$, we then obtain
    \begin{equation*}
        \f{\pr }{\pr v}(r_{\SS}^k f)=\O r_{\SS}^k b.
    \end{equation*}
  Incorporating with Gr\"onwall's inequality, we thus deduce \eqref{M=0}.
  
  Next, we proceed to prove \eqref{C^{N}_k estimate} by conducting induction on $M\in\mathbb{Z}_{\ge 0}$ with $N$ being fixed. Note that for the scenario $M=0$, the desired estimate has already been established in \eqref{M=0}.  
  
  Assume that \eqref{C^{N}_k estimate} holds for $M\leq N-3$. It suffices to bound terms of the form
	\begin{equation*}
	r_{\SS}^{j+k} (\f{\partial}{\partial \vartheta})^{i}(\f{\partial}{\partial v})^{j}f(u_0, v, \vartheta)
	\end{equation*}
	with $|i|+j=M+1$.
\vspace{2mm}

If $j=0$, applying $(\partial/ \partial \vartheta)^{i}$ to \eqref{gronwall lemma eqn}, we get
\begin{equation}\label{Eqn: pr th phi bi}
\O^{-1}\f{\partial}{\partial v}(\f{\partial}{\partial \vartheta})^{i}f+\f{k}{2}\trch (\f{\partial}{\partial \vartheta})^{i}f= (\f{\partial}{\partial \vartheta})^{i}b-\f{k}{2}\sum_{l< i}\left(\begin{array}{c}
i\\l
\end{array}\right) (\f{\partial}{\partial \vartheta})^{i-l} \trch (\f{\partial}{\partial \vartheta})^{l} f\eqqcolon b_i.
\end{equation}
Utilizing the inductive assumption and $\trch \in C^{N-2}_1(v)$, we then derive
\begin{equation*}
|r_{\SS}^{m}b_i|\lesssim  \max\limits_{|i|\leq M}\sup\limits_{\vartheta \in \mathbb{S}^2}|(\f{\partial}{\partial \vartheta})^{i}f(v_0, \vartheta)|+\|b\|_{C^{M+1}_{m}}\eqqcolon K.
\end{equation*}
Meanwhile, employing \eqref{M=0} for \eqref{Eqn: pr th phi bi} yields
\begin{equation}\label{Est:pr th phi bi}
|r_{\SS}^{k}(\f{\partial}{\partial \vartheta})^{i}f|\lesssim \max\limits_{|i|\leq M+1}\sup\limits_{\vartheta \in \mathbb{S}^2}|(\f{\partial}{\partial \vartheta})^{i}f(v_0, \vartheta)|+\int_{v_0}^v |r_{\SS}^k b_i|dv'.
\end{equation}
Observe that
\begin{equation}\label{Est:int b}
    \int_{v_0}^v |r_{\SS}^k b_i|dv'\le K\int_{v_0}^v r_{\SS}^{k-m}dv'\ls \left\{\begin{array}{ll}
         K   \quad &\text{if} \quad k<m-1,  \\
        K\log r_{\SS}   \quad &\text{if} \quad k=m-1,\\
         Kr_{\SS}^{k-m+1}   \quad &\text{if} \quad k>m-1.
    \end{array}\right.
\end{equation}
Combining with \eqref{Est:pr th phi bi}, for $k\neq m-1$, we thus deduce
\begin{equation*}
    |r_{\SS}^{\min(k, m-1)}(\f{\partial}{\partial \vartheta})^{i}f|\lesssim \max\limits_{|i|\leq M+1}\sup\limits_{\vartheta \in \mathbb{S}^2}|(\f{\partial}{\partial \vartheta})^{i}f(v_0, \vartheta)|+\|b\|_{C^{M+1}_{m}(v)}.
\end{equation*}
When $k=m-1$, due to the log-loss in \eqref{Est:int b}, we only obtain a weaker bound\footnote{This inequality is still true when $k-1$ is replaced by a number less than $k$. However, choosing $k-1$ suffices for us to implement the estimations for all geometric quantities.}
\begin{equation*}
    |r_{\SS}^{k-1}(\f{\partial}{\partial \vartheta})^{i}f|\lesssim \max\limits_{|i|\leq M+1}\sup\limits_{\vartheta \in \mathbb{S}^2}|(\f{\partial}{\partial \vartheta})^{i}f(v_0, \vartheta)|+\|b\|_{C^{M+1}_{m}(v)}.
\end{equation*}

For $j\geq 1$, we can write 
\begin{equation*}
(\f{\partial}{\partial v})^{j}f=(\f{\partial}{\partial v})^{j-1} \f{\partial}{\partial v}f.
\end{equation*}
Substituting this into \eqref{gronwall lemma eqn}, we deduce
\begin{align*}
(\f{\partial}{\partial v})^{j}f=&(\f{\partial}{\partial v})^{j-1} \O(b-\f{k}{2} \trch f) \\
=&(\f{\partial}{\partial v})^{j-1} (\O b)-\f{k}{2}\sum_{n=0}^{j-1}\left(\begin{array}{c}
j-1\\n
\end{array}\right) (\f{\partial}{\partial v})^{j-1-n} (\O \trch) (\f{\partial}{\partial v})^n f.
\end{align*}
Applying $(\partial/ \partial \vartheta)^{i}$ to the above equation thus yields
\begin{equation*}
(\f{\partial}{\partial \vartheta})^{i}(\f{\partial}{\partial v})^{j}f=(\f{\partial}{\partial \vartheta})^{i}(\f{\partial}{\partial v})^{j-1} (\O b)-\f{k}{2}\sum_{l\leq i}\sum_{n=0}^{j-1}\left(\begin{array}{c}
i\\l
\end{array}\right)\left(\begin{array}{c}
j-1\\n
\end{array}\right) (\f{\partial}{\partial \vartheta})^{i-l}(\f{\partial}{\partial v})^{j-1-n} (\O \trch) (\f{\partial}{\partial \vartheta})^{l}(\f{\partial}{\partial v})^n f.
\end{equation*}
Here note that all the terms on the right only invoke derivatives of $f$ in $v$ and $\th$ variables of order at most $j-1+|i|\le M$. Therefore, in light of the inductive assumption and $\trch \in C^{N-2}_1(v)$ again, for the scenario when $k<m-1$, we conclude
\begin{align*}
&|r_{\SS}^{k+j}(\f{\partial}{\partial \vartheta})^{i}(\f{\partial}{\partial v})^{j}f|\\\lesssim& |r_{\SS}^{k+1+j-1}(\f{\partial}{\partial \vartheta})^{i}(\f{\partial}{\partial v})^{j-1} (\O b)|+\sum_{l\leq i}\sum_{n=0}^{j-1} |r_{\SS}^{1+j-1-n}(\f{\partial}{\partial \vartheta})^{i-l}(\f{\partial}{\partial v})^{j-1-n} (\O \trch)|\cdot|r_{\SS}^{k+n} (\f{\partial}{\partial \vartheta})^{l}(\f{\partial}{\partial v})^n f| \\
\lesssim& \|\O b\|_{C^{M}_{k+1}(v)}+\|\O \trch\|_{C^{M}_{1}}\|f\|_{C^{M}_{k}(v)} 
\lesssim \max\limits_{|i|\leq M+1}\sup\limits_{\vartheta \in \mathbb{S}^2}|(\f{\partial}{\partial \vartheta})^{i}f(v_0, \vartheta)|+\|b\|_{C^{M+1}_{m}(v)}. 
\end{align*}
The scenario $k\ge m-1$ can be treated similarly to the case $j=0$ as above. We thus complete the proof of \eqref{C^{N}_k estimate}.
\end{proof}

We proceed to define a seed energy on $S_{u_0, v_0}$ by
\begin{equation*}
\mathbb{E}^{N}_{seed}=\max\limits_{|i|\leq N}\sup\limits_{\vartheta\in \mathbb{S}^2}\| (\f{\partial}{\partial \vartheta})^{i} (\log\f{\phi}{|u_0|},\widetilde{\Ga},\widetilde{R})(u_0, v_0, \vartheta)\| \qquad \text{with} \quad \Ga\neq \chih, \omega,\omegab, \ R\neq \a.
\end{equation*}
Here $\widetilde{\Ga}=\Ga-\Ga_{\SS}$ and $\widetilde{\Psi}=\Psi-\Psi_{\SS}$ represent the difference of the geometric quantities and their corresponding Schwarzschild values provided in \eqref{background sch value}. 

Denoting $N_1=N+10$ with $N\in \mathbb{N}$ being the same integer as in \eqref{Est:out ID} in \Cref{Subsec:admiKerr}, we then give smallness condition that guarantees existence of global initial data along $H_{u_0}$.
\begin{proposition}\label{global existence for initial data}
	Assuming $\mathbb{E}^{N_1}_{seed}\leq \epsilon_1$, $|r(u_0, v_0)/|u_0|-1|\le \epsilon_1$ and $\|\hat{g}-\mathring{g}\|_{C^{N_1}_{1.5+\de_B/2}}\leq \epsilon_1$ with $N_1\ge 10$ and  with sufficiently small $\epsilon_1$, then we can take $v_1=\infty$, i.e., the characteristic initial data can be globally prescribed along $H_{u_0}$. Furthermore, the following estimates hold 
    \begin{equation}\label{Est: phi trch}
\|\f{\phi}{r_{\SS}}-1\|_{C^{N_1-2}_{0}}+\|\trch-\trch_{\SS}\|_{C^{N_1-2}_{2}}\lesssim \epsilon_1.
\end{equation} 
\end{proposition}
\begin{proof}
	It suffices to show that $1/2\le \phi\ls r_{\SS}(v)$ for all $v\geq v_0$. We prove this through a bootstrap argument. 
    
    Assume there exists $v_2>v_0$ such that, for all $v_0\le v\le v_2$, there holds
    \begin{equation}\label{BA:phi}
    \f12\le \phi\le \f{2r_{\SS}(v)}{r(u_0, v_0)}, \qquad \|\trch-\trch_{\SS}\|_{C^{N_1-2}_{2}(v_2)}\leq 2B \epsilon_1.
    \end{equation}
Here $B>0$ is a large constant that will be determined later.
    
In view of the definition of $\trch$, we derive the below transport equation for $\phi$:
\begin{equation}\label{Apx:Eqn:phi rs}
\f{\partial}{\partial v}\log \f{\phi}{r_{\SS}}=\f{1}{2}\Omega \trch-\f{1}{2}\Omega \trch_{\SS}=\f{1}{2}\Omega\widetilde{\trch}.
\end{equation}
Combining with \Cref{gronwall lemma1}, this renders 
\begin{equation*}
\begin{split}
\|\log\f{\phi}{r_{\SS}}\|_{C^{N_1-2}_0(v_2)}\lesssim \mathbb{E}^{N_1}_{seed}+|\f{r(u_0, v_0)}{|u_0|}-1|+\|\trch-\trch_{\SS}\|_{C^{N_1-2}_{2}(v_2)}\lesssim (1+2B_1)\epsilon_1.
\end{split}
\end{equation*}
In particular, for $\epsilon_1$ being sufficiently small (depending on $B$), we thus improve both upper bound and lower bound of $\phi$ in \eqref{BA:phi} and get
\begin{equation*}
\f34\le\f{3}{4}\c\f{r_{\SS}(v)}{r(u_0, v_0)}\le\phi \leq \f{3}{2}\c\f{r_{\SS}(v)}{r(u_0, v_0)}.
\end{equation*}

The remaining part is to improve the second inequality in \eqref{BA:phi} and to obtain 
\begin{equation*}
    \|\trch-\trch_{\SS}\|_{C^{N_1-2}_{2}(v_2)}\le B\epsilon_1.
\end{equation*}
Substituting the expression of $\trch$ into \eqref{2nd order ODE} yields
\begin{equation*}
\f{\partial }{\partial v}(\Omega^{-1} \trch)=-\f{1}{2}(\trch)^2-2e.
\end{equation*}
Denoting $\widetilde{\trch}=\trch-\trch_{\SS}$, we then derive the evolution equation for $\widetilde{\trch}$:
\begin{equation}\label{Eqn:pr v trch}
	\begin{split}
\f{\partial }{\partial v}(\Omega^{-1}\widetilde{\trch})&=\f{\partial }{\partial v}(\Omega^{-1}\trch-\Omega^{-1}\trch_{\SS}) =-2e-\f{1}{2}(\trch)^2 +\f{1}{2}(\trch_{\SS})^2 
=-2e-\f{1}{2}(\widetilde{\trch})^2-\trch_{\SS} \widetilde{\trch},
\end{split}
\end{equation}
where we use the equality
  \begin{equation*}
        \f{\partial }{\partial v}(\O^{-1}\trch_{\SS})=\f{\partial }{\partial v}(\f{2}{r_{\SS}})=-\f{2}{r_{\SS}^2}(1-\f{2m_0}{r_{\SS}})=-\f12 (\trch_{\SS})^2.
    \end{equation*}
Moving the term $\trch_{\SS}\widetilde{\trch}$ to the left and multiplying $\O^{-1}$, we then convert \eqref{Eqn:pr v trch} into the form 
\begin{equation*}
    \O^{-1}\f{\partial }{\partial v}(\Omega^{-1}\widetilde{\trch})+\trch_{\SS} (\O^{-1}\widetilde{\trch})=-\O^{-1}\qty(2e+\f12(\widetilde{\trch})^2).
\end{equation*}
Observe that from \Cref{Est Product} we have
\begin{equation}\label{Est for e}
    \|e\|_{C^{N_1-2}_{5+\de_B}}\lesssim \|\pr_v \hat{g}\|_{C^{N_1-2}_{2.5+\de_B/2}}^2\les \|\hat{g}-\mathring{g}\|_{C^{N_1-1}_{2.5+\de_B/2}}^2 \leq \epsilon_1^2.
\end{equation}
Therefore, employing \Cref{gronwall lemma1} with $k=2, m=4$, along with \Cref{Est Product}, we deduce
\begin{align*}
    \|\Omega^{-1} \widetilde{\trch}\|_{C^{N_1-2}_2(v_2)}\le& C\qty(\mathbb{E}^{N_1}_{seed}+\|\O^{-1}\|_{C^{N_1-2}_{0}(v_2)}\|e\|_{C^{N_1-2}_{4}(v_2)}+ \|\O\|_{C^{N_1-2}_{0}(v_2)}\|\Omega^{-1}\widetilde{\trch}\|^2_{C^{N_1-2}_{2}(v_2)})\\
    \le& C(1+4B^2\epsilon_1)\epsilon_1.
\end{align*}
Together with \Cref{Est Product} this yields
\begin{equation*}
\begin{split}
\|\widetilde{\trch}\|_{C^{N_1-2}_2(v_2)}\leq C \|\Omega^{-1} \widetilde{\trch}\|_{C^{N_1-2}_2(v_2)} \|\Omega \|_{C^{N_1-2}_0} 
\le C(1+4B^2\epsilon_1)\epsilon_1.
\end{split}
\end{equation*}
Choosing $B\ge2C$ and $4B^2\epsilon_1\le 1$, we thus improve the bootstrap assumption for $\widetilde{\trch}$. The desired estimate \eqref{Est: phi trch} hence follows as we close the above bootstrap argument.
\end{proof}

The above proposition immediately implies
\begin{corollary}\label{gronwall lemma}
    The same conclusion of \Cref{gronwall lemma1} holds if replacing $\trch_{\SS}$ by $\trch$ in \eqref{gronwall lemma eqn}.
\end{corollary}
\begin{proof}
    Notice that the evolution equation for $\phi$ can be rewritten as
    \begin{equation*}
       \O^{-1} \f{\pr \phi}{\pr v}+\f{k}{2}\trch_{\SS}\phi=b-\widetilde{\trch} \phi.
    \end{equation*}
  Incorporating with Gr\"onwall's inequality and employing the estimate of $\widetilde{\trch}$ in \Cref{global existence for initial data}, we derive the desired result.
\end{proof}

By virtue of \eqref{hatg}, \eqref{3.1} and \Cref{global existence for initial data}, we also obtain
\begin{proposition}\label{3.2}
	Under the same condition of \Cref{global existence for initial data}, we have
	\begin{equation}\label{3.2 ineq}
	\|g-r_{\SS}^2 \mathring{g}\|_{C^{N_1-2}_{-2}}+\|\bfK-\f{1}{r^2_{\SS} }\|_{C^{N_1-4}_{2}}+\|\chih\|_{C^{N_1-1}_{0.5+\de_B/2}}+\|\widetilde{\trch}\|_{C^{N_1-2}_2}+\|\a\|_{C^{N_1-2}_{1.5+\de_B/2}}\lesssim \epsilon_1.
	\end{equation}
    Here $\bfK$ is the Gauss curvature of $S_{u_0, v}$.
\end{proposition}
\begin{proof}
    Here we only establish the estimate of $\a$ and $\bfK$, while the rest can be readily deduced from \eqref{hatg}, \eqref{3.1} and \Cref{global existence for initial data}. Employing \eqref{3.1}, we obtain
    \begin{equation*}
        \O\a_{ab}=\f12\phi^2\qty(\O \trch \f{\partial \hat{g}_{ab}}{\partial v}+ \f{\partial^2 \hat{g}_{ab}}{\partial v^2}).
    \end{equation*}
    Incorporating with \Cref{Est Product}, the estimate of $\trch$ in \Cref{global existence for initial data} and the condition $\|\hat{g}-\mathring{g}\|_{C^{N_1}_{1.5+\de_B/2}}\leq \epsilon_1$, we thus derive
    \begin{equation*}
        \|\a\|_{C^{N_1-2}_{1.5+\de_B/2}}\lesssim \|\phi\|^2_{C^{N_1-2}_{-1}}\qty(\|\trch\|_{C^{N_1-2}_{1}}\c \|\hat{g}-\mathring{g}\|_{C^{N_1-1}_{2.5+\de_B/2}}+\|\hat{g}-\mathring{g}\|_{C^{N_1}_{3.5+\de_B/2}})\lesssim \epsilon_1.
    \end{equation*}

    As for the Gauss curvature $\bfK$ of $(S_{u_0, v}, g)$, noting that $g=\phi^2 \hat{g}$ on $S_{u_0, v}$, we hence have
    \begin{equation}\label{Eqn:K conformal}
        \bfK=\phi^{-2}\qty(\hat{\bfK}-\hat{\De}\log{\phi}),
    \end{equation}
    where $\hat{\bfK}$ and $\hat{\De}$ represent the Gauss curvature and the Laplace--Beltrami operator of $(S_{u_0, v}, \hat{g})$.

    Therefore, using  the assumption $\|\hat{g}-\mathring{g}\|_{C^{N_1}_{1.5+\de_B/2}}\leq \epsilon_1$ and \eqref{Est: phi trch} in \Cref{global existence for initial data} again, we obtain the estimate of $\bfK$ as claimed.
\end{proof}

Denote $|\cdot|_{g}$ to be the norm of $S_{u_0, v}$-tensor with respect to the induced metric $g|_{S_{u_0 ,v}}$ and let $r=r(u, v)$ be the area radius of $S_{u, v}$. Given a $T^q_p$ type tensor $\xi$ on $S_{u_0, v}$, we remark it follows from \Cref{3.2} that 
\begin{equation}\label{equiv relation}
r=r_{\SS}+O(\epsilon_1) r_{\SS}\qquad \text{and} \qquad|\xi|_{g}\sim r_{\SS}^{q-p}\|\xi\|. \\[2mm]
\end{equation}

\subsection{Desired Estimates for Geometric Quantities Except \texorpdfstring{$\trchb$}{}}

This subsection is devoted to establishing below desired weighted estimates along the initial outgoing hypersurface $H_{u_0}$ after the short pulse for curvature components and for Ricci coefficients except $\trchb$.
\begin{proposition}\label{initial outgoing data}
		Assuming $\mathbb{E}^{N_1}_{seed}\leq \epsilon_1$, $|r(u_0, v_0)/|u_0|-1|\le \epsilon_1$ and $\|\hat{g}-\mathring{g}\|_{C^{N_1}_{1.5+\de_B/2}}\leq \epsilon_1$  with $N_1\geq 13$ and with sufficiently small $\epsilon_1$, then there holds
	\begin{equation*}
	\widetilde{\mathbb{E}}^{N_1-10}_{0,out}=\widetilde{\mathbb{E}}^{N_1-10}_{0,out} \lesssim \epsilon_1.
	\end{equation*}
    Here the outgoing initial energy norm $\widetilde{\mathbb{E}}^N_{0, out}$ is defined as
\begin{equation*}
\begin{split}
&\widetilde{\mathbb{E}}^N_{0,out}\coloneqq \sup\limits_{v\in[v_0,\infty)}\|(r\nab)^{\le N}(r^{\f{7+\de_B}{2}} \a, r^{\f{7+\de_B}{2}}\beta, r^3(\rho-\rho_{\SS}), r^3 \sigma, r^2\bb, r\ab)\|_{L^\infty(S_{u_0, v})}\\
&+\sup\limits_{v\in[v_0,\infty)}\|(r\nab)^{\le N}\qty(r^2 \chih, r\chibh, r^2(\trch-\trch_{\SS}), r(\trchb-\trchb_{\SS}), r^2 (\eta, \etab, \zeta), r^3(\omega-\omega_{\SS}), r^{-2}(g-\f{r_{\SS}^2}{|u_0|^2}\mathring{g}))\|_{L^\infty(S_{u_0, v})},
\end{split}
\end{equation*}
where $\nab$ is the covariant derivative on $S_{u_0 ,v}$ with respect to $(S_{u_0 ,v}, g)$ and the $L^{\infty}$ norms of tensors are taken with respect to $|\c|_g$.
\begin{remark}
    The decay rate of renormalized $\trchb$ in $\widetilde{\mathbb{E}}^{N}_{0,out}$  is weaker than the one as required in $\mathbb{E}^N_{0,out}$ in \eqref{Eqn:out ID1} for the admissible Kerr black hole formation initial data. This is due to the forward construction of characteristic initial data along $H_{u_0}$. The issue is addressed and overcome in the next subsection.
    \end{remark}
\end{proposition}
\begin{proof}
We start with the estimation of $\zeta$. Note that from $\nab\O=\nab\O_{\SS}=0$ and 
\begin{equation*}
    \eta=\zeta+\nab(\log\O), \qquad \etab=-\zeta+\nab(\log\O),
\end{equation*}
we have $\eta=-\etab=\zeta$. Consequently, the torsion $\zeta$ satisfies the following linear first-order ordinary differential equation in stereographic coordinates on $\mathbb{S}^2$:
\begin{equation*}
\O^{-1}\f{\partial \zeta_a}{\partial v}+\trch \zeta_a=\xi_a,
\end{equation*}
where $\xi=\div\chih-\f12\nab \trch$. Note that by \Cref{3.2}, we have $\|\xi\|_{C^{N_1-3}_{2}}\lesssim \epsilon_1$. 

As a direct consequence of \Cref{gronwall lemma}, with the selection of  $k=2$ and $m=2$, we obtain the estimate for $\zeta$:
\begin{equation}\label{initial estimate zeta}
\|\zeta\|_{C^{N_1-3}_{1}}\lesssim \mathbb{E}^{N_1-3}_{seed}(\zeta)+\|\xi\|_{C^{N_1-3}_{2}} \lesssim \epsilon_1.
\end{equation}

We then move to bound $\trchb$. Recall $\trchb$ satisfies the following propagation equation 
\begin{equation}\label{trchb eqn}
\O^{-1} \f{\partial (\O\trchb)}{\partial v}+\trch  (\O\trchb)=2 \O\qty(-\bfK-\div\zeta+|\zeta|^2_{g})\eqqcolon 2\O\la,
\end{equation}
For $\widetilde{\trchb}=\trchb-\trchb_{\SS}$, the equation \eqref{trchb eqn} can be converted into
\begin{equation}\label{renormaltrchb eqn}
\O^{-1}\f{\partial (\O\widetilde{\trchb})}{\partial v}+\trch  (\O\widetilde{\trchb})=2\O\qty((\lambda+\f{1}{r_{\SS}^2})-\O\trchb_{\SS}\widetilde{\trch})\eqqcolon2\O\lambda'.
\end{equation}
Using \Cref{3.2} and \eqref{initial estimate zeta} we infer that $\|\lambda'\|_{C^{N_1-4}_2}\lesssim \epsilon_1$. Applying \Cref{gronwall lemma} for $\O\trchb$ with $k=2$ and $m=2$, we then deduce
\begin{equation*}
\|\O\widetilde{\trchb}\|_{C^{N_1-4}_1}\lesssim \mathbb{E}^{N_1-4}_{seed}(\trchb)+\|\lambda'\|_{C^{N_1-4}_2}\lesssim \epsilon_1,
\end{equation*}
which renders
\begin{equation}\label{initial estimate trchb}
\|\widetilde{\trchb}\|_{C^{N_1-4}_1}\lesssim \epsilon_1. \\[3mm]
\end{equation}

We turn to $\chibh$. The transport equation for $\chibh$ in stereographic chart reads 
\begin{equation*}
\O^{-1}\f{\partial (\O\chibh_{ab})}{\partial v}-\f{1}{2}\trch (\O\chibh_{ab})-(\chih\c\O \chibh) g_{ab}=\O\qty(-\nab \hat{\otimes}\zeta+\zeta \hat{\otimes}\zeta-\f{1}{2}\trchb \chih)_{ab}\eqqcolon\O \theta_{ab}.
\end{equation*}
By \Cref{3.2}, \eqref{initial estimate zeta} and \eqref{initial estimate trchb}, we get $\|\theta\|_{C^{N_1-4}_{2}}\lesssim \epsilon_1$ and $\|\chih \c (\O \chibh)\|_{C^{N_1-4}_{2}}\lesssim \|\O \chibh\|_{C^{N_1-4}_{-1}}\c\epsilon_1$.
Thus, applying \Cref{gronwall lemma} for $\O \chibh$ with $k=-1$ and $m=2$, we deduce
\begin{equation*}
\|\O \chibh\|_{C^{N_1-4}_{-1}}\leq C(\mathbb{E}^{N_1-4}_{seed}(\chibh)+\|\O \chibh\|_{C^{N_1-4}_{-1}}\c\epsilon_1).
\end{equation*}
Choosing $\epsilon_1<1/2C$, we then obtain
\begin{equation*}
\|\O \chibh\|_{C^{N_1-4}_{-1}}\lesssim \mathbb{E}^{N_1-4}_{seed}(\chibh)\lesssim \epsilon_1,
\end{equation*}
which yields
\begin{equation}\label{initial estimate chibh}
\| \chibh\|_{C^{N_1-4}_{-1}}\lesssim \epsilon_1. \\[2mm]
\end{equation}

Next, recall that $\beta$ obeys the below transport equation:
\begin{equation*}
\O^{-1}\f{\partial (\O^{-1} \b_{a})}{\partial v}+\f{3}{2} \trch (\O^{-1} \b_a)-\chih_{a}{}^{b} \cdot (\O^{-1}\b_b)=\qty(\div\a+(\etab+2\zeta)\c \a)_a\eqqcolon\kappa_a.
\end{equation*}
Utilizing \Cref{Est Product}, \Cref{3.2} and \eqref{initial estimate zeta}, we then obtain
\begin{equation*}
\|\kappa\|_{C^{N_1-4}_{3.5+\de_B/2}}\lesssim \epsilon_1, \qquad \|\chih\cdot (\O^{-1}\b)\|_{C^{N_1-4}_{3.5+\de_B/2}}\lesssim \|\O^{-1}\b\|_{C^{N_1-4}_{2.5+\de_B/2}}\c \epsilon_1.
\end{equation*}
Applying \Cref{gronwall lemma} for $\O^{-1} \b_{a}$ with $k=3$ and $m=3.5+\de_B/2$, we derive
\begin{equation*}
\|\O^{-1}\b\|_{C^{N_1-4}_{2.5+\de_B/2}}\leq  C(\mathbb{E}^{N_1-4}_{seed}(\b)+\|\O^{-1} \b\|_{C^{N_1-4}_{2.5+\de_B/2}}\c\epsilon_1).
\end{equation*}
Choosing $\epsilon_1 <1/2C$, we deduce
\begin{equation*}
\|\O^{-1}\b\|_{C^{N_1-4}_{2.5+\de_B/2}}\lesssim \mathbb{E}^{N_1-4}_{seed}(\b)\lesssim \epsilon_1,
\end{equation*}
from which we infer
\begin{equation}\label{initial estimate beta}
\|\b\|_{C^{N_1-4}_{2.5+\de_B/2}}\lesssim  \epsilon_1. \\[2mm]
\end{equation}

We proceed to treat $\rho$. Note $\rho$ satisfies 
\begin{equation}\label{Eqn:rho}
\O^{-1}\f{\partial \rho}{\partial v}+\f{3}{2} \trch \rho=\div\b+(2\etab+\zeta)\c \b-\f12 \chibh\c\a\eqqcolon\mu.
\end{equation}
In view of 
\begin{equation*}
    \O^{-1}\f{\partial \rho_{\SS}}{\partial v}+\f{3}{2} \trch_{\SS} \rho_{\SS}=0,
\end{equation*}
we transfer \eqref{Eqn:rho} into
\begin{equation*}
    \O^{-1}\f{\partial \widetilde{\rho}}{\partial v}+\f{3}{2} \trch \widetilde{\rho}=-\f32 \widetilde{\trch}\rho_{\SS}+\mu\eqqcolon \mu'.
\end{equation*}
By \Cref{3.2}, \eqref{initial estimate trchb}, \eqref{initial estimate chibh} and \eqref{initial estimate beta}, we obtain
\begin{equation*}
\|\mu'\|_{C^{N_1-5}_{4.5+\de_B/2}}\lesssim \epsilon_1. 
\end{equation*}
Employing \Cref{gronwall lemma} for $\widetilde{\rho}$ with $k=3$ and $m=4.5+\de_B/2$, we get
\begin{equation}\label{initial estimate rho}
\|\widetilde{\rho}\|_{C^{N_1-5}_{3}}\leq  \mathbb{E}^{N_1-5}_{seed}(\widetilde{\rho})\lesssim \epsilon_1. \\[2mm]
\end{equation}

Next, we consider $\sigma$ and its equation is given by 
\begin{equation*}
-\curl \zeta=\curl\etab=\f{1}{2}\chih \wedge \chibh-\sigma.
\end{equation*}
Using \Cref{3.2}, \eqref{initial estimate zeta}, \eqref{initial estimate chibh} we deduce
\begin{equation}\label{initial estimate sigma}
\|\sigma\|_{C^{N_1-4}_3}\lesssim\epsilon_1. \\[2mm]
\end{equation}

We appeal to the null Bianchi equations to derive the estimates for $\ab$ and $\bb$. In stereographic coordinates on $\mathbb{S}^2$, the null Bianchi equation for $\bb$ can be written in the form
\begin{equation*}
\O^{-1}\f{\partial (\O \bb_{a})}{\partial v}+\f{1}{2} \trch (\O \bb_a)-\chih_{a}{}^{B} \cdot (\O\bb_B)=\O\qty(-\nab\rho+\dual\nab\sigma+3\zeta \rho-3{}^\ast\zeta \sigma+2\chibh \cdot \beta)\eqqcolon \O\underline{\kappa}_a.
\end{equation*}
It follows from \Cref{3.2}, \eqref{initial estimate zeta}, \eqref{initial estimate rho}, \eqref{initial estimate sigma} that 
\begin{equation*}
\|\underline{\kappa}\|_{C^{N_1-6}_{3}}\lesssim \epsilon_1, \qquad \|\chih \cdot (\O\bb)\|_{C^{N_1-6}_{3}}\lesssim \|\O\bb\|_{C^{N_1-5}_{1}} \c \epsilon_1.
\end{equation*}
Employing \Cref{gronwall lemma} for $\O \bb_{a}$ with $k=1$ and $m=3$, we get
\begin{equation*}
\|\O\bb\|_{C^{N_1-6}_{1}}\leq  C(\mathbb{E}^{N_1-5}_{seed}(\bb)+\|\O \bb\|_{C^{N_1-6}_{1}}\c\epsilon_1).
\end{equation*}
Picking $\epsilon_1 <1/2C$, we arrive at
\begin{equation*}
\|\O\bb\|_{C^{N_1-6}_{1}}\lesssim \mathbb{E}^{N_1-5}_{seed}(\bb)\lesssim \epsilon_1,
\end{equation*}
which gives
\begin{equation}\label{initial estimate bb}
\|\bb\|_{C^{N_1-6}_{1}}\lesssim  \epsilon_1. \\[2mm]
\end{equation}

Recall that $\ab$ also obeys a transport equation, which in stereographic chart reads
\begin{equation*}
\O^{-1}\f{\partial (\O^{2}\ab_{ab})}{\partial v}-\f{1}{2}\trch (\O^{2}\ab_{ab})-(\chih \c \O^{2}\ab)g_{ab}=\O^2\qty(-\nab \hat{\otimes}\bb+5\zeta \hat{\otimes}\bb-3\chibh\rho+3{}^\ast \chibh \sigma)\eqqcolon\O^2\tau_{ab}.
\end{equation*}
Applying \Cref{3.2}, \eqref{initial estimate zeta}, \eqref{initial estimate chibh}, \eqref{initial estimate rho}, \eqref{initial estimate sigma}, \eqref{initial estimate bb}, we obtain
\begin{equation*}
\|\tau\|_{C^{N_1-7}_{1}}\lesssim \epsilon_1, \qquad \|\chih\c (\O^{2}\ab)\|_{C^{N_1-7}_{1}} \lesssim \|\O^{2}\ab\|_{C^{N_1-7}_{-1}}\c\epsilon_1.
\end{equation*}
By virtue of \Cref{gronwall lemma} for $\O^{2}\ab$ with $k=-1$ and $m=1$, we then deduce
\begin{equation*}
\|\O^{2}\ab\|_{C^{N_1-7}_{-1}} \leq C(\mathbb{E}^{N_1-7}_{seed}(\ab)+\|\O^{2}\ab\|_{C^{N_1-7}_{-1}}\c\epsilon_1).
\end{equation*}
Taking $\epsilon_1<1/2C$, we hence conclude
\begin{equation*}
\|\O^{2}\ab\|_{C^{N_1-7}_{-1}} \lesssim \mathbb{E}^{N_1-7}_{seed}(\ab) \lesssim \epsilon_1.
\end{equation*}
This implies
\begin{equation*}
\|\ab\|_{C^{N_1-7}_{-1}} \lesssim \epsilon_1. \\[2mm]
\end{equation*}

Also, note that we already have
\begin{equation*}
\O_{\SS}\omega=-\f12 \pr_v(\log \Omega)=-\f14 \c \f{2m_0}{r_{\SS}^2}=-\f{m_0}{2r_{\SS}^2}=\O_{\SS}\omega_{\SS}.
\end{equation*} 
Summarizing all the above weighted estimates we obtained for Ricci coefficients and curvature components, together with \eqref{equiv relation}, we then finish the proof of \Cref{initial outgoing data}.
\end{proof}

\subsection{Improved Decaying Estimate for Renormalized \texorpdfstring{$\trchb$}{}}
In this subsection, we aim to construct appropriate conformal metric $\hat{g}=\hat{g}(u_0, v, \th)$ along $u=u_0$, such that the renormalized $\trchb$ exhibits better decay in $r$ compared to the one that appears in $\widetilde{\mathbb{E}}^N_{0, out}$. Here we require that the conformal metric $\hat{g}$ belongs to the following class
\begin{equation*}
    \mathcal{C}^{N_1}_{\epsilon_1}\coloneqq \qty{\hat{g}\in C^{N_1}_{1.5+\de_B/2}: \ \det \hat{g}=\det \mathring{g}, \ \|\hat{g}-\mathring{g}\|_{C^{N_1}_{1.5+\de_B/2}}\le \epsilon_1}.
\end{equation*}

First notice that from the Gauss equation on the 2-sphere $S_{u_0, v}$ we have
\begin{equation*}
     \bfK=-\rho-\f14 \tr\chi \tr\chib+\f12 \chih\c\chibh.
\end{equation*}
Consequently, it follows from \Cref{initial outgoing data} that\footnote{Here all estimates are understood to also hold true for higher derivatives $(r_{\SS}\nab)^{i}$ up to $N$-th order.}
\begin{equation*}
    \bfK-\f{1}{r_{\SS}^2}=-\widetilde{\rho}-\f14\tr\chi_{\SS}\widetilde{\trchb}-\f14\tr\chib_{\SS}\widetilde{\trch}+\f12 \chih\c\chibh=-\f14\tr\chi_{\SS}\widetilde{\trchb}+O(\f{\epsilon_1}{r_{\SS}^3}).
\end{equation*}
This gives
\begin{equation*}
    \widetilde{\trchb}=O(r_{\SS})\c(\bfK-\f{1}{r_{\SS}^2})+O(\epsilon_1 r_{\SS}^{-2}).
\end{equation*}
Thus, to derive the desired decaying estimate for $\widetilde{\trchb}$, we only need to control renormalized $\bfK$ as below
\begin{equation}\label{Est:expect decay K}
    \bfK-\f{1}{r_{\SS}^2}=O(\epsilon_1 r_{\SS}^{-3}).
\end{equation}

To this end, recall from \eqref{Eqn:K conformal} 
\begin{equation}\label{Eqn:K conformal1}
        \bfK=\phi^{-2}\qty(\hat{\bfK}-\hat{\De}\log{\phi}).
    \end{equation}
    Here $\hat{\bfK}$ and $\hat{\De}$ denote the Gauss curvature and the Laplace--Beltrami operator of $(S_{u_0, v}, \hat{g})$.

Using the assumption that $\hat{g}=\mathring{g}+O(\epsilon_1 r_{\SS}^{-\f{3+\de_B}{2}})$ and \eqref{Est: phi trch} in \Cref{global existence for initial data}, there holds
\begin{equation*}
\hat{\bfK}=1+O(\epsilon_1 r_{\SS}^{-\f{3+\de_B}{2}}), \qquad \hat{\De}\log{\phi}=\mathring{\De} \log\phi+O(\epsilon_1 r_{\SS}^{-\f{3+\de_B}{2}})
\end{equation*}
with $\mathring{\De}$  being the Laplace--Beltrami operator on $(\ms, \mathring{g})$.

Inserting these into \eqref{Eqn:K conformal1}, we then obtain\footnote{Notice that $r_{\SS}=r_{\SS}(v)$ is constant on $S_{u_0, v}$.}
\begin{equation*}
    \bfK-\f{1}{r_{\SS}^2}=O(r_{\SS}^{-2})\c\qty((\f{\phi}{r_{\SS}})^2-1)+O(r_{\SS}^{-2})\c\mathring{\De}\log \f{\phi}{r_{\SS}}+O(\epsilon_1 r^{-\f{7+\de_B}{2}}_{\SS}),
\end{equation*}
which implies \eqref{Est:expect decay K} provided that
\begin{equation}\label{Cond:phi over rs}
    \f{\phi}{r_{\SS}}=1+O(\f{\epsilon_1}{r_{\SS}}).
\end{equation}
Observe that by virtue of \eqref{Apx:Eqn:phi rs} and \eqref{Est: phi trch}, i.e.,
\begin{equation*}
    \f{\partial}{\partial v}\log \f{\phi}{r_{\SS}}=\f{1}{2}\Omega\widetilde{\trch}=O(\epsilon_1 r_{\SS}^{-2}),
\end{equation*}
we can further relax \eqref{Cond:phi over rs} by only requiring $\lim\limits_{v\to \infty} \phi/r_{\SS}=1$.
\begin{remark}
    Note that the linearization of geometric quantities in this section is conducted slightly differently from that in \eqref{Eqn:linearizeSchw} from \Cref{Subsec:admiKerr}. Once \eqref{Est:expect decay K}  is achieved, the Gauss–Bonnet theorem implies that the area radius of $S_{u_0, v}$ satisfies $r=r_{\SS}+O(\epsilon_1)$. As a result, the errors between these two linearizations decay faster and can be neglected.
\end{remark}

Now we state the main conclusion of this subsection:
\begin{proposition}\label{Prop:mainglue}
    With $\hat{g}\in \mathcal{C}^{N_1}_{\epsilon_1}([v_0, \infty))$ freely prescribed along $H_{u_0}\cap\qty{v\ge v_0}$ and with $\pr_v^{n}(\hat{g}-\mathring{g})(v_0, \th)=0$ for all $n\ge0$, we can always find $\hat{g}\in\mathcal{C}^{N_1}_{\epsilon_1}([0, v_0])$ that smoothly extends $\hat{g}\big|_{H_{u_0}\cap [v_0, \infty)}$, such that the equation \eqref{2nd order ODE} admits the unique solution $\phi$ in $[0, \infty)\times \ms$ satisfying 
    \begin{equation*}
         \lim\limits_{v\to \infty} \f{\phi(v,\vartheta)}{r_{\SS}}=1 \qquad \text{for all} \quad \vartheta\in\ms.
    \end{equation*}
\end{proposition}
\begin{remark}
    The condition $\pr_v^{n}(\hat{g}-\mathring{g})(v_0, \vartheta)=0$ for all $n\ge0$ is not crucial. In fact, for any freely prescribed $\hat{g}$ on $H_{u_0}\cap\qty{v\ge v_0+1}$, we can smoothly extend $\hat{g}$ to $[v_0, v_0+1]\times \ms$ so that $\hat{g}(v_0)$ is equal to $\mathring{g}$ up to any order of derivatives in $v$.
\end{remark}

To handle the second order equation \eqref{2nd order ODE}, we introduce the change of variable $r_{\SS}=r_{\SS}(v)$ to eliminate the first-order term. Denoting $\phi'\coloneqq\pr_{r_{\SS}}\phi=\O^{-2}\pr_v \phi=(1-\f{2m_0}{r_{\SS}})^{-1}\pr_v \phi$, a straightforward computation yields
\begin{equation*}
    \phi''+\O^{-4}e\phi=\O^{-4}\qty(\f{\partial^2 \phi}{\partial v^2}-\f{2\partial (\log \Omega)}{\partial v} \f{\partial \phi}{\partial v}+e\phi)=0.
\end{equation*}
The initial conditions of \eqref{2nd order ODE} now become
\begin{equation}\label{Apx:Eqn:phiID}
    \phi_0\coloneqq\phi\big|_{S_{u_0, 0}}=|u_0|, \qquad \phi'_0\coloneqq \phi'\big|_{S_{u_0, 0}}=(\O^{-2}\f{\pr\phi}{\pr v})\big|_{S_{u_0, 0}}=(1-\f{2m_0}{|u_0|})^{-2}\eqqcolon\O_0^{-2}.
\end{equation}

The proof of \Cref{Prop:mainglue} is separated into two parts. We first solve  \eqref{2nd order ODE} in the late stage for $v\in[v_0, \infty)$ by imposing the boundary conditions for $\phi$ at $v=v_0$ and $v=\infty$.
\begin{lemma}\label{Apx:Lem:hatg1}
    For any $\hat{g}\in \mathcal{C}^{N_1}_{\epsilon_1}([v_0, \infty))$ and $\phi_{\de}\in C^{N_1}(\ms)$ obeying $\|\phi_{\de}- r_{\SS}(v_0)\|_{C^{N_1-1}(\ms)}\le \epsilon_1$ with $0<\epsilon_1\ll 1$, there exists a unique solution $\phi$ to \eqref{2nd order ODE} for $v\ge v_0$ and $\vartheta\in \ms$, such that
    \begin{equation*}
       \phi(v_0, \vartheta)=\phi_{\de}(\vartheta), \qquad \lim\limits_{v\to \infty} \f{\phi(v,\vartheta)}{r_{\SS}}=1 \qquad \text{for all} \quad \vartheta\in\ms.
    \end{equation*}
    Moreover, $\phi'_{\de}(\vartheta)\coloneqq\phi'(v_0, \vartheta)$ satisfies $1\le \phi'_{\de}\le 1+C\epsilon_1^2$.
\end{lemma}
\begin{proof}
    This proof is based on the Banach fixed point theorem. Consider the function space 
    \begin{equation*}
        X\coloneqq \qty{\phi\in C^0\qty([v_0, \infty);C^{N_1-1}(\ms)): \ \sup_{v\ge v_0} \|\phi(v,\c)-r_{\SS}(v)\|_{C^{N_1-1}(\ms)}\le 2\epsilon_1}.
    \end{equation*}
    For any $\psi \in C^0([v_0, \infty); C^{N_1-1}(\ms))$, we define the map $F: C^0([v_0, \infty);C^{N_1-1}(\ms))\to C^0([v_0, \infty);C^{N_1-1}(\ms))$ by letting
    \begin{equation}\label{Def:F map}
       \phi(r_{\SS}, \vartheta)=F(\psi)(r_{\SS}, \vartheta)\coloneqq \phi_{\de}(\vartheta)+r_{\SS}-r_{\de}+\int_{r_{\de}}^{r_{\SS}}\int_{s}^{\infty} (\O^{-4}e\psi)(t, \vartheta) dt ds.
    \end{equation}
Here $r_{\de}\coloneqq r_{\SS}(v_0)=r_{\SS}(\de)$. Note that by differentiating \eqref{Def:F map} twice in $r_{\SS}$, we directly get
\begin{equation}\label{Apx:Eqn:phi psi}
\phi'= 1+\int_{r_{\SS}}^{\infty} (\O^{-4}e\psi)(s, \vartheta)ds, \qquad
    \phi''=-\O^{-4}\psi.
\end{equation}
Since $\hat{g}\in \mathcal{C}^N_{\epsilon_1}([v_0, \infty))$, there holds
\begin{equation}\label{Apx:Est:e}
    \|e\|_{C^{N_1-1}(\ms)}=\f{1}{8}\|(\hat{g}^{-1})^{ac}(\hat{g}^{-1})^{bd}\f{\partial \hat{g}_{ab}}{\partial v}\f{\partial \hat{g}_{cd}}{\partial v}\|_{C^{N_1-1}(\ms)}\les \f{\epsilon_1^2}{r_{\SS}^{5+\de_B}}.
\end{equation}
As a result, noting that $\psi\in  C^0([v_0, \infty); C^{N_1-1}(\ms))$, we infer
\begin{equation*}
    \phi(r_{\de})=\phi_{\de}, \qquad \quad\lim\limits_{r_{\SS}\to \infty} \f{\phi}{r_{\SS}}=\lim\limits_{r_{\SS}\to \infty}\phi'=1.
\end{equation*}
Thus, for the rest it suffices to prove that $F$ has a unique fixed point.

We first establish that $F(X)\subseteq X$. Picking any $\psi \in X$, for all $v\ge v_0$ we have
\begin{align*}
    \|F(\psi)(v, \c)-r_{\SS}(v)\|_{C^{N_1-1}(\ms)}\le& \|\phi_{\de}-r_{\de}\|_{C^{N_1-1}(\ms)}+\int_{r_{\de}}^{r_{\SS}}\int_{s}^{\infty} \|\O^{-4}e\psi(t, \c)\|_{C^{N_1-1}(\ms)} dt ds \\
    \le& \epsilon_1+\int_{r_{\de}}^{r_{\SS}}\int_{s}^{\infty} \f{C\epsilon_1^2}{t^{5+\de_B}}\c (t+2\epsilon_1) dt ds\\
    \le& \epsilon_1+C\epsilon_1^2 \int_{r_{\de}}^{r_{\SS}} (s^{-3-\de_B}+s^{-4-\de_B}) ds=(1+C\epsilon_1)\epsilon_1\le 2\epsilon_1
\end{align*}
provided that $C\epsilon_1\le 1$. This implies that $F(\psi)\in X$.

We turn to show that $F$ is a contraction. For any given $\psi_1,\psi_2\in C^0\qty([v_0, \infty);C^{N_1-1}(\ms))$, in view of the definition of $F$ and \eqref{Apx:Est:e}, for any $v\ge v_0$ we deduce
\begin{align*}
    \|F(\psi_1)(v,\c)-F(\psi_2)(v,\c)\|_{C^{N_1-1}(\ms)}\le&\int_{r_{\de}}^{r_{\SS}}\int_{s}^{\infty}\|\O^{-4}e(\psi_1-\psi_2)(t,\c)\|_{C^{N_1-1}(\ms)}dt ds\\
    \le& \int_{r_{\de}}^{r_{\SS}}\int_{s}^{\infty}\f{C\epsilon_1^2}{t^{5+\de_B}}dt ds\c \|\psi_1-\psi_2\|_{C^0([v_0,\infty);C^{N_1-1}(\ms))}\\
    \le& C\epsilon_1^2\|\psi_1-\psi_2\|_{C^0([v_0,\infty);C^{N_1-1}(\ms))}\le \f12 \|\psi_1-\psi_2\|_{C^0([v_0,\infty);C^{N_1-1}(\ms))}.
\end{align*}
Choosing $C\epsilon_1^2\le \f12$, we infer that $F$ is contractive. 

Therefore, employing the Banach fixed point theorem, the map $F$ admits a unique fixed point $\phi\in X$ satisfying
\begin{equation*}
    \phi''+\O^{-4}e\phi=0 \qquad \text{and} \qquad \phi(v_0, \vartheta)=\phi_{\de}(\vartheta), \quad\lim\limits_{v\to \infty} \f{\phi(v,\vartheta)}{r_{\SS}}=1.
\end{equation*}
Furthermore, in view of \eqref{Apx:Eqn:phi psi}, \eqref{Apx:Est:e} and noting that $\phi\in X$, we deduce
\begin{equation*}
    1\le \phi'_{\de}=\phi'(r_{\de})=1+\int_{r_{\de}}^{\infty} \O^{-4}e\phi(s,\vartheta)ds\le 1+C\epsilon_1^2
\end{equation*}
as desired.
\end{proof}

We proceed to construct $\hat{g}$ in the short pulse region for $v\in[0, v_0]$. To this end, we smoothly connect $\phi$ from $[v_0, \infty)$ as derived in \Cref{Apx:Lem:hatg1} with its initial data \eqref{Apx:Eqn:phiID} at $v=0$. Then we set $e=-\O^4 \phi^{-1}\phi''$ and recover $\hat{g}$ from $e$.  It is worthwhile to mention that there is one topological obstruction in the last step. Recalling that the metric $\hat{g}$ has the same volume form as $\mathring{g}$, we hence obtain the continuous symmetric 2-tensor $\pr_v g_{ab}$ on $(\ms, \hat{g})$ is trace-free. According to the hairy ball theorem, $\pr_v g_{ab}$ must vanish at certain point on $\ms$, and so is $e$. To ensure that $e$ has zero on $\ms$ for each $v$, we need the following lemma:
\begin{lemma}\label{Apx:Lem:hatg2}
    Let $g_0, g_1\in C^{N_1-1}(\ms)$ be two functions with $0< g_1<g_0$. Then there exists a non-negative function $f=f(x,\vartheta)$ with $\sqrt{f}\in C^{\infty}([0, 1]; C^{N_1-1}(\ms))$ such that 
    \begin{enumerate}
        \item We have $f^{(n)}(1, \vartheta)=0$ for all $n\ge0$;
        \item For any $x\in[0, 1]$, $f(x, \c)$ vanishes in two small antipodal discs $D(x), -D(x)$ on $\ms$;
        \item The below equalities hold
    \begin{equation}\label{Apx:Eqn:mathch g0 g1}
        \int_0^{1} f(x, \vartheta)dx=g_0(\vartheta), \qquad \int_0^{1}\int_{0}^{x} f(y, \vartheta)dy dx=g_1(\vartheta).
    \end{equation}
    \end{enumerate}
\end{lemma}
\begin{proof}
    Choose a smooth, non-negative cut-off function $\rho(x)$ in $[0, \infty)$ such that $\phi(x)=1$ for $x\in[0, \f12]$, $0<\rho\le 1$ in $[\f12, 1)$ and $\phi\equiv 0$ when $x\ge1$. Also pick a smooth, non-negative cut-off function $\chi(x,\c)$ on $\ms$ that is vanishing in $B_{\de}(\vth_0(x))\cup B_{\de}(-\vth_0(x))$ and is identical to $1$ outside $B_{2\de}(\vth_0(x))\cup B_{2\de}(-\vth_0(x))$ with $0<\de\ll1$. Here $B_r(\vth)$ denotes the disk of radius $r$ centered at the $\vth$ on $\ms$. We let $-\vth$ denote the antipodal point of $\vth$, and $\vartheta_0(x)=(\th(x), \varphi(x))=(\pi/2, 2\pi x)$ represents the equator of $\ms$ in spherical coordinates $(\th,\varphi)$. Consider $f=f(x, \vartheta)$ in the below form
    \begin{equation*}
        f(x, \vartheta)=\rho^2(x)\chi^2(x, \vth)\Big(a(\vartheta)\psi_0(x)+b(\vartheta)\psi_1(x)\Big),
    \end{equation*}
    where $\psi_0(x), \psi_1(x)\ge0$ are auxiliary functions that will be determined later.

   Plugging the ansatz of $f(x, \vartheta)$ into \eqref{Apx:Eqn:mathch g0 g1}, we obtain the system of equations for $a(\vartheta),b(\vartheta)$ as follows:
    \begin{align*}
A_0 a+A_1 b\coloneqq& \qty(\int_0^1 \rho^2\chi^2\psi_0 dx)a+\qty(\int_0^1 \rho^2\chi^2\psi_1 dx)=g_0, \\
B_0 a+B_1 b\coloneqq& \qty(\int_0^1 \rho^2\chi^2\psi_0\c(1-x) dx)a+\qty(\int_0^1 \rho^2\chi^2\psi_1\c(1-x) dx)=g_1.
    \end{align*}
    Here we utilize the fact that 
    \begin{equation*}
        \int_0^{1}\int_{0}^{x} f(y, \vartheta)dy dx=\int_0^1 \int_y^1 f(y,\vartheta)dx dy=\int_0^1 f(y,\vartheta)(1-y)dy.
    \end{equation*}

    By solving the above equations, we get
    \begin{equation*}
        a=D^{-1}(B_1g_0-A_1 g_1), \qquad b=D^{-1}(-B_0 g_0+A_0 g_1) \qquad \text{with} \quad D\coloneqq A_0B_1-A_1B_0.
    \end{equation*}
    Thus, to guarantee that both $a(\vartheta), b(\vartheta)$ are positive, noting that $A_i, B_i>0$ for $i=0,1$, we only need to require
    \begin{equation}\label{Apx:Cond:BoverA}
      \f{B_0}{A_0}< \f{g_1}{g_0}< \f{B_1}{A_1}.
    \end{equation}

    Now we construct $\psi_0, \psi_1$ in $[0, 1]$ that fulfill \eqref{Apx:Cond:BoverA}. From $0<g_1<g_0$ we can choose $0<\si_1\ll 1$ so that
    \begin{equation*}
        0<7\si_1\le \min\limits_{\ms} \f{g_1}{g_0}\le \max\limits_{\ms} \f{g_1}{g_0} \le 1-7\si_1<1.
    \end{equation*}
    Then we pick 
    \begin{equation*}
        \psi_0(x)=\rho(\f{1-x}{\si_1}), \qquad \psi_1(x)=\rho(\f{x}{\si_1})+\si_1^2.
    \end{equation*}
    Consequently, a direct computation implies\footnote{Here $\text{dist}(\c, \c)$ denotes the geodesic distance on $\ms$.}
  \begin{align*}
B_0 &= \int_{1-\si_1}^1 \rho^2(x)\chi^2(x,\vth)\rho\left(\frac{1 - x}{\si_1}\right)(1 - x) dx \\
&= \int_0^{\si_1} \rho^2(1 - x)\chi^2(1-x, \vth)\rho\left(\frac{x}{\si_1}\right)x dx \le \si_1 \int_0^{\si_1} \rho^2(1 - x)\chi^2(1 - x, \vth)\rho\left(\frac{x}{\si_1}\right) dx = \si_1 A_0, \\
A_1 - B_1 &= \int_0^1 \rho^2\chi^2\psi_1 x dx \le \int_0^{\si_1} \rho^2(x)\chi^2(x,\vth)\rho\left(\frac{x}{\si_1}\right) x dx + \si_1^2 \le \int_0^{\si_1} x dx + \si_1^2 = \frac{3}{2} \si_1^2, \\
A_1 &\ge \int_0^{\frac{\si_1}{2}} \rho^2(x)\chi^2(x,\vth)\rho\left(\frac{x}{\si_1}\right) dx \ge m\left(\left\{x \in \left[0, \frac{\si_1}{2}\right] : \text{dist}(\vth, \vth_0(x)) \ge 2\de_1 \right\}\right) \\
&\ge \frac{\si_1}{2} - m\left(\left\{x: \, \text{dist}(\vth, \vth_0(x)) \le 2\de_1 \right\} \right)\ge \frac{\si_1}{2}-m(\qty{x: \, |2\pi x-\varphi|\le 2\de_1}) \ge \frac{\si_1}{2} - \frac{2\de_1}{\pi}.
\end{align*}
    Therefore, with the choice of $\de_1=\pi\si_1/4$ we conclude
    \begin{equation*}
        \f{B_0}{A_0}\le \si_1<7\si_1  \le \f{g_1}{g_0}\le 1-7\si_1< 1-6\si_1\le 1-\f{A_1-B_1}{A_1}=\f{B_1}{A_1}.
    \end{equation*}
    This gives that 
    \begin{equation*}
f= \rho^2\chi\qty(a\psi_0+b\psi_1)\ge0, \qquad    \sqrt{f}=\rho\chi\sqrt{a\psi_0+b\psi_1}\in C^{\infty}([0, 1]; C^{N_1-1}(\ms))
    \end{equation*}
    as stated.
\end{proof}

We are ready to prove \Cref{Prop:mainglue}. First, applying \Cref{Apx:Lem:hatg1}, for any $\phi_{\de}(\vartheta)=r_{\de}+O(\epsilon_1)$ we can find a unique $\phi=\phi(v,\vartheta)$ in $[v_0, \infty)\times\ms$ satisfying
\begin{equation}\label{Apx:late phi}
    \phi''+\O^{-4}e\phi=0 \qquad \text{with} \qquad \phi(r_{\de})=\phi_{\de}, \quad \phi'(r_{\de})=\phi'_{\de}\in[1, 1+C\epsilon_1^2], \quad \lim\limits_{r_{\SS}\to\infty} \f{\phi}{r_{\SS}}=1.
\end{equation}

Next we select $\phi_{\de}(\vartheta)=r_{\SS}+\epsilon_1(r_{\de}-|u_0|)$. By prescribing $\phi''(r_{\SS}, \vartheta)=\tilde{f}(r_{\SS}, \vartheta)$ in $[|u_0|, r_{\de}]\times\ms$, we get
\begin{align}
    \phi'=&\phi'_0+\int_{|u_0|}^{r_{\SS}} \tilde{f}(s)ds, \qquad \phi=\phi_0+\phi'_0(r_{\SS}-|u_0|)+\int_{|u_0|}^{r_{\SS}}\int_{|u_0|}^{s} \tilde{f}(t)dt ds, \nonumber\\
    e=&-\O^{4}\phi^{-1}\phi''=-\O^{4}\phi^{-1} \tilde{f}. \label{Apx:Eqn:e express}
\end{align}
Injecting these into the boundary conditions for $\phi$ at $r_{\SS}=|u_0|$ and $r_{\SS}=r_{\de}$, we get the below two integral equations
\begin{align}\label{Apx:Eqn:tildef}
    \int_{|u_0|}^{r_{\de}}\tilde{f}(s,\vartheta) ds=\phi'_{\de}-\phi'_0, \qquad
    \int_{|u_0|}^{r_{\de}}\int_{|u_0|}^{s}\tilde{f}(t, \vartheta) dt ds=\phi_{\de}-\phi_0-\phi_0'(r_{\SS}-|u_0|).
\end{align}
To employ \Cref{Apx:Lem:hatg2}, for $|u_0|\le r_{\SS}\le r_{\de}$ we introduce
\begin{equation*}
    f(x, \vartheta)=-\tilde{f}(r_{\SS},\vartheta) \qquad \text{with} \quad x=\f{r_{\SS}-|u_0|}{r_{\de}-|u_0|}.
\end{equation*}
Substituting $\tilde{f}(r_{\SS},\vartheta)$ with $f(x, \vartheta)$ in \eqref{Apx:Eqn:tildef}, we derive 
\begin{equation}\label{Apx:Eqn:f2}
\begin{aligned}
    \int_{0}^{1}f(x,\vartheta) dx=&-\f{\phi'_{\de}-\phi'_0}{r_{\de}-|u_0|}\eqqcolon g_0(\vartheta), \\
    \int_{0}^{1}\int_0^{x}f(y,\vartheta) dydx=&-\f{\phi_{\de}-\phi_0-\phi_0'(r_{\SS}-|u_0|)}{(r_{\SS}-|u_0|)^2}\eqqcolon g_1(\vartheta).
\end{aligned}
\end{equation}
By virtue of \eqref{Apx:Eqn:phiID}, \eqref{Apx:late phi} and our choice of $\phi_{\de}$, together with the fact that $0<\O_0^2=1-\f{2m_0}{|u_0|}<1$, we hence obtain
\begin{align*}
    0<&\f{\O_0^{-2}-1-C\epsilon_1^2}{r_{\de}-|u_0|}\le g_0\le \f{\O_0^{-2}-1}{r_{\de}-|u_0|}, \\
    g_1=&-\f{r_{\de}+\epsilon_1(r_{\SS}-|u_0|)-|u_0|-\O_0^{-2}(r_{\SS}-|u_0|)}{(r_{\SS}-|u_0|)^2}=\f{\O_0^{-2}-1-\epsilon_1}{r_{\SS}-|u_0|}>0, \\
    \f{g_1}{g_0}\le&\f{\O_0^{-2}-1-\epsilon_1}{\O_0^{-2}-1-C\epsilon_1^2}=\f{1-c\epsilon_1}{1-C\epsilon_1^2}\le 1-c'\epsilon_1<1,
\end{align*}
provided that $\epsilon_1\ll 1$. 

Therefore, using \Cref{Apx:Lem:hatg2}, there exists $f=f(x, \vartheta)\ge 0$ with $\sqrt{f}\in C^{\infty}([0, 1]; C^{N_1-1}(\ms))$ and $f^{(n)}(1, \vartheta)=0$ for all $n\ge0$, so that \eqref{Apx:Eqn:f2} holds true. Translating these back to $\tilde{f}(r_{\SS}, \vartheta)$, we deduce that $\tilde{f}\le0$, $|\tilde{f}|^{\f12}\in C^{\infty}([|u_0|, r_{\de}]; C^{N_1-1}(\ms))$ and
\begin{equation*}
 \tilde{f}^{(n)}(r_{\de}, \vartheta)=0 \qquad \text{for any} \quad n\ge0.
\end{equation*}
This further implies that 
\begin{equation*}
    e^{(n)}(r_{\de}, \vartheta)=0 \qquad \text{for any} \quad n\ge0,
\end{equation*}
which matches with $e=e(r_{\SS}, \vartheta)$ when $r_{\SS}\ge r_{\de}$.

Finally, to construct $\hat{g}=\hat{g}(r_{\SS}, \vth)$, we let
\begin{equation}\label{Eqn:hatgexpress}
    \hat{g}_{ab}(r_{\SS}, \vth)=(m^T)_{a}{}^{c}\mathring{g}_{cd}(\vth)m^d{}_{b}.
\end{equation}
Here $m_{ab}=m_{ab}(r_{\SS}, \vth)$ is an unknown positive-definite matrix with $\det (m^a{}_b)=1$, while we raise and lower indices with respect to the round metric $\mathring{g}$. Then we have $m^{a}{}_b=\exp (\Psi^{a}{}_b)$ for a symmetric, trace-free (with respect to $\mathring{g}$) matrix $\Psi_{ab}=\Psi_{ab}(r_{\SS}, \vth)$. 

Via a direct calculation we derive
\begin{equation*}
 \hat{g}'_{ab}=(m^T)'_{a}{}^{c}\mathring{g}_{cd}m^d{}_{b}+(m^T)_{a}{}^{c}\mathring{g}_{cd}m'^d{}_{b}=(m^T)_{a}{}^{c}\qty((\Psi^T)'_{c}{}^{e}\mathring{g}_{ed}+\mathring{g}_{ce}\Psi'^{e}{}_d)m^d{}_{b}=2(m^T)_{a}{}^{c}\Psi'_{cd}m^d{}_{b}.
\end{equation*}
Combining with \eqref{Eqn:hatgexpress}, this further yields
\begin{align*}
    e=\f18 (\hat{g}^{-1})^{aa'}(\hat{g}^{-1})^{bb'}\f{\partial \hat{g}_{ab}}{\partial v}\f{\partial \hat{g}_{a'b'}}{\partial v}=&\f12  (\hat{g}^{-1})^{aa'}(\hat{g}^{-1})^{bb'}\c\O^4 (m^T)_{a}{}^{c}\Psi'_{cd}m^d{}_{b}(m^T)_{a'}{}^{c'}\Psi'_{c'd'}m^{d'}{}_{b'}\\
    =&\f12\O^4(m^{-1})^{a}{}_{c}\Psi'^{c}{}_{d'}m^{d'}{}_{b'}(m^{-1})^{b'}{}_{c'}\Psi'^{c'}{}_{d}m^{d}{}_{a}\\
    =&\f12\O^4(m^{-1})^{a}{}_{c}(\Psi'^2)^{c}{}_{d}m^{d}{}_a=\f12\O^4(\Psi'^2)^a{}_a.
\end{align*}
Here we utilize the fact that
\begin{equation*}
    (\hat{g}^{-1})^{aa'}(m^T)_{a'}{}^{c'}\Psi'_{c'd'}m^{d'}{}_{b'}=(m^{-1})^{a}{}_{c}(\mathring{g}^{-1})^{cd}\qty((m^{T})^{-1})_d{}^{a'}(m^T)_{a'}{}^{c'}\Psi'_{c'd'}m^{d'}{}_{b'}=(m^{-1})^{a}{}_{c}\Psi'^{c}{}_{d'}m^{d'}{}_{b'}.
\end{equation*}
Thus, back to \eqref{Apx:Eqn:e express}, it is equivalent to
\begin{equation*}
    \f12(\Psi'^2)^a{}_{a}=-\phi^{-1} \tilde{f}.
\end{equation*}

Now, in spherical coordinates $(\th,\varphi)$ on $(\ms, \mathring{g})$, we define a symmetric, traceless 2-tensor $T$ by
\begin{equation*}
    T\coloneqq d\th^2-\sin^2\th d\varphi^2.
\end{equation*}
Recall that for any $r_{\SS}\in [|u_0|, r_{\de}]$, $\t{f}(r_{\SS}, \vth)$ vanishes in two small antipodal discs on $\ms$. Denote the centers of these two discs by $\vth_0(r_{\SS}), -\vth_0(r_{\SS})$. Then we can choose a smooth isometry $\Phi_{r_{\SS}}:\ms\to\ms$ so that $\Phi_{r_{\SS}}$ maps $\vth_0(r_{\SS}), -\vth_0(r_{\SS})$ to the North pole and the South pole of $\ms$ respectively and $\Phi_{r_{\SS}}$ also smoothly depends on $r_{\SS}$ in $ [|u_0|, r_{\de}]$.

Picking $T_{r_{\SS}}=(\Psi_{r_{\SS}}^{-1})^* T$, we have that $T_{r_{\SS}}$ is smooth except at $\vth_0(r_{\SS}), -\vth_0(r_{\SS})$. Then we let
\begin{equation*}
    \Psi^a{}_{b}=-\int_{r_{\SS}}^{r_{\de}} \phi^{-\f12}|\t{f}|^{\f12} (T_{r_{\SS}})^{a}{}_{b} ds,
\end{equation*}
which is smooth in $(r_{\SS},\vth)$ owing to the smoothness of $|\t{f}|^{\f12} T_{r_{\SS}}$.

Therefore, by noting $\t{f}\le0$ and $(T^2_{r_{\SS}})^{a}{}_{a}=(T^2)^{a}{}_{a}=2$, we conclude the desired equality
\begin{equation*}
    (\Psi'^2)^a{}_{a}=-\phi^{-1} \t{f} (T^2_{r_{\SS}})^{a}{}_{a}=-2\phi^{-1} \t{f}.
\end{equation*}
This completes the proof of \Cref{Prop:mainglue}.
\begin{remark}
    By tracking the process as conducted in this subsection, in the short pulse regime, i.e., $v\in [0, v_0]=[0, \de]$, we have that the size of $e=\f12|\chih|_{g}^2$ obeys
    \begin{equation*}
        e\sim \f{|\t{f}|}{|u_0|}\sim \f{|g_0|+|g_1|}{|u_0|}\sim \f{\O^{-2}_{0}-1}{(r_{\de}-|u_0|)|u_0|}\sim \f{m_0}{\de|u_0|^2}=\f{a}{|u_0|^2}.
    \end{equation*}
    In light of $\phi=\phi_0+O(\de), \O^2=\O^2_0+O(\de)$ for $v\in [0, \de]$ and $\phi'_{\de}=1+O(\epsilon_1)$, it also holds 
    \begin{equation*}
        \int_0^{\de} |\chih|_{g}^2 dv=2\int_0^{\de} edv=-2\int_{|u_0|}^{r_{\de}} \O^{2} \phi^{-1}\phi''dr_{\SS}=-2\O_0^{2} \phi_0^{-1}(\phi'_{\de}-\phi'_0)+O(\de)=\f{4m_0}{|u_0|^2}+O(\epsilon_1).
    \end{equation*}
\end{remark}
\begin{remark}
    In the proof of \Cref{Apx:Lem:hatg2}, due to the construction of $\psi_0$ and $\psi_1$ in the expression of $f$, there might be an additional factor $\epsilon_1^{-1}$ involved for controlling the derivatives of $e, \chih$ in the short-pulse region. However, this does not affect the hyperbolic argument in this region, as we can treat $\epsilon_1>0$ as a fixed small constant and choose the scale-critical short-pulse parameter $A\gg e^{\epsilon_1^{-1}}$. Consequently, the seed energy of the renormalized quantities at $S_{u_0, v_0}$ still satisfies
    \begin{equation*}
        \mathbb{E}^{N_1}_{seed}\le C_{\epsilon_1} \de A^{\f12}= C_{\epsilon_1} A^{-\f12}\le \epsilon_1.
    \end{equation*}
\end{remark}

 \section{The Pretorius--Israel Coordinate System in \texorpdfstring{$\intM$}{}}\label{Appendix:coordinate in Mint}
In this section, we derive an expression of the optical functions $\tu_0, \tub_0$ in the exact Kerr spacetime $Kerr(a, m)$ with $0\le|a|<m$. When $a=0$, as in the Schwarzschild coordinates $(t, r, \th,\varphi)$ we can set 
\begin{equation*}
    \tu_0=t-r-2m\ln |r-2m|,\qquad \tub_0=t+r+2m\ln |r-2m|.
\end{equation*}
When $a\neq 0$, without loss of generality, we assume that $a>0$. Recall that in Boyer-Lindquist coordinates, the Kerr metric takes the form of
\begin{equation*}
    \bfg_{a, m}=-\f{|q|^2\Delta}{\Si^2}dt^2+\f{\Si^2 \sin^2 \theta}{|q|^2}\Big(d\phi-\f{2amr}{\Si^2} dt \Big)^2+\f{|q|^2}{\Delta} dr^2+|q|^2 d\theta^2,
\end{equation*}
where
\begin{equation*}
    q\coloneqq r+ia\cos \theta, \qquad  \Delta\coloneqq r^2-2mr+a^2, \qquad \Si^2\coloneqq(r^2+a^2)^2-a^2 \sin^2 \theta \Delta.
\end{equation*}

Our goal is to construct a pair of optical functions, namely, 
\begin{equation}\label{Eqn:tu0 tub0}
    \tu_0=t-r_{*}(r, \theta),\qquad \tub_0=t+r_{*}(r, \theta),
\end{equation}
with $r_{*}(r, \theta)$ to be solved such that
\begin{equation*}
    \bfg_{a, m}^{\a \b} \pr_{\a} \tu_0 \pr_{\b} \tu_0=\bfg_{a, m}^{\a \b} \pr_{\a} \tub_0 \pr_{\b} \tub_0=-\f{\Si^2}{|q|^2\Delta}+\f{\Delta}{|q|^2} (\pr_{r} r_{*})^2+\f{1}{|q|^2} (\pr_{\theta} r_{*})^2=0.
\end{equation*}
Inspired by Pretorius--Israel \cite{P-I} and Dafermos--Luk \cite{D-L:Kerrint}, we define 
\begin{equation*}
    F(\ths; r, \th)=\int_{\ths}^{\th} \f{d \th'}{a\sqrt{\sin^2 \ths-\sin^2 \th'}}+\sgn(r-r_+)\int_{r_+}^{r} \f{dr'}{\Si(r', \ths)} \qquad \text{for} \quad 0< \th \le \ths < \pi/2, \quad r\in [r_-, \infty).
\end{equation*}
Here $r_{\pm}\coloneqq m\pm\sqrt{m^2-a^2}$, $\Si^2(r', \ths)\coloneqq(r^{\prime 2}+a^2)^2-a^2 \sin^2 \ths \Delta'$ and $\Delta'\coloneqq r^{\prime 2}-2mr^{\prime}+a^2$. 
\begin{remark}
   Here we aim to construct $\tu_0, \tub_0$ that is initialized on the event horizon $r=r_+$ throughout the entire spacetime. Note that the construction in \cite{P-I} originates from the future null infinity $r=\infty$, while in \cite{D-L:Kerrint}, Dafermos and Luk constructed $\tub_0$ initialized on $r=r_+$ only within the interior region $\qty{r_-\le r\le r_+}$.
\end{remark}
\vspace{2mm}

Utilizing the implicit function theorem, we first establish
\begin{lemma}\label{Apx:existence of ths}
    For any $\th\in (0, \pi/2)$ and $r\in [r_-, \infty)$, there exists a unique $\ths=\ths(r, \th)\in [\th, \pi/2)$ such that 
    \begin{equation*}
        F(\ths(\rthe); r, \th)=0.
    \end{equation*}
    Furthermore, the function $\ths=\ths(r, \th)$ is continuous for $(\th, r)\in (0, \pi/2)\times [r_-, \infty)$.
\end{lemma}
\begin{proof}
Firstly, for the case $r=r_+$, we have
\begin{equation*}
    F(\ths; r_+, \th)=\int_{\ths}^{\th} \f{d \th'}{a\sqrt{\sin^2 \ths-\sin^2 \th'}}
\end{equation*}
if and only if $\ths=\th$. This implies that $\ths(r_+, \th)=\th$ uniquely solves
$F(\ths; r_+, \th)=0$.

Regarding the scenario $r\neq r_+$, we then deal with two cases $r\in [r_-, r_+)$ and $r\in (r_+, \infty)$ separately.
\begin{enumerate}
    \item \textbf{The case $r\in [r_-, r_+)$:} Introducing the change of variables $x=\f{\sin \th'}{\sin \ths}$, we obtain
\begin{equation*}
    \int_{0}^{\ths} \f{d \th'}{a\sqrt{\sin^2 \ths-\sin^2 \th'}}=\int_{0}^{1}\f{dx}{a\sqrt{(1-x^2)(1-\sin^2 \ths x^2)}}.
\end{equation*}
Denote $\la=\sin^2 \ths$ and $\Si'=\Si (r', \ths)$. Since $\la=\sin^2 \ths$ is strictly increasing for $\ths\in (0, \pi/2)$, 
we can reformulate $F$ as
\begin{equation*}
     G(\la; r, \th)\coloneqq F(\ths; \rthe)=\int_{0}^{\th} \f{d \th'}{a\sqrt{\lambda-\sin^2 \th'}}-\int_{0}^{1}\f{dx}{a\sqrt{(1-x^2)(1-\la x^2)}}+\int_{r}^{r_+} \f{dr'}{\Si'}.
\end{equation*}

Note that for any fixed $\th\in (0, \pi/2)$ and $r\in [r_-, \infty)$, $ G(\la; r, \th)$ is a continuous function of $\la$. By virtue of
\begin{equation*}
    \lim\limits_{\la\to 1}  G(\la; r, \th)=-\infty, \qquad G(\sin^2 \th; \rthe)=F(\th; \rthe)=\int_{r}^{r_+} \f{dr'}{\Si'}>0,
\end{equation*}
we have that there exists $\la\in (\sin^2 \th, 1)$ such that $G(\la; \rthe)=0$. 

To establish the uniqueness of $\la$, we claim that $G(\la; \rthe)$ is strictly decreasing in $\la$, when $\la\in (\sin^2 \th, 1)$. Taking $\pr_{\lambda}$ on both sides, we deduce
\begin{equation*}
\begin{aligned}
    \pr_{\lambda} G=&-\f12 \Big( \int_{0}^{\th} \f{d \th'}{a(\la-\sin^2 \th')^{\f{3}{2}}}+ \int_{0}^{1}\f{x^2 dx}{a\sqrt{1-x^2}(1-\la x^2)^{\f32}}-\int_{r}^{r_+} \f{a^2 \Delta' dr'}{\Si^{\prime 3}}\Big) \\
    =&-\f12(I+II+III).
\end{aligned}
\end{equation*}
Clearly $I, III\ge 0$.\footnote{Note that $\De'=(r-r_+)(r-r_-)\le 0$ when $r\in[r_-, r_+)$.}  Since $1-\lambda x^2 \in [1-\la, 1]$, we also have
\begin{equation}
    II\ge \int_{0}^{1}\f{x^2 dx}{a\sqrt{1-x^2}}=\f{\pi}{4a}.
\end{equation} 
These together yield
\begin{equation*}
    \pr_{\la} G \le -\f{\pi}{8a}<0.
\end{equation*}
The monotonicity of $G(\la; \rthe)$ thus follows. Due to the one-to-one correspondence between $\la=\sin^2 \th$ and $\th$ for $\th\in [0, \pi/2]$, we hence conclude the uniqueness of $\ths=\ths(r, \th)$, which satisfies $F(\ths; \rthe)=0$.

\item \textbf{The case $r\in (r_+, \infty)$:} Proceeding similarly as above, we obtain that for $G(\la;r, \th)= F(\ths;r, \th)$, it holds
\begin{align*}
     \lim\limits_{\la\to 1}  G(\la; r, \th)=-\infty, \qquad G(\sin^2 \th; \rthe)=F(\th; \rthe)=\int_{r_+}^{r} \f{dr'}{\Si'}>0.
     \end{align*}
    We also have
     \begin{align*}
      \pr_{\lambda} G=&-\f12 \Big( \int_{0}^{\th} \f{d \th'}{a(\la-\sin^2 \th')^{\f{3}{2}}}+ \int_{0}^{1}\f{x^2 dx}{a\sqrt{1-x^2}(1-\la x^2)^{\f32}}-\int_{r_+}^{r} \f{a^2 \Delta' dr'}{\Si^{\prime 3}}\Big) \\
      =&-\f12(I+II+III).
\end{align*}
Note that we still have $I\ge 0$, while $III$ becomes negative when $r\in(r_+, \infty)$. In order to prove $\pr_{\lambda} G<0$,  in below we move to show that the sum $II+III$ possesses a positive lower bound. 

In view of the fact $1-\lambda x^2 \in [1-\la, 1]$, we obtain the same lower bound for $II$:
\begin{equation}\label{est of II}
    II\ge \int_{0}^{1}\f{x^2 dx}{a\sqrt{1-x^2}}=\f{\pi}{4a}.
\end{equation} 
To control $III$, we notice that 
\begin{equation}\label{est of III}
\begin{split}
    -III= \int_{r_+}^{r} \f{a^2 \Delta' dr'}{\Si^{' 3}} 
       \le&  \int_{r_+}^{\infty}\f{a^2 (r'-r_+)(r'-r_-)dr'}{r^{\prime 6}} \\
       \le& \int_{r_+}^{\infty}\f{a^2 dr'}{r^{\prime 4} }\le \f{a^2}{3r_+^{3}}<\f{1}{3a},
\end{split}
\end{equation}
where in the second inequality we use the fact that, for all $r'\in [r_+, r]$,
\begin{equation*}
    \Si'^{2}=(r'^2+a^2)^2-a^2 \la \Delta'\ge (r'^2+a^2)^2-a^2 \Delta'=r'^4+2a^2 r'^2+2ma^2r'>r'^4.
\end{equation*}
Combining \eqref{est of II} and \eqref{est of III}, we arrive at the desired
\begin{equation*}
    \pr_{\la} G \le -\f12(\f{\pi}{4a}-\f{1}{3a})=-\f{3\pi -4}{24 a}<0.
\end{equation*}
\end{enumerate}

Next we turn to establish that $\ths=\ths(r,\th)$ is continuous in $r\in (r_-, \infty)$ and $\th\in(0, \pi/2)$. The continuity of $\ths=\ths(r,\th)$ on the complement of $\qty{r=r_+}$ follows directly by applying the implicit function theorem for $F(\ths;r, \th)=0$ in the respective sub-regions.

To check the continuity of $\ths=\ths(r,\th)$ or $\la=\la(r,\th)$ across the $r=r_+$,  we examine the below equation for $\ths(r,\th)$:
\begin{equation}\label{Apx:eqn F}
      \int_{\th}^{\ths}  \f{d \th'}{a\sqrt{\sin^2 \ths-\sin^2 \th'}}=\abs{\int_{r}^{r_+} \f{dr'}{\Si'}}\sim |r-r_+|.
\end{equation}
Letting $x=\f{\sin \th'}{\sin \ths}$, we estimate its LHS and get
    \begin{equation}\label{Apx: Est LHS1}
        \begin{split}
             \int_{\th}^{\ths}  \f{d \th'}{a\sqrt{\sin^2 \ths-\sin^2 \th'}}=&\int_{\f{\sin \th}{\sin \ths}}^1 \f{dx}{a\sqrt{(1-x^2)(1-x^2\sin^2 \ths)}}
        \\
        \ge& \int_{\f{\sin \th}{\sin \ths}}^1  \f{dx}{a\sqrt{1-x^2}}=\f{1}{a}\l\f{\pi}{2}-\arcsin(\f{\sin \th}{\sin \ths})\r\eqqcolon \f{\gamma}{a}.
        \end{split}
    \end{equation}
Along with \eqref{Apx:eqn F} we now have
\begin{equation*}
    0\le \gamma \lesssim a|r-r_+| \qquad \text{if} \quad |r-r_+|\ll 1.
\end{equation*}
Consequently it yields 
\begin{equation*}
    \sin\ths-\sin\th=\f{\sin\th}{\cos\gamma}-\sin\th=\f{2\sin\th}{\cos\gamma}\sin^2\f{\gamma}{2}\lesssim |r-r_+|^2.
\end{equation*}
This indicates that
\begin{equation*}
    \lim\limits_{r\to r_+} \la(r,\th)=\sin^2\th.
\end{equation*}
We hence complete the proof of this lemma.
\end{proof}

We further define $\ths(\rthe)=\pi-\ths(r, \pi-\th)$ for $\th\in (\pi/2, \pi)$ and set $\ths(r, 0)=\ths(r, \pi)=0, \,\ \ths(r, \pi/2)=\pi/2$. To prove the continuity of $\ths(r, \th)$ at $\th=0, \pi/2, \pi$, we appeal to deriving the following estimate:
\begin{lemma}\label{Apx: est for ths}
    For any $r\in [m, \infty)$ and $\th\in(0, \pi/2)\cup (\pi/2, \pi)$, there holds
    \begin{equation}\label{Apx eqn: est for ths}
        \f{\sin \ths}{\sin \th}+\f{\cos \th}{\cos \ths}\le C,
    \end{equation}
    with $C>0$ being independent of $m, a$.
\end{lemma}
\begin{proof}
    It is sufficient to show that \eqref{Apx eqn: est for ths} is valid when $\th\in (0, \pi/2)$. Recall the implicit definition of $\ths=\ths(\rthe)$:
    \begin{equation}\label{Apx: ths eqn}
        \int_{\th}^{\ths}  \f{d \th'}{a\sqrt{\sin^2 \ths-\sin^2 \th'}}=\abs{\int_{r}^{r_+} \f{dr'}{\Si'}}.
    \end{equation}
    Note that here we have $\Si'^2=(r'^2+a^2)^2-a^2 \sin^2 \ths \Delta'\ge (r'^2+a^2)^2$ when $r'\in [r_-, r_+]$ and $\Si'\ge r'^2$ for all $r'\in [r_+, \infty)$. Consequently, the term on the right of \eqref{Apx: ths eqn} obeys 
    \begin{equation}\label{Apx: Est RHS}
    \begin{split}
        \abs{\int_{r}^{r_+} \f{dr'}{\Si'}}\le& \max\qty(\int_{r_-}^{r_+} \f{dr'}{r'^2+a^2}, \  \int_{r_+}^{\infty} \f{dr'}{r'^2})\\
        \le&  \f{1}{a} \max\qty(\arctan (\f{r_+}{a})-\arctan (\f{m}{a}), \ \f{a}{r_+})\eqqcolon\f{c_{m, a}}{a}.
    \end{split}
    \end{equation}
   Since $0<a<m$ and $r_+=m+\sqrt{m^2-a^2}>m$, we deduce that
   \begin{equation*}
       0<c_{m, a}\le \max\qty(\f{\pi}{2}-\f{\pi}{4}, \ 1)=1.
   \end{equation*}
    
    We then proceed to estimate the term on the left of \eqref{Apx: ths eqn}. When $\th$ is near $\pi/2$, by the change of variables $x=\f{\cos\th'}{\cos\th}$, there holds
    \begin{equation}\label{Apx: Est LHS2}
        \begin{split}
               \int_{\th}^{\ths}  \f{d \th'}{a\sqrt{\sin^2 \ths-\sin^2 \th'}}=& \int_{\th}^{\ths}  \f{d \th'}{a\sqrt{\cos^2 \th'-\cos^2 \ths}} \\
         =&\int_{1}^{\f{\cos \th}{\cos \ths}}  \f{d x}{a\sqrt{(x^2-1)(1-x^2\cos^2 \ths)}} \\
         \ge& \int_{1}^{\f{\cos \th}{\cos \ths}}  \f{d x}{a\sqrt{x^2-1}}=\f{1}{a}\ln\l \f{\cos \th}{\cos \ths}+\sqrt{(\f{\cos \th}{\cos \ths})^2-1} \r.
        \end{split}
    \end{equation}
    Inserting \eqref{Apx: Est LHS1}, \eqref{Apx: Est RHS}, \eqref{Apx: Est LHS2} into \eqref{Apx: ths eqn}, we now arrive at
    \begin{align*}
         \f{\sin \th}{\sin \ths}\ge& \sin(\f{\pi}{2}-c_{m, a})=\cos c_{m, a}\ge \cos 1,\\
         \f{\cos \th}{\cos \ths}\le& \f{e^{c_{m, a}}+e^{-c_{m, a}}}{2}=\cosh c_{m, a}\le \cosh 1.
    \end{align*}
    Taking $C=\sec 1+\cosh 1$, we thus complete the proof of this lemma.
\end{proof}
\begin{remark}
    The estimate \eqref{Apx eqn: est for ths} does not hold for $r\in [r_-, m]$. In fact, we have
    \begin{equation*}
        \lim\limits_{a\to 0} \sup\limits_{\th\in(0, \pi/2)}\f{\sin \ths(r_-, \th)}{\sin \th}=\infty.
    \end{equation*}
\end{remark}
\begin{remark}\label{Apx:rmk:D L}
    In \cite{D-L:Kerrint} Dafermos--Luk proved that $\rs=\rs(r,\th)$ and $\ths=\ths(r,
    \th)$ are smooth for $r\in (r_-, r_+)$
    and $\th\in[0, \pi]$.\footnote{The left endpoint $r=r_-$ is excluded because $\ths=\ths(r,\th)$ is singular near $r=r_-$ when $a=0$. The smoothness on $\th\in[0, \pi]$ refers to the natural differential structure on $\ms$, with $\th$ serving as the polar angle.} Even through $\rs(r,\th)$ is singular at $r=r_+$, we can accordingly replace $\rs$ with 
    \begin{equation*}
         \t{r}=e^{\kappa \rs}+r_+=g(r,\th)\c(r-r_+)+r_+ \qquad \text{with} \quad \kappa=\f{r_+-r_-}{2mr_+}, \ \ g\sim \f{1}{r},
    \end{equation*}
    which remains smooth as $r$ approaches $r_+$. Their method in \cite{D-L:Kerrint} can be further extended to prove the smoothness of $\t{r}$ and $\ths$ in the range when $r>r_+$, as the involved integrals from $r_+$ to $\infty$ are bounded. This property has already been utilized in the proofs of \Cref{Apx:existence of ths} and \Cref{Apx: est for ths}. In addition, we note that when restricting to the region $r\in[m, \infty)$, the derivatives of $\t{r}$ and $\ths$ exhibit uniform bounds that do not diverge as $a\to 0$. 
\end{remark}

With the conclusions of \Cref{Apx:existence of ths} and \Cref{Apx: est for ths}, we further define 
\begin{equation*}
    \rho(r, \th, \ths)\coloneqq\int_{r}^{r_+}\f{r'^2+a^2-\Si'}{\Delta'} dr'-\sgn(r-r_+)\int_{\ths}^{\th} a\sqrt{\sin^2 \ths-\sin^2 \th'} d\th'.
\end{equation*}
Thanks to 
\begin{equation*}
    \f{r'^2+a^2-\Si'}{\Delta'}=\f{a^2 \sin^2 \ths }{r'^2+a^2+\Si'} \le \f{a^2 \sin^2 \ths }{r'^2+a^2},
\end{equation*}
we have the integrability of the first term. Thus $\rho(r, \th, \ths)$ is well-defined for all $ 0\le \th \le \ths \le \pi/2$ and $r\in [r_-, \infty)$. We then set
\begin{equation*}
    r_{*}(\rthe)=f(r)+f_{*}(r, \th)=f(r)+\rho(r, \th, \ths(\rthe)) \qquad \text{with} \quad f'(r)=\f{r^2+a^2}{\Delta},
\end{equation*}
which is continuous in $[0, \pi]\times (r_-, \infty)$ since $\ths(r_+, \th)=\th$. Consequently, a direct computation yields
\begin{equation}\label{Apx:eqn:pr rs}
\begin{aligned}
    \pr_{r} \rs=&\f{r^2+a^2}{\Delta}-\f{r^2+a^2-\Sis}{\Delta}+\pr_{r} \ths\int_{r}^{r_+}\f{a^2 \sin \ths \cos \ths}{\Si'} dr'-\sgn(r-r_+)\pr_{r} \ths \int_{\ths}^{\th} \f{a\sin \ths \cos \ths}{\sqrt{\sin^2 \ths-\sin^2 \th'} }d\th' \\
    =&\f{\Sis}{\Delta}-\sgn(r-r_+)a^2\pr_{r} \ths \sin \ths \cos \ths \cdot F(\ths(\rthe); \rthe)=\f{\Sis}{\Delta}, \\
    \pr_{\th} \rs=&a\bs+\pr_{\th} \ths\int_{r}^{r_+}\f{a^2 \sin \ths \cos \ths}{\Si'} dr'-\sgn(r-r_+)\pr_{\th} \ths \int_{\ths}^{\th} \f{a\sin \ths \cos \ths}{\sqrt{\sin^2 \ths-\sin^2 \th'} }d\th' \\
    =&a\bs-\sgn(r-r_+)a^2\pr_{\th} \ths \sin \ths \cos \ths \cdot F(\ths(\rthe); \rthe)=a\bs,
\end{aligned}
\end{equation}
where $\Sis\coloneqq\Si(r, \ths)$, $\bs\coloneqq -\sgn(r-r_+)\sqrt{\sin^2 \ths-\sin^2 \th}$. In the derivation of \eqref{Apx:eqn:pr rs} we also use the fact that $F(\ths(\rthe); \rthe)=0$ from the definition of $\ths(\rthe)$. With the aid of \eqref{Eqn:tu0 tub0} and \eqref{Apx:eqn:pr rs}, now we can easily check that
\begin{align*}
    \bfg_{a, m}^{\a \b} \pr_{\a} \tu_0 \pr_{\b} \tu_0=\bfg_{a, m}^{\a \b} \pr_{\a} \tub_0 \pr_{\b} \tub_0 
    =&-\f{\Si^2}{|q|^2\Delta}+\f{\Delta}{|q|^2} (\pr_{r} r_{*})^2+\f{1}{|q|^2} (\pr_{\theta} r_{*})^2 \\
    =&-\f{\Si^2}{|q|^2\Delta}+\f{\Delta}{|q|^2}\cdot \f{(r^2+a^2)^2-a^2\sin^2\ths \Delta}{\Delta^2}+\f{a^2(\sin^2 \ths-\sin^2 \th)}{|q|^2}=0.  
\end{align*}

\vspace{2mm}
Employing \eqref{Eqn:tu0 tub0}, we can also express the incoming optical function $\tub_0$ as 
\begin{equation*}
    \tub_0=t+\rs(r,\th)=t+f(r)+\fs(r,\th)=\ub+\fs(r,\th).
\end{equation*}
Thus, in view of \Cref{Apx:rmk:D L}, we have that $\tub_0=\tub(\ub, r, \th)$ is smooth in $(\ub, r, \th,\varphi)$ coordinates for $r\in [m, r_+)\cup (r_+, \infty)$. We then turn to show the smoothness of $\tub_0=\tub(\ub, r, \th)$ when $r$ passes through $r=r_+$.
\begin{lemma}
    The incoming optical function $\tub_0=\tub(\ub, r,\th)$ is smooth near $r=r_+$.
\end{lemma}
\begin{proof}
    Instead of computing the derivatives of $\tub_0$ directly, we regard $\tub$ as the solution of the eikonal equation
    \begin{equation}\label{Apx:incoming eikonal eqn}
        \bfg_{a, m}^{\a \b} \pr_{\a} \tub \pr_{\b} \tub=0 \qquad \text{with} \quad \tub=\tub_0 \quad \text{along} \quad r=r_+.
    \end{equation}
    Our aim here is to show that the solution to the eikonal equation \eqref{incoming eikonal eqn} is unique and smooth near $r=r_+$. If this is true, the desired result readily follows.
    
    Defining $F(x, p)\coloneqq \bfg^{\a \b}_{a, m}p_{\a} p_{\b}$, we then proceed similarly as in the proof of \Cref{Lem:est for tub} and reformulate \eqref{incoming eikonal eqn} as\footnote{Although the Kerr metric $\bfg_{a,m}$ degenerates near $\th=0, \f{\pi}{2}$ in $(\ub,r,\th,\varphi)$ coordinates, we can replace $(\th,\varphi)$ with the regular coordinate system $(x^1, x^2)=(\sin\th\cos\varphi, \sin\th\sin\varphi)$ for $\th\in [0, \f{\pi}{3}]\cup [\f{2\pi}{3}, \pi]$, and the entire argument in this proof still works.}
   \begin{equation*}
       F(x, D\tub)=0 \qquad \text{with} \quad   x=(\ub, r,\th,\varphi) \quad \text{and} \quad  D\tub\coloneqq(\pr_{\ub} \tub, \pr_{r} \tub, \pr_{\th} \tub, \pr_{\varphi} \tub).
   \end{equation*}
    With initial data prescribed in \eqref{Apx:incoming eikonal eqn}, we can further transform the equation above to the below Hamiltonian system for the characteristic curve $x=x(s)$ with $p=\f{dx(s)}{ds}$:
   \begin{equation}\label{Apx:Eqn: Character for hb}
      \left\{ \begin{aligned}
           &\f{d}{ds} p_{\a}=D_{x^{\a}}F(x, p), \\
           &\f{d}{ds} x^{\a}=-D_{p_{\a}} F(x, p)=-2\bfg^{\a \b}_{a,m}p_{\b}, \\
       \end{aligned} \right.
   \end{equation}
with  initial conditions
\begin{equation*}
     x(0)=(\ub, r_+, \th, \varphi), \quad p(0)=D\tub_0(\ub, r_+, \th, \varphi).
\end{equation*}
Now the solution $\tub$ to \eqref{Apx:incoming eikonal eqn} can be written as
\begin{equation*}
    \tub(x(s))=\tub(x(0))-\int_{0}^{s} D_{p_{\a}} F(x(s'), p(s')) p_{\a} (s') ds'=\ub-2\int_{0}^{s} \bfg^{\a \b}_{a,m}p_{\a} p_{\b} ds'.
\end{equation*}
Here we use the fact that
\begin{equation*}
    \ths(r_+,\th)=\th \qquad \text{and}\qquad \tub_0=\ub+\rho(r_+, \th, \ths(r_+,\th))=\ub \qquad \text{on} \quad r=r_+.
\end{equation*}

Applying \eqref{Apx:eqn:pr rs} and noting that $\ths(r_+,\th)=\th$, $\De\big|_{r=r_+}=0$, along $r=r_+$ we hence deduce
\begin{align}
    \pr_{\ub} \tub_0=&1, \qquad \pr_{\th}\tub_0=\pr_{\th}\fs=\pr_{r}\rs=-\sgn(r_+-r_+)a\sqrt{\sin^2\ths(r_+,\th)-\sin^2\th}=0, \label{Eqn:pr th tub0}\\
    \quad \pr_{r}\tub_0=&\pr_{r}\fs=\pr_{r}\rs-f'(r)=-\lim\limits_{r\to r_+}\f{r^2+a^2-\Sis}{\De}=-\lim\limits_{r\to r_+}\f{a^2 \sin^2\ths}{r^2+a^2+\Sis}=-\f{a^2\sin^2\th}{2(r_+^2+a^2)}. \nonumber
\end{align}
This infers that the initial datum
\begin{equation*}
    p(0)=D\tub_0(\ub, r_+, \th, \varphi)=(1, -\f{a^2\sin^2\th}{2(r_+^2+a^2)}, 0, 0)
\end{equation*}
is smooth in $(\ub,\th,\varphi)$ along the event horizon $r=r_+$. 

Conducting analogously the process as in the proof of \Cref{Lem:est for tub}, we are thus able to establish the desired local existence, uniqueness and smoothness for the solution $x=x(s,\ub,\th,\varphi)$ to \eqref{Apx:Eqn: Character for hb}. Furthermore, the corresponding flow map
$\Phi=x(s,\ub,\th,\varphi)$ is invertible. These together yield the smoothness of $\tub$ near $r=r_+$, and $\tub$ must be identical to $\tub_0$ around $r=r_+$.
\end{proof}

With above arguments, we conclude that $\tub_0=\tub_0(\ub,r,\th)$ is smooth for $(\ub,r,\th)\in\mathbb{R}\times [m, \infty)\times [0, \pi]$. As a consequence, we derive
\begin{lemma}\label{Apx:tla limit}
 For any $\th\in(0, \pi/2)\cup (\pi/2, \pi)$, the following limit equality holds
 \begin{align*}
     \lim\limits_{r\to r_+} \f{\pr_{\th}\la-\sin(2\th)}{\sqrt{\sin^2\ths-\sin^2\th}}=0.
 \end{align*}
 \end{lemma}
 \begin{proof}
     In view of \eqref{Apx:eqn:pr rs}, we have
     \begin{equation*}
         \pr_{\th} \tub_0=\pr_{\th}\rs=-\sgn(r-r_+) \sqrt{\sin^2\ths-\sin^2\th}. 
     \end{equation*}
     Differentiating this equation in $\th$ gives
     \begin{equation*}
         \pr^2_{\th} \tub_0=-\sgn(r-r_+)\f{\pr_{\th}\la-\sin(2\th)}{2 \sqrt{\sin^2\ths-\sin^2\th}} \qquad \text{for any} \quad r\neq r_+, \  \th\not\in \qty{0, \f{\pi}{2}, \pi}.
     \end{equation*}

Meanwhile, from \eqref{Eqn:pr th tub0} we also have $\pr^2_{\th} \tub_0\equiv 0$ along $r=r_+$. Since $\tub_0$ is smooth near $r=r_+$, we must have
     \begin{equation*}
         \lim\limits_{r\to r_+} \f{\pr_{\th}\la-\sin(2\th)}{\sqrt{\sin^2\ths-\sin^2\th}}=0
     \end{equation*}
     as stated.
 \end{proof}
We note that \Cref{Apx:tla limit} will be used in \Cref{Appendix:Null expansion}.

\section{Properties of Bondi Mass along Future Null Infinity}\label{Appendix:property Bondi mass}
In this section, we establish the Bondi mass loss formula and calculate the limit of Bondi mass at the timelike infinity. Proceeding similarly as in Section 8.2 of \cite{KS:main}, we first construct a double null foliation in $\Ext\MM$ as follows:
 \begin{itemize}
        \item We solve for the outgoing optical function $\tu$ and the incoming optical function $\Ext\tub$, which satisfies
            \begin{equation*}
         \bfg^{\a \b}\pr_{\a} \tu \,  \pr_{\b} \tu=0 \qquad \text{and}\qquad \bfg^{\a \b}\pr_{\a} \Ext\tub \,  \pr_{\b} \Ext\tub=0.
    \end{equation*}
    \item Let $\tH_{\tu}$ be the level set of $\tu$ and denote $\tS_{\tu, \Ext\tub}$ as the intersection of $\tH_{\tu}$ and the constant $\Ext\tub$ hypersurface. Then we introduce the corresponding null frame\footnote{Throughout the proof of this proposition, we use tildes to denote the null frame with respect to the double null foliation constructed in the exterior region $\Ext\MM$.} 
\begin{equation}\label{Eqn:def double null}
     \te_1, \te_2\in T\tS_{\tu, \Ext\tub}, \qquad \te_3=-2\O\bfD \Ext\tub, \qquad \te_4=-2\O \bfD u \qquad \text{with} \quad \O^{-2}\coloneqq-\f12\bfg(\bfD\tu, \bfD\Ext\tub).
\end{equation}
\item We assign a local chart $(\Ext\tth^1, \Ext\tth^2)$ on $\tS_{\tu, \tub}$ such that
$\te_4(\Ext\tth^a)=0$ with $a=1,2$.
    \end{itemize}
\vspace{2mm}

With the double-null foliation established, we move to derive the precise hyperbolic estimates with respect to the newly constructed null frame $(\te_{\mu})$.
\begin{lemma}\label{Appendix:hyper est new}
Relative to the null frame $(\te_1, \te_2, \te_3, \te_4)$, adapted to the double null foliation $(\tu, \Ext\tub, \Ext\tth^1, \Ext\tth^2)$ we have
    \begin{equation}\label{Est in double null frame}
    \begin{aligned}
    \tr\t{\chi}=&\f{2}{r}+r^{-1}\Ga_b^{(2)}, &\qquad  \tr\t{\chib}=&-\f{2}{r}\qty(1-\f{2m_{\infty}}{r})+O(r^{-2})+\Ga_b^{(2)}, \\
        \t{\chih}=&O(r^{-3})+r^{-1}\Ga_b^{(2)}, &\qquad  \t{\chibh}=&O(r^{-3})+\Ga_b^{(2)}, \\
        \t{\eta}=&O(r^{-2})+r^{-1}\Ga_b^{(2)},&\qquad
        \t{\rho}=&-\f{2m_{\infty}}{r^3}+O(r^{-4})+r^{-2}\Ga_b^{(1)}.
    \end{aligned}
\end{equation}
Here $\Ga_g$ and $\Ga_b$ satisfy the below estimates with $\fd\coloneqq \qty{\nab_3, r\nab_4, r\nab}$ and $0<\de_{dec}\ll 1$,
\begin{equation*}
    |\fd^{\le k_{small}}\, \Ga_g|\lesssim \epsilon_0 r^{-2} u^{-\f12-\de_{dec}}, \qquad |\fd^{\le k_{small}}\, \Ga_b|\lesssim \epsilon_0 r^{-1} u^{-\f12-\de_{dec}},
\end{equation*}
and we denote $\Ga_g^{(s)}=\fd^{\le s} \Ga_g$ and $\Ga_b^{(s)}=\fd^{\le s} \Ga_b$.
\end{lemma}
\begin{remark}
    Notice that for $\t{\eta}$, we have better decaying in $r$ compared to the estimate of $\eta$ in \eqref{KS hyperbolic Est ext}. And we have $\xib=0$ within the double null foliation. These two facts play an important role in the derivation of the Bondi mass loss formula.
\end{remark}
\begin{proof}
    In view of hyperbolic estimates from \cite{KS:main}, within the exterior region $\Ext\MM$, relative to the outgoing PG structure with the coordinate system $(u, r, \th, \varphi)$ and the corresponding null frame $(e_1, e_2, e_3, e_4)$, we have\footnote{Note that $e_4(u)=e_4(\th)=e_4(\varphi)=\nab(r)=0, e_4(r)=1$ and $\omega=\xi=0$ are ensured by the conditions of outgoing PG structures.}
\begin{equation}\label{KS hyperbolic Est ext}
    \begin{aligned}
    e_3(u)=&2+O(r^{-2})+\Ga_b, &\quad e_3(r)=&-(1-\f{2m_{\infty}}{r})+O(r^{-2})+r\Ga_b,\\
    \nab(u),\nab(\th)&, \nab(\varphi)=O(r^{-1})+\Ga_b, &\quad e_3(\th)=&\Ga_b, \qquad e_3(\varphi)=O(r^{-2})+\Ga_b,\\
        \trch=&\f{2}{r}+O(r^{-3})+\Ga_g,& \quad \trchb=&-\f{2}{r}(1-\f{2m_{\infty}}{r})+O(r^{-3})+\Ga_g,\\
        \chih=&\Ga_g, \quad\chibh, \xib=\Ga_b,  & \qquad \zeta=&-\etab=O(r^{-2})+\Ga_g, \quad \eta,\omegab=O(r^{-2})+\Ga_b, \\
        \rho=&-\f{2m_{\infty}}{r^3}+O(r^{-4})+r^{-1}\Ga_g, &\qquad \sigma=&O(r^{-4})+r^{-1}\Ga_g,  \quad \a,\b=r^{-1}\Ga_g, \quad \bb,\ab=r^{-1}\Ga_b.
    \end{aligned}
\end{equation}

Following the process as in Section 8.2 of \cite{KS:main}, we now can solve the eikonal equations
\begin{equation*}
   \bfg^{\a \b}\pr_{\a} \tu \,  \pr_{\b} \tu=0 \qquad \text{and}\qquad \bfg^{\a \b}\pr_{\a} \Ext\tub \,  \pr_{\b} \Ext\tub=0 \qquad \text{in} \quad \Ext\MM.
\end{equation*}
And the optical functions $\tu$ and $\Ext\tub$ obey the following asymptotic formulas
\begin{equation*}
    \tu=u-\f{a^2\sin^2\th}{2r}+O(r^{-3})+\Ga_b, \qquad \Ext\tub=u+2\int \f{r^2+a^2}{\De} dr+\f{a^2\sin^2\th}{2r}+O(r^{-3})+\Ga_b.
\end{equation*}
As a consequence, for $r\ge r_0\gg 1$ we have
\begin{equation}\label{Est for der tu tub}
\begin{aligned}
    e_3(\tu)=&2+O(r^{-2})+\Ga_b^{(1)}, \qquad e_4(\tu)=O(r^{-2})+r^{-1}\Ga_b^{(1)}, \qquad\ \ \qquad \nab(\tu)=O(r^{-1})+r^{-1}\Ga_b^{(1)}, \\
    e_4(\Ext\tub)=&\f{2(r^2+a^2)}{\De}+O(r^{-2})+r^{-1}\Ga_b^{(1)}=2+O(r^{-1})+r^{-1}\Ga_b^{(1)},  \qquad \nab(\Ext\tu)=O(r^{-1})+r^{-1}\Ga_b^{(1)},\\
    e_3(\Ext\tub)=&2+O(r^{-2})+\Ga_b+\f{2(r^2+a^2)}{\De}\qty(-(1-\f{2m_{\infty}}{r})+r\Ga_b+O(r^{-2}))+O(r^{-2})+\Ga_b^{(1)}\\
    =&O(r^{-1})+\Ga_b^{(1)}.
\end{aligned}
\end{equation}
These further yield
\begin{equation*}
    -2\O^{-2}=\bfg(\bfD\tu, \bfD \Ext\tub)=-\f12 e_3(\tu) e_4(\Ext\tub)-\f12 e_3(\Ext\tub) e_4(\tu)+\nab(\tu)\c \nab(\Ext\tub)=-2+O(r^{-1})+\Ga_b^{(1)},
\end{equation*}
or equivalently,
\begin{equation}\label{Est for der O}
    \O=1+O(r^{-1})+\Ga_b^{(1)}.
\end{equation}

To derive precise hyperbolic estimates for geometric quantities adapted to the newly constructed double null foliation, we proceed to study the change of frame from $(e_1, e_2, e_3, e_4)$ to $(\te_1, \te_2, \te_3, \te_4)$. Denote the corresponding transition coefficients by $(f, \fb, \la)$. In light of \Cref{Lem:frame transform}, the following relations hold
\begin{align*}
     &\te_4=\lambda \left(e_4+f^b e_b+\f14|f|^2e_3\right), \\
     &\te_a=\Big(\delta_a^b+\f12 \fb_a f^b\Big)e_b+\f12 \fb_a e_4+(\f12 f_a+\f{1}{8}|f|^2 \fb_a)e_3, \\
            &\te_3=\lambda^{-1}\Big[ \Big(1+\f12 f\cdot \fb+\f{1}{16}|f|^2|\fb|^2\Big)e_3+\Big(\fb^b+\f{1}{4}|\fb|^2 f^b\Big)e_b+\f14 |\fb|^2 e_4 \Big].
\end{align*}
Note also by \eqref{Eqn:def double null}, we have
\begin{align*}
    \te_4=&-2\O\qty(-\f12e_3(\tu)e_4-\f12e_4(\tu)e_3+e^a(\tu)e_a), \\
     \te_3=&-2\O\qty(-\f12e_3(\Ext\tub)e_4-\f12e_4(\Ext\tub)e_3+e^a(\Ext\tub)e_a).
\end{align*}
Comparing with coefficients in front of each $e_{\mu}$, we thus obtain the expression of  $(f, \fb, \la)$ as
\begin{align*}
    \la=\O e_3(\tu), \qquad f=-\f{2\nab (\tu)}{e_3(\tu)}, \qquad \fb=-2\O^2\qty(e_3(\tu)\nab(\Ext\tub)-e_3(\Ext\tub)\nab(\tu)).
\end{align*}
Inserting the estimates for derivatives of $\tu,\Ext\tub$ from \eqref{Est for der tu tub} into above, we infer that
\begin{equation}\label{Appendix:Est for f fb la}
    f=O(r^{-1})+r^{-1}\Ga_b^{(1)}, \qquad \fb=O(r^{-1})+r^{-1}\Ga_b^{(1)}, \qquad \la=2+O(r^{-1})+\Ga_b^{(1)}.
\end{equation}
Incorporating with the transformation formulas provided in \Cref{Lem:transformation formula}, together with employing \eqref{KS hyperbolic Est ext} and \eqref{Est for der O}, we hence derive
    \begin{align*}
        \t{\chih}=&\la\qty(\chih+\tnab f+\Ga\c \sum_{i=1}^3(f, \fb)^{i})=O(r^{-3})+r^{-1}\Ga_b^{(2)}, \\  \t{\chibh}=&\la^{-1}\qty(\chih+\tnab \fb+\Ga\c \sum_{i=1}^3(f, \fb)^{i})=O(r^{-3})+\Ga_b^{(2)}, 
\end{align*}
    \begin{align*}
        \t{\zeta}=&\zeta-\tnab(\log \la)+\Ga\c\sum_{i=1}^3(f, \fb)^{i}=O(r^{-2})+r^{-1}\Ga_b^{(2)},
        \qquad \t{\eta}=\t{\zeta}+\tnab\log\O=O(r^{-2})+r^{-1}\Ga_b^{(2)},\\
        \t{\rho}=&\rho+(\b, \bb)\c(f, \fb)+(\rho,\sigma)(f, \fb)^2+(\a,\b,\ab,\bb)(f, \fb)^3=-\f{2m_{\infty}}{r^3}+O(r^{-4})+r^{-2}\Ga_b^{(1)}
    \end{align*}
as desired.
\end{proof}

We also need to control $\t{r}-r$, where $\t{r}$ is the area radius of $\tS_{\tu, \Ext\tub}$. Note that in $(u, r, \th, \varphi)$ coordinates adapted to the PG structure within $\Ext\MM$, we have
    \begin{equation*}
        e_{\mu}=e_{\mu}(u)\pr_{u}+e_{\mu}(r) \pr_r+e_{\mu}(\th)\pr_{\th}+e_{\mu}(\varphi)\pr_{\varphi} \qquad \text{with} \quad \mu=1, 2, 3, 4.
    \end{equation*}
Combining with hyperbolic estimates in \eqref{KS hyperbolic Est ext}, this allows us to express the corresponding coordinate derivatives in terms of the null frame $\{\te_1, \te_2, \te_3, \te_4 \}$ as follows:
\begin{equation}\label{Appendix:coor der}
\begin{aligned}
    \pr_{u}=&\f12(e_3+e_4)+\qty(O(r^{-1})+\Ga_b)(e_3, re_4, re_a), &\quad \pr_r=&e_4, \\
    \pr_{\th}=&re_1+\qty(O(r^{-1})+\Ga_b)(e_3, re_4, re_a), &\quad \pr_{\th}=&r\sin\th e_2+\qty(O(r^{-1})+\Ga_b)(e_3, re_4, re_a).
\end{aligned}
\end{equation}
As a result, together with using \eqref{Est for der tu tub}, we can estimate the Jacobian matrices for the pair $(\tu, \Ext\tub)$ with regard to $(u, r)$ and $(\th, \varphi)$:
\begin{equation*}
    \f{\pr(\tu, \Ext\tub)}{\pr(u, r)}=\mqty(1& \\ 1& 2)+O(r^{-1})+\Ga_b^{(1)}, \qquad \f{\pr(\tu, \Ext\tub)}{\pr(\th, \varphi)}=O(r^{-1})+\Ga_b^{(1)}.
\end{equation*}
Therefore, by applying the implicit function theorem for
\begin{equation*}
    \tu=\tu(u, r, \th, \varphi), \qquad \Ext\tub=\Ext\tub(u, r, \th, \varphi),
\end{equation*}
we can find two smooth functions $U=U(\tu, \Ext\tub, \th, \varphi), R=R(\tu, \Ext\tub, \th, \varphi)$ such that
\begin{equation*}
     \tu=\tu(U, R, \th, \varphi), \qquad \Ext\tub=\Ext\tub(U, R, \th, \varphi).
\end{equation*}
Furthermore, by differentiating above equations in $\th, \varphi$ variables, we obtain
\begin{align}\label{Est for pr U R pr th}
    \f{\pr(U, R)}{\pr(\th, \varphi)}=&\qty[\f{\pr(\tu, \Ext\tub)}{\pr(u, r)}]^{-1} \c \f{\pr(\tu, \Ext\tub)}{\pr(\th, \varphi)}=O(r^{-1})+\Ga_b^{(1)}.
\end{align}

We then proceed to view the 2-sphere $\tS_{\tu, \Ext\tub}=\qty{u=U(\tu, \Ext\tub, \th, \varphi), r=R(\tu, \Ext\tub, \th, \varphi)}$ as a graph over $(\th, \varphi)\in\mathbb{S}^2$. Utilizing \eqref{Appendix:coor der} and \eqref{Est for pr U R pr th}, we can hence compute its induced metric and it holds
\begin{align*}
        \t{g}_{\th \th}=&\bfg(\pr_{\th} U \, \pr_{u}+\pr_{\th} R\pr_r+\pr_{\th} , \pr_{\th} U \, \pr_{u}+\pr_{\th} R+\pr_{\th})
        =r^2+O(1)+r^2\Ga_b^{(1)},\\
        \t{g}_{\th \varphi}=&\bfg(\pr_{\th} U \, \pr_{u}+\pr_{\th} R\pr_r+\pr_{\th}, \pr_{\varphi} U \, \pr_{u}+\pr_{\varphi} R \pr_r+\pr_{\varphi})=O(r^{-1})+r^2\Ga_b^{(1)},\\
        \t{g}_{\varphi \varphi}=&\bfg( \pr_{\varphi} U \, \pr_{u}+\pr_{\varphi} R \pr_r+\pr_{\varphi},  \pr_{\varphi} U \, \pr_{u}+\pr_{\varphi} R \pr_r+\pr_{\varphi})=r^2\sin^2\th+O(1)+r^2\Ga_b^{(1)}.
    \end{align*}
Thus, the determinant of $\t{g}$ satisfies
\begin{equation}\label{Apx:Est det tg}
    \det\t{g}=\t{g}_{\th \th}\t{g}_{\varphi \varphi}-\t{g}_{\th \varphi}^2=r^4\sin^2\th+O(r^2)+r^4\Ga_b^{(1)}.
\end{equation}
This yields
\begin{align*}
    4\pi\t{r}^2=|\tS_{\tu, \Ext\tub}|=\int_{0}^{\pi}\int_{0}^{2\pi}  \sqrt{\det\t{g}} \ d\th d\varphi 
    =&\int_{0}^{\pi}\int_{0}^{2\pi}  \qty(r^2\sin\th+O(1)+r^2 \Ga_b^{(1)}) \ d\th  d\varphi \\
    =&4\pi r^2+O(1)+r^2 \Ga_b^{(1)}.
\end{align*}
It further implies
\begin{equation}\label{Est tr-r}
    \t{r}=r+O(r^{-1})+r\Ga_b^{(1)}. \\[2mm]
\end{equation}

Denote $M_B(\tu)$ as the Bondi mass along $\tH_{\tu}$ and let $m_{\infty}$ be the final mass of perturbed Kerr spacetime. Now we are ready to establish
\begin{proposition}\label{Prop:Bondi}
    Along the future null infinity $\II^+$, the Bondi mass $M_B(\tu)$ is non-increasing in $\tu$. Moreover, the limit of $M_B(\tu)$ at $\tu=\infty$ exists and it satisfies
    \begin{equation*}
        \lim\limits_{\tu\to \infty} M_B(\tu)=m_{\infty}.
    \end{equation*}
\end{proposition}
\begin{proof}
Define the Hawking mass of the 2-sphere $\tS_{\tu, \Ext\tub}$ as
\begin{equation*}
    m_H(\tu, \Ext\tub)=\f{\t{r}}{2}\l 1+\f{1}{16\pi}\int_{\tS_{\tu, \Ext\tub}} \tr\t{\chi}\tr\t{\chib} \r.
\end{equation*}
 
By mimicking the computations in Section 3.8.2 provided in \cite{KS:main} and by employing the hyperbolic estimates obtained in \Cref{Appendix:hyper est new} and \eqref{Est tr-r}, we deduce
\begin{align}
     \pr_{\Ext\tub} m_H=&O(r^{-2}), \label{Eqn:pr ub mH2} \\ 
     m_H=&m_{\infty}\l 1+O({r}^{-1})+O(\epsilon_0\tu^{-\f12-\de_{dec}}) \r. \label{Eqn:mH2}
\end{align}
By applying \eqref{KS hyperbolic Est ext}, \eqref{Est for der O}, \eqref{Appendix:Est for f fb la}, we also have 
\begin{equation}\label{Est:pr tub r}
    \pr_{\Ext\tub} r=2+O(r^{-2})+\Ga_b^{(1)}\sim 1.
\end{equation}
From the integrability of ${r}^{-2}$ in $\Ext\tub$, we obtain that $\lim\limits_{\Ext\tub\to \infty} m_H(\tu, \Ext\tub)$ exists, whose value is defined to be the Bondi mass $M_B(\tu)$ along $\tH_{\tu}$. Integrating \eqref{Eqn:pr ub mH2} backwards from $\infty$ to $\Ext\tub$, we then derive
\begin{equation}\label{ineq Bondi2}
\begin{aligned}
      |m_H(\tu, \Ext\tub)-M_B(\tu)|\lesssim& \f{1}{\t{r}}. 
\end{aligned}
\end{equation}
Together with \eqref{Eqn:mH2}, this renders
\begin{equation*}
    M_B(\tu)=m_{\infty}\l 1+O(\tu^{-\f12-\de_{dec}}) \r. 
\end{equation*}
By sending $\tu\to\infty$, we hence get
\begin{equation*}
    M_B(\infty)=m_{\infty}. \\[2mm]
\end{equation*}

We proceed to prove the Bondi mass loss formula. Firstly, as conducting in Section 3.8.2 in \cite{KS:main}, we deduce
\begin{equation}\label{Appendix:limit chibh}
    \t{\Thetab}\coloneqq\lim\limits_{\tH_{\tu}, \Ext\tub\to \infty}\t{r}\t{\chibh} \quad \text{exists} 
\end{equation}
and
\begin{equation*}
    \tnab_3(\tr\t{\chi} \tr\t{\chib})=-\tr\t{\chi}(\tr\t{\chib})^2+2\tr\t{\chib}\,\t{\rho}+2\tr\t{\chib}\,\t{\div}\t{\eta}-\tr\t{\chi}|\t{\chih}|^2+\tr\t{\chib}\qty(2|\t{\eta}|^2-\t{\chih}\c\t{\chibh}).
\end{equation*}
Using the identity
\begin{equation*}
    \O\te_3(\int_{\tS_{\tu, \Ext\tub}} h)=\int_{\tS_{\tu, \Ext\tub}} \O\qty(\te_3(h)+\tr\t{\chib}\, h) \qquad \text{with} \quad h \quad \text{being a scalar function}
\end{equation*}
and inserting hyperbolic estimates stated in \Cref{Appendix:hyper est new} and \eqref{Est tr-r}, we then derive\footnote{We sometimes use $\tS$ to represent $\tS_{\tu, \Ext\tub}$ for simplicity.}
\begin{align*}
    \O\te_3(\int_{\tS} \tr\t{\chi} \tr\t{\chib})=&\int_{\tS}\O\qty(2\tr\t{\chib}\,\t{\rho}+2\tr\t{\chib}\,\t{\div}\t{\eta}-\tr\t{\chi}|\t{\chibh}|^2+\tr\t{\chib}\qty(2|\t{\eta}|^2-\t{\chih}\c\t{\chibh})) \\
    =&\int_{\tS} \O\qty(-\tr\t{\chi}|\t{\chibh}|^2+O(\t{r}^{-4}))=-\int_{\tS} \O\tr\t{\chi}|\t{\chibh}|^2+O(\t{r}^{-2}).
\end{align*}

Further noting that
\begin{equation*}
    \f{2m_H}{\t{r}}=1+\f{1}{16\pi}\int_{\tS} \tr\t{\chi} \tr\t{\chib},
\end{equation*}
we can hence estimate $\te_3(m_{H})$ as below:
\begin{equation}\label{Est:pr tub mH}
\begin{aligned}
   \pr_{\Ext\tub} m_H=\O\te_3(m_{H})=&\O\te_3(\f{\t{r}}{2})\c \f{2m_H}{\t{r}}+\f{\t{r}}{2}\c \qty(-\f{1}{16\pi}\int_{\tS} \O\tr\t{\chi}|\t{\chibh}|^2+O(\t{r}^{-2}))\\
   =&-\f{\t{r}}{32\pi}\int_{\tS}\O\tr\t{\chi}|\t{\chibh}|^2+ O(\t{r}^{-1}).
\end{aligned}
\end{equation}
Here we use \eqref{Eqn:mH2} and the fact that
\begin{equation*}
    \O\te_3(\t{r})=\f{\t{r}}{2}\c \f{1}{|\tS|}\int_{\tS} \O\tr\t{\chi}=O(1).
\end{equation*}
Along $\tH_{\tu}$, from \eqref{Est:pr tub r} and \eqref{Est tr-r} we have $\t{r}\to \infty, \f{\t{r}}{r}\to 1$ as $\Ext\tub$ tends to infinity. Therefore, letting $\Ext\tub\to \infty$ in \eqref{Est:pr tub mH} along $\tH_{\tu}$, by virtue of \Cref{Appendix:hyper est new}, \eqref{Est for der O}, \eqref{Apx:Est det tg} and \eqref{Appendix:limit chibh}, we infer
\begin{equation*}
    \pr_{\Ext\tub}M_B=-\f14 \oint_{\mathbb{S}^2} |\t{\Thetab}|^2\le 0.
\end{equation*}
This finishes the proof of this proposition.
\end{proof}

\section{Miranda-Talenti Type Inequality on \texorpdfstring{$\mathbb{S}^2$}{}}
Following the arguments in \cite{D-P} on $\mathbb{R}^2$, in this appendix we establish a sharp Miranda--Talenti type inequality on $\mathbb{S}^2$. 
\begin{lemma}\label{Apx:M-T ineq on S2}
    Define the stereographic coordinates of $\mathbb{S}^2$ on the north pole chart and on the south pole chart by
    \begin{align*}
        &F^+: (\th^1, \th^2)\in B_{\rho} \mapsto \l\f{2\th^1}{1+(\th^1)^2+(\th^2)^2}, \f{2\th^2}{1+(\th^1)^2+(\th^2)^2}, \f{1-(\th^1)^2-(\th^2)^2}{1+(\th^1)^2+(\th^2)^2} \r,\\
         &F^-: (\th^1, \th^2)\in B_{\rho} \mapsto \l\f{2\th^1}{1+(\th^1)^2+(\th^2)^2}, \f{2\th^2}{1+(\th^1)^2+(\th^2)^2}, \f{-1+(\th^1)^2+(\th^2)^2}{1+(\th^1)^2+(\th^2)^2} \r,
    \end{align*}
    with $\rho>1$ and  $B_{\rho}=\{(\th^1, \th^2): \ (\th^1)^2+(\th^2)^2<\rho^2 \}$. Then for any function $u\in H^2(\mathbb{S}^2)$, we have
    \begin{equation}\label{Apx:MTineq}
        \int_{B_{1}} |D^2 u(F^+(z))|^2 dz+\int_{B_{1}} |D^2 u(F^-(z))|^2 dz\le \int_{B_{1}} |\De u(F^+(z))|^2 dz+\int_{B_{1}} |\De u(F^-(z))|^2 dz.
    \end{equation}
    Here $D^2$ and $\De$ represent the Hessian and Laplacian operators with respect to the Cartesian coordinates in $z=(\th^1, \th^2)$.
\end{lemma}
\begin{remark}
    The above inequality \eqref{Apx:MTineq} is equivalent to \eqref{Ineq:MT}, i.e., $\|\De R\|_{L^2(\ms)}\ge \|D^2 R\|_{L^2(\ms)}$. Inequality \eqref{Ineq:MT} is crucially used in the proof of \Cref{R1-R2 C 1 apha} in \Cref{Subsec:C1aest}.
\end{remark}
\begin{proof}
    We first consider any function $v$ defined on $\pr B_1$. Define $\de$ to be the gradient operator on $\pr B_1$ and set
    \begin{equation*}
        \de v\coloneqq(v_1, v_2).
    \end{equation*}
    Within the polar coordinates $(R, \Theta)$ in $B_1$, it is easy to verify
    \begin{align*}
        \de v=\f{\pr v}{\pr \Theta} (-\sin \Theta, \cos \Theta).
    \end{align*}
    This further gives
    \begin{equation}\label{Appendixeqn:derivative}
        \begin{aligned}
        v_1=&-\f{\pr v}{\pr \Theta}\sin \Theta, \qquad v_2=\f{\pr v}{\pr \Theta}\cos \Theta, \\
        v_{11}=&\sin \Theta \l\f{\pr^2 v}{\pr \Theta^2}\sin \Theta+\f{\pr v}{\pr \Theta}\cos \Theta \r, \\
        v_{22}=&\cos \Theta \l\f{\pr^2 v}{\pr \Theta^2}\cos \Theta-\f{\pr v}{\pr \Theta}\sin \Theta \r.
    \end{aligned}
    \end{equation}
As conducting in to \cite{D-P}, for any function $v\in H^2(B_1)$, we have the following equality holds
\begin{equation}\label{Apprndixeqn:MT}
    \int_{B_1} (|\De v|^2-|D^2 v|^2)dz=-\int_{\pr B_1} \l\de v_0\c \de v-v_0\sum_{i=1}^2 v_{ii}+B(\de v, \de v)+v_0^2 H \r d\sigma,
\end{equation}
where $v_0=\f{\pr v}{\pr n}$ denotes the normal derivative on $\pr B_1$, $B$ and $H$ represent the second fundamental form and the mean curvature on $\pr B_1$ with respect to the inward pointing unit normal, and $d\sigma$ is the surface element of $\pr B_1$. 

Note that along $\pr B_1$, we have
\begin{equation*}
    B(\de v, \de v)=-\|\de v\|^2, \qquad H=-1.
\end{equation*}
Hence from \eqref{Apprndixeqn:MT} we deduce
\begin{equation*}
    \int_{B_1} (|\De v|^2-|D^2 v|^2)dz\ge -\int_{\pr B_1} \l\de v_0\c \de v-v_0\sum_{r=1}^2 v_{rr} \r d\sigma \qquad \text{for all} 
    \quad v\in H^2(B_1).
\end{equation*}
Picking $v=u^{\pm}\coloneqq u\circ F^{\pm}$, we further derive
\begin{equation}\label{Appendixeqn:MT inter}
\begin{aligned}
     &\int_{B_1} (|\De u^+|^2+|\De u^-|^2-|D^2 u^+|^2-|D^2 u^-|^2)dz \\
     \ge& -\int_{\pr B_1} \l\de u^+_0\c \de u^+ -u^+_0\sum_{i=1}^2 u^+_{ii}+\de u^-_0\c \de u^- -u^-_0\sum_{i=1}^2 u^-_{ii} \r d\sigma.
\end{aligned}
\end{equation}
Notice that in spherical coordinates\footnote{This coordinate is regular near the equator $\th=\pi/2$.} $(\th, \varphi)\in \mathbb{S}^2$, with $\th\in [0, \pi]$ and $\varphi\in [0, 2\pi]$, the stereographic maps of the north chart and the south chart on $\mathbb{S}^2$ take the forms of
\begin{align*}
    F^+(R, \Theta)=(\pi-2\arctan \f1R, \Theta),\qquad F^-(R, \Theta)=(2\arctan \f1R, \Theta).
\end{align*}
Therefore, in view of \eqref{Appendixeqn:derivative} and the fact that $u^{\pm}_0=\f{\pr u^{\pm}}{\pr R}$, we hence obtain 
\begin{equation*}
    u^+_0=-u^-_0, \qquad \de u^{+}_0=-\de u^{-}_0, \qquad \de u^+=\de u^{-}, \qquad u^+_{ii}=u^-_{ii} \qquad \text{on} \quad \pr B_1=\{R=1 \}.
\end{equation*}
Inserting these into \eqref{Appendixeqn:MT inter}, we have that the right hand side of \eqref{Appendixeqn:MT inter} is zero and therefore conclude
\begin{equation*}
    \int_{B_1} (|\De u^+|^2+|\De u^-|^2-|D^2 u^+|^2-|D^2 u^-|^2)dz
     \ge 0
\end{equation*}
as stated.
\end{proof}


\begin{thebibliography}{99} 
\bibitem{Allen:2025fhj}
B.~Allen, E.~Bryden, D.~Kazaras and M.~Khuri, \textit{Proof of the Spacetime Penrose Inequality With Suboptimal Constant in the Asymptotically Flat and Asymptotically Hyperboloidal Regimes}, preprint (2025), arXiv:2504.10641.

\bibitem{An:Trapped} X.~An, \textit{Formation of Trapped Surfaces from Past Null Infinity}, preprint (2012), arXiv:1207.5271. 

\bibitem{An:AH} X.~An, \textit{Emergence of Apparent Horizon in Gravitational Collapse}, Ann. PDE, 6 (2020): 1-89. 

\bibitem{An:scale} X.~An, \textit{A scale-critical trapped surface formation criterion: a new proof via signature for decay rates}, Ann. PDE 8, 3 (2022).

\bibitem{Annakedsingularity} X.~An, \textit{Naked singularity censoring with anisotropic apparent horizon}, Ann. of Math, 201(3), (2025), 775-908.

\bibitem{An-Han} X.~An, Q.~Han, \textit{Anisotropic dynamical horizons arising in gravitational collapse}, arXiv preprint arXiv:2010.12524 (2020).

\bibitem{An-He} X.~An, T.~He,  \textit{Dynamics of apparent horizon and a null comparison principle}, Ann. PDE \textbf{10} (2), 15 (2024).

\bibitem{An-He-Shen} X.~An, T.~He, D.~Shen, \textit{Angular momentum memory effect}, arXiv:2403.11133 (2024).

\bibitem{A-L} X.~An, J.~Luk, \textit{Trapped surfaces in vacuum arising dynamically from mild incoming radiation}, Adv. Theor. Math. Phys., 21 (2017), 1-120.

\bibitem{Andersson:stability} L.~Andersson, M.~Mars, W.~Simon, \textit{Stability of marginally outer trapped surfaces and existence of marginally outer trapped tubes}. Adv. Theor. Math. Phys., 12.4 (2008): 853-888.

\bibitem{A-Met} L.~Andersson, J.~Metzger, \textit{The area of horizons and the trapped region}, Comm. Math. Phys., \textbf{290} (3):941–972, 2009.

\bibitem{A-K1} A.~Ashtekar, B.~Krishnan, \textit{Isolated and dynamical horizons and their applications}, Living Reviews in Relativity, 2004, 7(1): 1-91.


\bibitem{A-K2} A.~Ashtekar, B.~Krishnan, \textit{Dynamical horizons and their properties}, Phys. Rev. D, 68, 104030–1–25, (2003). 

\bibitem{At-Le} A.~Athanasiou, M.~Lesourd, \textit{Construction of Cauchy data for the dynamical formation of apparent horizons and the Penrose Inequality}, Adv.Theor.Math.Phys. \textbf{26} (2022) 8.


\bibitem{Bray} H. L.~Bray,  \textit{Proof of the Riemannian Penrose inequality using the positive mass theorem}, J. Diff. Geom., \textbf{59} (2), 177-267 (2001).

\bibitem{B-K} H. L.~Bray, M.~Khuri, \textit{P.D.E.'s which imply the Penrose conjecture}, Asian. J. Math. Vol. 15, No.4, pp. 557-610, December 2011. 

\bibitem{B-L} H. L.~Bray, D. A.~Lee, \textit{On the Riemannian Penrose inequality in dimensions less than eight}, Duke Mathematical Journal 148 (2009).

\bibitem{B-W} S.~Brendle, M. T. ~Wang, \textit{A Gibbons–Penrose Inequality for Surfaces in Schwarzschild Spacetime}, Commun. Math. Phys. 330, 33–43 (2014).

\bibitem{B-H-W} S.~Brendle, P. K.~Hung, M. T.~Wang, \textit{A Minkowski Inequality for Hypersurfaces in the Anti-de Sitter-Schwarzschild Manifold}, Commun.Pure Appl.Math. 69 (2016) 124-144.

\bibitem{C-N} G.~Caciotta, F.~Nicol\`o, \textit{Non Linear Perturbations of Kerr Spacetime in External Regions and the Peeling Decay}, Annales Henri Poincar\'e \textbf{11} (3) (2010), pp. 433–497.

\bibitem{Cam1} S.~Campanato, \textit{A Cordes type Condition for nonlinear non-variational Systems}, Rend. Accad. Naz. Sci. Detta XL, V Ser. 13, No 1,(1989), 307-321.

\bibitem{Cam2} S.~Campanato, \textit{Nonvariational basic parabolic systems of second order}, Atti Accad. Naz. Lincei, Cl. Sci. Fis. Mat. Nat., IX Ser., Rend. Lincei,
Mat. Appl. 2, No.2, (1991), 129-136.

\bibitem{Chen-K:EH} X.~Chen, S.~Klainerman, \emph{Regularity of the Future Event Horizon in Perturbations of Kerr}, arXiv preprint arXiv:2409.05700 (2024).

\bibitem{Chr:book} D.~Christodoulou, \textit{The Formation of Black Holes in General Relativity}, Monographs in Mathematics, European Mathematical Soc. (2009). 

\bibitem{Chr-Kl} D.~Christodoulou, S.~Klainerman, \textit{The global nonlinear stability of the Minkowski space}, Princeton mathematical series 41, (1993).

\bibitem{Dafermos} M.~Dafermos, \textit{The formation of black holes in general relativity}, Astrisque 352 (2013).


\bibitem{D-L:Kerrint} M.~Dafermos, J.~Luk, \textit{The interior of dynamical vacuum black holes I: the $ C^0$-stability of the Kerr Cauchy horizon}, arXiv:1710.01722 (2017). To appear in Ann. of Math.

\bibitem{D-H-R-T} M.~Dafermos, G.~Holzegel, I.~Rodnianski, M.~Taylor,  \textit{The non-linear stability of the Schwarzschild family of black holes},   arXiv:2104.08222 (2021).

\bibitem{Dain} S.~Dain, \textit{Proof of the angular momentum-mass inequality for axisymmetric black holes}, J. Diff. Geom. 79, 33-67 (2008).

\bibitem{D-P} G.~Devillanova, F.~Pugliese, \textit{A variant on Miranda-Talenti estimate}, Le Matematiche \textbf{54} (1) (1999): 91-97.

\bibitem{E} M.~Eichmair, \textit{The plateau problem for marginally trapped surfaces}, J. Differential Geom., \textbf{83} (3):551–584, 2009.

\bibitem{Evans}  L. C.~Evans, \textit{Partial differential equations}, Vol. 19. American Mathematical Society, (2022).

\bibitem{F-S-T} A. J.~Fang, J.~Szeftel,  A.~Touati, \textit{Spacelike initial data for black hole stability}, arXiv:2405.02071 (2024).

\bibitem{G-T} D.~Gilbarg, N. S.~Trudinger, \textit{Elliptic Partial Differential Equations of Second Order}, Grundlehren, Vol. 224, Springer-Verlag, Berlin, 1983.

\bibitem{KS:formula} E.~Giorgi, S. Klainerman, J. Szeftel, \textit{A General Formalism for the Stability of Kerr}, arXiv preprint arXiv:2002.02740 (2020).

\bibitem{GKS} E.~Giorgi, S.~Klainerman, J.~Szeftel, \emph{Wave equations estimates and the nonlinear stability of slowly rotating Kerr black holes},  Pure Appl. Math. Q.. \textbf{20} (7) (2024).

\bibitem{H-K-W-X} Q.~Han, M.~Khuri, G.~Weinstein and J.~Xiong, \emph{The Mass-Angular Momentum Inequality for Multiple Black Holes}, arXiv:2501.15093 (2025).

\bibitem{H-L:ellipticpdetextbook} Q.~Han, F. H.~Lin, \textit{Elliptic partial differential equations (2nd ed.)}, Courant lecture notes, Vol. 1, (2011).

\bibitem{Hawking} S.W.~Hawking, \emph{Black holes in general relativity}, Comm. Math. Phys 25 (1972): 152-166.

\bibitem{Hintz}  P.~Hintz,  \textit{Horizons of some asymptotically stationary spacetimes}, arXiv:2411.12568 (2024).

\bibitem{H-V}  P.~Hintz, A.~Vasy, \textit{The global non-linear stability of the Kerr–de Sitter family of black holes}, Acta Math, \textbf{220} (2018): 1-206.

\bibitem{H-I} G.~Huisken, T.~Ilmanen, \textit{The inverse mean curvature flow and the Riemannian Penrose inequality}, J. Diff. Geom. \textbf{59} (3), 353-437 (2001).

\bibitem{K-U:gluing} C.~Kehle, R.~Unger, \textit{Event horizon gluing and black hole formation in vacuum: the very slowly rotating case}, Adv. Math. 452 (2024): 109816.

\bibitem{Khuri} M.~Khuri, \textit{A Penrose-Like Inequality for General Initial Data Sets}, Commun. Math. Phys. 290, 779–788 (2009).

\bibitem{KNI:book} S.~Klainerman,   F.~Nicolo, \textit{The evolution problem in General Relativity}, Progress in Mathematical Physics, Birkha\"user (2003).

\bibitem{KN:peeling} S.~Klainerman,   F.~Nicolo, \textit{Peeling properties of asymptotically flat solutions to the Einstein vacuum equations}, Class. Quantum Grav. 20 (2003), 3215-3257.


\bibitem{K-L-R} S.~Klainerman, J.~Luk, I.~Rodnianski, \textit{A fully anisotropic mechanism for formation of trapped surfaces in vacuum}, Invent. Math. 198 (2014), no.1, 1-26.

\bibitem{KR:causal} S.~Klainerman, I.~Rodnianski, 
\textit{Causal geometry of Einstein-Vacuum 
spacetimes with finite curvature flux}, Invent. Math. 159 (2005), 437-529.  


\bibitem{KR:LP} S.~Klainerman,  I.~Rodnianski,
\textit{A geometric approach to the Littlewood-Paley theory},
Geom. Funct. Anal. 16, (2006) no. 1, 126-163. 

\bibitem{KR:Scarred} S.~Klainerman, I.~Rodnianski, 
\textit{On emerging scarred surfaces for the Einstein vacuum equations}, Discrete Contin. Dyn. Syst. , 28 (2010), no. 3, 1007-1031.

\bibitem{KR:Trapped} S.~Klainerman, I.~Rodnianski, 
\textit{On the formation of trapped surfaces}, Acta Math. 208 (2012), no.2, 211-333.

\bibitem{KSW} S.~Klainerman, D.~Shen, J.~Wan, \textit{A canonical foliation on null infinity in perturbations of Kerr}, arXiv:2412.20119 (2024).

\bibitem{K-S} S.~Klainerman, J.~Szeftel, \textit{Global nonlinear stability of Schwarzschild spacetime under polarized perturbations}, (AMS-210). Vol. 210. Princeton University Press, 2020.

\bibitem{KS:Kerr1} S.~Klainerman, J.~Szeftel, \emph{Construction of GCM spheres in perturbations of Kerr}, Ann. PDE \textbf{8} (2), Art. 17 (2022).

\bibitem{KS:Kerr2} S.~Klainerman, J.~Szeftel, \textit{Effective results in uniformization and intrinsic GCM spheres in perturbations of Kerr}, Ann. PDE \textbf{8} (2), Art. 18 (2022).

\bibitem{KS:main} S.~Klainerman, J.~Szeftel, \textit{Kerr stability for small angular momentum}, Pure Appl. Math. Q. \textbf{19} (3) (2023), 791--1678.

\bibitem{Le} P.~Le, \textit{The intersection of a hyperplane with a lightcone in the Minkowski spacetime}, J. Differential Geom., \textbf{109}, (2018), no.3, 497-507. 

\bibitem{L-M} J.~Li, H.~Mei, \textit{A construction of collapsing spacetimes in vacuum}, Commun. Math. Phys. \textbf{378} (2020): 1343-1389.

\bibitem{L-Y} J.~Li, P.~Yu, \textit{Construction of Cauchy data of vacuum Einstein field equations evolving to black holes}, Ann. of Math. 181 (2015), no.2, 699-768.

\bibitem{L-Z} J.~Li, X.P.~Zhu,   \textit{On the local extension of the future null infinity. Journal of Differential Geometry}, \textbf{110} (1), 73-133 (2018).

\bibitem{Luk} J.~Luk,
\textit{On the local existence for the characteristic initial value problem in general relativity}, Int. Mat. Res. Notices 20, (2012), 4625-4678.

\bibitem{Luk2} J.~Luk, \textit{Weak null singularity in general relativity}, Journal of AMS. 31 (2018) 1-63.

\bibitem{L-R:Propagation} J.~Luk, I.~Rodnianski, 
\textit{Local propagation of impulsive gravitational waves}, Comm. Pure Appl. Math., 68 (2015), no.4, 511-624.

\bibitem{L-R} J.~Luk, I.~Rodnianski, 
\textit{Nonlinear interactions of impulsive gravitational waves for the vacuum Einstein equations}, Cambridge Journal of Math. 5 (4): 435-570, 2017.

\bibitem{M-M-S} E.~Malec, M.~Mars, W.~Simon, \textit{On the Penrose Inequality for General Horizons}, Phys. Rev. Lett. 88, 121102, 2002. 

\bibitem{Mars} M.~Mars, \textit{Present Status of the Penrose Inequality}, Class. Quant. Grav. 26:193001, 2009.

\bibitem{Penrose} R.~Penrose, 
\textit{Gravitational collapse and space-time singularities}, Phys. Rev. Lett. 14 (1965), 57-59.

\bibitem{Penrosenaked} R.~Penrose, \textit{Naked Singularities}, Annals of the New York Academy of Sciences 224 (1), 1973.


\bibitem{P-I} F.~Pretorius, W. Israel, \textit{Quasispherical light cones of the Kerr geometry}, Class. Quantum Grav., 15:2289–2301 (1998).

\bibitem{R-T} M.~Reiterer, E. Trubowitz, \textit{Strongly focused gravitational waves}, Comm. Math. Phys. 307 (2011), no. 2, 275-313.

\bibitem{R-S} I.~Rodnianski, Y. Shlapentokh-Rothman, \textit{The asymptotically self-similar regime for the Einstein vacuum equations}, Geom. Funct. Anal. Vol. 28 (2018) 755–878.

\bibitem{Shen} D.~Shen, \emph{Construction of GCM hypersurfaces in perturbations of Kerr}, Ann. PDE \textbf{9} (1), Art. 11 (2023).

\bibitem{Shen:Minkext} D.~Shen, \emph{Stability of Minkowski spacetime in exterior regions}, Pure Appl. Math. Q. \textbf{20} (2), 757–868, 2024.

\bibitem{Shen:Kerr} D.~Shen, \textit{Kerr Stability in External Regions}, Ann. PDE \textbf{10} (1), 9 (2024).

\bibitem{Soft} L.G.~Softova, \emph{An integral estimate for the gradient for a class of nonlinear elliptic equations in the plane}, Z. Anal. Anwen 17(1) (1998): 57-66.

\bibitem{W} W.~Wahl, \textit{\"Uber quasilineare elliptische Differentialgleichungen in der Ebene}, manuscripta mathematica 8.1 (1973).

\bibitem{Yu1} P.~Yu, \textit{Energy estimates and gravitational collapse}, Comm. Math. Phys. 317 (2013), no. 2, 275-316.

\bibitem{Yu2} P.~Yu, \textit{Dynamical formation of black holes due to the condensation of matter field}, preprint (2011), arXiv: 1105.5898.  

\end{thebibliography}
\end{document}